\documentclass[preprint]{aastex}
\usepackage[dvipsnames]{xcolor}
\usepackage{natbib,lscape}
\usepackage{url}    

\newcommand{\HST}{{\sl HST}}

\newcommand{\Msun}{\mbox{$M_{\sun}$}}

\newcommand{\Lsun}{\mbox{$L_{\sun}$}}

\newcommand{\Mjup}{\mbox{$M_{\rm Jup}$}}

\newcommand{\degree}{\mbox{$^{\circ}$}}

\newcommand{\etal}{et al.}
\newcommand{\eg}{e.g.}
\newcommand{\ie}{i.e.}

\newcommand{\IRAS}{{\sl IRAS}}

\newcommand{\kms}{\hbox{km~s$^{-1}$}}
\newcommand{\rchisq}{\mbox{$\tilde\chi^2$}}

\newcommand{\htwoo}{{\hbox{H$_2$O}}}   
\newcommand{\meth}{{\hbox{CH$_4$}}}   

\newcommand{\Ks}{\mbox{$K_S$}}
\newcommand{\degs}{\mbox{$^{\circ}$}}

\newcommand{\Teff}{\mbox{$T_{\rm eff}$}}
\newcommand{\logg}{\mbox{$\log(g)$}}

\newcommand{\Lp}{\mbox{${L^\prime}$}}

\newcommand{\bPic}{\hbox{$\beta$}~Pic}

\newcommand{\WISE}{{\sl WISE}}

\newcommand{\vtan}{\mbox{$V_{\rm tan}$}}

\newcommand{\Kn}{\mbox{$K_{\rm H2}$}}

\newcommand{\fldg}{\hbox{\sc fld-g}}
\newcommand{\intg}{\hbox{\sc int-g}}
\newcommand{\vlg}{\hbox{\sc vl-g}}

\newcommand{\masyr}{\hbox{mas\,yr$^{-1}$}}

\newcommand{\perpix}{\mbox{pixel$^{-1}$}}
\newcommand{\Vtan}{\mbox{$V_{\rm tan}$}}

\newcommand{\rchisqXYZ}{\mbox{$\tilde\chi^2_{XYZ}$}}
\newcommand{\rchisqUVW}{\mbox{$\tilde\chi^2_{UVW}$}}

\slugcomment{ApJ, accepted October 5, 2016}
\shorttitle{Parallaxes of Young Ultracool Dwarfs}
\shortauthors{Liu et al.}

\begin{document}

\title{The Hawaii Infrared Parallax Program.\\
II. Young Ultracool Field Dwarfs\altaffilmark{1,2}}


\author{Michael C. Liu,\altaffilmark{3,4} \email{mliu@ifa.hawaii.edu}
Trent J. Dupuy,\altaffilmark{5}
Katelyn N. Allers\altaffilmark{4,6}
}

\altaffiltext{1}{This work is largely based on observations obtained
  with WIRCam, a joint project of CFHT, Taiwan, Korea, Canada, France,
  and the Canada-France-Hawaii Telescope (CFHT) which is operated by the
  National Research Council (NRC) of Canada, the Institut National des
  Sciences de l'Univers of the Centre National de la Recherche
  Scientifique of France, and the University of Hawaii.}
\altaffiltext{2}{Some of the data presented herein were obtained at the
  W.M. Keck Observatory, which is operated as a scientific partnership
  among the California Institute of Technology, the University of
  California, and the National Aeronautics and Space Administration. The
  Observatory was made possible by the generous financial support of the
  W.M. Keck Foundation.}
\altaffiltext{3}{Institute for Astronomy, University of Hawaii, 2680
  Woodlawn Drive, Honolulu HI 96822}
\altaffiltext{4}{Visiting Astronomer at the Infrared Telescope Facility,
  which is operated by the University of Hawaii under Cooperative
  Agreement no. NNX-08AE38A with the National Aeronautics and Space
  Administration, Science Mission Directorate, Planetary Astronomy
  Program.}
\altaffiltext{5}{The University of Texas at Austin, Department of
  Astronomy, 2515 Speedway C1400, Austin, TX 78712, USA}
\altaffiltext{6}{Department of Physics and Astronomy, Bucknell
  University, Lewisburg, PA 17837}

\begin{abstract} 
  We present a large, uniform analysis of young
  ($\approx$10--150~Myr) ultracool dwarfs, based on new high-precision
  infrared (IR) parallaxes for 68~objects.  We find that low-gravity
  (\vlg) late-M and L~dwarfs form a continuous sequence in IR
  color-magnitude diagrams, separate from the field population and from
  current theoretical models. These \vlg\ objects also appear distinct
  from young substellar (brown dwarf and exoplanet) companions,
  suggesting the two populations may have a different range of physical
  properties.  In contrast, at the L/T transition, young, old, and
  spectrally peculiar objects all span a relatively narrow range in
  near-IR absolute magnitudes.  At a given spectral type, the IR
  absolute magnitudes of young objects can be offset from ordinary field
  dwarfs, with the largest offsets occurring in the $Y$ and $J$ bands
  for late-M dwarfs (brighter than the field) and mid/late-L dwarfs
  (fainter than the field).  Overall, low-gravity (\vlg) objects have
  the most uniform photometric behavior while intermediate-gravity
  (\intg) objects are more diverse, suggesting a third governing
  parameter beyond spectral type and gravity class.  We examine the
  moving group membership for all young ultracool dwarfs with
  parallaxes, changing the status of 23~objects (including 8~previously
  identified planetary-mass candidates) and fortifying the status of
  another 28~objects.  We use our resulting age-calibrated sample to
  establish empirical young isochrones and show a declining frequency of
  \vlg\ objects relative to \intg\ objects with increasing age.  Notable
  individual objects in our sample include high-velocity
  ($\gtrsim$100~\kms) \intg\ objects; very red late-L dwarfs with high
  surface gravities; candidate disk-bearing members of the MBM20 cloud
  and $\beta$~Pic moving group; and very young distant interlopers.
  Finally, we provide a comprehensive summary of the absolute magnitudes
  and spectral classifications of young ultracool dwarfs, using a
  combined sample of 102~objects found in the field and as substellar
  companions to young stars.
\end{abstract}

\keywords{brown dwarfs --- proper motions --- parallaxes --- solar neighborhood}

\section{Introduction}

Observational studies of brown dwarfs have been inspired by two
complementary perspectives. One has been the desire to expand the
parameter space of classical stellar astrophysics. Brown dwarfs are the
lowest mass products of the star formation process and thus represent
the terminus of the mass function that has been well-mapped for
hydrogen-burning stars \citep[e.g.][]{2005ASSL..327...41C}. Similarly,
the ultracool temperatures of field brown dwarfs extend spectral
classification from the late-M~dwarfs into the new regimes of the L, T,
and Y~dwarfs ($\Teff\approx2500$ to 300~K; e.g., \citealp{kirk05},
\citealp{2011ApJ...743...50C}). There has been sustained progress over
the last $\approx$15~years in exploring ever lower masses and
temperatures, in large part thanks to wide-area digital sky surveys
sensitive to such cool objects \citep[e.g.][]{1999A&AS..135...41D,
  1999ApJ...519..802K, 2006AJ....131.2722C, 2010MNRAS.406.1885B, 2010A&A...522A.112R,
  2011ApJ...740L..32L, 2011ApJS..197...19K, 2015ApJ...814..118B}.

The second perspective has been the connection between brown dwarfs and
gas-giant planets, both planets in our own solar system and around other
stars. This has been an equally powerful inspiration --- indeed, the
interpretation of the first unambiguous brown dwarf Gl~229B
\citep{1995Natur.378..463N} was guided by its spectral similarity to the
planet Jupiter \citep{1995Sci...270.1478O}. However, in contrast to the
numerous connections with stellar astronomy, observations in common
between the brown dwarf and exoplanet populations have been far more
limited.
The underlying cause is that magnitudes, colors, and spectra are the
most accessible measurements for brown dwarfs but the least accessible
ones for exoplanets. 
In addition, for the case of planets in our own solar system, the gap
between them and brown dwarfs is substantial. Jupiter (124\,K;
\citealp{1981JGR....86.8705H}) is much cooler than the coldest known
brown dwarf WISE~J085510.83$-$071442.5 ($\approx$250\,K;
\citealp{2014ApJ...786L..18L}), which is itself much cooler than the
next coolest known brown dwarfs at $\approx$350--400\,K
\citep{2012ApJ...744..135L, 2013Sci...341.1492D}. This range of
effective temperature corresponds to a factor of $\sim$100 in
luminosity.

Direct detection of young gas-giant exoplanets is now strengthening the
link between the exoplanet and brown dwarf populations, enriching our
understanding of both classes of objects. Young ($\lesssim$100~Myr)
planetary-mass ($\lesssim$13~\Mjup) companions have been discovered over
a wide range of separations, luminosities, and mass ratios with respect
to the host star \citep{2005A&A...438L..29C, 2006ApJ...649..894L,
  2008ApJ...689L.153L, 2009A&A...493L..21L, 2010Sci...329...57L,
  2011ApJ...726..113I, 2012ApJ...758L...2J, 2013ApJ...763L..32C,
  2013ApJ...774...11K, 2013ApJ...772L..15R, 2014ApJ...780L...4B,
  2015Sci...350...64M, 2015ApJ...804...96G}, possibly suggesting a range
of origins for these systems. The inferred temperatures for these
objects amply overlap the known brown dwarf census, but the young ages
mean these companions have significantly lower surface gravities. The
HR~8799 system has been particularly illuminating as it is the only
directly imaged multi-planet system \citep{2008Sci...322.1348M,
  2010Natur.468.1080M}, and the planets possess extreme photospheric
properties compared to previously known substellar objects.  Their
absolute magnitudes and inferred effective temperatures
($\Teff\approx1000$~K) overlap field T~dwarfs, which are cloud-poor and
methane-rich. But the planets' infrared photometry and spectra indicate
the presence of very cloudy photospheres and weak methane, stemming from
the lower surface gravity \citep{2008Sci...322.1348M,
  2010ApJ...716..417H, bowler10-hr8799b, 2011ApJ...729..128C,
  2011ApJ...733...65B, 2012ApJ...753...14S, 2014ApJ...792...17S}.
Theoretical work \citep[e.g.,][]{2011ApJ...735L..39B, madhu11-8799,
  2012ApJ...754..135M, 2013ApJ...778...97L, 2014ApJ...797...41Z} has
helped provide physical context for direct imaging observations. But we
are still faced with the conundrum of how to integrate young gas-giant
planets and field brown dwarfs into a common understanding of substellar
evolution.

A promising approach to understanding such low-gravity ultracool
atmospheres is to identify the analogs of young exoplanets among the
field brown dwarf census.
The initial mass functions in the young star-forming clusters appear to
go down to a few Jupiter masses \citep[e.g.][]{2008MNRAS.383.1385L}, and
thus such low-mass objects should be found in the field after departing
their birth sites. For objects of spectral type~L
($\Teff\approx1300-2300$~K; \citealp{2004AJ....127.3516G,
  stephens09-irac}), a field object of typical age ($\approx$1--5~Gyr)
has a mass of $\approx$40--80~\Mjup\ according to evolutionary models
\citep{2000ApJ...542..464C, 2008ApJ...689.1327S}. But for the same
temperature, an age of 10--100~Myr corresponds to an object of
6--40~\Mjup, encompassing the planetary-mass regime
($\lesssim$13~\Mjup). Such young field objects will have larger radii
and, combined with their lower masses, this means a reduction in surface
gravity by a factor of $\sim$10 compared to older field objects.
Evolutionary models predict that brown dwarf radii become nearly
constant at ages of $\gtrsim$300~Myr, which sets an approximate upper
age limit for field objects displaying signatures of low gravity.

Young objects are expected to be a small minority population in the
field and can be identified primarily in three ways. (1)~The first young
field L~dwarf, G196-3B, was found in seeing-limited imaging as a wide
companion to a star \citep{1998Sci...282.1309R}. Such imaging surveys
have now uncovered $\approx$20 brown dwarfs as bound companions to young
($\approx$10--300~Myr) field stars in seeing-limited data
\citep{2001AJ....121.3235K, 2006PASP..118..671R, 2010MNRAS.405.1140G,
  2010A&A...521L..54H, 2014ApJ...787....5N, 2015ApJ...804...96G},
space-based imaging \citep{1999ApJ...512L..69L, 2000ApJ...541..390L,
  2007ApJ...654..570L}, and ground-based adaptive optics imaging
\citep[][]{2004ApJ...617.1330M, 2005A&A...438L..29C,
  2005A&A...430.1027C, 2006ApJ...651.1166M, 2010ApJ...720L..82B,
  2011ApJ...729..139W, 2012ApJ...750...53N, 2013A&A...553L...5D,
  2013ApJ...774...55B, 2013ApJ...763L..32C, 2015MNRAS.453.2378M}.
(2)~Young objects can be distinguished from older objects as their lower
surface gravities are manifested as differences in their optical and
near-infrared (IR) spectra compared to field objects
\citep[e.g.][]{1999AJ....118.2466M, 2001MNRAS.326..695L}.
Among the spectroscopic studies of hundreds of ultracool dwarfs from
wide-field surveys, a small fraction of young late-M and L~dwarfs have
been identified in this fashion \citep[e.g.][]{2008AJ....136.1290R,
  2008ApJ...689.1295K, 2009AJ....137.3345C, 2013ApJ...772...79A,
  2013ApJ...777L..20L, 2015ApJ...814..118B}, with implied masses down to
$\approx$7~\Mjup.
(3) Finally, there have been significant recent efforts to find young
ultracool dwarfs in the nearest young moving groups through kinematic
and stellar-activity criteria \citep[e.g.,][]{2009ApJ...699..649S,
  2011ApJ...727....6S, 2012ApJ...758...56S, 2011ApJ...727...62R,
  2012AJ....143...80S, 2013ApJ...762...88M, 2014ApJ...783..121G,
  2014AJ....147..146K, 2015ApJ...798...73G, 2016ApJ...821..120A}. For
such objects, their ages can be assigned using the ages of the higher
mass stellar members derived from theoretical isochrones and/or lithium
depletion boundaries.

Parallactic distances for young brown dwarfs are critical for directly
establishing basic properties like absolute magnitudes and bolometric
magnitudes. Given their younger ages, it may not be appropriate to
derive photometric distances using relations between absolute magnitude
and spectral type established for field objects
\citep[e.g.][]{2004AJ....127.2948V, 2012ApJS..201...19D}.\footnote{In
  principle, spectroscopic distance estimates can be computed from
  fitting model atmospheres to the observed spectra and assuming a
  radius \citep{2009ApJ...706.1114B}, though current models have not yet
  been validated for this purpose, even for old (high-gravity) field
  objects \citep{2011arXiv1103.0014L, 2011ApJ...743...50C}.} In
addition, high-quality parallaxes and proper motions allow for more
stringent assessments of membership in young moving groups than possible
with spectrophotometric distances alone.
Previous work has produced parallaxes for relatively small samples of
low-gravity ultracool dwarfs \citep{2012ApJ...752...56F,
  2013AN....334...85L, 2014A&A...568A...6Z}. This work has resulted in
discordant conclusions regarding spectrophotometric trends, such as
whether young L~dwarfs are brighter, fainter, or similar to field
objects at $JHK$ bands. To date, with the exception of late-M~dwarfs in
the TW~Hydrae Association (TWA), only a handful of free-floating
ultracool dwarfs have membership in a young moving group validated by a
measured parallax and radial velocity \citep{2013AJ....145....2F,
  2013AN....334...85L, 2013ApJ...777L..20L, 2015ApJ...799..203G,
  2015ApJ...808L..20G}. Furthermore, the codification of near-IR
spectral types and gravity indicators for M~and L~dwarfs by
\citet{2013ApJ...772...79A} enables a rigorous investigation of the
spectrophotometric properties of low-gravity ultracool dwarfs using a
uniform classification system.

We present here a large sample of high-precision parallaxes for young ultracool
field dwarfs, based on our ongoing astrometric monitoring program at the
Canada-France-Hawaii Telescope (CFHT). 
The immediate aims of this effort are to provide distances for these
objects and to compare their spectrophotometric properties to the field
population.
Preliminary results from this work for a smaller set of objects were
presented in \citet{2013AN....334...85L}, where we showed for the first
time that young field objects form a separate sequence on the near-IR
color-magnitude diagram, residing redder and/or brighter than the field
population.
We revisit this work with a much larger sample of parallaxes,
supplemented with a uniform analysis of near-IR spectral type and
gravity classifications using the methods of
\citet{2013ApJ...772...79A}. We also perform a uniform assessment of
membership in canonical young moving groups for the entire known sample
of young ultracool dwarfs with parallaxes.
The broader goal of our work is to critically examine this intriguing
class of brown dwarfs as empirical analogs for directly imaged
exoplanets.

\section{Observations \label{sec:observations}}

Since 2007, we have been carrying out a high-precision parallax program
at the 3.6-meter Canada-France-Hawaii Telescope (CFHT) using the
facility wide-field IR camera WIRCam \citep{2004SPIE.5492..978P}. CFHT
offers an excellent platform for parallax work, given its combination of
large aperture, excellent seeing, and queue scheduling, though to our
knowledge it had not been used to measure IR parallaxes prior to our
program. Our initial motivation was to determine accurate distances to
ultracool binaries, which can be used to directly measure dynamical
masses when parallaxes are combined with visual orbit determinations
\citep[e.g.][]{liu08-2m1534orbit, 2009ApJ...692..729D,
  2011ASPC..448..111D, 2015ApJ...805...56D}. This goal compelled us to
establish observing protocols and data reduction methods to achieve the
best possible accuracy, since the total dynamical mass scales with the
cube of the binary distance via Kepler's Third Law.

After a few seasons of observations, it became apparent that the
CFHT/WIRCam was incredibly effective for IR parallaxes. Thus, we
expanded the scope of our program to other interesting classes of
ultracool dwarfs. As described in \citet{2012ApJS..201...19D}, our
measurements are as good as have ever been achieved in the near-IR,
producing parallaxes with typical uncertainties of 1.3~mas and as good
as 0.7~mas, but for objects $\approx$2--3~mags fainter than have been
measured by previous parallax programs. This combination of faint
limiting magnitudes \underline{and} high precision naturally lends
itself to two broad classes of targets: (1) extremely low-luminosity
late-type brown dwarfs, as illustrated by our IR parallax to the
$J=19.7$~mag T9~binary CFBDS~J1458+1013AB (\citealp{2011arXiv1103.0014L}
and updated in \citealp{2012ApJS..201...19D}), and (2) more distant
objects, where the same small uncertainties needed for nearby binaries
can benefit science cases that need only moderate S/N distances. The
latter is relevant for studying young field brown dwarfs. Since such
objects are a small minority population (\eg,
\citealp{2008ApJ...689.1295K} find that 8\%~$\pm$~2\% of field L~dwarf
are younger than $\approx$100~Myr), their typical distances will be
larger than those of older field objects. Furthermore, the known stellar
members of even the youngest and nearest moving groups can extended to
distances of $\approx$60~pc \citep[e.g.][]{2008hsf2.book..757T}. This is
quite distant compared to previous brown dwarf parallax programs --- in
comparison, among $\ge$L4~dwarfs, no objects had high precision
($\le3$\%) parallaxes beyond 13~pc prior to our CFHT effort.

\subsection{Sample \label{sec:sample}} 

Our ongoing CFHT program is monitoring candidate young field objects
with spectral types of M6 and later that have been identified from a
variety of sources.
This paper presents results for a sample of 67~such objects (and
1 object with a spectral type of M4.5), chosen for publication solely
based on a sufficient number of observing epochs and time baseline to
determine a robust parallax and proper motion. These objects fall into
four subsets (Table~\ref{table:sample}):

\begin{itemize}

\item {\em Low-gravity objects (44 objects):} We selected targets noted
  as having lower surface gravity than typical field objects based on
  their optical and/or near-IR spectra. Most of these objects were
  originally identified by 2MASS-based searches for ultracool dwarfs in
  the solar neighborhood \citep[e.g.,][]{2003AJ....126.2421C,
    2007AJ....133..439C, 2008AJ....136.1290R, 2008ApJ...689.1295K}, with
  low-gravity signatures identified in their discovery spectra or
  followup studies \citep[e.g.][]{2009AJ....137.3345C,
    2013ApJ...772...79A}.
  We included the young L~dwarf binary SDSS~J2249+0044AB, which has signs
  of low gravity in the near-IR spectrum presented by
  \citet{2004ApJ...607..499N} and has been resolved into a tight binary
  by \citet{2010ApJ...715..561A}. We also included the extremely red young
  L~dwarf PSO~J318.5$-$22 discovered from our ultracool dwarf search
  using Pan-STARRS; this parallax has already been published in
  \citet{2013ApJ...777L..20L}, and a slightly revised parallax and
  proper motion based on a larger dataset are presented here.

  Also, we added two L~dwarf companions to young stars, G~196-3B
  \citep{1998Sci...282.1309R} and LP~261-75B
  \citep{2006PASP..118..671R}. The former was included for its known
  low-gravity spectrum \citep[e.g.][]{1998Sci...282.1309R,
    2004ApJ...600.1020M}. The latter was included as its primary star
  (LP~261-75A, NLTT~22741) is young ($\approx$100-200~Myr
  [\citealp{2006PASP..118..671R}], $\approx$40--300~Myr
  [\citealp{2009ApJ...699..649S}]), though in the end we classified the
  near-IR spectrum of LP~261-7B as field gravity (see Appendix). In
  fact, the M4.5 primary LP~261-75A was the target of our CFHT
  observations, so we used a narrow-band filter to observe this
  system. The same data also yielded a parallax for LP~261-75B, albeit
  with 2$\times$ larger uncertainties than for the primary.

  In total, our 44~young field objects in 42 systems represent a
  4$\times$ larger sample than targeted by previous parallax studies of
  such objects \citep{2012ApJ...752...56F, 2014A&A...568A...6Z}.

\item {\em Candidate stellar association members (12 objects):} We targeted
  a smaller sample of ultracool dwarfs proposed as members of open
  clusters and moving groups, for the purpose of assessing their
  membership and/or establishing them as age-calibrated benchmark
  objects. These candidates have been identified based on their common
  kinematics with the stellar members of these groups. As these were
  drawn from different sources, the candidacy criteria differs in
  construction and stringency.

  The two sources from the TW Hya Association (TWA; $\approx$10~Myr
  [\citealp[e.g.,][]{2013ApJ...762..118W, 2014A&A...563A.121D}]) had
  already been strongly confirmed as members based on their kinematics
  \citep{2005ApJ...634.1385M, 2008A&A...489..825T} and corroborated to
  be young based on their low-gravity spectra. We included one T~dwarf
  candidate in the Hyades cluster ($\approx$600--800~Myr;
  \citealp{1998A&A...331...81P, 2015ApJ...807...58B}) identified in a
  proper motion search by \citet{2008A&A...481..661B}. 

  Our remaining objects were flagged by \citet{2010A&A...512A..37S} as
  candidates of the nearby moving groups studied by
  \citet{2001MNRAS.328...45M}: the Pleiades (1~object; $130\pm20$~Myr
  [\citealp{2004ApJ...614..386B}]), Ursa Majoris (2~objects;
  $\approx$400--500~Myr [\citealp{2003AJ....125.1980K,
    2015ApJ...813...58J}]), and Hyades (4~objects;
  $\approx$600--800~Myr) moving groups.  Note that these are not
    the open clusters themselves, but rather more dispersed stars
    claimed to be coeval with the clusters.

  This sample contains 12 objects in 10 systems, with
  DENIS~J1441$-$0945AB and DENIS J2200$-$3038AB being binaries.  For
  the latter, we were able to measure parallaxes to each individual
  component (see Appendix).

\item {\em Extremely red ultracool dwarfs (8 objects):} A small number
  of field objects have extremely red near-IR colors and peculiar
  near-IR spectra. The original archetype for this genre is
  2MASS~J2244+2043, which has an ordinary L6.5-type optical spectrum but
  a very unusual near-IR spectrum, distinguished by its much redder
  color, stronger CO, and more peaked $H$-band continuum shape compared
  to other field objects \citep{2000AJ....120..447K,
    2003ApJ...596..561M}.  \citet{2008ApJ...689.1295K} and
  \citet{2013ApJ...772...79A} have concluded this is in fact a young
  field object, as opposed to being a ordinary (high gravity) object
  with extreme cloud properties. However, since its discovery, about a
  dozen similarly red objects have been identified that do not appear to
  be low gravity (\citealp{2008ApJ...686..528L} [though see Appendix for
  discussion of 2MASS~J2148+4003]; \citealp{2010ApJS..190..100K,
    2011ApJS..197...19K, 2013ApJS..205....6M, 2013PASP..125..809T,
    2014MNRAS.439..372M}). Instead their extreme colors are thought to
  be due to high gravity (\ie, old age) and/or metallicity. Parallaxes
  for eight of these objects are presented here, as these objects are
  valuable laboratories for understanding the extrema of ultracool
  atmospheres.

\item {\em Strong H$\alpha$ emitters (3+1 objects):} Strong line
  emission from L~dwarfs is uncommon. Unlike for earlier M-type objects,
  H$\alpha$ emission is not necessarily a diagnostic of youth
  \citep[e.g.][]{2015AJ....149..158S}. Nevertheless we included three
  objects with strong H$\alpha$ emission in our program to help better
  diagnose these objects, including one object with a close T7.5
  companion (2MASS~J1315$-$2649AB; \citealp{2011ApJ...739...49B}).  Note
  that a fourth object known to have strong and variable H$\alpha$
  emission, 2MASS~J1022+5825 \citep{2007AJ....133.2258S}, was already
  included in our low-gravity sample ($\beta$ in the optical but \fldg\
  in the near-IR).

\end{itemize}

Table~\ref{table:sample} details the full sample of 67~ultracool objects
in 63 systems, including youth properties. All but one of our targets
have near-IR spectra, mostly from published work and a few obtained by
us (Section~\ref{sec:spectra}). (The exception is the young M4.5 primary
LP~261-75A, whose early spectral type removes it from our subsequent
analysis.)  This complete spectroscopic dataset allows the entire sample
analyzed in this paper, both objects with parallaxes from our CFHT
program and those from the literature (Section~\ref{sec:photom}), to be
spectrally classified in a homogeneous fashion using the
\citet{2013ApJ...772...79A} system (Section~\ref{sec:spectra}).

\subsection{CFHT Parallaxes \label{sec:parallaxes}}

Our analysis methods for obtaining high-precision astrometry from
CFHT/WIRCam images are described in detail in Section~2 of
\citet{2012ApJS..201...19D}. Briefly, we obtain many dithered images
(average of 21) at each observing epoch, thereby achieving a
median standard error per epoch of 3.6~mas for targets presented here.
Targets are observed on each WIRCam run if they fulfill strict airmass
constraints, needed to reduce differential chromatic refraction to a
level below the astrometric noise.
We began observing the majority of the sample presented here in
mid-2009, except for SDSS~J2249+0044AB which was started in mid-2008 and
several objects that we added to our program during 2010--2013.  

Images were obtained in $J$~band for most of our targets, while brighter
targets at risk of saturating in the WIRCam minimum integration time
(5~s) were observed in a narrow $K$-band filter (\Kn) centered at
2.122~\micron. Table~\ref{table:sample} summarizes our target list and
the details of our observations for each target such as the seeing
(median of 0.63\arcsec, full range of 0.50--0.85\arcsec), number
of epochs, time baseline, number of reference stars (S/N$ > 10$), and
the mean parallax for the reference stars determined from the
Besan\c{c}on model of the Galaxy \citep{2003A&A...409..523R} as
described in Section~2.4.2 of \citet{2012ApJS..201...19D}. In addition,
we now derive proper motion corrections also from the Besan\c{c}on model
of the Galaxy in a similar manner as the parallax correction, using the
median proper motion of modeled stars as the correction and computing
the uncertainty in this correction through a Monte Carlo simulation of
sampling variance given our finite number of reference stars. These
proper motion corrections are small, with a median amplitude of
2.5\,\masyr, but necessary for accurately computing space motions.

We derived astrometric parameters and their uncertainties from our data
using a Markov Chain Monte Carlo fit. Figure~\ref{fig:plx} plots the
results, and the parallax and proper motion values are given in
Table~\ref{table:plx}, along with the $\chi^2$ of the best-fit
solutions. In all cases $\chi^2$ is commensurate with the degrees of
freedom and thus validates the accuracy of the astrometric errors used
in our analysis, which are simply the standard error of the position
measured across multiple dithers at a given epoch. We report parallaxes
and proper motions both as relative values (\ie, the direct output of
our MCMC analysis) and absolute values (\ie, corrected for the parallax
and proper motion of our reference grid). The median absolute parallax
uncertainty for the sample presented here is 1.4~mas (5.0\% uncertainty
in distance), with a best precision of 0.8~mas (1.0\%) and 90\% have
errors of 1.0--2.1~mas.  Figure~\ref{fig:j-eplx} summarizes all
parallaxes measured to date for ultracool dwarfs by us and other
programs, illustrating CFHT's excellent combination of small parallax
errors and faint limiting magnitudes.

Our astrometry pipeline automatically detects any object in the field
with a significant parallax by comparing the $\chi^2$ of astrometric
solutions where the parallax was fixed to zero or freely fitted.
Normally this simply allows for easy identification of our targets,
but occasionally we find other objects in the field with significant
parallaxes. We found two such objects in the sample presented here,
LP~415-22\footnote{LP~415-22 is listed in SIMBAD as a Hyades 
  member, but our proper motion and parallax rule out this association
  for any assumed radial velocity, in agreement with previous work
  using only the proper motion \citep{1969AJ.....74....2V}.} ($\pi_{\rm
  abs} = 13.0\pm1.6$\,mas, $M_K = 6.49\pm0.26$\,mag) and
2MASS~J07521548+1614237 ($\pi_{\rm abs} = 13.2\pm1.4$\,mas, $M_K =
8.76\pm0.23$\,mag), and their astrometric solutions are included in
Table~\ref{table:plx}. Based on the photometric relationship with
spectral type from \citet{2007AJ....134.2340K}, we estimate spectral
types of M3 for LP~415-22 and M6 for 2MASS~J07521548+1614237.

\subsection{Near-IR Spectroscopy and Classification \label{sec:spectra}}

\subsubsection{New Near-IR Spectroscopy}

We obtained near-IR spectra for a few of our targets in order to assign
spectral types and assess surface gravity. CFHT-Hy-20 and
DENIS~J1441$-$0945AB did not have any previous comparable
spectra. 2MASS~J0619$-$2903, 2MASS~J2140+3655, and WISE~J2335+4511 have
spectra published in \citet{2013ApJ...772...79A},
\citet{2010ApJS..190..100K} and \citet{2013PASP..125..809T},
respectively, but with only modest S/N, so we re-observed these objects
to better classify them.

We obtained high S/N spectra of these five objects using the NASA
Infrared Telescope Facilty on the summit of Mauna Kea, Hawaii. We used
the facility spectrograph SpeX \citep{2003PASP..115..362R} in prism mode
with the 0.5\arcsec\ slit, which provided an average spectral resolution
($R\equiv\Delta\lambda/\lambda$) of $\approx$150. We dithered in an ABBA
pattern on the science target to enable sky subtraction during the
reduction process. We obtained calibration frames (arcs and flats) and
observed an A0~V star contemporaneously for telluric calibration of each
object. All spectra were reduced using version 4.0 of the SpeXtool
software package \citep{2003PASP..115..389V,2004PASP..116..362C}.
Table~\ref{table:spex} summarizes our observations.

\subsubsection{Spectral Typing and Gravity
  Classification \label{sec:typing}}

For consistency, we classify the near-IR spectra of all M5--L7 dwarfs in
our sample using the system developed by \citet[][hereinafter
AL13]{2013ApJ...772...79A}. The AL13 classification system includes a
gravity-independent determination of spectral type, followed by a
classification of the surface gravity of the object. First, a
combination of visual comparison to field standards as well as spectral
indices is used to determine the near-IR spectral type.  These generally
agree with optical spectral types to better than 1 subtype.  Then given
an object's near-IR spectral type, indices and equivalent widths for
gravity-sensitive features (VO, FeH, \ion{K}{1}, \ion{Na}{1}, and the
$H$-band continuum shape) are used to classify the gravity of the object
as \vlg\ (very low), \intg\ (intermediate), or \fldg\ (field). As noted
in AL13, the gravity classifications of \vlg, \intg, and \fldg\ roughly
correspond to ages of $\lesssim$30~Myr, $\sim$30--200~Myr, and
$\gtrsim$200~Myr, respectively.  The near-IR gravity classifications of
\vlg, \intg, and \fldg\ are designed to agree with the
\citet{2009AJ....137.3345C} optical classifications of $\gamma$,
$\beta$, and $\alpha$, respectively. The majority of our sample have
classifications already published in the AL13 system. For the remaining
objects, we determine the spectral type and gravity
(Table~\ref{table:spectra}). Table~\ref{table:sample} gives the near-IR
spectral types and gravity classifications on the AL13 system for our
entire sample of M5--L7 objects. Unless stated otherwise, these types
and classifications are used throughout the paper.

The AL13 method of spectral typing is only valid for objects with
spectral types of L7 and earlier. To determine near-IR spectral types
for the three late-L and early-T dwarfs in our sample (WISE~J0206+2640,
CFHT-Hy-20 and WISE~J0754+7909), we compute the set of flux indices
defined by \citet{2006ApJ...637.1067B} and assign spectral types to each
resulting index based on the \citet{burg2006-lt} polynomial fits
(Table~\ref{table:spectra-lateL+T}).
Each object has four of the indices with values suitable for this
purpose. We average the spectral types from the four usable indices and
compute an overall index-based type, with the RMS of the spectral types
quoted as the uncertainty.
We also visually compared these objects to spectral standards
from \citet{2006ApJ...637.1067B}. (For the T3 subclass, the original
standard 2MASS~J1209$-$10 has since been found to be a binary with
estimated component types of T2+T7.5, so we also compared with the T3
dwarf SDSS~J1206+28 [\citealp{2010liu-2m1209}].)  
There is no established system for identifying low-gravity features for
late-L and early-T dwarfs,\footnote{For mid/late-T~dwarfs, the ratios of
  the near-IR continuum flux peaks are known to be sensitive to
  metallicity and gravity \citep[e.g.][]{2006liu-hd3651b,
    burningham08-T8.5-benchmark}, and \citet{2014ApJ...787....5N}
  suggest a similar effect for the young T3.5 companion GU~Psc~b.} so
this information is not available for such objects in our sample. The
Appendix discusses the typing results for each of these three objects.

\subsection{Keck Laser Guide Star Adaptive Optics Imaging \label{sec:keck-ao}}

We obtained resolved imaging of DENIS~J1441.6$-$0945AB on
2014~March~14~UT using the facility camera NIRC2 in concert with the
laser guide star adaptive optics (LGS AO) system on the Keck~II
telescope \citep{2004SPIE.5490..321B, 2006PASP..118..297W,
  2006PASP..118..310V}. We used the wide camera mode of NIRC2, with a
pixel scale of $39.686\pm0.008$\,mas\,\perpix\ (H.\ Fu, priv.\ comm.),
and standard Mauna Kea Observatories (MKO) filters $J$, $H$, and $K$
\citep{2002PASP..114..169S, 2002PASP..114..180T} in addition to the
NIRC2 $Y$-band filter (see Appendix of \citealp{2012ApJ...758...57L}).
Our procedure for reducing and analyzing NIRC2 imaging data of binaries
is described in detail in our previous work
\citep[\eg][]{liu08-2m1534orbit, 2009ApJ...692..729D,
  2009ApJ...699..168D,
  2010ApJ...721.1725D}. 
In brief, we measure binary parameters using a
three-component Gaussian representation of the point-spread function and
use the RMS of the dithered images to determine our measurement
uncertainties. We measured flux ratios of
$\Delta{Y_{\rm NIRC2}} = 0.32\pm0.06$\,mag,
$\Delta{J_{\rm MKO}} = 0.250\pm0.016$\,mag,
$\Delta{H_{\rm MKO}} = 0.232\pm0.014$\,mag, and
$\Delta{K_{\rm MKO}} = 0.200\pm0.010$\,mag.  Given the nearly equal
magnitudes and colors of the components, we assume that within the
measurement errors here $\Delta{Y_{\rm NIRC2}} = \Delta{Y_{\rm MKO}}$,
$\Delta{J_{\rm MKO}} = \Delta{J_{\rm 2MASS}}$,
$\Delta{H_{\rm MKO}} = \Delta{H_{\rm 2MASS}}$, and
$\Delta{K_{\rm MKO}} = \Delta{K_{S,{\rm 2MASS}}}$. The astrometry we
measured was consistent between filters, and after correcting for
distortion\footnote{NIRC2 wide camera distortion solution computed by
  H.\ Fu: \url{http://herschel.uci.edu/fu/idl/nirc2wide/}.} the weighted
averages are $314.2\pm1.1$\,mas for the separation and
$328\fdg12\pm0\fdg13$ for the PA.

The vast majority of our field ultracool sample (both our CFHT targets
and the literature objects described in Section~\ref{sec:otherobjects})
have been imaged at high angular resolution, either with \HST\
\citep[e.g.][]{2003AJ....126.1526B, 2003AJ....125.3302G} or with Keck
LGS AO by us.  Also, young companion objects included in our study have
largely been identified with AO imaging.  For the objects lacking high
angular resolution imaging, the low intrinsic binary frequency among
field ultracool dwarfs \citep[e.g.][]{2013ARA&A..51..269D-duchene}
suggests few undetected binaries, especially when accounting for the
decline in binary frequency with stellar mass.  (The young ages of our
targets mean they are lower mass than old field objects of similar
spectral type.)  Overall, unresolved binarity is unlikely to be an
important factor in our analysis.

\subsection{Photometry \label{sec:photom}}

\subsubsection{Our CFHT Sample}

The vast majority of our targets have photometry in the 2MASS Point
Source Catalog \citep{2003tmc..book.....C}, and several have published
photometry on the Mauna Kea Observatories filter system
\citep{2002PASP..114..169S, 2002PASP..114..180T}, primarily from Data Release~10
(DR10) of the UKIRT Infrared Deep Sky Survey (UKIDSS;
\citealp{2007MNRAS.379.1599L}). All targets also have photometry in
the AllWISE Source Catalog that merges the cryogenic \WISE\ mission
survey data \citep{2010AJ....140.1868W} with the post-cryogenic
NEOWISE survey \citep{2011ApJ...731...53M}. Only LP~261-75B is not
resolved from its primary in the \WISE\ data. In order to create a
homogeneous collection of near-infrared photometry, we have
supplemented these published results with synthesized photometry. All
of our targets have spectra published either here or elsewhere, and we
use these to compute 2MASS--MKO photometric conversions for each
object as well as synthetic $Y-J$ colors. As described in
\citet{2012ApJS..201...19D}, we assume the synthesized colors within a
given bandpass (e.g., $J_{\rm 2MASS}-J_{\rm MKO}$) are errorless and assume an
error of 0.05~mag for the synthesized $Y-J$ colors.  The young binary
SDSS~J2249+0044AB has published $K$-band flux ratios in both 2MASS and
MKO systems but $J$- and $H$-band flux ratios reported only in the MKO
system \citep[$\Delta{J}_{\rm MKO} = 1.024\pm0.016$\,mag and
$\Delta{H}_{\rm MKO} = 0.953\pm0.009$\,mag;][]{2010ApJ...715..561A}.  We
used the spectral decomposition method of \citet{2012ApJS..201...19D} to
derive 2MASS flux ratios of $\Delta{J}_{\rm 2MASS} = 1.04\pm0.03$\,mag
and $\Delta{H}_{\rm 2MASS} = 0.95\pm0.04$\,mag.

\subsubsection{Other Objects from the Literature \label{sec:otherobjects}}

We have also compiled photometry for all field objects with published
parallaxes that have possible evidence for low-gravity spectra, youth
($\approx$10--200\,Myr), or being unusual redness. This compilation of
objects from the literature includes 11 field objects and 20 systems
with (one or more) ultracool companions.\footnote{Four well-known young
  ultracool companions are not included in our compilation as their
  primary stars lack parallaxes:
  GSC 08047$-$00232~B (M9.5$\pm$1, Tuc-Hor member;
  \citealp{2003A&A...404..157C, 2004A&A...420..647N,
    2005A&A...430.1027C}),
  1RXS~J235133.3+312720 (L0$^{+2}_{-1}$, probable AB~Dor member;
  \citealp{2012ApJ...753..142B}), 
  2MASS J01225093$-$2439505~B (L3.7$\pm$1.0, probable AB~Dor member;
  \citealp{2013ApJ...774...55B, 2015ApJ...805L..10H}),
  and GU~Psc~b (T3.5$\pm$1, probable AB~Dor member;
  \citealp{2014ApJ...787....5N}).}
We do not include objects in star-forming regions such as Taurus and
Sco-Cen, as they are younger than the sample being considered here.
Combined with our parallax sample of 67~ultracool objects in 63 systems
(excluding the young M4.5 primary LP~261-75A), our compilation includes
a total of 102~ultracool dwarfs in 93~systems.  (For objects with
parallaxes from multiple sources, we choose the one with the lowest
uncertainty.)  We supplemented the published photometry for this
extended sample in the same way as our main sample, using published
spectra to compute synthesized photometry when available. For objects
without published spectra, we estimated 2MASS--MKO photometric
conversions from other objects with similar spectral types or absolute
magnitudes. We expect these estimates to be reasonably accurate as,
e.g., over our entire sample the differences in these corrections vary
by $\pm$0.06\,mag or less in any one band. These estimates allow us to
have an almost complete collection of $JHK$ photometry for our combined
sample, only lacking $J$-band photometry for HD~948B and HR~8799e and
$K$-band photometry for 51~Eri~b.

Table~\ref{table:photom} gives the 2MASS ($JH\Ks$), MKO ($YJHK$), and
AllWISE ($W1$ and $W2$) photometry for our combined sample along with
the parallax we use for each object. Table~\ref{table:absmag} gives the
resulting absolute magnitudes.

\subsubsection{Objects Not Included in Our Analysis \label{sec:exclude}}

We discuss other potential young objects with parallaxes in the
literature that we excluded from our analysis for the various reasons
given here.
EROS-MP~J0032$-$4405 \citep[L0$\gamma$ optical/L0~\intg\
near-IR;][]{2009AJ....137.3345C, 2013ApJ...772...79A} has two published
parallaxes that are highly discrepant (1.9$\sigma$, 78\% different),
$38.4\pm4.8$\,mas from \citet{2012ApJ...752...56F} and $21.6\pm7.2$\,mas
from \citet{2013AJ....146..161M}. These gives absolute magnitudes that
are 1.25\,mag different, and rather than perform our analysis contingent
on both scenarios we simply exclude this object.
2MASSW~J0103320+193536 \citep[L6$\beta$;][]{2000AJ....120..447K,
  2012ApJ...752...56F} has a parallax from \citet{2012ApJ...752...56F}.
However, its near-IR gravity status is uncertain, as
\citet{2013ApJ...772...79A} type this object as L6~\intg\ but
\citet{2015ApJS..219...33G} typed it as L6p with no clear sign of low
gravity.
%
TVLM831-154910 has a parallax from \citet{1995AJ....110.3014T} and has
recently been typed as M7:$\beta$~\vlg\ and identified as a $\beta$~Pic
moving group candidate member by \citet{2015ApJS..219...33G}. An updated
parallax from Pan-STARRS-1 (E.\ Magnier 2015, priv.\ comm.) is highly
discrepant with the published value, making the distance and membership
indeterminate.
%
NLTT~13728, 2MASS~J19303829$-$1335083, and G44-9 are M6 dwarfs that
have parallaxes from \citet{2012ApJ...758...56S} but lack infrared
spectra.
2MASS~J06524851$-$5741376AB is a binary discovered by
\citet{2012A&A...548A..33C} with a S/N~=~10 parallax from
\citet{2012ApJ...752...56F} who also typed it as M8$\beta$. This binary
lacks resolved spectral information for the individual components,
though the $J\Ks\Lp$ flux ratios for the system are modest
($\approx$0.3~mag) suggesting comparable spectral types. If they were
late-M dwarfs, both components would be $\sim$1\,mag fainter than any
such objects in our sample, suggesting this is an atypical system or
else the parallax has a significant systematic error.
2MASS~J07123786$-$6155528 \citep[L1$\beta$;][]{2009AJ....137.3345C} has
a S/N~=~2.5 parallax from \citet{2012ApJ...752...56F}.
The primary star of the VHS~1256$-$12AB binary (M7.5~\intg\ +
  L7~\vlg; \citealp{2015ApJ...804...96G}) has been found to be a binary
  by \citet{2016ApJ...818L..12S}, who note that the resolved components
  would be extremely faint compared to any other known M7.5~objects if
  the parallax from \citet{2015ApJ...804...96G} is correct.  Also, a
  preliminary parallax from Pan-STARRS1 (E.~Magnier 2015, priv.\ comm.)
  disagrees substantially with the published parallax.  So we exclude
  all three components of this system from our sample.
LSPM~J1314+1320AB has a parallax and optical spectral type of M7 from
\citet{2009AJ....137.4109L} and was discovered to be a binary by
\citet{2006MNRAS.368.1917L}. \citet{2014ApJ...783...27S} present
resolved photometry and integrated-light near-infrared spectral evidence
of low gravity. Gravity classifications are lacking for the individual
components, so this object is not included in our work here.  A detailed
analysis of this system, including its component dynamical masses, is in
\citet{2016ApJ...827...23D}.
SIPS~J2045$-$6332 \citep[M9/L1;][]{2007AJ....133.2258S,
  2013AJ....146..161M} has a parallax from
\citet{2014AJ....147...94D}, but the only published youth signatures
are a triangular $H$-band spectrum from \citet{2013AJ....146..161M}
and lithium from \citet{2014MNRAS.439.3890G}.
2MASS~J21011544+1756586AB is a binary with a parallax from
\citet{2004AJ....127.2948V} that has been noted to have a triangular
$H$-band spectrum and red colors \citep{2014ApJ...783..121G}. It has no
resolved spectra or near-infrared photometry. We classify the
integrated-light near-IR spectrum from \citet{2010ApJ...710.1142B} as L7
and lacking any strong indication of youth, though the spectral
resolution ($R\sim120$) of the data is not particularly sensitive to
gravity for such a late-type object (Figure~\ref{fig:2m2101}).
CFBDSIR~J214947.2$-$040308.9 (T7) was discovered by
\citet{2012A&A...548A..26D} as a potential low-gravity object in the
AB~Dor moving group.  However, a recent parallax measurement rejects
this membership (P.~Delorme \etal, submitted), and thus the
interpretation of its unusual near-IR spectrum as low gravity is now
uncertain.

\section{Comparison to Published Distances \label{sec:compare}}

\subsection{Parallactic Distances \label{sec:compare-parallaxes}}

A number of objects in our sample have independent measurements of their
parallaxes in the literature. Comparing these published values to each
other and to our CFHT results allows us to vet the quality of all of
these measurements. There are two programs that have produced parallaxes
for significant subsets of our sample. \citet{2012ApJ...752...56F,
  2013AJ....145....2F} have published parallaxes for 10 of our targets
based on SMARTS/ANDICAM and CTIO Blanco/ISPI observations. In addition,
our sample includes all 10 parallaxes from \citet{2014A&A...568A...6Z},
which are based on combining astrometry from Calar Alto/OMEGA2000,
NOT/NOTCam and WHT/LIRIS. Finally, 14 objects in our sample have
published parallaxes in several other papers \citep{2002AJ....124.1170D,
  2004AJ....127.2948V, 2006AJ....132.1234C, 2008A&A...489..825T,
  2009AJ....137..402G, 2011AJ....141...54A, 2012ApJ...758...56S,
  2013ApJ...762..118W, 2014A&A...563A.121D, 2015ApJ...799..203G}. Given
that there is some overlap among these various literature sources, the
total number of objects in our sample with some published parallax is
26.

Figure~\ref{fig:compare} shows our new parallaxes compared to published
values. Our results broadly agree with published values as most of our
parallaxes are consistent at $\le$1$\sigma$ with previous results, and
90\% are consistent at $\le$2.5$\sigma$. However, there are a number of
more extreme ($\approx$4--6$\sigma$) outliers. As can be seen from the
histograms of parallaxes differences in $\sigma$ in
Figure~\ref{fig:compare}, all of the most extreme outliers are from the
programs of \citet{2012ApJ...752...56F} and \citet{2014A&A...568A...6Z}.
We now consider all of the $>$2$\sigma$ outliers
individually,\footnote{Note that we consider proper motion discrepancies
  in units of \masyr\ rather than $\sigma$ since most authors do not
  report absolute proper motions, and even absolute proper motions may
  contain systematic errors associated with imperfect knowledge of the
  reference frame.} and in the end we decide to use our parallaxes for
all these cases (except for 2MASS~J1139$-$3159).

\textit{2MASSW~J0033239$-$152131.} \citet{2014A&A...568A...6Z} report an
absolute parallax of $24.8\pm2.5$\,mas, and we find a value 5.8$\sigma$
different, $42.8\pm1.8$\,mas. The total proper motion of this object is
$300.6\pm2.3$\,\masyr, and our RA and Dec proper motions both agree to
better than 4\,\masyr\ with those reported by
\citet{2014A&A...568A...6Z}. In this case, our parallax seems to resolve
an anomaly that \citet{2014A&A...568A...6Z} found when placing
2MASSW~J0033$-$1521 on their H-R diagram that used a spectral type-based
\Teff\ and parallax-based luminosity. They found that this object
appeared to have an age of $<$10\,Myr and thereby a mass well below the
lithium fusion limit, but it has a clear non-detection of lithium
absorption \citep[$<$1~\AA;][]{2009AJ....137.3345C}. It also does not
display evidence of very low surface gravity (i.e., $\gamma$, $\delta$,
or \textsc{vl-g}) as might be expected for such a young object, rather
it is classified as L4$\beta$ \citep{2009AJ....137.3345C} in the optical
and L1~\textsc{fld-g} \citep{2013ApJ...772...79A} in the near-infrared
(see Appendix). \citet{2014A&A...568A...6Z} note that their parallax
could be the cause of the problem, as a closer distance would reconcile
the H-R diagram position and lack of lithium, and our parallax does
indicate a much closer distance. We thereore use our parallax for this
object.

\textit{2MASSW~J0045214+163445.} \citet{2014A&A...568A...6Z} report an
absolute parallax of $57.3\pm2.0$\,mas, and we find a value 3.6$\sigma$
different, $65.9\pm1.3$\,mas. Our proper motions agree reasonably well
within their quoted errors ($\approx$16\,\masyr\ or 1.5$\sigma$). Our
correction from relative to absolute parallax is only $1.56\pm0.20$\,mas
and thus not sufficient to explain why our parallax is 8.6\,mas larger.
We share all 10 objects from \citet{2014A&A...568A...6Z} in common with
our program, and as noted above there is at least one other case
(2MASS~J0033$-$1521) where the \citet{2014A&A...568A...6Z} parallax is
likely to be in error. In contrast, the 15 parallax measurements we have
in common with programs other than \citet{2012ApJ...752...56F} and
\citet{2014A&A...568A...6Z} have no $>$2$\sigma$ outliers. We therefore
use our own parallax for 2MASSW~J0045+1634.

\textit{GJ~3276 (a.k.a.\ 2MASS~J04221413+1530525).}
\citet{2012ApJ...752...56F} report an absolute parallax of
$24.8\pm3.1$\,mas and a relative proper motion of
$(-17.2\pm2.7, 7.4\pm2.6)$\,\masyr. However, we find only a marginally
significant relative parallax detection of $2.6\pm1.0$\,mas, from which
we compute an absolute parallax of $3.9\pm1.0$\,mas. We also do not
detect a significant relative proper motion, finding
$(1.4\pm1.5, 1.8\pm1.4)$\,\masyr. Thus our parallax is 6.4$\sigma$
discrepant from \citet{2012ApJ...752...56F}, and the total proper motion
is different by 16\,\masyr. We believe this is unlikely to be a case of
mistaken identity, as we observe no other stars near GJ~3276 in our CFHT
imaging with $\approx$0$\farcs$5 seeing, and there are no other objects
in the field with a similar parallax or proper motion. We automatically
detect any sources with significant parallaxes in each field, and this
field did have one such detection (LP~415-22, 4$\farcm$1 away from
GJ~3276), but its parallax ($13.2\pm1.5$\,mas) and total proper motion
($112.5\pm1.7$\,\masyr) are also inconsistent with
\citet{2012ApJ...752...56F}.  \citet{2013ApJ...772...79A} note that this
the spectrum of GJ~3276 is particularly red for its spectral type and
report a reddening of $A_V=4.6$\,mag. They also note that it is only
$\sim$10$\degree$ from the Taurus star-forming region on the sky and
thus may be behind or embedded in interstellar material. Either of these
scenarios is consistent with our parallax, which places it at or behind
the distance of Taurus within 2$\sigma$. Such reddening would be more
challenging to explain using the \citet{2012ApJ...752...56F} parallactic
distance of $\approx$40\,pc. Thus, we favor our parallax for this
object, but given its low S/N we do not end up including it in the
analysis that follows.

\textit{2MASS~J05012406$-$0010452.} \citet{2012ApJ...752...56F} report
an absolute parallax of $76.4\pm4.8$\,mas, and we find a value
5.6$\sigma$ different, $48.4\pm1.4$\,mas. Like GJ~3276 above, our value
is $\approx$20\,mas smaller than the \citet{2012ApJ...752...56F} value.
Our two proper motions agree reasonably well ($<$10\,\masyr) for this
object that has a total motion of 236\,\masyr.
\citet{2014A&A...568A...6Z} also report a parallax for this object,
$51.0\pm3.7$\,mas, which agrees with ours at 0.7$\sigma$, and proper
motions that also agree well ($\lesssim$1\,\masyr).
\citet{2014A&A...568A...6Z} note this parallax discrepancy and show that
2MASS~J0501$-$0010 would be unique among objects of its spectral
type, i.e., much fainter than the field sequence or other low-gravity
objects, if the \citet{2012ApJ...752...56F} parallax is used. Therefore,
we use our parallax, consistent with but more precise than that of
\citet{2014A&A...568A...6Z}.

\textit{LP261-75B (2MASS J09510549+3558021).}
\citet{2004AJ....127.2948V} reported a preliminary absolute parallax of
$16.1\pm7.4$\,mas, and we find a value 2.1$\sigma$ different,
$31.6\pm1.3$\,mas. Our proper motions agree well (3--6\,\masyr) for this
object with a total motion of 189\,\masyr. An updated parallax distance
of $33.0^{+2.8}_{-2.4}$\,pc from Vrba et al.\ is reported in
\citet{2013ApJ...774...55B}, which is in excellent agreement with our
parallax distance. Therefore, we do not consider this object a
$>$2$\sigma$ outlier in the end, and we use our parallax in the
following analysis.

\textit{2MASSW~J1139511$-$315921 (a.k.a.\ TWA~26).}
\citet{2012ApJ...752...56F} report an absolute parallax of
$35.1\pm4.3$\,mas, and we find a value 2.5$\sigma$ different,
$23.7\pm1.6$\,mas. Our proper motions are also somewhat discrepant at
10--20\,\masyr. \citet{2013ApJ...762..118W} and
\citet{2014A&A...563A.121D} also report parallaxes for this object of
$23.8\pm2.6$\,mas and $26.2\pm1.1$\,mas, respectively, which agree with
ours at 0.1$\sigma$ and 1.3$\sigma$ (as does the
\citealp{2005ApJ...634.1385M} kinematic parallax of $25.0\pm2.5$~mas).
Their proper motions also agree with ours at 6\,\masyr.  We use
  the Ducourant \etal\ result, given its somewhat higher precision than
  our CFHT result and the fact that this group has measured parallaxes
  for several other TWA members using the same methods.

\textit{2MASSW~J2208136+292121.} \citet{2014A&A...568A...6Z} report an
absolute parallax of $21.2\pm0.7$\,mas, and we find a value 2.2$\sigma$
different, $25.1\pm1.6$\,mas. This marginal discrepancy is largely
driven by the very small quoted parallax error of
\citet{2014A&A...568A...6Z}. Our proper motions agree reasonably well
($<$10\,\masyr) for this object with a total motion of 96\,\masyr. Our
correction from relative to absolute parallax is $0.85\pm0.04$\,mas,
which is slightly larger than the quoted parallax error from
\citet{2014A&A...568A...6Z} and slightly smaller than their adopted
correction of 1.0\,mas. Given that 2 of 10 objects from
\citet{2014A&A...568A...6Z} are $\approx$4--6$\sigma$
($\approx$10--20\,mas) outliers, we are reluctant to take their very
small quoted parallax uncertainty of 0.7\,mas for 2MASS~J2208+2921
at face value. Therefore, we use our own parallax value in the following
analysis.

To summarize, our parallax sample overlaps with 10 out of 15 young
objects from \citet{2012ApJ...752...56F,
  2013AJ....145....2F},\footnote{Of the 5 non-overlapping
  \citet{2012ApJ...752...56F} objects, we use two in our analysis
  (2MASS~J02212859$-$6831400 and DENIS~J205754.1$-$025229), as their
  near-IR absolute magnitudes are consistent with the objects from our
  CFHT sample. The remaining 3 parallaxes appear to be problematic
  (Section~\ref{sec:exclude}) and thus are excluded from our analysis.}
the entire sample of 10 objects from \citet{2014A&A...568A...6Z}, and 15
measurements from various other programs (Table~\ref{table:compare}).
The only $>$2$\sigma$ outliers in this sample come from
\citet{2012ApJ...752...56F} and \citet{2014A&A...568A...6Z}, and for the
three largest discrepancies ($>$5$\sigma$) our new parallaxes seem to
resolve standing problems with the interpretation of those objects. One
of the other $>$2$\sigma$ discrepant objects, 2MASS~J1139$-$3159, has
two independent parallaxes in the literature that agree with ours. For
all of these overlapping cases except 2MASS~J1139$-$3159, in the end we
choose to use our CFHT parallaxes, which has the added benefit of
internal consistency for the handling of systematic effects such as the
correction from relative to absolute parallax and proper motion.

\subsection{Kinematic Distances}

Given prior knowledge of the expected space motions of stars, such as
membership in co-moving associations, distance estimates are possible
for objects that have some kinematic information (proper motion and/or
radial velocity). Such kinematic distance estimates exist in the
literature for a number of our targets, either because they have been
proposed as members of young moving groups or through a kinematic model
of the solar neighborhood. The most homogeneous and complete collection
of such kinematic distances for our ultracool dwarf sample comes from
the BANYAN~II analysis \citep{2013ApJ...762...88M, 2014ApJ...783..121G}.
(Note that subsequent BANYAN~II analysis in
\citealp{2015ApJS..219...33G} does not publish independent kinematic
distances.)
We now compare our trigonometric parallax distances to these kinematic
distance estimates. We also compare our TWA parallaxes to the kinematic
distances from \citet{2005ApJ...634.1385M}. Table~\ref{table:dkin}
presents the complete list of overlapping objects.

Overall, the kinematic distances are reasonably consistent with parallax
distances within the quoted uncertainties, with 50\% of BANYAN~II
distances agreeing within 1.2$\sigma$ and 90\% agreeing within
2.3$\sigma$.
In terms of absolute fractional difference, 50\% of BANYAN~II kinematic
distances are within 20\% of the parallax distance, with the other half
being more discrepant. (Note that the typical parallax error is much
smaller than this: the median fractional error for the CFHT sample is
6\%.)
In fact, for the best studied groups (TWA, $\beta$~Pic, AB~Dor,
Tuc-Hor), there are no $>$2$\sigma$ outliers out of a sample of 11
objects. However, outliers are present in the larger sample, and we find
that SIMP~J2154$-$1055 is significantly more distant (2.9$\sigma$, 28\%)
than predicted from its kinematic distance based on Argus membership
\citep{2014ApJ...792L..17G}. Other candidate group members with
$>$2$\sigma$ or $>$50\% discrepancies are 2MASS~J0045+1634 (Argus) and
2MASS~J0619$-$2903 (Columba). All four TWA and Tuc-Hor proposed members
that have CFHT parallaxes are consistent at $<$1$\sigma$ and 5\%--10\%
with kinematic distances from \citet{2005ApJ...634.1385M} and
\citet{2014ApJ...783..121G}.

For objects identified to belong to the ``young field'' population by
\citet{2014ApJ...783..121G}, there is a significant preference for
kinematic distances to be larger than parallax distances
(Figure~\ref{fig:dkin}). The median offset is +11\% with a dispersion
of $\pm$37\%. In terms of the absolute value of the fractional
difference, the median offset for the young field sample is 25\%.
The rate of highly discrepant BANYAN~II kinematic distances
($>$2.5$\sigma$, $>$60\% fractional offset) is 10\% for the whole
sample. The largest of these discrepancies seem to only occur for the
young field population. For example, we find that 2MASS~J0435$-$1414
is at $88^{+9}_{-12}$\,pc while \citet{2014ApJ...783..121G} report
$10.5\pm1.6$\,pc. Another large difference in the opposite direction
is for 2MASS~J1711+2326 that we place at $32.7^{+1.7}_{-1.9}$\,pc but
for which \citet{2014ApJ...783..121G} estimate to be at
$53^{+7}_{-6}$\,pc.

In summary, we find that BANYAN~II kinematic distances are generally
not as accurate as their quoted precision of 8\%
\citep{2014ApJ...783..121G}, with 50\% of kinematic distances being
$\geq$20\% different from parallactic distances.  Kinematic distances
for young field objects seem to be systematically overestimated by
11\%, with an RMS dispersion of 37\% with respect to the parallactic
distances, and have a failure rate (i.e., values that are
$>$2.5$\sigma$, $>$60\% discrepant) of $\approx$10\%.

\section{Results \label{sec:results}}

\subsection{Absolute Magnitudes \label{sec:absmags}}

Figures~\ref{fig:absmag-2mass}, \ref{fig:absmag-mko},
and~\ref{fig:absmag-wise} show the absolute magnitudes of our sample in
the standard 2MASS, MKO, and \WISE\ infrared bandpasses as a function of
\citet{2013ApJ...772...79A} near-IR spectral type and gravity
classifications, applicable for objects down to L7. (Later-type
  objects are discussion in Section~\ref{sec:LTtransition}.)  While
there are many possible criteria for partitioning our sample, near-IR
spectral type and gravity classification are the only spectroscopic
properties available for all our CFHT targets. Also, surface gravity is
the key defining characteristic of this young sample, so it is natural
to examine how reduced gravity impacts the photometry of ultracool
dwarfs.

The figures also show the loci of field ultracool dwarfs for comparison.
Because our sample is predominantly composed of late-M and L dwarfs,
which change monotonically in absolute magnitude with spectral type, we
derive new linear relations to represent the M6--L8 field population,
instead of using the higher-order polynomials from
\citet{2012ApJS..201...19D} which were derived for a much larger range
in spectral type. (In particular, the near-IR magnitudes become
non-monotonic in the L/T transition. This leads to curvature in the
polynomial fits, and thus the higher-order fits do not track the
behavior of the earlier L~dwarfs as well as linear fits.) Also, our
complete young sample has near-IR spectral types, and thus we would like
to compare with a field relation based solely on near-IR types (unlike
the relations of \citealp{2012ApJS..201...19D}, which used optical types
for M and L dwarfs when available, though the effect of their inclusion
is small).
For our field data, we used the IDL structure posted at the Database of
Ultracool
Parallaxes,\footnote{\url{http://www.as.utexas.edu/~tdupuy/plx},
  maintained by T.\ Dupuy and last updated 2013 September 9.} which
includes the comprehensive summary of \citet{2012ApJS..201...19D} along
with more recent published parallaxes and photometry for ultracool
dwarfs (M6 and later). In order to use only normal field objects with
high-quality absolute magnitudes, we excluded anything without
\texttt{null} in the IDL structure's \texttt{flag} tag (which excludes,
\eg, subdwarfs and young objects), with a peculiar optical or infrared
spectral type noted in the \texttt{osptstr} or \texttt{isptstr} tags, or
with a ${\rm S/N} < 10$ parallax. We then used the least absolute
deviation method (\texttt{LADFIT} in IDL) to compute robust linear fits
of absolute magnitude as a function of infrared spectral type for
objects having spectral types of M6 to L8.  Table~\ref{table:coeff}
reports the coefficients of these fits along with the RMS about the
fit. Note that the RMS does vary with spectral type range so a single
number is not fully representative (\eg, Figure~27 of
\citealp{2012ApJS..201...19D} and discussion below.)

\subsubsection{General Trends in Absolute Magnitudes with Spectral Type
  and Gravity}

Our data reveal three main trends in the behavior of the near-IR
($YJHK$) absolute magnitudes.
\begin{enumerate}

\item In a given bandpass, the loci of young objects relative to the
  field follows a consistent pattern, with young late-M dwarfs being
  brighter than the field, young early-L ($\approx$L0-L2) dwarfs being
  comparable, and young mid/late-L ($\gtrsim$L3) dwarfs being fainter
  than the field sequence. This is in accord with the preliminary
  $J$-band results presented in \citet{2013AN....334...85L}.

\item The offset of the young objects relative to the field sequence
  (brighter at earlier types, fainter at later types) is more dramatic
  at bluer wavelengths. At $Y$~and $J$~bands, the young late-M's and
  late-L's can be $\approx$2~magnitudes offset from the field sequence,
  whereas at $K$~band most L~dwarfs are comparable to the field.

\item The low-gravity (\vlg) population is more distinct from the field
  objects than the intermediate-gravity (\intg) population, especially
  for the late-L dwarfs, though individual young objects of both gravity
  classifications do overlap. To highlight the differences between each
  other and the field, Figures~\ref{fig:absmag-2mass},
  \ref{fig:absmag-mko}, and~\ref{fig:absmag-wise} include robust linear
  fits to the \vlg\ and \intg\ samples, and Table~\ref{table:coeff}
  gives the fitted coefficients.

\end{enumerate}

The first two trends have been noted in earlier studies of young
ultracool dwarfs using smaller samples and more heterogenous spectra
classifications, though not as conclusively.
(1)~Based on the companions to young ($\approx$10--500~Myr) stars known
at the time, \citet{2011ApJ...729..139W} showed that young late-L
dwarfs were always fainter than the field sequence, but that young
early/mid-L dwarfs could be brighter or fainter, broadly consistent with
our results presented here. The spectral types available for their
companion sample was of variable quality, perhaps enhancing the scatter
in the absolute-magnitude relations.
(2)~\citet{2013AJ....145....2F} found that most young L~dwarfs are
fainter than the field sequence in $J$~band, by up to 1~mag, based on
their own IR parallaxes for a comparable sample of young field objects.
This result is not confirmed with our CFHT sample: while some of the
young L~dwarfs show such a large offset, many are comparable to the
field sequence.\footnote{Note that the summary plots in Wahhaj \etal\
  and Faherty \etal\ both show the young L0~companion AB~Pic~b
  \citep{2005A&A...438L..29C} as being significantly fainter in $J$~band
  than the field sequence. However, improved photometry by
  \citet{2013ApJ...777..160B} produces a brighter flux for AB~Pic~b,
  leading to a less anomalous appearance.}
(3)~Finally, \citet{2014A&A...568A...6Z} found most of the young
early/mid L~dwarfs in their IR parallax sample are comparable or
brighter than the field sequence (\ie, the opposite of Faherty \etal),
whereas our CFHT results do not find brighter objects except among the
earliest L~dwarfs. 
To summarize, while young late-L dwarfs have long been recognized as
being fainter at $J$~band than the field, previous studies of young
early/mid-L dwarfs have been inconsistent. The large sample and high
precision of our CFHT sample now affords a clear view of the
spectrophotometric trends.

Recently, \citet{2015ApJ...810..158F} reported a similar trend in the
$J$-band magnitudes of young late-M to late-L dwarfs as seen in our CFHT
sample. (Note that a significant fraction of their sample relied on
kinematic distances, rather than parallaxes.) They found that their
young sequence and field sequence cross at a spectral type of L0,
whereas our \vlg\ and field sequences cross at a spectral type of L2.
This small discrepancy may be due to our separate analysis of \vlg\ and
\intg\ samples whereas \citeauthor{2015ApJ...810..158F} consider all
young objects as a single sample, or due to slightly different
definitions of the field sequence (see below). In addition,
\citeauthor{2015ApJ...810..158F} find young objects have brighter $W2$
magnitudes than the field sequence at all spectral types, in accord with
our results.

In the \WISE\ mid-IR bandpasses, the young targets appear generally to
be brighter than the field sequence in Figure~\ref{fig:absmag-wise},
especially among the young early/mid-L dwarfs. There may be a more clear
distinction between the loci of \intg\ and \vlg\ L~dwarfs than seen at
shorter wavelengths, with the \vlg\ objects being systematically
brighter in the mid-IR than \intg, and similarly between \intg\ and
\fldg.

Many of our objects have optical spectra, so
Figure~\ref{fig:absmag-optical} shows the absolute-magnitude trends
using optical spectral types from the literature. Many of the objects
also have optical gravity classifications, or at least reported
signatures of youth in the literature
\citep[e.g.][]{2008AJ....136.1290R, 2008ApJ...689.1295K,
  2009AJ....137.3345C, 2009ApJ...699..649S, 2010ApJS..190..100K}. The
overall behavior of the young objects relative to the field sequence is
the same for optical and near-IR spectral types. (Note that optical
types for L~dwarfs only go down to L8, while near-IR types include L9.)
The scatter in the young L~dwarf sequence is smaller for optical types,
likely reflecting the greater spectral heterogeneity of L~dwarfs in the
near-IR (\eg, see discussion in \citealp{kirk05} and
\citealp{2013ApJ...772...79A}).

\subsubsection{Dispersion in Absolute Magnitudes}

In conjunction with the aforementioned general trends, there is
significant scatter in the absolute magnitudes of young objects. This
phenomenon is real and not simply due to parallax uncertainties.
Among the late-M dwarfs, the \vlg\ objects show relatively uniform
magnitudes, perhaps in part because 5 out of 7 of them are TWA members,
and thus have common ages. (The six TWA objects in our sample are all
the brightest \vlg\ objects in $J$-band for their spectral type, ranging
from M6 [TWA~8B] to L0 [2MASS~J1245$-$4429 a.k.a.\ TWA~29].)
The scatter is far more notable for the \intg\ late-M dwarfs, where
objects can be comparable to the field locus or up to
$\approx$2~magnitudes brighter at $YJHK$. 

In fact, while the sample is small, the \intg\ late-M's seem to
bifurcate in the magnitude-vs-spectral type plots and near-IR
color-magnitude diagrams (Section~\ref{sec:cmd}), into a pair of M8's
brighter than field objects (2MASS~J1411$-$2119 and 2MASSI~J0019+4614)
and a fainter pair of M7.5's comparable to field objects (SO~J0253+1652
[a.k.a. Teegarden's Star] and 2MASS~J0714+3702).
Furthermore, this bifurcation is also reflected in the lithium
abundances,
with the two brighter objects having lithium absorption in their optical
spectra \citep{2009ApJ...705.1416R} while the one fainter object with
an optical spectrum does not show lithium
\citep[SO~J0253+1652;][]{2004AJ....128.2460H}.
And thus the bifurcation of our \intg\ late-M sample may reflect the
conflation of a younger population (\eg, 2MASSI~J0019+4614 is a member
of the $\approx$125~Myr AB~Dor group) and an older one.
An additional consideration is that the M7--M8 range is where
  \citet{2002MNRAS.335L..79D} have suggested the mass-luminosity
  relation changes due to the onset of dust formation, as inferred from
  the luminosity function of young clusters \citep[see
  also][]{2003MNRAS.342.1241P, 2005MNRAS.361.1323C}. So the \intg\
  bifurcation may also reflect dust opacity variations. (Intriguingly,
  the \vlg\ objects seem to show the largest dispersion right at the
  M/L~transition.)
Altogether, more \intg\ late-M objects with parallaxes are needed to
refine our view of this sub-population.  

For young L~dwarfs, the scatter relative to the field population is less
than for the late-M's, though young L's can still be brighter or fainter
than the field.  The notable outlier is the companion
2MASS~J1207$-$3932b (L3~\vlg), which may reflect the difficulties of
assigning a single number, namely spectral type, to this spectrally
extreme object (\eg, see Figure~2 of \citealp{2013ApJ...777L..20L}).
In this vein, it is worth remembering that differences between the young
and old populations also reflect the challenges of near-IR spectral
typing for dusty ultracool dwarfs (\eg, see Section~4.1 of
\citealp{2013ApJ...772...79A}). Some of the scatter may be due to
spectral type uncertainties, rather than the intrinsic dispersion of
absolute magnitudes in objects of the same spectral type.

While the \vlg\ and \intg\ populations show clear trends, perhaps a
comparably interesting result is how the near-IR absolute magnitudes are
only partly influenced by surface gravity.  A sizeable number of young
objects do {\em not} have significantly different absolute magnitudes
than the field sample. Similarly, for objects of comparable near-IR
spectral type, the offsets from the field sequence only partly correlate
with gravity classification. Some \intg\ objects can be just as faint or
bright relative to the field as \vlg\ objects, except for the late-L
dwarfs, where the \vlg\ objects can achieve fainter magnitudes. Spectral
type (a proxy for temperature) and gravity (a proxy for age) do not seem
to be the only factors driving the absolute magnitudes, at least given
our current ability to assign types and gravities. (Note this is not the
case for the color-magnitude diagram, where \vlg\ objects show a clearly
defined sequence, as discussed in Section~\ref{sec:cmd}.)

\subsubsection{Absolute Magnitudes of Unusually Red Field-Gravity Objects}

Figures~\ref{fig:absmag-2mass}, \ref{fig:absmag-mko},
and~\ref{fig:absmag-wise} also show the behavior of field objects with
unusually red colors but \fldg\ gravities, such as 2MASS~J1821+1414 and
2MASS~J2148+4003.  Interestingly, such objects mostly fall within the
absolute magnitude range seen in field objects. Thus while these objects
have absolute magnitudes and colors that overlap low-gravity objects
(see also Section~\ref{sec:cmd}), they are not unusually bright (for the
case of the late-M dwarfs) or faint (for the case of the mid/late-L
dwarfs) compared to objects of similar spectral type, unlike the
low-gravity objects.
In this respect, absolute magnitudes for late-M and late-L dwarfs may
help to distinguish dusty objects from low-gravity ones, in cases where
the spectra are insufficient or unavailable to assess surface gravity.

Figure~\ref{fig:absmag-misc} shows the \fldg\ objects and objects having
strong H$\alpha$ emission from our CFHT sample.
For the most part, the \fldg\ objects appear similar to the field locus,
including the young companion LP~261-75B and peculiar object
SDSS~J1025+3212. The most discrepant \fldg\ object compared to the field
locus is 2MASSW~J0033$-$1521, which is L4$\beta$ in optical but
L1~\fldg\ in NIR. This is 0.8~mag fainter at $J$-band than the field
locus, making it notably fainter than even the low-gravity L1 dwarfs,
though its absolute magnitudes would be field-like if it were classified
as L4.
Among the H$\alpha$ emitters, the two objects brighter than the field
locus both have low-gravity signatures: 2MASS~J0335+2342 is a \vlg\
object, and LP~423-31 is classified as \fldg\ in the near-IR but has
signs of low gravity in its optical spectrum
\citep{2009ApJ...699..649S}. The two H$\alpha$ emitters without any
low-gravity signatures (2MASS~J0407+15464 and 2MASS~J1707+4301) have
comparable magnitudes to the field.

\subsubsection{Polynomial Relations} 

A common use of absolute magnitude-spectral type relations is to
estimate photometric distances. As seen in our plots, the field and
young sequences differ depending on bandpass, spectral type, and
gravity, but there are some regimes where they overlap. For instance,
the $H$-band absolute magnitudes of early-L dwarfs ($\approx$L0--L2) are
comparable for field and young objects, likewise for $K$~band with mid-L
dwarfs ($\approx$L3--L6), and for $W1$~band with late-L dwarfs
($\approx$L5--L8), though there still exist young outliers. Such trends
may be useful for estimating distances for ultracool dwarfs where
spectral types are available but gravity information is not, \eg, typing
from modest S/N spectra or purely photometric estimates. Unfortunately,
for the late-M dwarfs, the young and field sequences are distinct so
they are not amenable to this approach. Similarly, if gravity
information is known, then specific choices of bandpasses for certain
spectral types may be quite useful for photometric distance estimates,
\eg, the low-gravity L~dwarfs show remarkably small scatter in their
$W1$-band magnitudes, more so than it would appear from simply quoting
the RMS about the robust linear fit in Table~\ref{table:coeff}.

Figure~\ref{fig:absmag-compare} compares polynomial fits for absolute
magnitude versus near-IR spectral type for the field sequence from
\citet{2012ApJS..201...19D}; for field late-M and L~dwarfs from this
paper, \citet{2015ApJS..219...33G}, and \citet{2015ApJ...810..158F}; for
the \intg\ and \vlg\ samples from this paper; and for the young field
sample from \citet{2015ApJS..219...33G}. We caution that the Gagne
\etal\ relations are derived from a sample that has some objects with
parallactic distances but most with kinematic distances derived from
their Bayesian classification method (described here in
Section~\ref{sec:membership}).  Overall, the differences between the
polynomial relations can be very significant between our low-gravity
relations and those of Gagne \etal, depending on the spectral types and
filters of interest.

\subsection{Color-Magnitude Diagrams \label{sec:cmd}}

As another comparison of our sample to the field population, we examine
several infrared color--magnitude diagrams (CMDs) based on near-IR
($YJHK$) and mid-IR ($W1$ and $W2$) data with objects distinguished by
their gravity classifications (Figures~\ref{fig:cmd1}
and~\ref{fig:cmd2}). For the field population, we use the same data
described in the previous section. Note that our CMDs contain more
objects than the plots of absolute magnitude versus spectral type in the
previous section, since some objects have measured photometry but have
not been classified spectroscopically.  For brevity, we plot near-IR
photometry on the MKO system only, though our compilation is complete
for both 2MASS and MKO systems.  All CMD plots span 9.5\,mag in absolute
magnitude for ready comparison.

\subsubsection{General CMD Trends with Gravity}

The most obvious feature of all the CMDs is that \vlg\ objects are
displaced from the field sequence, being brighter and/or redder than
field objects.  In fact, the distinction between the respective \vlg\
and field sequences is remarkable, with only a very small fraction of
interlopers.  The Appendix discusses the individual interlopers in
detail.  In short, four of them (2MASS~J0407+1546, LP~423-31,
2MASS~J0835$-$0819, and 2MASS~J2208+2921) appear to have CMD positions
that are discrepant with their gravity classifications, though three of
these have hints of low-gravity spectral signatures (all except
2MASS~J2208+2921).  One interloper (2MASS~J1935$-$2846) appears to have
a mid-IR excess.

In contrast, objects classified as \intg\ share CMD positions with both
the field and \vlg\ objects. There seems to be a slight preference for
the early-type \intg\ objects, brighter than
$M_{J,{\rm MKO}} \sim 12$\,mag, to sit on the field sequence (in
  accord with \citealp{2008ApJ...689.1327S} models which show surface
  gravity has a minimal influence on the CMD positions of the hottest
  ultracool dwarfs), while fainter \intg\ objects are redder and
intermediate between the field and \vlg\ sequences. However, there are
also counterexamples of this trend, with a few bright \intg\ objects
being closer to the \vlg\ sequence or faint \intg\ objects being
consistent with the field population.

Figure~\ref{fig:cmd-comp} provides a master summary of our low-gravity
sequences, the field population, the unusually red field objects, and
the locus of young ultracool companions of stars, with many individual
objects labeled. We plot $J$-band absolute magnitude versus $J-K$ color
because we have nearly complete data in these bands, and this captures
more dynamic range in the observed properties than other CMDs. We have
also plotted linear fits to the three sequences shown here (\vlg, \intg,
and the field).  We used the same least absolute deviation method
(\texttt{LADFIT} in IDL) as in Section~\ref{sec:absmags} to fit
$(J-K)_{\rm MKO}$ color as a function of absolute magnitude, since
absolute magnitude is the more precisely measured quantity and has the
larger dynamic range.  We exclude 2MASS~J1207$-$3932 from the fits given
its extreme position.  These linear fits quantify the behavior discussed
above for the three sequences, and the coefficients and RMS about the
fits are in Table~\ref{table:coeff}.

Figure~\ref{fig:cmd-vlgseq} examines how the CMD behavior described
above relates to spectral type, plotting our \vlg\ sample and the field
population, as well as the linear fits between color and absolute
magnitude. We also show the mean absolute magnitude at each spectral
type from M7 to L7 using the linear relations derived in
Section~\ref{sec:absmags} for both the \vlg\ and field samples, thereby
indicating the direction in which objects of the same spectral type but
different gravity classifications move on the CMD.  Late-M dwarfs are
much brighter at nearly the same color at lower gravity.
This behavior gradually changes from late-M to early-L such that low
gravity objects of the same spectral become more similar in absolute
magnitude but increasingly redder in $J-K$. At spectral type L3 and
beyond, low-gravity counterparts are on average fainter in $J$ band
and even redder in $J-K$. At the latest L~types, low gravity objects
are fainter in both $J$ band (by $\sim$1.6\,mag) and $K$ band (by
$\sim$0.8\,mag) and correspondingly $\sim$0.8\,mag redder in $J-K$.
Thus, the overall effect of decreasing surface gravity on the CMD
positions of ultracool dwarfs follows a complex pattern, with the
changes in the absolute magnitudes and colors depending on the
spectral type

\subsubsection{Young Companions versus Young Field Objects}

One of the prime motivations for studying young ultracool dwarfs
  is to use these objects as analogs for young gas-giant exoplanets.
  The overlapping temperatures, ages, and masses of young ultracool
  dwarfs and young imaged companions make this analogy appealing.  At
  the same time, we expect that field objects and companions typically
  form via different mechanisms, \eg, gravitational collapse of a gas
  cloud as compared to formation within a circumstellar disk.,
  respectively, which could result in different thermal evolution,
  metallicities, elemental mixtures, etc.  High-quality data enables an
  empirically driven comparison of free-floating objects with substellar
  companions.  In this aspect, our CMD plots reveal the following:

\begin{enumerate}

\item At the bright end, the young companion and young field samples
  overlap on the CMD.  The brightest companions (HR~7329B, PZ~Tel~B,
  and TWA~5B) are consistent with the \vlg\ sequence, mainly because
  they are brighter than anything of similar spectral type and/or
  color.

\item Most of the fainter young companions are more consistent with the
  old field sequence than the \vlg\ sequence.
  There are exceptions among the early/mid-L companions, as
    CD-35~2722B (\intg), AB~Pic~b (\vlg), and G~196-3B (\vlg) all
    concide with the \vlg\ sequence, and 2MASS~J0103$-$5515~b might as
    well given its (larger) photometric uncertainties.  In addition, as
    is well known, the extreme companions HR~8799b and
    2MASS~J1207$-$3932 fall off the field sequence.

  \indent Overall, the late-M to late-L companions appear to be bluer
  and/or fainter than the young field sample, which suggests that the
  companions and the field objects span a different range in physical
  properties. These two samples cover similar age ranges and spectral
  types, including objects in the same moving groups
  (Section~\ref{sec:membership}), and thus should have similar masses
  and temperatures given their similar absolute magnitudes.  Thus,
  difference in other physical properties, \eg, formation history or
  composition, may be relevant here.

  Figure~\ref{fig:cmd-comp-vs-VLG} highlights the difference between the
  \vlg\ sequence and the young companions, using a robust linear fit as
  in our other CMD plots. The construction of the two samples are
  heterogenous, which deters a more rigorous comparison between the two.
  But the overall pattern is intriguing --- the \vlg\ sequence is
  notably offset from the young companion sequence.  Note that most of
  the free-floating objects were identified as low-gravity in {\em
    optical} spectra (Section~\ref{sec:sample}), and thus their redder
  colors are not due to a selection effect.  Given the large number of
  young field objects with parallaxes, the next significant step in such
  a comparison will likely rely on more discoveries to grow the
  companion sample.\footnote{The paucity of young companions with
    $M(J)\approx13-15$~mag on the CMDs is interesting, though we note
    that the young mid-L~dwarf companion 2MASS~J0122$-$2439~B
    \citep{2013ApJ...774...55B} has $M(J_{\rm MKO})=14.0\pm0.3$~mag
    based on a photometric distance. Its $(J-K)_{\rm MKO}$ color is
    $2.28\pm0.15$~mag, placing it in the \vlg\ sequence.}

\item Companions in the L/T transition are beyond the spectral type and
  color range where we have defined our linear fits, but the companions
  broadly appear to be consistent with the field population on the CMD.
  Finally, the late-T companions Ross~458C and Gl~504b are notably
  redder than the field T~dwarf CMD sequence, likely due to the gravity
  dependence of collisionally induced H$_2$ absorption
  \citep[e.g.][]{2011MNRAS.414.3590B}.

\end{enumerate}

\subsubsection{Unusually Red Field-Gravity Objects \label{sec:veryred}}

The eight \fldg\ objects identified as unusually red behave in a way
distinct from either of the field or low-gravity samples
(Figure~\ref{fig:cmd-redseq}).  The four earlier-type red objects
(M9--L5) are indistinguishable from the field sequence on all our CMDs.
2MASS~J2224$-$0158 (L3~\fldg) is the reddest of these, but it is only
marginally inconsistent with the field sequence within its color
errors. In contrast, the late-type red objects 2MASS~J2148+4003
(L6~\fldg) and WISE~J2335+4511 (L7p~\fldg) intermingle with the bottom
of our \vlg\ sequence on CMDs where the color includes $K$~band, having
colors and magnitudes similar to such objects as 2MASS~J0355+1133
(L3~\textsc{vl-g}), 2MASS~J2244+2043 (L6~\vlg), and WISE~J0047+6803
(L7~\intg). This intermingling is suprising, given that work to
  date on late-L dwarfs has focused on low-gravity being the prime mover
  to generate extremely red colors and faint magnitudes (see
  Section~\ref{sec:LTtransition}).  Perhaps these two red \fldg\ late-L
  dwarfs have enhanced metallicity compared to most objects in the
  field, a possibility that could be examined in the future with
  higher-resolution spectra.

In the mid-infrared $W1-W2$ versus $M_{W2}$ plot, the red \fldg\ objects
both appear to be intermediate between the field and \vlg\
sequence. Only in $Y-J$ or $Y-H$ CMDs do they distinguish themselves,
with 2MASS~J2148+4003 sitting with the \vlg\ sequence and
WISE~J2335+4511 being consistent with the field. Finally, the
latest-type unusually red objects WISE~J0206+2640 (L8 (red)) and
WISE~J0754+7909 (T2p (red)) appear in locations on all CMDs that seem to
be consistent with the field sequence, which we discuss further
  in Section~\ref{sec:LTtransition}.

\subsection{Tangential Velocities \label{sec:vtan}}

Given our measured parallaxes and proper motions, we can readily compute
the projected tangential velocities (\Vtan) for all of our targets.
Objects that have not acquired significant velocities relative to their
natal material should have small tangential velocities with
  respect to the local standard of rest (LSR) and overall a small
  velocity dispersion as a sample.
This allows for an interesting probe of our low-gravity sample, since we
do not expect our objects to be old enough to have acquired large space
motions. While tangential velocities cannot be used to rule out large
space motions, if an object does have a large space motion we might
detect it through \Vtan.

While it is common practice to quote
  directly observed tangential velocities ($\Vtan \equiv \mu/\pi$), we
  also compute here tangential velocities corrected for the Sun's space
  motion with respect to the LSR, adopting the recent determination of
  $(U,V,W)_{\sun} = (-11.10, 12.24, 7.25)$\,\kms\ from
  \citet{2010MNRAS.403.1829S}. For a given position on the sky it is
  straightforward to convert $(U,V,W)_{\sun}$ into projected component
  velocities in RA and Dec, and with a parallactic distance these can be
  converted into proper motion in angular units. Table~\ref{table:vtan}
  reports our uncorrected tangential velocities, our derived corrections
  to proper motion in RA and Dec, and the final tangential velocities of
  our objects in the LSR frame ($V_{\rm tan}^{\prime}$). We compute all
  of these quantities in a Monte Carlo fashion, accounting for the
  uncertainties in parallax and proper motion.

Two objects with signatures of low gravity display raw (\ie, not
LSR corrected) tangential velocities of $\gtrsim$90\,\kms. 2MASS
J1022+5825 (L1$\beta$/L1~\fldg) is an object from our parallax program
for which we measure
$\mu_{\rm abs} = 1\farcs0775\pm0\farcs0031$\,yr$^{-1}$ and
$\pi_{\rm abs} = 0\farcs0526\pm0\farcs0031$, resulting in
$\Vtan = 97\pm6$\,\kms ($V_{\rm tan}^{\prime} = 82\pm6$\,\kms).
SO~J0253+1652 (M7.5~\intg; a.k.a., Teegarden's Star) has a parallax of
$\pi_{\rm abs} = 0\farcs2593\pm0\farcs0009$ from
\citet{2009AJ....137..402G} and proper motion of
$\mu_{\rm abs} = 5\farcs129\pm0\farcs029$\,yr$^{-1}$ from
\citet{2010AJ....139.2440R}, resulting in $\Vtan = 93.8\pm0.6$\,\kms\
($V_{\rm tan}^{\prime} = 79.2\pm0.6$\,\kms). In fact, both of
these objects have radial velocities (Table~\ref{table:uvw}) showing
that 2MASS~J1022+5825 and SO~J0253+1652 have total space motions of
$99\pm4$\,\kms\ and $116.0\pm0.5$\,\kms, respectively (not LSR
corrected).  Such velocities are inconsistent with young stars,
\eg, the high probability ($>$75\%) moving group members from
  \citet{2008hsf2.book..757T} have a median of 24~\kms\ and a 95\%
  confidence limit of 18--34~\kms.
The existence of these two objects demonstrates that one of two
possibilities must be true: (1)~objects with low gravity spectral
signatures are indeed young but are sometimes born with, or quickly
acquire, large space motions; or (2)~the optical/near-IR spectral
signatures being used to identify low-gravity objects are sometimes
manifested by much older objects. If the former is true, then high
velocity cannot be used to distinguish between objects that are red
because they are dusty and those that are red because they are low
gravity, as has been suggested previously (\eg, see discussion of
2MASS~J2148+4003 in the Appendix). We note that these kinematic outliers
only occur among objects with intermediate gravities, not the \vlg\
objects, which thus might prefer the latter possibility, namely that
other effects besides lower gravity can lead to an \intg\ spectrum.
However, visual inspection of the near-IR SpeX prism spectra for
  our two high-velocity objects \citep{2013ApJ...772...79A,
    2008arXiv0803.0295B} do not reveal any obvious discrepancies
  compared to other \intg\ objects of similar spectral type but lower
  tangential velocity.

Figure~\ref{fig:vtan} shows all of the tangential velocity measurements
for objects with signatures of low gravity as noted in
Table~\ref{table:absmag}, including those without formal gravity
classifications, along with the unusually red \fldg\ objects. With the
exception of the two highest velocity objects discussed above, the
low-gravity objects have a median $V_{\rm tan}^{\prime}$ of 10\,\kms\
and standard deviation of 8\,\kms.  This distribution is a
somewhat tighter than has been noted previously in the literature for
low-gravity objects \citep[e.g.,][]{2012ApJ...752...56F,
  2013PASP..125..809T}. It is also tighter than standard
  deviations typically quoted for normal field dwarfs of similar
spectral type \citep[$\approx$20\,\kms; e.g.,][]{2007AJ....133.2258S,
  2010AJ....139.1808S, 2009AJ....137....1F, 2012ApJ...752...56F}. These
results are consistent with the notion that low-gravity objects have
younger ages.\footnote{Note that our text and the references
    cited here refer to the standard deviation of the tangential
    velocities.  As highlighted by \citet{2009ApJ...705.1416R} and
    \citet{2010A&A...512A..37S}, this is distinct from the velocity
    dispersion, a term which commonly refers to the quadrature sum of
    the standard deviations in the three components of space velocity
    ($U$, $V$, and $W$).}

Unusually red \fldg\ objects tend to have higher tangential velocities,
with a median $V_{\rm tan}^{\prime}$ of 29\,\kms\ and a
standard deviation of 15\,\kms.  We caution however that
these objects have sometimes been discovered in proper motion searches
\citep[e.g.,][]{2008ApJ...686..528L, 2010ApJS..190..100K} and thus could
harbor a selection bias toward higher velocities.\footnote{We
    note that one red object, 2MASS~J21481633+4003594 (L6/L6~\fldg), was
    previously thought to have a relatively high tangential velocity
    \citep[\Vtan = 62\,\kms;][]{2008ApJ...686..528L}, but our parallax
    and proper motion give a lower velocity ($34.4\pm0.5$\,\kms; see
    Appendix), consistent with the bulk of the low-gravity objects.}
The unusually red objects with the highest velocities are the two
earliest-type ones (2MASS~J2224$-$0158 [L4.5/L3~\fldg] and
  2MASS~J1331+3407 [L0/L0~\fldg] with
  $V_{\rm tan}^{\prime} = 41.4\pm0.7$\,\kms and $40.1\pm2.4$\,\kms,
  respectively) and the latest-type one (WISE~J0754+7909 [T2p (red)]
with $56.0\pm1.0$\,\kms).
As noted in Section~\ref{sec:cmd}, all three of these red objects are
consistent with the field sequence on CMDs. In contrast, the two
unusually red objects that seem to follow the \vlg\ sequence on CMDs
have velocities marginally consistent with the low-gravity sample
($V_{\rm tan}^{\prime} = 25.1\pm0.5$\,\kms\ for 2MASS~J2148+4003 and
$V_{\rm tan}^{\prime} = 24.8\pm0.4$\,\kms\ for WISE~J2335+4511).

\subsection{Membership \label{sec:membership}}

\subsubsection{Young Moving Groups}

Assessing the membership of our young sample in nearby young moving
groups (YMGs) is a prime result from our new parallaxes. Placing young
brown dwarfs in such groups allows their ages to be established, by
tying to the stellar YMG members. Establishing membership also allows a
more complete census of the low-mass population in YMGs. We consider
several approaches.

\begin{enumerate}

\item {\em Kinematic BANYAN~II analysis:} We used the BANYAN~II online
  tool \citep{2013ApJ...762...88M,2014ApJ...783..121G}, which relies on
  a model of the kinematic and spatial distribution of the solar
  neighborhood, assumed to be composed of seven YMGs (Argus [ARG],
  Columba [COL], \bPic\ [BPMG], AB~Dor [ABDMG], Carina [CAR], TW Hyd
  [TWA], and Tuc-Hor [THA]) and two components of the field population
  (old [O.FLD] and young [Y.FLD]). Each of these nine components is
  represented by triaxial gaussians in their space motions ($UVW$) and
  positions ($XYZ$), where the gaussians' principal axes are not
  necessarily aligned with the $UVWXYZ$ coordinate axes, and then
  weighted by the expected number of objects, based on a log-normal
  initial mass function that has been scaled to the number of known
  stellar members in these YMGs.

  Given the observed proper motion, parallax, and (when available)
  radial velocity for an ultracool dwarf, this online tool uses Bayesian
  classification to compute the probability for membership in each of
  the nine components. (See \citealp{2014ApJ...783..121G} and
  \citealp{2015ApJ...798...73G} for an in-depth discussion about the
  classification and resulting probabilities.) Note that the BANYAN
  model only contains groups within 100~pc and the field population only
  extends to 200~pc, so the model is not appropriate for analysis of our
  most distant objects (GJ~3276, 2MASS~J0557$-$1359, and
  2MASS~J0619$-$2903).

  For objects with clear spectroscopic signatures of youth, we use the
  online tool's ``Young Field'' option, which removes the old field
  component from consideration, leaving just the young field population
  to represent field objects $<$1~Gyr old. Note that for objects with
  discrepant information between their optical and near-IR gravity
  classifications,\footnote{For all objects with discrepant optical and
    near-IR gravity classifications (2MASS~J0033$-$1521
    [L4$\beta$/L1~\fldg], 2MASS~J0253+3206 [M7~low-$g$/M6~\fldg],
    LP~423-31 [M7~low-$g$/M6~\fldg], 2MASS~J1022+0200
    [M9$\beta$/M9~\fldg], and 2MASS~J1022+5825
    [L1$\beta$/L1~\textsc{fld-g}], the optical spectrum is low gravity
    while the IR spectrum is classified as \fldg. This suggests that
    optical spectra are somewhat more discriminating for gravity.}  we
  run the tool both with the Young Field option and without. The
  resulting probabilities for field membership are unchanged, though the
  objects are assigned to young field or old field depending on the age
  assumed.

  Table~\ref{table:membership} shows the resulting Bayesian
  probabilities for all the field objects, computed using the astrometry
  compiled in Table~\ref{table:absmag}, and supplemented with radial
  velocities from the literature given in Table~\ref{table:uvw}. As
  discussed in \citet{2014ApJ...783..121G}, bona fide YMG members within
  1$\sigma$ of their group's six-dimensional $UVWXYZ$ location have
  membership probabilities of $\gtrsim$95\%, and more peripherial
  members ($\approx$1.0--2.5$\sigma$) have probabilities of
  $\approx$10--95\%, based on a full Bayesian analysis using kinematic
  and photometric information (described below). Using this as a
  benchmark, we report all membership outcomes with a probability
  $>$10\% from our kinematics-only BANYAN~II analysis. Most objects are
  uniquely matched to a single YMG or the field population.

\item {\em Kinematic+SED BANYAN~II analysis:} For most of our objects,
  membership results are available from the full BANYAN~II modeling by
  \citet{2014ApJ...783..121G} and \citet{2015ApJS..219...33G}. While the
  BANYAN~II online tool uses only kinematics, the full BANYAN~II model
  evaluates YMG membership for ultracool dwarfs using both kinematic and
  spectrophotometric information, the latter based on the 2MASS and
  \WISE\ colors and absolute magnitudes of old and young ultracool
  dwarfs. For this reason, the probabilities that compute from the
  BANYAN~II online tool cannot be directly compared to those from the
  full BANYAN~II analysis of Gagne \etal. The full BANYAN~II model
  evaluates 4 kinematic criteria and 2 photometric ones, and thus
  kinematics still plays a large role in the membership calculation but
  not an exclusive one.

  While BANYAN~II can efficiently identify low-gravity ultracool dwarfs
  \citep{2015ApJS..219...33G}, there are known deficiencies that deter
  us from relying solely on its results.  The first is its kinematic
  model of the solar neighborhood may not be complete as there may be
  (1) stellar YMG members that are currently
  unidentified, which would alter the triaxial gaussians used by
  BANYAN~II, 
  (2) uncertainties in the reality of all seven adopted YMGs, \eg, the
  exact nature of Argus is disputed \citep{2015MNRAS.454..593B}, and
  (3) new YMGs in the solar neighborhood that remain to be discovered.
  The second important deficiency is the photometric aspect of
  BANYAN~II, which relies on the census of young ultracool dwarfs known
  at the time of its construction. Our CFHT work has expanded this
  census by a factor of $\gtrsim$4, most notably growing the L~dwarf
  sample, and has significantly revised parallaxes for some objects used
  by BANYAN~II. Thus, the definition of the sequence used by the full
  BANYAN~II analysis may need to be updated. Finally, Gagne \etal\
  require objects to show low-gravity features to be considered YMG
  members. However, we consider this to be too strict, given the
  diversity of gravity classifications seen in YMG brown dwarfs and the
  lack of firm constraints on the upper age at which gravity signatures
  persist \citep{2013ApJ...772...79A, 2016ApJ...821..120A} --- this is
  especially relevant for 2MASS~J2351+3010 discussed below.

  In short, while the full BANYAN~II analysis provides a powerful tool
  for assessing membership, we do not consider it to be a complete
  representation of the nearby young ultracool population.

\item{\em Comparison to known stellar members:} As a simple check, we
  compare our objects' $UVWXYZ$ position with the nearby young groups,
  using reduced chi-squared (\rchisq) as a metric, namely in velocity
  position:
  \begin{equation}
    \rchisqUVW = \frac{1}{3} \left( 
      \frac{(U_{obj} - U_{ymg})^2}{\sigma^2_U} + 
      \frac{(V_{obj} - V_{ymg})^2}{\sigma^2_V} + 
      \frac{(W_{obj} - W_{ymg})^2}{\sigma^2_W}
    \right)  
  \end{equation}
  where $\sigma$ is the quadrature sum of the measurement uncertainty
  for an individual object and uncertainties in the group kinematics
  (both centroid uncertainty and group dispersion). \rchisqXYZ\ is
  calculated similarly.
  We adopt a cutoff value of 4 in \rchisqXYZ\ and \rchisqUVW, as this
  encompasses most ($\approx$95\%) of the known members.

  Thus, \rchisq\ values provides a simple assessment of whether an
  object's membership proposed by BANYAN~II is plausible. Obviously,
  this is an incomplete view since it does not consider false positives,
  namely that an object from the young field population happens to have
  a very similar $UVW$ and/or $XYZ$ with the YMGs being considered (and
  likewise for intermixing of members from different YMGs). For this
  reason, we chiefly use \rchisq\ to rule out cases of marginal
  membership.

  We calculate the $UVW$ velocities for our sample in a right-handed
  coordinate system. Given the uncertainty for the parallax, proper
  motion and radial velocity of each object, we track uncertainties
  using a Monte Carlo approach. Table \ref{table:uvw} lists literature
  radial velocities as well as our calculated $UVW$ velocities and total
  velocity for the subset of our sample having measured radial
  velocities. For this \rchisq\ consideration, we exclude our targets
  with RV uncertainties of $>$3~\kms, as these will have unsuitably
  small \rchisqUVW\ values.

  For objects without RVs, we calculate the radial velocity that gives
  the smallest velocity offset to each group and the accompanying
  minimum possible \rchisqUVW\ value, as a plausibility check on
  membership. Objects that never reach a small enough velocity distance,
  or corresponding have minimum possible \rchisqUVW\ values that are
  large, can be refuted as members even in the absence of RV
  information.

  For the YMG properties, we use \citet{2008hsf2.book..757T} as a
  starting point and incorporate more recent updates as appropriate.
  Table~\ref{table:ymg} summarizes our adopted groups. Note that
  BANYAN~II uses rotated ellipses to define the galactic position and
  velocity of each group, thus we cannot directly compare their group
  definitions to ours. As a check,
  we computed the mean and standard deviation of $UVWXYZ$ for group
  members from the BANYAN~II lists \citep{2014ApJ...783..121G}. With the
  exception of the Argus group, we find that our adopted group $UVWXYZ$
  agree with those determined from BANYAN~II members to within the
  uncertainties. For Argus, our adopted $Y$ for the group
  ($-115.1\pm35.5$~pc) does not agree with the mean $Y$ value for
  objects in the BANYAN~II Argus membership list ($-21.7\pm26.7$~pc).
  This discrepancy could be due to contamination of the membership lists
  for Argus, as recently noted by \citet{2015MNRAS.454..593B}.

\end{enumerate}

In the end, we adopt a holistic approach in assessing the membership of
our young sample. At one extreme, for objects with complete
spatial-kinematic data (both parallaxes and radial velocities) as well
as full agreement among our three membership methods (kinematics-only
BANYAN~II, full kinematics+SED BANYAN~II, and \rchisq\ confirmation),
there is no ambiguity in the result. Similarly, objects with parallaxes
only, without RVs, but with agreement between BANYAN~II analyses are
also considered highly probable members, especially if the minimum RV
distance between the object and the known members is plausible
(equivalently, the minimum possible \rchisqUVW\ value is less than our
nominal cutoff of $\lesssim4$). And likewise, some objects flagged by
Gagne \etal\ as strong members from their full BANYAN~II analysis but
that lacked parallactic distances now have measured parallaxes from our
work, thereby strengthening the case for membership.

Where the three membership methods disagree, we scrutinize the differing
results along with the available data to make an assessment. For
instance, when our kinematic BANYAN~II analysis has a different outcome
than the Gagne \etal\ full BANYAN~II analysis, we examine the
observations and the Bayesian calculations. Our new parallaxes often
help to resolve the disagreement. The full (kinematic+SED) BANYAN~II
analyses for many objects were done without parallaxes, producing
membership assignments along with statistical estimates of the
distances. These statistical distances, and thus the associated
membership claims, can be evaluated in light of our CFHT parallactic
distances. (Likewise, some objects' parallaxes have changed
significantly from published work to our CFHT measurements; see
Section~\ref{sec:compare-parallaxes}.) Similarly, new CFHT distances for
objects can establish their $UVWXYZ$ positions as being highly
discrepant with known stellar members based on \rchisq, even if the
BANYAN~II analyses (either our kinematics-only one or the published
kinematics+SED one) points to membership.

Table~\ref{table:membership} provides our membership results. Young
companions are not included here, as their membership is assessed in
their discovery papers. Individual objects that warrant discussion are
given in the Appendix, primarily those where the different membership
methods give discrepant results or where our CFHT parallaxes lead to a
revision in the membership assigned by previous work. A few particularly
notable findings are:

\begin{itemize}

\item {\em Argus \fldg\ members:} we identify two possible Argus members
  with \fldg\ near-IR gravities (2MASS~J0030$-$1450 and
  2MASS~J2351+3010), suggesting an older age than previously claimed
  ($\approx$40~Myr; \citealp{2008hsf2.book..757T}), if the group
  actually a physical one (see \citealp{2015MNRAS.454..593B}).  Another
  Argus candidate found by \citet{2014ApJ...792L..17G},
  SIMP~J2154$-$1055, is refuted as a member based on our new parallax.

\item {\em Disk-bearing brown dwarfs in the \bPic\ moving group:} we
  support the membership of the accreting object 2MASS~J0335+2342
  proposed by \citet{2012ApJ...758...56S}, and we identify
  2MASS~J1935$-$2846 as a potential member with a mid-IR excess
  indicative of a disk. These are the two longest-lived (24~Myr;
  \citealp{2015MNRAS.454..593B}) brown dwarf disks known to date.

\item {\em Distant young ultracool dwarfs with disks:} Four M5--M7
  dwarfs originally thought to be nearby young objects turn out to be
  very young ($\sim$Myr) objects at distances of $\approx$100--300~pc:
  GJ~3276 (a.k.a.\ 2MASS~J0422+1530), 2MASS~J0435$-$1414,
  2MASS~J0557$-$1359, and 2MASS~J0619$-$2903. All of them have
  signatures of having circumstellar disks, most notably
  2MASS~J0435$-$1414 which is detected at far-IR, sub-millimeter, and
  millimeter wavelengths.  Based on their spectral types, they straddle
  the stellar/substellar mass boundary.

\item {\em An intermediate-age very red L~dwarf}: 2MASS~J2148+4003 is a
  notable very red L6~\fldg\ object claimed to be an old object by
  \citet{2008ApJ...686..528L} based on its \vtan. As discussed in the
  Appendix, we find its \vtan\ value is actually only half the
  Looper~\etal\ value, and a BANYAN~II-based analysis suggests it may be
  a young field object.  When combined with its \fldg\ gravity, these
  results suggest this may be an intermediate-age object.

\item {\em Refuted planetary-mass members of YMGs:} A number of
  candidate planetary-mass ($\lesssim$13~\Mjup) objects in young moving
  groups have been identified by the BANYAN~II searches:
  2MASS~J0033$-$1521 (9--11~\Mjup), 2MASS~J0253+3206 (13--15~\Mjup),
  2MASS~J2148+4003 (6--7~\Mjup), and 2MASS J2351+3010 (9--11~\Mjup) from
  \citet{2014ApJ...783..121G};
  2MASS~J0030$-$1450 ($10.8^{+0.4}_{-0.6}$~\Mjup), 2MASS~J0501$-$0010
  ($10.2_{-1.0}^{+0.8}$~\Mjup), and 2MASS~J2213$-$2136
  ($13.5\pm0.3$~\Mjup) from \citet{2015ApJS..219...33G}; 
  and SIMP~J2154$-$1055 ($10.3_{-0.3}^{+0.7}$~\Mjup) from
  \citet{2014ApJ...792L..17G}.
  In all these cases, we weaken or eliminate the case for planetary-mass
  status, either by assigning field membership or finding the near-IR
  spectra are not low-gravity. Our analysis does preserve one young
  planetary-mass candidate from \citet{2014ApJ...783..121G}, the \bPic\
  candidate 2MASS~J2208+2921 ($12.9_{-0.1}^{+0.3}$~\Mjup). These are all
  discussed individually in the Appendix.

\end{itemize}

Table~\ref{table:changes-membership} provides a concise summary of our
membership results relative to the published BANYAN~II analyses. We adopt
two simple categories: (1)~23~objects for which our results have changed the
membership assignments, and (2)~28~objects for which our results have
improved/fortified the existing assignments.
For the former, we include all objects identified as YMG member
candidates which end up with changed memberships. This count includes
objects previously identified as only marginal candidates based on their
low BANYAN~II membership probabilities and/or high contamination rates,
under the rationale that such candidates would still warrant followup
observations whereas we conclude they are unlikely to be worth pursuing.
For the latter, most are objects with BANYAN~II membership
determinations done without parallaxes where our CFHT results reinforce
the BANYAN~II assignments. There is inevitable ambiguity for a handful
of objects when using this simple 2-bin categorization, so those
interested in the details of specific objects should consult the
Appendix and Table~\ref{table:membership}.

Broadly speaking, objects with $\gtrsim80\%$ BANYAN~II probabilities to
reside in a YMG are confirmed by our work, 
except for the Argus (1~object) and Columba (1~object) candidates, which
we refute and assign to the field. Objects with $\lesssim60\%$
probabilities are usually refuted as members and assigned to the field.
These numbers are a very rough summary --- the reasons for the
membership changes depend on the individual objects and their specific
datasets.

For several objects, our parallaxes improve/fortify the previous YMG
membership claims based on BANYAN~II analyses without parallaxes
\citep{2014ApJ...783..121G, 2015ApJ...798...73G}. These are
2MASS~J0019+4614 (ABDMG), 2MASS~J0045+1634 (ARG), WISE~J0047+6803 (ABDMG;
see also \citealp{2015ApJ...799..203G}), 2MASS~J0117$-$3403 (THA),
2MASS~J0241$-$0326 (THA), SDSS~J0443+0002 (BPMG), 2MASS~J0518$-$2756
(COL), 2MASS~J0536$-$1920 (COL), and 2MASS~J2244+2043 (ABDMG).

Similarly, our parallaxes have added three objects for which YMG
  memberships are now secured based on both parallax and RV data:
  2MASSI~J0019+4614 (ABDMG), 2MASS~J0045+1634 (ARG), and SDSS~0443+0002
  (BPMG). (These had previously been noted as members based on BANYAN~II
  statistical distances and RVs, and hence are also listed in the
  previous paragraph.)  We have also tentatively added a fourth YMG
  member with both parallax and RV data, 2MASS~J0608$-$2753 (COL).  This
  object was previously considered a marginal candidate by BANYAN~II,
  and our improvement in the parallax has strengthened the case (see
  Appendix).

Finally, Figures~\ref{fig:cmd-ymg} and~\ref{fig:cmd-ymg2}
present the color-magnitude diagrams again, this time with the YMG
memberships indicated.  These results provide empirical isochrones for
substellar evolution as a function of age and spectral type.  We use
these in Section~\ref{sec:models} to test current theoretical isochrones.


\subsubsection{Older Stellar Kinematic Groups}

Several field L~dwarfs from \citet{2010A&A...512A..37S} were included in
our sample, as they were reported to be candidate members of older
($\gtrsim$100~Myr) moving groups, namely the Hyades, Pleaides, Ursa
Major streams \citep[e.g.][]{2001MNRAS.328...45M, 2004ARA&A..42..685Z}.
These are more kinematically distributed then the aforementioned young
moving groups and lack a definite spatial concentration. Accordingly
\citet{2010A&A...512A..37S} adopted a rather large velocity dispersion
(7--10~\kms) combined with photometric distances (and an unrealistically
small uncertainty of 1~pc) to select candidates.

With our CFHT parallaxes, the $UVW$ values of these candidates is
consistent with group membership, as judged by \rchisqUVW\ and using the
updated group definitions from \citet{2014A&A...567A..52K}
(Table~\ref{table:ymg}). However, the large velocity extent of these
older groups means that a given object can overlap more than one group.
Moreover, the reality of these groups and the purity of their membership
lists are still being debated \citep[e.g.][]{2010ApJ...717..617B,
  2012A&A...547A..13T}. In the absence of a quantitative method to
assign membership and the lack of any matches to younger groups, we
simply assign these objects to the general field population and
propagate their original assignments from \citet{2010A&A...512A..37S}.

\subsection{Age Calibration of the Ultracool Gravity Sequence \label{sec:gravity}}

Low-gravity spectral signatures in optical and near-IR spectra of
ultracool dwarfs have been valuable for distilling the rare subset of
young field brown dwarfs from the general field population. Theory
predicts that the radii of substellar objects cease contracting by
$\sim$300~Myr \citep[e.g.][]{bur01}, but accurate predictions about the
time-evolution of gravity-sensitive features leading up to this
cessation are beyond current models. Thus, empirical calibration of the
spectral evolution is needed, based on ultracool dwarfs in stellar
associations/groups and as companions to stars. Such a calibration would
also be prized for diagnosing ages of field ultracool objects
unaffiliated with any stellar group as well as ultracool companions
around host star with uncertain ages.

For late-M and L~dwarfs, evidence to date supports the expected
$\sim$300~Myr timescale disappearance of low-gravity features
\citep[e.g.][]{2008ApJ...689.1295K, 2013ApJ...772...79A,
  2015MNRAS.454.4476S}, with ample evidence that objects at
$\sim$100~Myr still have distinct gravity signatures
\citep[e.g.][]{1999AJ....118.2466M, 2004ApJ...600.1020M,
  2016ApJ...821..120A}.
At optical wavelengths, \citet{2008ApJ...689.1295K} also showed that
low-gravity signatures for late-M dwarfs are distinct between
$\sim$10~Myr and $\sim$100~Myr objects, and \citet{2009AJ....137.3345C}
speculated that a similar distinction for young L~dwarfs was possible
for their $\beta$/$\gamma$ gravity classifications.
At near-IR wavelengths, \citet{2013ApJ...772...79A} used a sparse sample
of YMG members and candidates to estimate \vlg\ and \intg\
classifications occur for $\sim$10--30~Myr and $\sim$50--200~Myr,
respectively. At the youngest ages ($\approx$10~Myr), they found \vlg\
gravity classifications are ubiquitous, but intermediate ages could
display objects with both \vlg\ and \intg\ classifications (see also
\citealp{2013MmSAI..84.1089A}), in accord with evolutionary models (\eg,
Figures~4 and~5 of \citealp{2008ApJ...689.1327S}).  Surprisingly, they
also found that brown dwarfs of the same age and spectral type could
have differing gravity classifications, as demonstrated by the AB~Dor
members CD-35~2722B (L3~\intg) and 2MASS~J0355+1133 (L3~\vlg).
Finally, \citet{2015ApJS..219...33G} found that the strength of
gravity-sensitive near-IR spectral features do generally correlate with
age, but with too much scatter to diagnose individual objects.

Our membership assignments and uniform near-IR gravity classifications
afford an opportunity to revisit this issue.
Figure~\ref{fig:ymg-gravities} shows the distribution of \vlg\ and
\intg\ objects in different young moving groups, as a function of group
age.  While the number of objects and age sampling is still rather
sparse, the data do indicate a decline in \vlg\ objects with age, with
\intg\ objects becoming prevalent by $\approx$150~Myr.
However, the transition from low to high gravity (\ie, \vlg\ to \intg\
to \fldg) does not purely depend on age, nor simply depend on spectral
type, as might be expected.  This is highlighted in the AB~Dor group,
where L~dwarfs are are both \vlg\ (2MASS~J0355+1133 [L3] and
2MASS~J2244+2043 [L6]) and \intg\ (CD-35~2722B [L3] and
DENISJ1425$-$3650 [L4]). The one AB~Dor late-M dwarf has \intg, but one
cannot reach conclusions based on one object.  (See also
\citealp{2016ApJ...821..120A}.)
Finally, the Argus group may be the best illustration of the diverse
spectral behavior, where the one secure group member is \vlg\
(2MASS~J0045+1634 [L2]) while the two other possible members are \fldg\
(2MASS~J0030$-$1450 [L6] and 2MASS~J2351+3010
[L5]).\footnote{\citet{2016ApJ...827...23D} have recently
    determined the individual component dynamical masses for the
    ultracool binary LSPM~J1314+1320AB, which has an integrated-light
    spectrum on the borderline of \fldg\ and \intg.  The measured masses
    and luminosities combined with evolutionary models yield an age
    estimate of 81$\pm$3~Myr, rather young compared to its spectroscopic
    gravity.}

\subsection{Comparison to Theoretical Models \label{sec:models}}

We now examine whether current model atmospheres can reproduce the locus
of low-gravity ultracool dwarfs on the CMD. This has previously been
addressed primarily for planetary-mass companions like HR~8799b
\citep[e.g.,][]{2011ApJ...733...65B, 2012ApJ...754..135M} and
2MASS~J1207$-$3932b \citep[e.g.,][]{2011ApJ...735L..39B}, and our new
large sample of low-gravity objects spans a much wider range in mass and
temperature.  Here we consider two sets of
models. \citet{2008ApJ...689.1327S} provide a model atmosphere grid over
a wide range of \Teff, \logg, and cloud properties based on the cloud
treatment of \citet{2001ApJ...556..872A}, which is parameterized by the
sedimentation efficiency $f_{\rm sed}$.  (Larger values of $f_{\rm sed}$
mean more efficient sedimentation, leading to faster particle growth,
larger particle sizes, and thinner clouds.)  This grid predicts the
range of colors and magnitudes that objects would have, regardless of
the particular prescription for cloud evolution as objects cool. We also
examine BT-Settl evolutionary model isochrones
(\citealp{2015A&A...577A..42B}; Allard et al.~2015, in preparation) that
adopt a particular prescription for the formation and sedimentation of
clouds as objects age \citep{2012RSPTA.370.2765A}.

Figure~\ref{fig:cmd-sm08} shows our \vlg\ CMD sequence, along with the
field population and HR~8799bcd. (HR~8799e is lacking $J$-band
photometry.) We have overplotted \citet{2008ApJ...689.1327S} model
atmospheres of different gravities and cloud thicknesses, from very
thick ($f_{\rm sed} = 1$) and red in $J-K$ to thin ($f_{\rm sed} = 3$)
and bluer. We plot models for $\Teff=1100$--2400\,K at $\logg=4.5$\,dex
and 5.5\,dex, while for the lowest gravity $\logg=4.0$\,dex models are
only complete over $\Teff=1100$--1500\,K.\footnote{Previous
    comparison of these models by \citet{2012ApJ...752...56F} for a
    sample of four low-gravity L~dwarfs seemed to show (1)~the model
    locus overlapping with the data, and (2)~the low-gravity models
    being too red relative to to their data (see their Figure~12).
    Their two results are the opposite of our results, likely due to
    issues with their plotted sample. Two out of the four have
    problematic parallaxes (2MASS~J0032$-$4405 and 2MASS~J0501$-$0010;
    see Sections~\ref{sec:exclude} and~\ref{sec:compare-parallaxes},
    respectively); another is only intermediate gravity
    [2MASS~J2322$-$3133]; and the fourth has either intermediate or
    field gravity (2MASS~J0103+1935; see Section~\ref{sec:exclude}).}

We find that the vast majority of our \vlg\ objects do not overlap with
the \citet{2008ApJ...689.1327S} model atmosphere locus. This is because
the models do not reach sufficiently red colors at a given absolute
magnitude (or equivalently are too faint at a given color). The
discrepancy is especially severe for the earlier type objects, the
late-M and early-L dwarfs.
and also occurs when using $K$-band absolute magnitude for the
  CMD.  While the Saumon \& Marley models are similarly too blue (and/or
  faint) compared to the field population given the expected field
  gravities of $\logg\approx5$ (\eg, also see Figure~10 of
  \citealp{2008ApJ...689.1327S}), the discrepancy is more severe for the
  \vlg\ sample.  Finally, a similar discrepancy with the data is seen
  when considering the \citet{2012ApJ...754..135M} evolutionary models,
  which adopt a gravity dependence to the L/T transition
  (Figure~\ref{fig:cmd-m12}).

The cloud treatment is perhaps the most ready explanation for the
discrepancy on the CMD between the \vlg\ sequence and the models.  Since
Figure~\ref{fig:cmd-sm08} shows that reducing $f_{\rm sed}$ at fixed
temperature and gravity moves the models redder, one natural explanation
is that more extreme $f_{\rm sed}$ values are needed for young objects
than currently modeled, namely even smaller $f_{\rm sed}$ and
  thus smaller particles and thicker clouds.  Another alternative is
unphysically large radii (\ie, lower gravities) to boost the absolute
magnitudes at a given temperature.

A more speculative idea is that the cloud particle size distribution
differs in low-gravity objects compared to higher-gravity field
  objects, the latter of which have served as the notional references
  for model developments.  For instance, \citet{2001ApJ...556..872A}
  assume the size distribution is always log-normal, as detailed
  computations over the range of parameter space are prohibitive.
  In addition, brown dwarf radii decrease by a factor of $\approx$2--3
  from 10--100~Myr \citep[e.g.][]{bur01}, so perhaps the correspondingly
  longer rotation periods of young brown dwarfs change the dynamics of
  condensate formation and sedimentation.  Note that there is a large
  spread in natal brown dwarf rotation rates \citep[e.g., factor of
  $\approx$10;][]{2015ApJ...809L..29S}, which might lead to differences
  in the cloud properties.  These could manifest as differences in
  colors, magnitudes, and spectra for objects of similar mass, age, and
  metallicity (\eg, Sections~\ref{sec:absmags} and~\ref{sec:gravity}).
  However, the fact that the \vlg\ sequence is rather tight and distinct
  from the field sequence on the CMD hints that evolutionary changes
  play a larger role than initial conditions.

Figure~\ref{fig:cmd-btsettl} shows our \vlg\ CMD sequence alongside
BT-Settl isochrones ranging from 1\,Myr to 10\,Gyr. We also show the
confirmed and probable members of young moving groups compared to
isochrones encompassing the age ranges for these groups from
\citet{2015MNRAS.454..593B}. Overall, isochrones tend to have bluer
colors (or equivalently fainter $J$-band absolute magnitudes) than the
empirical sequences, and faint red L~dwarfs are not reproduced at all as
the models begin to turn to the blue due to cloud clearing at relatively
bright magnitudes. Notable exceptions to these trends are $\beta$~Pic~b
(L1), which is bluer than the 20--30\,Myr isochrones (albeit with a
large $J-K$ uncertainty), and the AB~Dor member DENIS~J1425$-$3650
(L4~\intg), which aligns well with the 120--200\,Myr isochrone.

\subsection{The L/T Transition at Low Gravity\label{sec:LTtransition}}

Our sample contains mostly M6--L7 dwarfs and thus well covers the
earlier spectral types of the L/T transition (which we loosely consider
as spectral types L6--T5 here).  In this arena, the mid/late-L~dwarfs
have been the focus on much attention, in light of the relevance to
young directly imaged planets such as 2MASS~J1207$-$3932b and
HR~8799bcde, which have redder and fainter IR fluxes than field objects.
Similarly, studies of benchmark brown dwarfs (those with known
  masses and/or ages) show the L/T transition occurs at lower effective
  temperatures for younger objects \citep[e.g.][]{2006ApJ...651.1166M,
    2009ApJ...699..168D, 2013ApJ...774...55B, 2013ApJ...777L..20L}.
Initial attempts to model the full evolution of the transition from
\citet{2008ApJ...689.1327S} indicated that the absolute magnitudes of
the transition (where the near-IR CMD locus crosses from red L~dwarfs to
blue T~dwarfs) should be brigher for young ages.  However, with the
addition of a gravity-dependent recipe for the transition, chosen to
match the HR~8799 planets, more recent evolutionary models by
\citet{2012ApJ...754..135M} point to fainter absolute magnitudes for the
transition at young ages.

Our combined sample also contains a handful of young objects at later
types and thus begins to probe the impact of reduced surface gravity
across the entire L/T transition.  Figure~\ref{fig:absmag-LT}
  again shows behavior of the absolute magnitude versus spectral type,
  but with a larger range in spectral types than previous plots to
  highlight the L/T objects.  The most straightforward result comes
from the T2.5 Hyades member CFHT-Hy-20, which at all IR bandpasses is
consistent with field objects. This is not surprising since evolutionary
models indicate that brown dwarf contraction has largely completed by
the $\approx$600--800~Myr age of the Hyades. Similarly, the young
($\approx$100--500~Myr) wide L/T companions known to date --- HD~203030B
(L7.5; \citealp{2006ApJ...651.1166M}), HN~Peg~B (T2.5;
\citealp{2007ApJ...654..570L}), and GU~Psc~b (T3.5, which only
  has a BANYAN statistical distance; \citealp{2014ApJ...787....5N}) ---
have positions on the IR color-magnitude diagrams consistent with the
field locus (\eg, Figure~\ref{fig:cmd-comp}).  Finally, the AB~Dor
member SDSS~J1110+0116 (T5.5; \citealp{2015ApJ...808L..20G}) also
coincides with the field locus. Thus, for L/T objects as
  young as $\approx$100~Myr, while the temperatures for a given spectral
  type may be age-dependent, the IR absolute magnitudes across the L/T
transition do not seem to strongly depend on age.

The two latest-type red objects in our sample, WISE~J0206+2640
(L9~(red)) and WISE~J0754+7909 (T2p~(red)), may be illuminating
landmarks for the transition. Their spectra are unusual compared to most
field objects of similar type, especially WISE~J0754+7909 (see details
in Appendix).  And yet their absolute magnitudes are largely consistent
with field objects, with the largest discrepancy being WISE~J0206+2640
having $\approx$0.5~mag fainter magnitudes at $Y$~and $J$~bands compared
to the field. Even given its very peculiar spectrum, WISE~J0754+7909 has
absolute magnitudes and color-magnitude positions in accord with the
field. Thus, while the spectral peculiarities of these two objects may
not be due to low gravity (indeed both have old field kinematics;
Table~\ref{table:changes-membership}), their consistency with field
objects may be circumstantial evidence that the IR magnitudes of the L/T
transition are robust over a range of physical properties.

One interesting question that we can address is whether the $J$-band
flux reversal that is a hallmark of L/T transition binaries in the field
population \citep[e.g.][]{2006astro.ph..5037L} would appear different at
lower gravities. The most extreme such flux reversal binaries known are
2MASS~J1404$-$3159AB \citep[$\Delta{J}=-0.54\pm0.08$\,mag,
$\Delta{K}=1.33\pm0.05$\,mag;][]{2008ApJ...685.1183L} and
SDSS~J1052+4422AB \citep[$\Delta{J}=-0.45\pm0.09$\,mag,
$\Delta{K}=0.52\pm0.05$\,mag;][]{2015ApJ...805...56D}. SDSS~J1052+4422AB
has a very precise evolutionary model-based age derived from its
dynamical masses and luminosities, clearly demonstrating that it is
field age and gravity \citep[$1.11^{+0.17}_{-0.20}$\,Gyr and
$\logg = 5.0$--5.2\,dex;][]{2015ApJ...805...56D}. Now that we have
parallaxes for members of young moving groups that span the L/T
transition, we can imagine hypothetical L/T transition binaries composed
of these coeval objects. For example, if the AB Dor members
WISEP~J0047+6803 (L7~\intg) and SDSS~J1110+0116 (T5.5) were paired
together in a binary, they would have $\Delta{J}=-0.37\pm0.11$\,mag,
$\Delta{K}=2.04\pm0.09$\,mag. If the primary were instead
2MASS~J2244+2043 (L6~\vlg), the flux ratios would be
$\Delta{J}=-0.47\pm0.09$\,mag, $\Delta{K}=1.89\pm0.09$\,mag.

Thus, it seems plausible that young $J$-band flux-reversal binaries have
comparable $J$-band flux ratios as field systems, with the caveat that
the current sample of young objects for such hypothetical pairings is
very small. However, the ``bump'' relative to $K$ band in these binaries
would be somewhat larger, i.e., their $K$-band ratios may be larger, as
low gravity late-L dwarfs are both fainter at $K$ and redder in $J-K$
than field objects of comparable spectral type, while the young mid-T
dwarf SDSS~J111010.01+011613 seems to lie close to or perhaps only
slightly below the field CMD sequence. Although little work has been
done on field $J$-band flux-reversal binaries at bandpasses bluer than
$J$, we note that the $Y$-band bump of a hypothetical AB~Dor L/T
transition binary composed of 2MASS~J2244+2043 and SDSS~J1110+0116 would
be even larger than at $J$-band ($\Delta{Y} = -0.71\pm0.09$\,mag), while
SDSS $z$-band photometry of these sources indicates no significant
$z$-band bump ($\Delta{z} = 0.18\pm0.14$\,mag).

\section{Conclusions \label{sec:conclusions}}

We have completed a large, uniform analysis of the spectrophotometric
properties of young ultracool dwarfs in the field, based on a populous
new sample of high precision distances from the Hawaii Infrared Parallax
Program combined with a uniform analysis of near-IR spectral types and
gravity classifications. Our parallax sample is a factor of 4$\times$
larger than previous studies of young field objects and with
significantly higher parallax precision. Objects with already published
parallaxes from other groups mostly agree with our (typically higher
precision) CFHT results, though we note that at least 6 out of the
15~parallaxes for young objects measured by \citet{2012ApJ...752...56F}
appear to be problematic. A few young objects previously thought to have
odd/extraordinary properties in fact are unremarkable when considering
our new distances.

Combined with previously published measurements (mostly substellar
companions to young stars), we have assembled a sample of 102
ultracool objects with parallaxes that allows us to clearly delineate
the spectrophotometric behavior of young ultracool dwarfs as a function
of spectral type, gravity, and age. The absolute magnitudes of young
field objects as a function of near-IR spectral type differ from those
of old field objects, with young late-M objects being brighter and
mid/late-L dwarfs being fainter in the near-IR ($YJHK$) bandpasses. The
brightness differences relative to the field are more dramatic for bluer
bandpasses.  In contrast, in the \WISE\ mid-IR bandpasses, young targets
generally appear brighter than the field sequence.

As delineated by the \cite{2013ApJ...772...79A} gravity classifications,
the \vlg\ and \intg\ populations have differing spectrophotometric
behavior. The \vlg\ population forms a well-defined sequence, both in
its absolute magnitudes as a function of spectral type and in near-IR
color-magnitude diagrams. For the latter, the \vlg\ sequences now
extended smoothly from late-M to the late-L spectral types, offset from
the field sequence to both redder colors and brighter absolute
magnitudes. Our work reveals that the effect of decreasing surface
gravity from field to \vlg\ follows an evolving pattern with spectral
type, making the IR absolute magnitudes of late-M and early-L dwarfs
brighter, then making the colors of mid-L dwarfs redder while not
affecting their IR absolute magnitudes, and then making the late-L
dwarfs both redder and fainter.

In contrast, the \intg\ population is more heterogeneous. Its absolute
magnitudes as a function of spectral type have more scatter, especially
among the late-M dwarfs. Limited data, based on the presence of lithium,
suggests this scatter may be age-related. The \intg\ sample is also
notable for having two kinematic outliers, with tangential velocities of
$>$99~\kms, far exceeding known young stars. 
Two natural explanations are that some young objects experience early
dynamical interactions that lead to significant acceleration (\eg,
ejection from their birth clouds; \citealp{2001AJ....122..432R}) or old
(high-gravity) objects can display spectra that mimic low gravity. We
note that the absence of any high-velocity \vlg\ objects disfavors the
former possibility. Though the current sample sizes and selection
effects limit firm conclusions, we notionally propose that older
high-gravity objects can show near-IR spectra that appear to be
intermediate in gravity (\intg) between typical young and old field
objects.

Along with these trends, there is notable dispersion among the
population as a whole that is not well-correlated with spectral type or
gravity classification, at least given current ability to diagnose these
two properties. There appears to be (at least) a third physical
parameter that governs the spectral energy distributions of young
ultracool dwarfs. Metallicity is unlikely to be such a parameter, given the
metallicity of local high-mass stars \citep{2012A&A...539A.143N} and YMG
stars \citep{2012MNRAS.427.2905B, 2013ApJ...766....6B,
  2014AJ....148...70M} appear to be basically solar
($\sigma\lesssim0.1$~dex in metallicity). Other possibilities might be
early formation history, rotation, or inclination.

We included some objects in our study with unusual properties possibly
connected to youthfulness. Very red objects with field gravities turn
out to have IR absolute magnitudes similar to ordinary field objects, as
do ultracool field dwarfs with strong H$\alpha$ emission. The
latest-type red \fldg\ objects do overlap the color-magnitude positions
of the young late-L dwarfs, meaning they might provide an interesting
case study for extreme cloud formation. The extreme redness of young
field L~dwarfs is believed to be due to extreme clouds, which in turn
are believed to be the consequence of reduced surface gravity. But we
find two field-gravity objects (2MASS~J2148+4003 and WISE~J2335+4511)
also demonstrate such extreme near-IR properties. {\em Thus, the current
  theoretical framework for the extreme red and faint magnitudes of
  late-L dwarfs is incomplete.}

Young field brown dwarfs are appealing in large part given their
potential utility as analogs for brown dwarf and planetary companions
around stars, given the observational difficulties of studying the
latter.  We find the photometric properties of the free-floating and
companion populations are quite similar. However, the young field
population may inhabit more extreme parameter space, as seen by the
brighter and/or redder locations of the young late-M and L~dwarfs in the
near-IR color magnitude diagrams relative to known young companions.
With the young field population now well-mapped, progress on this issue
will rely on discovery of more substellar companions to determine
whether the two populations do in fact differ in their physical range,
which might reflect differences in formation history, photospheric
processes, and/or composition.
In a similar vein, theoretical models for ultracool objects roughly
coincide with the data, but the observational properties are more
extreme than the models, with the young objects being redder and/or
brighter than the models especially among the young late-M and early-L
dwarfs (contrary to the results of \citealp{2012ApJ...752...56F}).

We have conducted a comprehensive membership analysis of all young field
objects with parallaxes. We establish several new members of young
moving groups, as well as strengthen, revise, or refute previously
published memberships of many others. Altogether, memberships assigned
by BANYAN~II for young field objects appear to be reliable when group
membership probabilities are $\gtrsim$80\% (except for the less-certain
Columba and Argus groups), whereas objects with lower group membership
probabilities appear in fact to be field objects. Similarly, objects
with high membership probabilities in well-established groups have
BANYAN~II kinematic distances that are comparable to actual parallactic
distances ($\lesssim10\%$ errors). For the remaining objects, either
with lower membership probabilities or assignments to the field
population, the BANYAN~II statistical distances have significantly
larger errors than reported. (We caution that our comparison with
BANYAN~II is based on our specific sample of objects, which does not
uniformly sample the full range of spectral types, photometric
properties, and moving groups.) Out of the nine planetary-mass
($\lesssim$13~\Mjup) objects from the BANYAN~II searches
\citep[e.g.][]{2015ApJS..219...33G} that coincide with our sample, we
weaken or eliminate the case for planetary-mass status for eight of
them.

With a robust sample of objects in young moving groups, our work
provides age-calibrated benchmarks for ultracool dwarfs. We find clear
evolution in the mix of IR gravity classifications, with
$\lesssim$20~Myr objects being entirely composed of \vlg\ objects and by
$\sim$100~Myr most objects being \intg. At the older ages, we note that
objects of the same spectral type can have different gravity
classifications even within the same moving group, as first suggested by
Allers \& Liu (2013). Our benchmark sample defines empirical
color-magnitude isochrones for substellar evolution from 10--150~Myr,
and shows that current BT-Settl evolutionary models are not a good match
to the data. The few young age-benchmarks in the L/T transition regime
(late-L to mid-T) suggest that the $J$-band brightening may be slightly
enhanced for young objects compared to the field. But the evolutionary
locus in the near-IR CMD does not appear to be significantly different,
namely young early/mid-T dwarfs, appear to have similar IR absolute
magnitudes as the field sequence and even spectrally peculiar objects.
Overall, the L/T transition on the CMD is relatively narrow in near-IR
absolute magnitudes ($\lesssim$1~mag).

As a by-product of such a large parallax sample, we have identified a
number of individual objects that warrant specialized followup.  Four
objects appear to be very young ($\lesssim$Myr), distant
($\gtrsim$200~pc) brown dwarfs (GJ~3276, 2MASS~J0435$-$1414,
2MASS~J0557$-$1359, and 2MASS~J0619$-$2903), at least one of which has a
well-detected circumstellar disk.  Another candidate disk-bearing object
might be a member of the \bPic\ moving group (2MASS~J1935$-$2846), which
would make it the oldest disk found to date around a brown dwarf.  The
two aforementioned high-velocity \intg\ objects would benefit from high
resolution spectroscopy to assess their gravity and abundances.

Our work enables several future lines of inquiry.  A larger sample of
parallaxes at later types than covered by our sample would help
delineate the L/T transition at low gravities.  A larger sample of
directly imaged substellar companions would help to clarify if the
companion population and young field populations have different
photometric properties, perhaps reflecting different origins.
Memberships for strong YMG candidates with parallaxes should be further
secured by measuring radial velocities.  Expanding the stellar census in
these groups will be an important complementary effort, energized by the
upcoming high-precision astrometric catalogs from {\em Gaia}.  The
culmination of such work will establish the full membership census from
the highest mass stars to the planetary-mass regime, creating a rich
empirical grid of stellar and substellar evolution to test our
understanding of how low-mass self-gravitating objects form and evolve.

\acknowledgments

We greatly appreciate the CFHT staff for their constant observing
support and dedication to delivering the highest-quality data.  We thank
Will Best and Kimberly Aller for assistance with the IRTF near-IR
spectroscopy. We thank Didier Saumon, Mark Marley, Adam Burrows, Bruce
Macintosh, Jonathan Fortney, Michael Line, Caroline Morley, and Ruth
Murray-Clay for insightful comments on this work.  We thank Jonathan
Gagn{\'e} for helpful discussions and for making published data and the
BANYAN~II model tools publicly available.
M. Liu thanks the University of California Observatories; the University
of California, Santa Cruz; and the Other Worlds Laboratory for
supporting a sabbatical to help complete this work.
This publication makes use of data products from the Wide-Field Infrared
Survey Explorer, which is a joint project of the University of
California, Los Angeles, and the Jet Propulsion Laboratory/California
Institute of Technology, funded by the National Aeronautics and Space
Administration.
Our research has employed the 2MASS data products; NASA's Astrophysical
Data System; the SIMBAD database operated at CDS, Strasbourg, France;
the M, L, T, and Y~dwarf compendium housed at {\tt \small
  DwarfArchives.org}; the Spex Prism Spectral Libraries maintained by
Adam Burgasser;
the Montreal Brown Dwarf and Exoplanet Spectral Library maintained by
Jonathan Gagn{\'e};
and the Database of Ultracool Parallaxes maintained by Trent Dupuy.
This research was supported by NSF grants AST-0507833, AST-0909222, and
AST-1518339 awarded to M.\ Liu.
Finally, the authors wish to recognize and acknowledge the very
significant cultural role and reverence that the summit of Mauna Kea has
always had within the indigenous hawaiian community. We are most
fortunate to have the opportunity to conduct observations from this
mountain.  

{\it Facilities:} \facility{CFHT (WIRCAM), IRTF (SpeX), Keck-2 (LGS/NIRC2)}

\appendix


\section{Notes on Individual Objects}

For each object, the optical and near-IR spectral types are listed in
parentheses, along with gravity classifications when available.
  Objects noted in the literature as having low gravity but without a
  formal gravity classification are indicated by ``low-$g$'' (also see
  Table~\ref{table:sample}).

\noindent{\bf 2MASSW~J0030300$-$145033 (L7/L6~\fldg):}
\citet{2015ApJS..219...33G} classify 2MASS~J0030$-$1450 as
L4--L6~$\beta$ based on the published near-IR spectrum of
\citet{2010ApJ...710.1142B}.\footnote{The spectral classification system
  of \citet{2015ApJS..219...33G} uses visual comparison to spectral
  templates to assign near-IR spectral types and gravities, whereas the
  AL13 system uses a combination of visual comparison and spectral
  indices. \citet{2015ApJS..219...33G} use gravity designations of
  $\beta$/$\gamma$/$\delta$ for near-IR spectra (the same notation as
  the \citealp{2009AJ....137.3345C} optical gravity classification),
  which are similar to (but not the same as) the near-IR \intg\ and
  \vlg\ gravity classifications of the AL13 system.} Using the same
spectrum, we assign a spectral type of L6 in the AL13 system to
2MASS~J0030$-$1450 (Table~\ref{table:spectra}). We note that our
spectral type is close to the optical L7 spectral type of this object
\citep{2000AJ....120..447K}. At low spectral resolution, the AL13 system
can not classify the gravities of objects with spectral types later than
L5. We can visually compare the spectrum of 2MASS~J0030$-$1450 to \vlg\
and \intg\ L6 objects (Figure~\ref{fig:L6comp}). 2MASS~J0030$-$1450 has
has deeper FeH and alkali features relative to the L6~\intg\ standard,
indicating that it likely has higher gravity. Overall, the features of
2MASS~J0030$-$1450 better match the field dwarf standard, so we adopt a
spectral type of L6~\fldg. Oddly, our calculated indices agree with
those of \citet{2015ApJS..219...33G}, but the overall classification
does not. We note that \citet{2015ApJS..219...33G} use their own
(template-based) spectral type when determining the AL13 gravity
classification of 2MASS~J0030$-$1450, but the AL13 gravity
classification is designed to incorporate AL13 spectral type
classifications which are derived from both index-based and visual
typing (as described in Section~\ref{sec:typing}). This likely explains
the discrepant gravity classifications.\\
\indent \citet{2015ApJS..219...33G} find that 2MASS~J0030$-$1450 is a
possible ARG member, with 26\% probability and 3\% contamination, based
on a parallactic distance of $27\pm3$~pc from
\citet{2004AJ....127.2948V}.  Their resulting mass estimate is
$10.8^{+0.4}_{-0.6}$~\Mjup, based on an age of 30--50~Myr. While we do
not provide a new parallax for this object, our reclassification of the
spectrum as field gravity would have led
\citeauthor{2015ApJS..219...33G} to remove this object from YMG
consideration. However, we retain it given (1)~the uncertain diversity
of gravity classifications within young groups
(Section~\ref{sec:gravity}) and (2)~the ambiguous age/nature of the
Argus moving group \citep{2015MNRAS.454..593B}. Both of these aspects
suggest a substantially higher mass for this object. \\
\indent Removing the assumption of youth used by
\citeauthor{2015ApJS..219...33G}, our kinematics-only BANYAN~II analysis
gives 16\% ARG, 20\% Y.FLD, and 64\% O.FLD. We note that the
membership lists of Argus differ substantially between BANYAN~II and
\citet{2008hsf2.book..757T}, with the former having a much larger
$UVWXYZ$ extent. Using the Torres \etal\ list, we find \rchisqUVW=1.0
and \rchisqXYZ=4.1, making 2MASS~J0030$-$1450 only a marginal
candidate. A radial velocity of 2.5~\kms\ would place the object at
its minimum velocity distance from the Argus group (based on Torres
\etal) of 4.9~\kms. We consider this a possible Argus member for now,
but a radial velocity for this object and a renewed evalution of the
Argus group are needed to clarify the situation.


\noindent{\bf 2MASSW~J0033239$-$152131 (L4$\beta$/L1~\fldg):} 
\citet{2008ApJ...689.1295K} characterized this as low gravity based on
relatively weak \ion{K}{1}, CaH, and FeH in its optical spectrum
compared to field objects.  However, they caution that its atypically
weak VO compared to other young L~dwarfs (which show enhanced VO
absorption) may indicate that its age is near the $\sim$100~Myr limit
where low-gravity indications are present. \citet{2013ApJ...772...79A}
find that this object has near-IR FeH, VO, \ion{K}{1} and $H$-band
continuum features consistent with field dwarfs and classify
2MASS~J0033$-$1521 as L1~\fldg.  Our parallax shows that the $JHK$
absolute magnitudes of this object are
comparable to other field objects. \\
\indent Based on a parallax-free analysis, \citet{2014ApJ...783..121G}
found this is a modest probability ARG member, with 32\% probability,
22\% contamination, and a statistical distance of
$17.3_{-2.0}^{+1.6}$~pc.  Their associated mass estimate is 9--11~\Mjup,
based on an age of 30--50~Myr. Our new parallactic distance of
$23.4_{-1.0}^{+0.9}$~pc disfavors this possibility, agreeing instead
with their $28.1_{-4.8}^{+4.4}$~pc statistical distance for Y.FLD. If we
assume youthfulness ($<$1~Gyr), our kinematics-only BANYAN~II analysis
reports 99\% Y.FLD. Thus we categorize this as a young field object.  If
we do not assume youthfulness, then BANYAN~II gives membership
probabilities of O.FLD(82\%) and Y.FLD(18\%).

\noindent {\bf 2MASSW~J0058425$-$065123 (L0/L1~\intg):}
\citet{2014ApJ...783..121G} identify this as a possible ABDMG (64\%) or
BPMG (32\%) member, based on a parallax from
\citet{2013AJ....146..161M}. Our kinematics-only BANYAN~II analysis
finds a similar bifurcation in membership probability using the same
published parallax, with ABDMG(75\%) and BPMG(22\%).  
 
\noindent{\bf 2MASS~J01262109+1428057 (L0/L1~\intg):} Based on a
parallax-free analysis, \citet{2014ApJ...783..121G} indicate this is a
low probability BPMG member, with 3\% probability, 77\% contamination,
and a statistical distance of $38.6_{-4.8}^{+6}$~pc. Our new parallactic
distance of $58.5_{-3.8}^{+4.4}$~pc disfavors this possibility, agreeing
instead with their $62.6_{-14}^{+18}$~pc statistical distance for
Y.FLD. Our kinematics-only BANYAN~II analysis reports 97\% Y.FLD with a
small residual probability BPMG (1.3\%) and COL (1.2\%). The minimum
$UVW$ distance from these groups are modest, 4.1~\kms\ and 2.9~\kms\ for
RVs of 4.5~\kms\ and 0.4~\kms, respectively, and the object seems to be
at the spatial periphery of the known members (\rchisqXYZ=5.5 and 2.2,
respectively). Young field membership seems preferred, but an RV
measurement would be useful.


\noindent{\bf WISEP J020625.26+264023.6 (---/L8 (red)):} This was
discovered and classified as ``L9p~(red)'' by
\citet{2011ApJS..197...19K}. We derive a spectral type of L8 based on
visual comparison with spectral standards and the index measurements
given in Table \ref{table:spectra-lateL+T}. The spectrum is slightly
redder than a normal L8, so we assign a spectral type of L8 (red).


\noindent {\bf 2MASS~J02212859$-$6831400 (M8$\beta$/---):}
\citet{2014ApJ...783..121G} find this is a young field object based on a
parallax from \citet{2012ApJ...752...56F}, as does our kinematics-only
BANYAN~II analysis. We note that its position is consistent with the known
BPMG members (\rchisqXYZ=1.3). A radial velocity of 14.2~\kms\ would
yield the smallest possible velocity distance from BPMG, 6.0~\kms, with
a corresponding \rchisqUVW=2.6. Thus, BPMG membership seems marginal at
best given the velocity distance.


\noindent{\bf 2MASSI~J0253597+320637 (M7 low-$g$/M6~\fldg):} Based on a
parallax-free analysis, \citet{2014ApJ...783..121G} indicate this is a
modest probability BPMG member, with 26\% probability, 30\%
contamination and a statistical distance of $35.8_{-2.4}^{+2.8}$~pc.
(Their tables report both a 21\% and 26\% membership probability. We
take the larger value as it is the one reported alongside the
contamination rate.) Their associated mass estimate is 13--15~\Mjup,
based on an age of 12--22~Myr. Our new parallactic distance of
$47.0_{-2.3}^{+2.1}$~pc disfavors this possibility, agreeing instead
with their $60.6_{-11.2}^{+12.8}$~pc statistical distance for Y.FLD.
Using an RV of $-36.3\pm0.8$~\kms\ (E. Shkolnik, priv. comm.), the $UVW$
position is inconsistent with any known young group. Assuming
youthfulness, our kinematics-only BANYAN~II analysis reports $>$99.9\%
Y.FLD. Thus we categorize this as a young field object. \\
\indent \citet{2003AJ....126.2421C} identify this as a young
  object based on its optical spectrum, which upon examination may show
  weak CaH, \ion{K}{1}, and \ion{Na}{1} though not conclusively so.
  \citet{2013ApJ...772...79A} classify the near-IR spectrum as \fldg.
  If we do not assume youthfulness in the BANYAN~II analysis, the result
  is 99.3\% O.FLD probability.


\noindent {\bf 2MASS~J03140344+1603056 (L0/L0~\fldg):}
  \citet{2010A&A...512A..37S} identify this object as a candidate member
  of the Ursa Majoris moving group. Its location on the sky, however,
  does not coincident with the core of UMa nucleus members. As discussed
  in \citet{2013ApJ...772...79A}, its near-IR spectrum does not show any
  signs of low gravity \citep{2013ApJ...772...79A}, which would not be
  unexpected at the age of UMa ($500\pm100$~Myr;
  \citealp{2005PASP..117..911K}).


\noindent{\bf 2MASSI~J0335020+234235 (M8.5 low-$g$/M7~\vlg):} This
object was first proposed to be a $\beta$~Pic member by
\citet{2012ApJ...758...56S}, using a parallax (23.6$\pm$1.3~mas) that
agrees well with ours (21.8$\pm$1.8~mas). \citet{2002AJ....124..519R}
detected \ion{Li}{1} absorption in this object, and
\citet{2009ApJ...699..649S} noted it as a low-gravity accreting source.
This makes it the oldest known accreting brown dwarf.


\noindent {\bf LP~944-20, a.k.a.\ 2MASS~J03393521$-$3525440
  (M9/L0~\fldg):} As discussed in \citet{2014ApJ...783..121G}, BANYAN~II
analysis indicates that ARG membership may be an option for this object,
which is also found from our kinematics-only analysis, but this object
most likely belongs to the Castor moving group, which is not included in
the standard BANYAN~II model.


\noindent {\bf 2MASS~J03552337+1133437 (L5$\gamma$/L3~\vlg):}
  This object has the reddest $J-K$ color in our sample, and both
  \citet{2009AJ....137.3345C} and \citet{2013ApJ...772...79A}
  characterize it as very low gravity. We find that this object is faint
  for its spectral type at $J$~band, but its $H$ and
  $K$-band absolute magnitudes are comparable to the field population.\\
  \indent At a distance of only $9.10\pm0.10$~pc, this object is the
  nearest known young brown dwarf. In addition, it is one of the very
  closest known young moving group members, with only the M2~dwarf
  BD+01~2447 ($7.07\pm0.02$~pc; \citealp{2007A&A...474..653V}) in the
  AB~Dor moving group being closer \citep{2008hsf2.book..757T} and the
  young M~dwarfs AT~Mic (10.7$\pm$0.4~pc) and AU~Mic ($9.91\pm0.10$~pc)
  in the \bPic\ moving group \citep{2001ApJ...562L..87Z} being
  comparable.


\noindent{\bf 2MASS~J04070752+1546457 (L3.5/L3~\fldg)}: This is a strong
($EW\sim60$~\AA) H$\alpha$ emitting L3.5 dwarf
\citep{2008AJ....136.1290R}, classified as \fldg\ by
\citet{2013ApJ...772...79A}. It has normal $J$-band absolute magnitudes
for its near-IR spectral type but slightly brighter in $K$ and $W2$ than
the field sequence (Figure~\ref{fig:absmag-misc}), On both near- and
mid-IR CMDs, this is makes it a clear \fldg\ interloper in the \vlg\
sequence, being unusually red in $J-K$, $H-{\rm W1}$, and
${\rm W1}-{\rm W2}$ given its absolute magnitude. \\
\indent \citet{2015ApJS..219...33G} note some marginal visual signs of
low gravity but do not associate it with any YMG. Our kinematics-only
BANYAN analysis finds 49\% Y.FLD and 40\% O.FLD, perhaps in accord with
possible mild youth seen in the near-IR spectrum.  For these spectral
subclasses, the IR absolute magnitudes of the young and field
populations are very similar, so they provide no age discrimination.  We
assign a generic ``field'' membership to this object, as we cannot
distinguish between the young and old possibilities.


\noindent {\bf GJ~3276, a.k.a.\ 2MASS~J04221413+1530525
  (M6$\gamma$/M6~\vlg):} \citet{2013ApJ...772...79A} note that its
near-IR spectrum appears reddened and that it has a mid-IR excess
indicative of a circumstellar disk, which combined with its sky position
points to possible membership with the Taurus star formation region
$\sim$10\degs\ away. Our CFHT parallax puts this at a distance
($240_{-40}^{+70}$~pc) that is marginally consistent with the young
stars in Taurus \citep[130--160~pc;][]{2009ApJ...698..242T}.  (See
Section~\ref{sec:compare} for a discussion of the highly discrepant
parallax from \citealp{2012ApJ...752...56F}.) The absolute magnitudes of
this object derived from our CFHT parallax are $\approx$5~mag brighter
than the field dwarf sequence and comparable to very young
($\lesssim$3~Myr old) mid-M dwarfs in the Taurus star-forming region
\citep[e.g.][]{2003ApJ...590..348L}.


\noindent{\bf CFHT-Hy-20 (---/T2.5):} The depth of this object's $JH$
and $HK$ water absorption bands are well-matched to the T2~standard
SDSSp~J1254$-$0122, though its $H$-band and $K$-band methane absorption
and $YJ$ water absorption are slightly deeper and would suggest a type
of T2.5 (Figure~\ref{fig:Hy20}). Our visual and index-based types are in
good agreement, and we assign a final near-IR spectral type of
T2.5. This agrees well with the T2 assigned by
\citet{2008A&A...481..661B}, based on their very low resolution ($R=50$)
near-IR spectrum of CFHT-Hy-20 compared to field
T~dwarfs observed with their same instrument.  \\
\indent CFHT-Hy-20 lies 3\degs\ from the center of the Hyades (\eg,
Figure~1 in \citealp{2008A&A...481..661B}). No radial velocity is
available, but it has a minimum $UVW$ distance of 2.5~\kms\ from the
Hyades. We measure a parallactic distance of $32.5_{-1.7}^{+1.5}$~pc,
which agrees well with the {\sl Hipparcos} distance of $46.34\pm0.27$~pc
given the cluster tidal radius of 10~pc \citep{1998A&A...331...81P}. We
therefore confirm this object as a Hyades member.


\noindent{\bf 2MASSI~J0435145$-$141446 (M6$\delta$/M7~\vlg)}: A
parallax-free analysis by \citet{2014ApJ...783..121G} indicates this is
a young field object ($>$99.9\% probability) with a statistical distance
of $10.5\pm1.6$~pc.  We measure a parallactic distance of
$87.7_{-11.8}^{+9.4}$~pc, with a kinematics-only BANYAN~II analysis
still indicating $>$99.9\% Y.FLD.
\citet{2003AJ....126.2421C} noted that the object lies in the direction
of the high-galactic latitude MBM20 (LDN~1642) cloud, which has an
estimated distance of
$124\pm19$~pc \citep{2014ApJ...786...29S}. Our parallax is 1.7$\sigma$
different than the cloud, placing the object in front of or plausibly
within the cloud. The radial velocity for the object (E.~Shkolnik, priv.
comm.) is notably different than the cloud's CO emission at
$\approx$0--1~\kms\ \citep{1991A&A...244..483L, 2003A&A...409..135R},
perhaps due to spectroscopic binarity or unusual kinematics. Our proper
motion measurement of $(\mu_\alpha,\ \mu_\delta) = (-0.6\pm1.9,\
+9.3\pm2.0)$~mas/yr is consistent with that of the \IRAS-excess star
EW~Eri associated with the cloud \citep{1987A&A...181..283S}, which has
a proper motion of ($-5.3\pm5.6,\ +7.1\pm5.5$~mas/yr) from UCAC4
\citep{2013AJ....145...44Z}. \\
\indent \citet{2013ApJ...772...79A} noted that this object's \WISE\
photometry (3.6--22~\micron) shows no sign of a disk, while
\citet{2014A&A...563A.125M} associate this object with a source detected
in far-IR, sub-millimeter, and millimeter data (100--1100~\micron).  The
overall (1.2--1100~\micron) spectral energy distribution points to a
Class~III source, namely one without a substantial circumstellar disk
but still possessing residual material at large separations. A modified
blackbody fit finds a dust color temperature of only $\approx$15~K.
\citet{2014A&A...563A.125M} remark that the object is clearly extended
at 160~\micron\ and overall conclude 2MASS~J0435$-$14 is an embedded
young object, consistent with the very bright absolute magnitudes
derived from our parallax. \\
\indent \citet{2013ApJ...772...79A} also noted that the spectrum of
this M7~\vlg\ dwarf appears to be highly reddened, with a rough estimate
of $A_V\sim7.4$ based on its observed $J-K$ color ($1.93\pm0.04$~mag in
2MASS) compared other young objects of similar spectral type.
Similarly, we also found that young late-M dwarfs in Taurus from
\citet{2009ApJ...697..824A} are a reasonable match to the near-IR
spectrum assuming a modest reddening ($A_V\sim4.5$).
However, we note that the \citet{2003AJ....126.2421C} optical spectrum
shows a normal continuum shape, which is inconsistent with reddening as
is the overall Class~III SED.
Instead, we suggest that the unusual near-IR spectrum and very red $J-K$
color may be manifestations of very low surface gravity, more extreme
than young field objects (Figure~\ref{fig:2m0435+2m0557}).  Low surface
gravity is also consistent with the very bright absolute magnitudes and
its optical gravity classication of $\delta$, the only such object in
our entire sample.  Given the low effective temperature, relatively
large bolometric luminosity ($\approx$0.05~\Lsun, from scaling the
Malinen \etal\ results to our parallax), and very low surface gravity,
this object appears to be a very young ($\lesssim$1~Myr) brown dwarf
with a residual circumstellar disk.


\noindent{\bf SDSS~J044337.60+000205.2 (M9 low-$g$/L0~\vlg):} 
\citet{2007AJ....133..439C} identified this object as having low-gravity
based on optical spectroscopy. It also shows strong signatures of
low-gravity in its near-IR spectrum \citep{2013ApJ...772...79A}.  Based
on a radial velocity of $17.1\pm2.0$\,\kms\ from
\citet{2009ApJ...705.1416R} and a SUPERBLINK proper motion of
($+48, -122$)\,\masyr, \citet{2012AJ....143...80S} reported
SDSS~J0443$+$0002 as a likely member of the AB~Dor moving group. Using
the same RV but a proper motion of ($+36\pm8, -98\pm8$)\,\masyr\ from
\citet{2009AJ....137....1F}, \citet{2014ApJ...783..121G} instead
concluded that SDSS~J0443$+$0002 is a strong $\beta$~Pic moving group
candidate (99.8\%) and predicted a statistical distance of
$25.7_{-2.4}^{+3.2}$\,pc, with \citet{2015ApJS..219...33G} providing a
mass estimate of $20.6^{+5.9}_{-3.8}$~\Mjup. Our work now provides a
parallax distance of $21.1_{-0.5}^{+0.4}$\,pc and a refined proper
motion of ($+53.6\pm1.3,-104.6\pm1.5$)\,\masyr. With this new
information we confirm SDSS~J0443$+$0002 as a member of $\beta$~Pic
(99.8\% probability from our kinematics-only BANYAN~II analysis).


\noindent{\bf 2MASS~J05012406$-$0010452 (L4$\gamma$/L3~\vlg):}
\citet{2015ApJS..219...33G} identify this as a possible COL (49\%) or
CAR (17\%) member, based on a parallactic distance of $13.1\pm0.8$~pc
from \citet{2012ApJ...752...56F}, with an associated mass estimate of
$10.2_{-1.0}^{+0.8}$~\Mjup. As discussed in
Section~\ref{sec:compare-parallaxes} and shown in
Figure~\ref{fig:compare}, their parallax is the among the most
discrepant of all the results with our CFHT measurements
($20.7\pm0.6$~pc) and also disagrees with \citet{2014A&A...568A...6Z}.
Our CFHT distance leads to brighter absolute magnitudes, and our
kinematics-only BANYAN~II analysis gives $>$99.9\% probability of young
field membership.  So our new parallax weakens the case for
2MASS~J0501$-$0010 being a planetary-mass object. \\
\indent We note that the membership lists of Columba and Carina differ
substantially between BANYAN~II and \citet{2008hsf2.book..757T}, with
the BANYAN~II lists having larger $UVWXYZ$ extents. Using our CFHT
parallax and the BANYAN~II membership lists, radial velocities of 19.2
and 18.4~\kms\ would would place the object at minimum velocity
distances of 8.5 and 8.2~\kms\ from the Columba and Carina group cores,
respectively, which is rather far given the dispersions of the known
members. We conclude that 2MASS~J0501$-$0010 is a young field object.


\citet{2007AJ....133..439C} originally identified this as a giant during
a search for nearby ultracool dwarfs, \citet{2009ApJ...699..649S} found
this object is actually a M7.0 dwarf and estimated a photometric
distance of 60~pc assuming an age of 10~Myr based on its strong lithium
absorption, low-gravity features in the optical, and possible H$\alpha$
accretion.  The Allers \& Liu (2013) near-IR classification of \vlg\ is
in accord with this young age.  Its 2MASS $JHK$ colors are consistent
with late-type dwarfs, rather than giants. \\
\indent Our parallax measurement is very marginal ($S/N=2.8$) and places
the object at $\approx$300~pc.  Our result agrees with a similarly poor
parallactic distance by \citet{2012ApJ...758...56S} of
$526\pm277$~pc. Based on the evolutionary models of
\citet{2000ApJ...542..464C}, such a high luminosity implies that this is
a very young ($\sim$1~Myr old) brown dwarf. Similarly, its
  $K$-band absolute magnitude of 4.4$\pm$0.8~mag is $\approx$1.5~mag
  brighter than the individual components of the very young
  ($\sim$1~Myr) M6.5~eclipsing binary 2MASS~J05352184$-$0546085 in Orion
  \citep{2006Natur.440..311S}.  A comparison of the near-IR spectrum of
  2MASS~J0557$-$1359 with the aforementioned 2MASS~J0435$-$1414 shows
  the two have a comparable surface gravities, both more extreme than
  young field objects (Figure~\ref{fig:2m0435+2m0557}). \\
\indent A parallax-free analysis by \citet{2014ApJ...783..121G}
indicated that 2MASS~J0557$-$1359 is a young field member (99.7\%) with
a statistical distance of $45_{-8}^{+12}$~pc. Our parallactic distance
of $290_{-110}^{+80}$~pc suggests that 2MASS~J0557$-$1359 is beyond the
200-pc spatial extent of the BANYAN~II field model.


\noindent{\bf 2MASSI~J0608528$-$275358 (M8.5$\gamma$/L0~\vlg):}
\citet{2010ApJ...715L.165R} proposed this is a \bPic~moving group member
based on their radial velocity measurement ($24.0\pm1.0$\,\kms) and an
ad~hoc photometric distance of $30\pm10$~pc. Based on a parallactic
distance of $31.3\pm3.5$~pc and proper motion of
($+8.9\pm3.5, +10.7\pm3.5$)\,\masyr\ from \citet{2012ApJ...752...56F},
\citet{2013ApJ...762...88M} determined 2MASS~J0608$-$2753 as a
\emph{bona fide} member of BPMG.  In the analysis of
\citet{2014ApJ...783..121G}, 2MASS~J0608$-$2753 was excluded from the
list of \emph{bona fide} members based on their proper motion
significance criterion ($S/N>5\sigma$).  \citet{2014ApJ...783..121G}
then indicated 2MASS~J0608$-$2753 as a peripheral COL candidate, with
4\% probability and 2\% contamination, finding that BPMG membership is
not viable in their model due to the RV measurement. Our new parallactic
distance of $40.0_{-2.2}^{+2.3}$~pc and proper motion of
($+12.4\pm0.6, +7.3\pm0.8$)~\masyr\ strengthens the case for COL
membership, with our kinematics-only BANYAN~II analysis giving a 33\%
membership probability. The BANYAN~II membership list for COL has a
significantly larger $UVWXYZ$ extent than that of
\citet{2008hsf2.book..757T}, but even the latter suggests COL is viable,
with \rchisqUVW=2.7, \rchisqXYZ=1.1, and a velocity distance of
3.2~\kms. Thus Columba membership seems
plausible but not definitive yet. \\
\indent We note that the \citet{2014MNRAS.445.2169M} group membership
suggests that \bPic\ membership is still a marginal option, with
\rchisqUVW=3.2, \rchisqXYZ=1.2, and a velocity distance of 5.2~\kms.


\noindent{\bf 2MASSI~J0619526$-$290359 (M6 low-$g$/M5 low-$g$):} 
\citet{2007AJ....133..439C} identified this object as having low-gravity
in the optical. \citet{2013ApJ...772...79A} classify 2MASS~J0619$-$2903
as M5 and therefore outside the spectral type range of their gravity
classification scheme. They deredden the spectrum ($A_V=6.5$) and note
that it appears low in gravity (Figure~\ref{fig:2m0619}), and that its
mid-IR photometry suggests a circumstellar disk.  A parallax-free
analysis by \citet{2014ApJ...783..121G} indicate this object is a modest
probability COL member, with a 81\% membership, 22\% contamination, a
statistical distance of $55.8_{-6}^{+5.6}$~pc, and a mass of
15--23~\Mjup\ based on an age of 20--40~Myr. We measure a parallactic
distance of $270_{-100}^{+270}$~pc making 2MASS~J0619$-$2903 likely
beyond the 200-pc extent of the BANYAN~II field model. The very bright
absolute magnitudes derived from our CFHT parallax support the notion
that this is a very young low-mass
star.\\
\indent Comparing our new 2015 spectrum with the 2008 one from
  \citet{2013ApJ...772...79A} shows that the object appears to be
  spectrally variable (Figure~\ref{fig:2m0619-2epochs}).  We classify
  the new spectrum as M6~\vlg, slightly later than the published M5
  type.  (The published classification did not include a gravity class,
  since the object was too early-type, but the first-epoch spectrum was
  visually identified as low gravity.)  The individual \htwoo\ spectral
  indices suggest a slightly later type in the second-epoch spectrum
  (Table~\ref{table:spectra}).


\noindent{\bf 2MASS~J07140394+3702459 (M8/M7.5~\intg):}
\citet{2015ApJS..219...33G} classify this as M7.5$\beta$ and assign
this to ARG with 88.9\% probability and 0.6\% contamination rate,
based on a parallax from \citet{2014ApJ...784..156D} and no radial
velocity information.  \citet{2015ApJ...798...73G} estimate a mass of
$20.6_{-1.7}^{+2.5}$~\Mjup\ based on an age of 30--50~Myr. Using the
published parallax and an RV from \citet{2013AJ....146..156D},
our kinematics-only BANYAN~II analysis robustly places the object in the
young field population ($>$99.9\%). In $UVWXYZ$, this object barely
overlaps the spatial positions of the most distant members of Argus
($\rchisqXYZ=4.8$) and is 120~pc from the group core. Also its $UVW$
position is $22.2\pm0.4$~\kms from the mean group velocity, placing it
far beyond the $\approx$1~\kms\ dispersion of the known members
($\rchisqUVW=299$). Thus we categorize this as a young field object
and hence expect a more uncertain (and likely higher) mass than
originally reported.


\noindent {\bf LP~423-31 (M7 low-$g$/M6~\fldg):} This object is
  relatively bright for its spectral type, and on $YJHK$ CMDs is a
  marginal case of a \fldg\ interloper in the \vlg\ sequence.  LP~423-31
  is classified as \fldg\ in the near-IR \citep{2013ApJ...772...79A} but
  has signs of low gravity in its optical spectrum
  \citep{2009ApJ...699..649S}.  Our BANYAN~II analysis gives $>$99.9\%
  Y.FLD or 89\% O.FLD membership depending on whether youthfulness is
  assumed or not, respectively.


\noindent{\bf WISE~J075430.95+790957.8 (---/T2p (red)):}
\citet{2013ApJS..205....6M} discovered and classified this object as
``extremely red,'' noting that its extremely unusual near-IR spectrum
made classification very difficult with available methods. We visually
compared this object to spectral standards, finding a spectral type of
T2 in the $J$-band and T1.5 in the $K$-band. The $H$-band spectrum,
however, is very unusual, with the central continuum plateau resembling
an L8--L9 spectrum but the surrounding water absorption bands being
closer to T2.  The spectral indices in Table~\ref{table:spectra-lateL+T}
indicate a spectral type of T$1.7\pm1.5$, while WISE~J0754+7909 is much
redder than a typical T2. Thus, we assign a spectral type of T2p (red)
to WISE~J0754+7909, with the caveat that it warrants more detailed
examination.


\noindent {\bf 2MASS~J08354256$-$0819237 (L5/---):}
This object is not in our CFHT sample, but it is in the sample of
parallaxes that we plot for the field population
(Sections~\ref{sec:absmags} and~\ref{sec:cmd}). It has a parallax from
\citet{2011AJ....141...54A} and is one of the nearest objects of its
spectral type ($8.5_{-0.7}^{+0.9}$~pc). It is also one of the reddest
with $(J-K)_{\rm MKO}=1.97\pm0.03$\,mag (and $M(J)=13.42\pm21$). Thus,
in our $YJHK$ CMDs it appears as a possible \vlg\ sequence
interloper. To our knowledge, this object has no properties associated
with low gravity or youth. Examination by us of its near-IR SpeX prism
spectrum \citep{2010ApJ...710.1142B} suggests the object's gravity is on
the borderline between \fldg\ and \intg.


\noindent {\bf LP~261-75B, a.k.a.\ 2MASSW~J09510549+35580021 (L6/L6~\fldg):}
\citet{2013ApJ...774...55B} classified the low resolution infrared
spectrum of LP~261-75B as L4.5~\fldg. Using the same spectrum, we
determine a later spectral type of L6 in AL13 system, in agreement with
its optical spectral type of L6 \citep{2000AJ....120..447K}. At low
spectral resolution, the AL13 system can not classify the gravity of an
L6 object, so we use visual comparison. Figure~\ref{fig:L6comp} compares
the spectrum of LP~261-75B to \vlg, \intg, and \fldg\ L6 standards as
well as the aforementioned 2MASS~J0030$-$1450. Overall, the spectrum of
LP~261-75B is very similar to that of 2MASS~J0030$-$1450 but has deeper
FeH and alkali features indicating likely higher gravity. The spectral
features of LP~261-75B are better matched to the \fldg\ template than
the \intg\ template, and thus we classify LP~261-75B as L6~\fldg. \\
\indent Based on a parallax-free analysis and no assumption of youth,
\citet{2015ApJS..219...33G} indicated this as a possible ABDMG member,
with 19\% probability, 31\% contamination, and a statistical distance of
32.5$_{-2.0}^{+1.6}$~pc. While our new parallactic distance for
LP~261-75A (31.6$_{-1.4}^{+1.2}$~pc) agrees well, our kinematics-only
BANYAN~II analysis reports $>$99.9\% Y.FLD membership. The $XYZ$ is
broadly consistent with known AB~Dor members ($\rchisqXYZ=1.9$), but the
$UVW$ is off by $15.2\pm1.1$~\kms,
much larger than the group dispersion ($\rchisqUVW=64$). Thus we
categorize this as a young field object.


\noindent {\bf G~196-3B, a.k.a.\ 2MASS~J10042066+5022596 (L3$\beta$/L3~\vlg):} 
As a well studied companion to a young M3 star that lacks a parallax
measurement, the age of G~196-3B has been estimated to be anywhere from
20--300\,Myr \citep{1998Sci...282.1309R, 2001AJ....121.3235K,
  2009ApJ...699..649S}.  Based on a parallax-free analysis,
\citet{2014ApJ...783..121G} placed this either in ABDMG or CAR,
depending on the adopted radial velocity. Considering our new parallax,
the $UVW$ agrees much better with CAR than ABDMG, but the reverse is
true for $XYZ$. Our kinematics-only BANYAN~II analysis favors young
field, which we adopt as the outcome.


\noindent{\bf 2MASS~J10220489+0200477 (M9$\beta$/M9~\fldg):} Based on a
low S/N parallax of $38\pm16$~pc from \citet{2012ApJ...752...56F} and a
radial velocity of $-7.9\pm4.8$~\kms\ derived from the literature,
\citet{2015ApJS..219...33G} indicated this is a peripheral ABDMG
candidate, with a 3\% probability and 6\% contamination. Using our new
parallactic distance of $28.9_{-1.8}^{+1.6}$~pc and a radial velocity of
$-8.4\pm5.0$~\kms\ from \citet{2008AJ....135..785W}, our kinematics-only
BANYAN~II analysis reports $>$99.9\% Y.FLD membership assuming
youthfulness or $>$99.9\% O.FLD membership without this
  assumption.


\noindent {\bf SDSS~J102552.43+321234.0 (---/L7~\fldg):} Upon its
discovery, \citet{2006AJ....131.2722C} reported that SDSS~J1025+3212 has
an unusual L7.5 spectrum, with scattered spectral indices indicating a
spectral type of L4.5--T2 and hints of variability based on two epochs
of spectra.  \citet{2008ApJ...689.1295K} listed SDSS~J1025+3212 as a
potential young object on the basis of weak \ion{K}{1} and H$_2$O
features. AL13 determined a spectral type of L7 for this object, but
could not reliably determine its gravity, as the AL13 system only
provides gravity classification for low-resolution spectra with spectral
types earlier than L6. Recently, two bona-fide young L7~objects have
been discovered, PSO~J318.5$-$22 \citep[L7~\vlg;][]{2013ApJ...777L..20L}
and WISE~J0047+6803 \citep[L7~\intg;][]{2015ApJ...799..203G}, which
allows for a visual assessment of the gravity of SDSS~J1025+3212
(Figure~\ref{fig:L7comp}). We do not find evidence of low gravity, as
the FeH absorption bands, alkali lines, and $H$-band continuum shape are
all consistent with \fldg. The $H$-cont index from AL13 also yields a
score of ``0'', which corresponds to \fldg. We assign a
spectral type of L7~\fldg. \\
\indent Based on a parallax-free analysis and adopting the
\citet{2008ApJ...689.1295K} interpretation of the near-IR spectrum as
being low gravity, \citet{2014ApJ...783..121G} assign young field
membership ($>$99.9\%) with a statistical distance of
$19.7_{-2.4}^{+2.8}$~pc. Using our \fldg\ gravity classification and
CFHT parallax ($26.8_{-0.8}^{+0.9}$~pc),
a kinematics-only BANYAN~II analysis finds a probability of 99.8\% of
belonging to the old field population. (If youth is assumed based on the
optical spectrum, then the result is $>$99.9\% membership for young
field.)  Its $UVW$ position is far from any known young group. We also
note that our parallax gives absolute magnitudes for SDSS~J1025+3212
(e.g., $M_J=14.75\pm0.09$\,mag) that are more consistent with field
objects of the same spectral type ($M_J=14.7$\,mag;
\citealp{2012ApJS..201...19D}) than with the low-gravity L7 dwarfs
PSO~J318.5$-$22 ($M_J=15.42\pm0.09$\,mag) and WISE~J0047+6803
($M_J=15.07\pm0.08$\,mag) Thus, this appears to be an old field object.


\noindent {\bf 2MASSW~J1139511$-$315921, a.k.a.\ TWA~26 (M9 low-$g$/M9~\vlg):}
\citet{2002ApJ...575..484G} identified this young brown dwarf as a
candidate TWA member based on its proximity to previously known members
and its low-gravity optical spectral features.
\citet{2003ApJ...593L.109M} found a radial velocity consistent with the
other TWA members, while \citet{2003MNRAS.342..837R} suggested that this
object is not a TWA member based on the (low-quality) proper motion
available at the time.  As discussed in
  Section~\ref{sec:compare}, our parallactic distance of
  $42.2^{+2.6}_{-3.0}$~pc agrees well with parallaxes from
  \citet{2013ApJ...762..118W} and \citet{2014A&A...563A.121D} as well as
  the \citet{2005ApJ...634.1385M} kinematic distance.  (All of these
  disagree with the \citealp{2012ApJ...752...56F} parallax.)  Thus, our
  membership analysis is also in accord with this previous work, showing
  unambigous TWA membership.  It is clearly overluminous in $JHK$
relative to field objects, consistent with its young age.


\noindent {\bf 2MASS~J13153094$-$2649513AB (L5.5/L6 \fldg)}:
  2MASS~J1315$-$2649 was first identified as a mid-L dwarf with strong,
  variable H$\alpha$ emission by \citet{2002ApJ...564L..89H}.  Based on
  a high-quality composite optical spectrum, \citet{2008ApJ...689.1295K}
  assigned a spectral type of L5.5.  \citet{2011ApJ...739...49B}
  resolved 2MASS~J1315$-$2649 into binary and determined component
  spectral types of L3.5$\pm$2.5 and T7$\pm$0.6 from $H$-band spectra.
  Using the composite near-IR spectrum of \citet{2011ApJ...739...49B},
  we classify 2MASS~J1315$-$2649AB as L6~\fldg\ on the
  \citet{2013ApJ...772...79A} system.  The composite near-IR spectrum
  shows subtle peculiarities, but these do not appear low-gravity in
  nature, and the $H$-cont index is consistent with field gravity.  As
  discussed by \citeauthor{2011ApJ...739...49B}, the peculiar spectrum
  is well-explained by unresolved L+T binarity.

\noindent {\bf 2MASS~J13313310+3407583 (L0/L0~\fldg):} Based on
  comparison to field dwarf standards, \citet{2010ApJS..190..100K}
  classify 2MASS~J1331+3407 as L0 from its optical spectrum and an
  ``L1~pec~(red)'' based on its near-IR spectrum.  Using the near-IR
  spectrum of \citet{2010ApJS..190..100K}, we classify 2MASS~J1331+3407
  as L0~\fldg.  The $(J-\Ks)_{\rm 2MASS}$ color of 2MASS~J1331+3407
  (1.448$\pm$0.035~mag) is slightly redder than the $J-\Ks$ color of L0
  field dwarfs \citep[median color~=~1.24~mag,
  RMS~=~0.18~mag;][]{2010AJ....139.1808S}.
  Likewise, the near-IR spectrum of 2MASS~J1331+3407 is slightly redder
  than a field L0 standard (\eg, Figure~10 of
  \citeauthor{2013ApJ...772...79A}), but otherwise shows no spectral
  peculiarities.


\noindent{\bf 2MASS~J14112131$-$2119503 (M9/M8 \intg):} A
kinematics-only BANYAN analysis using our new CFHT parallax favors the
young field population, as does the full BANYAN~II analysis by
\citet{2014ApJ...783..121G} done with a radial velocity from
\citet{2009ApJ...705.1416R} but without a parallax. We note that the
$UVWXYZ$ position of this object is close to TWA, with $\rchisqUVW=2.0$
and $\rchisqXYZ=2.0$. However, the known members of TWA have a
filamentary $XYZ$ distribution \citep[e.g.][]{2013ApJ...762..118W}, and
2MASS~J1411$-$2119 is not well aligned with this. Thus we favor
assignment to the young field population.


\noindent {\bf 2MASSI~J1615425+495321 (L4$\gamma$/L3~\vlg):}
\citet{2007AJ....133..439C} first noted low gravity features in the
optical spectrum of 2MASS~J1615+4953. \citet{2008ApJ...689.1295K}
tentatively concluded this object is low gravity, but caution that the
modest S/N of their optical spectra cannot discount the possibility that
the object is a very dusty, old field L~dwarf.
\citet{2013AJ....145....2F} assigned an optical gravity classification
of $\gamma$ to this L4 dwarf. \citet{2013ApJ...772...79A} determined
that the near-IR spectrum of this source is indicative of a very low
surface gravity (\vlg\ classification), not merely being dusty. Its
$J$-band absolute magnitude is notably fainter than field objects,
supporting its low-gravity classification.


\noindent {\bf DENIS~J170548.3$-$051645 (L0.5/L1~\fldg):}
This was identified as a candidate member of the Ursa Major moving
group by \citet{2010A&A...512A..37S},
which agrees with our kinematic determination.
This object is far removed on the sky from the core Ursa Major
members, being one of many group candidates from
\citet{2001MNRAS.328...45M} that are widely distributed.
\citet{2013ApJ...772...79A} find this object's near-IR spectrum is
consistent in gravity with field objects. \citet{2003AJ....125.1980K}
and \citet{2015ApJ...813...58J} estimate ages for the Ursa Major core
members of $500\pm100$~Myr and $414\pm23$~Myr, respectively, consistent
with the spectroscopic gravity.
The absolute magnitudes of DENIS~J1705$-$0516 are also consistent with
field objects, so its spectrophotometry and kinematics are in
accord. \\
\indent \citet{2006AJ....132..891R} identified a possible companion to
this object in \HST/NICMOS imaging from 2005~June~24~UT at a separation
of 1$\farcs$36, PA of 355\degree, and $\Delta{J}\approx4.1$\,mag. At a
recent CFHT epoch with good seeing (JD~2455988.17) we measured a source
with $\Delta{J}=3.82\pm0.06$\,mag at separation of
$2\farcs445\pm0\farcs007$ and PA of $332\fdg71\pm0\fdg15$. Accounting
for the proper motion of DENIS~J1705$-$0516, this source would have been
at a separation of $1\farcs413\pm0\farcs011$ and PA of
$346\fdg1\pm0\fdg4$ at the NICMOS epoch. (Note that our quoted errors do
not account for the uncertainty in the proper motion of the faint
source.) We therefore conclude that this possible companion is a
background object, which is seemingly at odds with a recent note by
\citet{2014AJ....147...94D} that they observe orbital motion in their
astrometry of DENIS~J1705$-$0516.


\noindent {\bf 2MASS~J18212815+1414010 (L4.5/L5p~\fldg):}
\citet{2008ApJ...686..528L} flag this as a potential young object,
based on its slightly red color (2MASS $J-\Ks=1.78\pm0.05$\,mag),
peculiar near-IR spectrum (including somewhat triangular $H$-band peak
and strong CO), and low tangential velocity of 10\,\kms\ (based on a
photometric distance of $9.8\pm1.3$\,pc). Using the low resolution
near-IR spectrum of \citet{2008ApJ...686..528L}, we determine a
near-IR spectral type of L5~\fldg\ and similarly assign a peculiar
designation to denote its red color and triangular $H$-band continuum
shape. \\
\indent Our astrometry gives a tangential velocity of
$15.20\pm0.17$\,\kms\ and a parallactic distance of
$9.56\pm0.09$\,pc. The photometric and parallactic distances agree well
because 2MASS~J1821+1414 has normal absolute magnitudes compared to the
field sequence. Also its position in the near-IR CMDs resides among the
field L~dwarfs. Similar to 2MASS~J0314+1603 and DENIS~J1705$-$0516, this
object has kinematics consistent with the Ursa Major moving group
but resides far from the core members on the sky. The
$\approx$400--500~Myr age of the group would be consistent with the
near-IR gravity classification. Thus, we conclude that 2MASS~J1821+1414
is likely to be an older dusty object, as opposed to a young one as
suggested by \citet{2008ApJ...686..528L}. This is consistent with the
notion that objects with triangular $H$-band spectra
are not necessarily young, as found by \citet{2013ApJ...772...79A}. \\
\indent Based on a parallax-free analysis using an radial velocity of
$9.8\pm0.8$\,\kms\ (\citealp{2010ApJ...723..684B};
\citealp{2010ApJS..190..100K}) and an assumption of low gravity,
\citet{2014ApJ...783..121G} assign this young field membership
($>$99.9\%) with a statistical distance of $13.3\pm2.8$\,pc. With our
new CFHT parallax and no assumption of youth, a kinematics-only
BANYAN~II analysis reports 82\% old field population and 18\% young
field population. Its $UVW$ places it far from any known young
group. Altogether we assign it to the old field population.


\noindent {\bf 2MASS~J19355595$-$2846343 (M9 low-$g$/M9~\vlg):}
A parallax-free BANYAN~II analysis of 2MASS J1935$-$2846 by
\citet{2014ApJ...783..121G} found a slim possibility for BPMG membership
(9\%), with a statistical distance of $57\pm5$\,pc and favoring the
hypothesis that the system is a binary. (Keck laser guide star adaptive
optics imaging by us finds the source to be single at 0.07\arcsec\
resolution at $K$~band.) For young field membership, they derive a
statistical distance of $62_{-10}^{+12}$\,pc. Our new CFHT parallax
($70.4_{-5.5}^{+6.5}$~pc) yields a smaller BPMG possibility (1.1\%) from
the kinematics-only BANYAN~II analysis, but shows that the $XYZ$
position of this object is quite compatible with known BPMG members
($\rchisqXYZ=2.0$). This BPMG probability increases modestly if the
radial velocity for the system (E.~Shkolnik 2015, priv.\ comm.) is
included in the analysis but is still small. We note that this object
lies beyond the $\approx$50-pc limiting distance of the original
\citet{2001ApJ...562L..87Z} search that identified the \bPic\ moving
group, and it is also farther than 42 out of the 48~members identified
by \citet{2008hsf2.book..757T}.  Thus the relatively low BPMG
probability assigned by the BANYAN~II model may simply reflect the
incomplete membership census for more distant stars. A radial velocity
of $-$9.3\,\kms\ would bring the object to 3.2\,\kms\ from the group
core. Altogether, we consider BPMG membership to be unlikely but still
viable --- further investigation is warranted.  More precise $UVWXYZ$
data for this object as well as a more expansive
census of the \bPic\ stellar members would help refine the membership. \\
\indent \citet{2010A&A...517A..53M} noted that 2MASS~J1935$-$2846 is
near the R~CrA star forming region on the sky, but our proper motion of
$(+27.3\pm0.9, -61.6\pm1.1)$~\masyr\ is inconsistent with the mean of
known members, $(+5.5, -27.0)$\,\masyr\ \citep{2000A&AS..146..323N}, and
our parallax distance of $70.4_{-5.5}^{+6.5}$~pc is much closer than the
region's best distance estimate of 130\,pc
\citep{2008hsf2.book..735N}. \\
\indent 2MASS~J1935$-$2846 also possesses a mid-IR excess, standing out
in Figure~\ref{fig:cmd2} as being both bright in $W2$ and red in $W1-W2$
. Only two other objects are redder at this absolute magnitude:
2MASS~J1207$-$3932 (TWA~27) and SSSPM~J1102$-$3431 (TWA~28). Both of
these are noted by \citet{2008ApJ...681.1584R} to have excess thermal
emission as short as 5\,\micron\ due to circumstellar disks that likely
explains their red colors on this plot. The similar mid-infrared
photometry of 2MASS~J1935$-$2846 suggests that it too possesses
circumstellar material. 
We note that the mid-IR excess of 2MASS~J1935$-$2846 lends
circumstantial support to the possibility of $\beta$~Pic membership,
since an older age would make a circumstellar disk more implausible.  If
confirmed, this would join the \bPic\ member 2MASS~J0335+2342 as the
oldest disks found to date around brown dwarfs.


\noindent{\bf 2MASS~J20135152$-$2806020 (M9 low-$g$/L0~\vlg):} 
\citet{2008AJ....136.1290R} classify this object as a low-gravity
M9~dwarf, while \citet{2008ApJ...689.1295K} indicate an optical
spectral type of M8--M9~III. Our parallax detection and the resulting
absolute magnitudes show that it is a dwarf. Near-IR spectra of the
object from \citet{2013ApJ...772...79A} also show that it is a
low-gravity dwarf and not a giant. \\
\indent A parallax-free BANYAN~II analysis by
\citet{2014ApJ...783..121G} identified this as a low probability BPMG
member, with a 43\% probability, a 71\% contamination rate, and a
statistical distance of $44_{-4}^{+5}$\,pc, as well as an estimated mass
of $15.7_{-0.6}^{+1.5}$\,\Mjup\ from \citet{2015ApJS..219...33G}.  Our
new CFHT parallax ($48\pm3$\,pc) also suggests BPMG membership (70\%
probability) and places the object comfortably among the known members
($\rchisqXYZ=1.0$). The velocity offset between 2MASS~J2013$-$2806 and
the group would be only 2.2~\kms\ for a radial velocity of $-$8.1\,\kms.
Like 2MASS~J1935$-$2806, we thus consider this a promising \bPic\
member, with radial velocity followup and a more complete group census
needed.  
%


\noindent {\bf 2MASS~J21403907+3655563 (---/M8p~\fldg):}
\citet{2010ApJS..190..100K} originally typed the near-IR spectrum of
2MASS~J2140+3655 as ``M8 pec'' and noted that it appears to be redder
than vB~10, the dwarf field M8 standard. We obtained a higher S/N
spectrum of 2MASS~J2140+3655 suitable for gravity classification on
2015~July~23~UT (Figure~\ref{fig:2mass2140}). The $H$-cont index of
Allers \& Liu (2013) indicates intermediate gravity, but the FeH$_z$ and
\ion{K}{1} indices are consistent with \fldg. We type 2MASS~J2140+3655
as M8~\fldg\ but also assign a ``p'' designation to its spectral
classification, to denote the peculiarity of having a triangular
$H$-band continuum shape while lacking any other low-gravity
indications.  Its IR absolute magnitudes are consistent with field
objects. \\
\indent In contrast to \citeauthor{2010ApJS..190..100K}, our new
spectrum does not appear redder than vB~10 but instead is slightly
bluer. To check for possible spectral variability, we obtained an
additional spectrum of 2MASS~J2140+3655 on 2015~September~25~UT
(Figure~\ref{fig:2mass2140-compare}). Our two spectra from 2015 are
consistent with each other, making the \citet{2010ApJS..190..100K}
spectrum the discrepant one. In addition, the 2MASS $J-\Ks$ color of
this object ($0.94\pm0.13$\,mag) agrees with the synthetic 2MASS $J-\Ks$
colors from our spectra (1.01\,mag and 0.93\,mag for our July and
September spectra, respectively) and is bluer than vB~10
($1.14\pm0.03$\,mag). Thus, 2MASS~J2140+3655 does not appear to be
systematically red for its spectral type. Assessing its potential
variability would benefit from more observations.


\noindent {\bf 2MASS~J21481633+4003594 (L6/L6~\fldg):}
\citet{2008ApJ...686..528L} discovered this object, noting that its red
color, peculiar near-IR spectrum, and field-gravity optical spectrum
indicate an unusually dusty photosphere. From their measured proper
motion ($1\farcs33\pm0\farcs24$, $75\fdg6\pm0\fdg3$), and estimated
distance of $9.9\pm1.3$\,pc, they estimated $\Vtan = 62$\,\kms\ and
suggested that such a value implied this object is not young but rather
possibily high metallicity.
Our proper motion of ($0\farcs9041\pm0\farcs0018$, $59\fdg63\pm0\fdg13$)
is 50\% (1.8$\sigma$) smaller than theirs, and our parallax distance of
$8.03^{+0.11}_{-0.12}$\,pc is 10\% (0.6$\sigma$) smaller. (The
\citealp{2008ApJ...686..528L} photometric distance turns out to be an
overestimate, because this object is faint at $J$~band compared to field
objects.) Thus, we find a significantly smaller
$\Vtan = 34.4\pm0.5$\,\kms, which is in fact consistent with the high
end of tangential velocities for our low-gravity sample
(Section~\ref{sec:vtan}). \\
\indent A parallax-free BANYAN~II analysis by
\citet{2014ApJ...783..121G} identified this as a modest ARG candidate,
with a probability of 48\%, a contamination rate of 37\%, a statistical
distance of $4.9_{-0.4}^{+0.4}$~pc, and an estimated mass of 6--7~\Mjup\
based on an age of 30--50~Myr. They also derived a statistical distance
of $8.1_{-1.2}^{+0.8}$~pc, under the assumption that the object is
young.  However, \citet{2013ApJ...772...79A} classify the gravity as
\fldg.  Our new CFHT parallactic distance ($8.03_{-0.08}^{+0.11}$~pc)
greatly favors field membership, 67\% young field and 33\% old field
based on no prior assumption about the object's youth.  The $UVW$ are
incompatible with any known young group. Thus, 2MASS~J2148+4003 might
have an intermediate age, old enough that spectroscopic signatures of
gravity have receded but whose kinematics are still indicative of
youth. Its mass is more uncertain and likely higher than estimated by
\citeauthor{2014ApJ...783..121G}


\noindent{\bf SIMP~J215434.5$-$105530.8 (---/L4~\intg):} Based on a
parallax-free analysis, \citet{2014ApJ...792L..17G} identify this as a
ARG candidate, with \citet{2015ApJ...798...73G} reporting a revised
59\% membership probability, 34\% contamination rate, a statistical
distance of $22.5\pm2.4$~pc, and an estimated mass of
$10.3_{-0.3}^{+0.7}$~\Mjup\ based on an age of 30--50~Myr.
We had independently identified this object as a low-gravity L~dwarf
during a search for nearby ultracool dwarfs using Pan-STARRS1
\citep[e.g.][]{2013ApJ...777L..20L} and had begun parallax
observations prior to its published discovery.
Using our resulting CFHT parallax ($30.7_{-0.9}^{+1.0}$~pc), a
kinematics-only BANYAN~II analysis leads to ARG membership probability
of 23\% and a spatial position that is just beyond the known members
($\rchisqXYZ=4.6$). Its smallest velocity offset from ARG would be
6.0~\kms\ for a radial velocity of $-$14.5~\kms, using the
\citet{2008hsf2.book..757T} group properties. Radial velocity is needed
for better assessment, but given the available data we favor young field
membership and thus likely a higher mass than originally estimated.


\noindent{\bf DENIS~J220002.0$-$303832AB (M9/M9~\fldg\ and L0/L0~\fldg):}
Resolved photometry and spectra are available for both components from
\citet{2006AJ....131.1007B}. \citet{2013ApJ...772...79A} analyzed
only the A~component, assigning M9~\fldg. We find the B~component is
L0~\fldg. \citet{2008AJ....136.1290R} report component optical spectral
types of M9 and L0, respectively. \\
\indent In our CFHT images we typically cleanly resolve this $1\farcs09$
binary.  Any images in which two sources were not detected by SExtractor
were not used in our astrometric analysis.  We report the individual
solutions in Table~\ref{table:plx}.  The individual parallaxes are
$41\pm4$\,mas and $35\pm4$\,mas, among the worst precision of our sample
due to the difficulty in obtaining centroids for such a close pair.  We
adopt a common parallax of $38\pm4$\,mas for the system.


\noindent {\bf 2MASSW~J2208136+292121 (L3$\gamma$/L3~\vlg):}
\citet{2008ApJ...689.1295K} estimated an age of $\sim$100~Myr for this
object, based on the similarity of its optical spectrum to the young
L-type companion G~196-3B, whose primary star has an estimated age of
20--300~Myr.\\
\indent This object is the only case where something classified as \vlg\
lies closer to the field CMD sequence than the \vlg\ CMD sequence,
although given the photometric and parallax uncertainties this is only a
marginally significant discrepancy ($\approx$2$\sigma$). Given a sample
of 28 \vlg\ objects on our CMD plots, such a $\approx$2$\sigma$ outlier
is not unexpected. In fact, if we use the spectrum of this object to
synthesize its $(J-\Ks)_{\rm 2MASS}$ color, we get a value 0.12\,mag
redder than the catalog value of $1.65\pm0.11$\,mag, which would reduce
the discrepancy to an $\approx$1$\sigma$. Therefore, we determine that
this object is not unusual for its gravity classification. \\
\indent Based on a parallax-free analysis, \citet{2014ApJ...792L..17G}
identify this as a modest BPMG candidate, with 10\% membership
probability,\footnote{Their text and tables report both a 7\% and 10\%
  probability. We take the larger value as it is the one reported
  alongside the contamination rate.} a 54\% contamination rate, and a
statistical distance of $35.4_{-3.2}^{+3.6}$~pc, along with an estimated
mass of $12.9_{-0.1}^{+0.3}$~\Mjup\ from \citet{2015ApJS..219...33G}
based on an age of 20--26~Myr. The young field hypothesis gives a
statistical distance of $55_{-10}^{+9}$~pc.  Using our new CFHT parallax
($39.8_{-2.7}^{+2.4}$~pc), a kinematics-only BANYAN~II analysis leads to
a BPMG membership probability of 18\% and a spatial position that is at
the outskirts of the known members ($\rchisqXYZ=2.4$).
A radial velocity of $-$12.4~\kms\ would give this object an offset of
only 0.7~\kms\ from the group core.
We consider this a promising \bPic\ member but not definitive. Similar
to 2MASS~J1935$-$2846 and 2MASS~J2013$-$2806, a radial velocity
measurement and a more complete group census are needed.

  
\noindent {\bf 2MASS~J22134491$-$2136079 (L0$\gamma$/L0~\vlg):}
\citet{2008ApJ...689.1295K} estimate an age significantly younger than
the Pleiades ($\approx$125~Myr), based on this object's spectral
similarity to the very young field L~dwarf 2MASS~J01415823$-$4633574,
for which they estimate an age of 1--50~Myr. They also noted the
proximity of this object on the sky to members of the \bPic\ moving
group.  Based on a parallax-free analysis, \citet{2014ApJ...783..121G}
identify this as a marginal BPMG candidate, with 3\% membership
probability, a 80\% contamination rate, and a statistical distance of
$45\pm3.6$~pc, along with an estimated mass of $13.5\pm0.3$~\Mjup\
from \citet{2015ApJS..219...33G} based on an age of 20--26~Myr. Their
young field hypothesis gives a statistical distance of
$63_{-8}^{+7}$~pc. \\
\indent Using our new CFHT parallax ($47.8_{-4.0}^{+4.8}$~pc), its
spatial position is at the outskirts of the known \bPic\ members
($\rchisqXYZ=3.1$), and a kinematics-only BANYAN~II analysis gives 88\%
probability of young field. Its smallest possible velocity offset from
BPMG would be a relatively large value of 6.0~\kms, for a radial
velocity of $-$3.1~\kms.
Altogether, we favor young field membership and thus the mass is more
uncertain and likely to be higher than reported by
\citeauthor{2015ApJS..219...33G}


\noindent {\bf 2MASSW~J2224438$-$015852 (L4.5/L3p~\fldg):}
Using the low resolution near-IR spectrum of \citep{2010ApJ...710.1142B},
we classify 2MASS~J2224$-$0158 as L3~\fldg. \citet{2005ApJ...623.1115C}
first noted that 2MASS~J2224$-$0158 has abnormally red colors for its
spectral type ($(J-\Ks)_{\rm 2MASS}=2.03\pm0.04$\,mag), thus we assign a ``p''
designation to denote this spectral peculiarity.


\noindent {\bf 2MASSW~J2244316+204343 (L6.5/L6~\vlg):} This object
serves as the archetype for extremely red L~dwarfs, with an ordinary
optical spectrum but a very peculiar near-IR one
\citep{2000AJ....120..447K, 2003ApJ...596..561M}. The very red near-IR
colors, weak FeH and alkali features in the $J$-band, and triangular
$H$-band spectra indicate this a low-gravity object
\citep{2013ApJ...772...79A} as opposed to an ordinary
(high-gravity) object with extreme cloud properties.
\citet{2008ApJ...689.1295K} suggest that the bland optical spectrum
arises from a fortuitous degeneracy of photospheric chemistry, whereby
the weakened alkali lines and stronger oxides due to reduced surface
gravity act to mimic an earlier type L~dwarf of higher gravity. As
noted in Section~\ref{sec:cmd}, its near-IR absolute magnitudes are
among the faintest in our sample (exceeded only by SDSS~J2249+0044B),
and on the CMD it appears to be an extension of the L~dwarf sequence
towards the very red planetary-mass objects 2MASS~J1207$-$3932b and
HR~8799b.  


\noindent {\bf SDSS~J224953.46+004404.6AB (---/L3~\intg\ and ---/L5~\intg):}
\citet{2010ApJ...715..561A} identified this as possible young object by
examining the near-IR spectrum from \citet{2004ApJ...607..499N} as well
as their own low-resolution spectrum. They then resolved this object
into a 0.32\arcsec\ binary with component spectral types of L3 and L5.
The agrees with the integrated-light optical spectrum typed as L3 by
\citet{2002AJ....123.3409H}. \\
\indent The secondary component of the binary has the faintest near-IR
absolute magnitudes of our entire
sample. \citeauthor{2010ApJ...715..561A}\ estimated a photometric distance
of $54\pm16$~pc by comparing to other young objects with distances,
which was also consistent with the photometric distance to a
wide-separation early-M dwarf comoving with the binary.  We find that
the object is closer, with a parallactic distance of
$39.2^{+2.3}_{-2.1}$\,pc. \\
\indent 
The closer parallax distance also means that SDSS~J2249+0044A and B are
less luminous than assumed by \citet{2010ApJ...715..561A}. As a result,
the estimated masses of the components are also smaller. For possible
ages of 20--300~Myr, we estimate masses of 0.007--0.027\,\Msun\ and
0.005--0.019\,\Msun\ for the A and B components, respectively, based on
the \citet{2008ApJ...689.1327S} $f_{sed}=2$ models.  

\noindent{\bf WISE~J233527.07+451140.9 (---/L7p~\fldg):}
WISE~J2335+4511 was discovered and classified as ``L9 pec (v.red)'' by
\citet{2013PASP..125..809T}, based on a fairly low S/N spectrum. Using
our higher S/N spectrum (Figure~\ref{fig:L7comp}), we determine a
spectral type of L7 for WISE~J2335+4511, based on visual
classification (L9$\pm$2 in the $J$-band and L7$\pm$1 in the $K$-band)
and the H$_2$OD index (L6.9$\pm$0.9). The only index in the Allers \&
Liu (2013) system applicable to low-resolution L7 spectra is the
$H$-cont index. The $H$-cont index value for WISE~J2335+4511 yields a
gravity score of ``0'', indicating \fldg. The gravity of this object
can be assessed by visual comparison to L7~\vlg\ and \intg\ spectra
(Figure \ref{fig:L7comp}). Overall, the spectrum of WISE~J2335+4511
appears intermediate between that of WISE~J0047+6803 (L7~\intg) and
the L7 field standard, showing subtle hints of low gravity in its FeH
and alkali features but not distinct enough from the field standard to
warrant an \intg\ classification.  Thus, we assign a final spectral
type of L7p~\fldg\ to WISE~J2335+4511, where the ``p'' derives from
its red color and subtle spectral peculiarities.  \\
\indent As discussed in Section~\ref{sec:cmd}, WISE~J2335+4511 also
displays the unusual behavior of appearing with the \vlg\ sequence on
CMDs where the color involves $K$-band but appearing with the field
sequence on $Y-J$ and $Y-H$ CMDs.  Thus, this object (along with the
aforementioned 2MASS~J2148+4003) demonstrates that {\em high gravity
  L~dwarfs can reside in the extremely red, faint portion of the CMD
  previously ascribed solely to low-gravity objects.} \\
\indent Based on a parallax-free analysis and no assumption of youth,
\citet{2014ApJ...783..121G} identify this as an old field candidate
(97\% probability) with statistical distance of $23.3_{-7.6}^{+9.6}$~pc,
with a slim probability (3\%) of being young field with a statistical
distance of $14.9_{-4.4}^{+4.8}$~pc. Our new CFHT parallax
($22.7_{-0.7}^{+0.7}$~pc) yields old field membership at 73\% with a
kinematics-only BANYAN~II analysis, and young field at 27\%.  Thus we
assign this to the general field population.


\noindent {\bf 2MASS~J23512200+3010540 (L5.5/L5~\fldg):} 
\citet{2010ApJS..190..100K} assign a near-IR type of L5~pec~(red), while
we assign a type of L5~\fldg\ to their spectrum without a peculiar flag.
Using a parallax-free analysis and no assumption of youth,
\citet{2014ApJ...783..121G} identify this as a modest ARG candidate,
with 47\% membership probability, a 63\% contamination rate, a
statistical distance of $20.9_{-2.4}^{+2.0}$~pc, and a mass of
9--11~\Mjup\ based on an age of 30--50~Myr. The corresponding statistical
distances for young and old field are $32.5\pm6$~pc and
$47.8_{-9.2}^{+8.8}$~pc, respectively. Based on its near-IR spectrum
having only weak evidence for low gravity (consistent with our
classification as \fldg), \citet{2015ApJS..219...33G} place this in the
field population, with 93\% probability and a 6\% contamination rate.
Using our new CFHT parallax ($24.3_{-0.9}^{+0.8}$~pc), its spatial
position is at the outskirts of the known members ($\rchisqXYZ=5.0$),
and a kinematics-only BANYAN~II analysis gives 89\% probability for ARG
membership.  Its smallest velocity offset from the group would be
3.2~\kms\ for a radial velocity of $-$3.4~\kms. Given the uncertain
state of this group's membership and thus its age
\citep{2015MNRAS.454..593B}, the presence of low-gravity spectral
features seems to be an inappropriate selection criterion at this point.
Thus we consider Argus membership to be viable despite the absence of
low-gravity spectra, with more data needed both for this object and the
group as whole.







\clearpage

\clearpage

\begin{landscape}
\begin{figure}

  \centerline{
    \includegraphics[width=2.0in,angle=0]{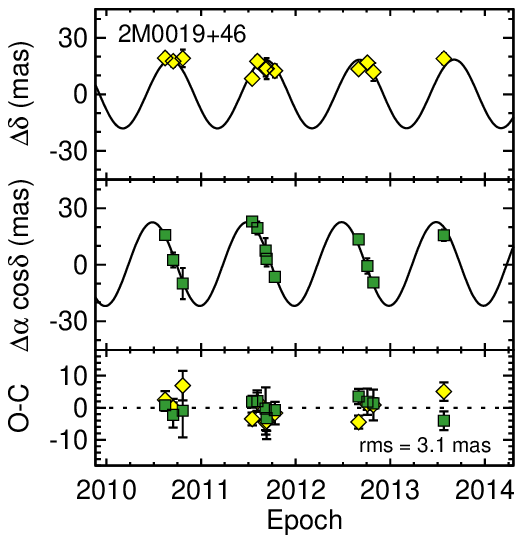}
    \includegraphics[width=2.0in,angle=0]{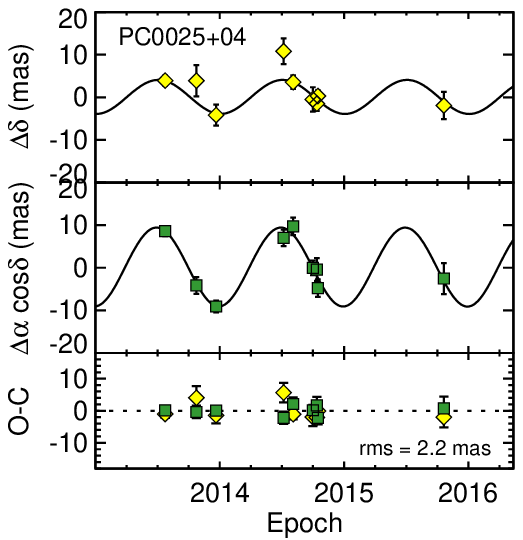}
    \includegraphics[width=2.0in,angle=0]{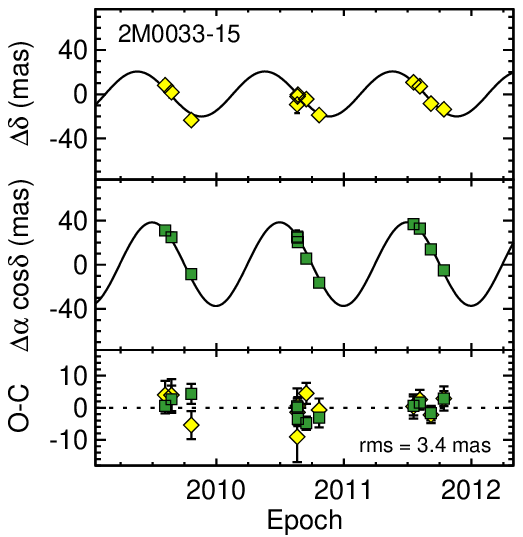}
    \includegraphics[width=2.0in,angle=0]{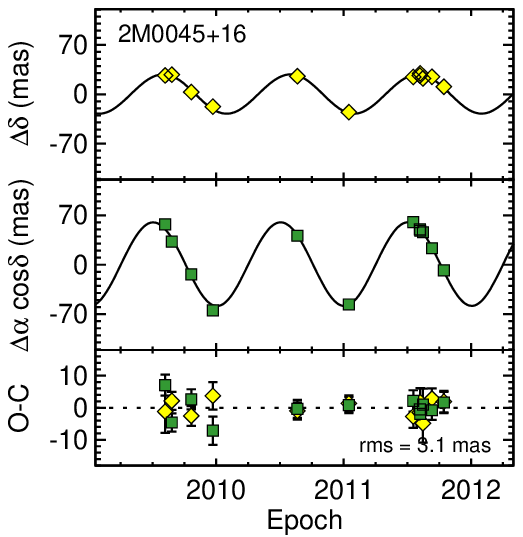}
  }
  \centerline{
    \includegraphics[width=2.0in,angle=0]{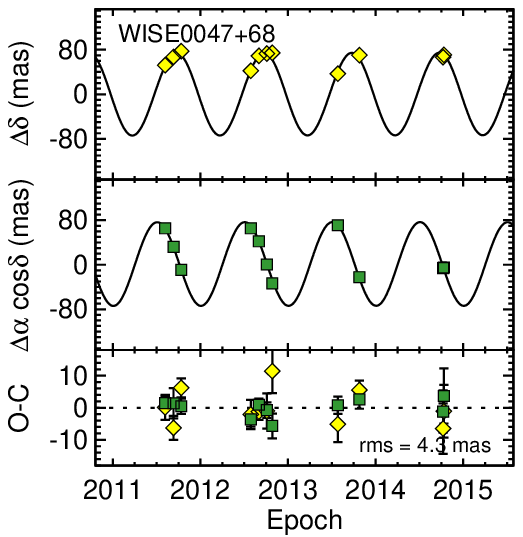}
    \includegraphics[width=2.0in,angle=0]{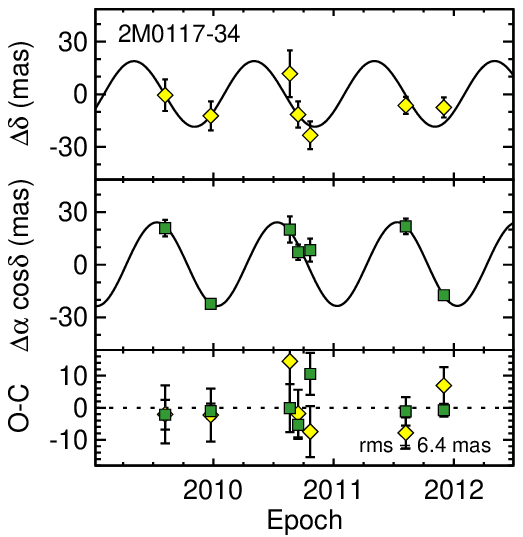}
    \includegraphics[width=2.0in,angle=0]{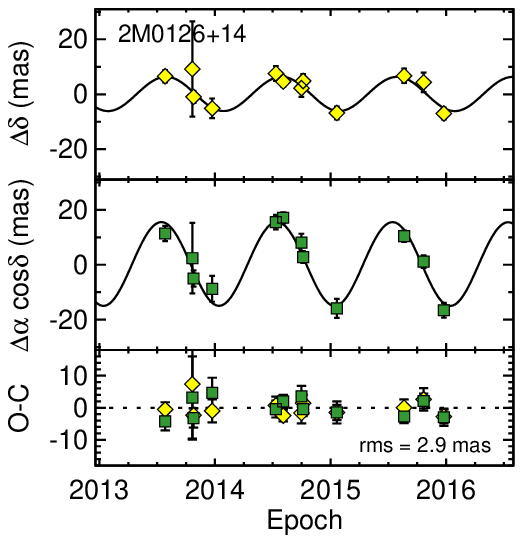}
    \includegraphics[width=2.0in,angle=0]{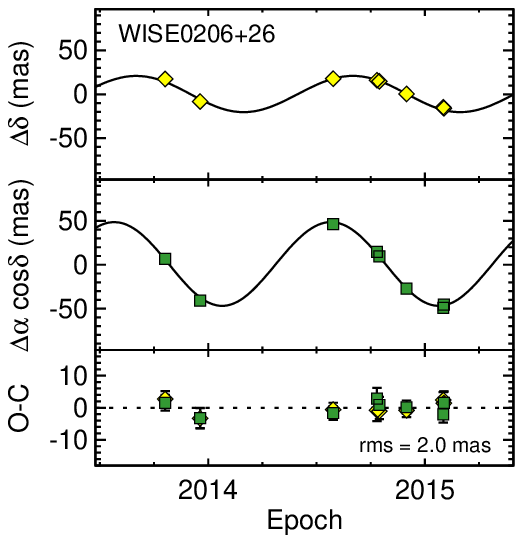}
  }
  \centerline{
    \includegraphics[width=2.0in,angle=0]{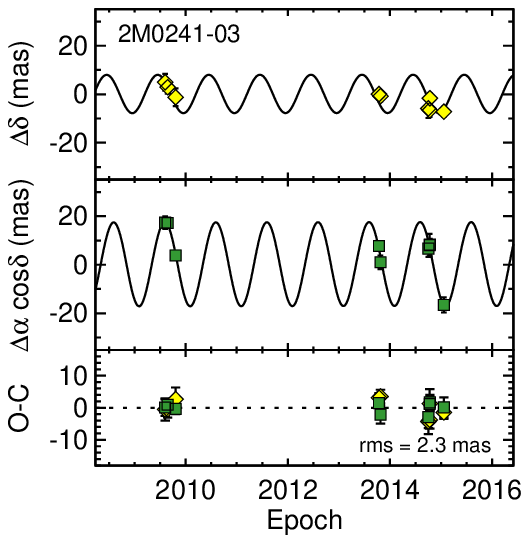}
    \includegraphics[width=2.0in,angle=0]{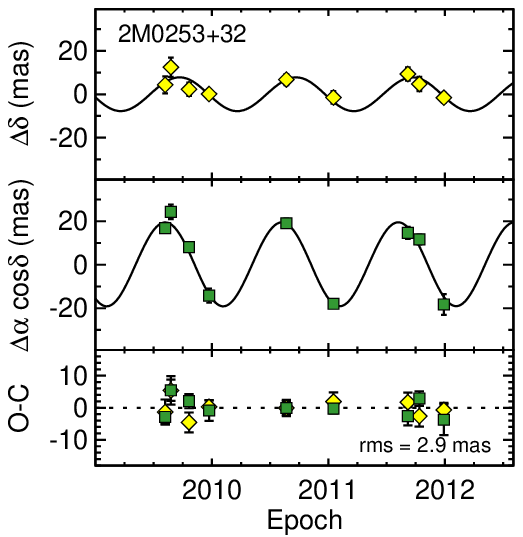}
    \includegraphics[width=2.0in,angle=0]{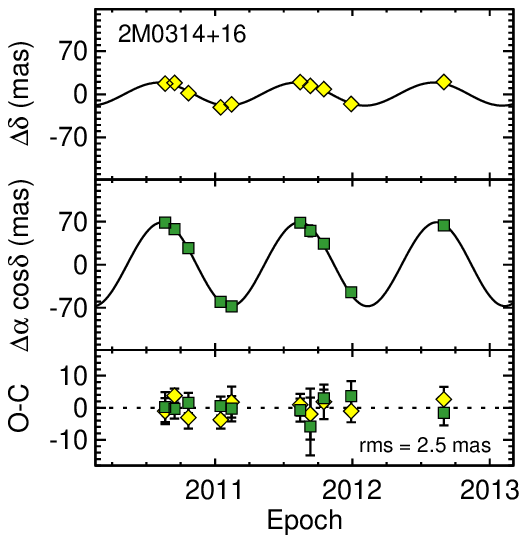}
    \includegraphics[width=2.0in,angle=0]{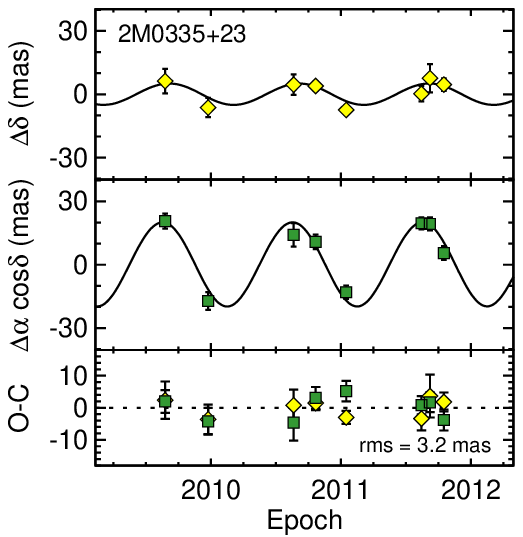}
  } 
  \caption{\normalsize For each object, the top and middle panels show
    relative astrometry in $\delta$ and $\alpha$, respectively, as a
    function of Julian year after subtracting the best-fit proper
    motion.  (This is for display purposes only; in our analysis we fit
    for both the proper motion and parallax simultaneously.)  The bottom
    panels show the residuals after subtracting both the parallax and
    proper motion. \label{fig:plx}}
\end{figure}

\setcounter{figure}{0}   
\begin{figure}
  \centerline{
    \includegraphics[width=2.0in,angle=0]{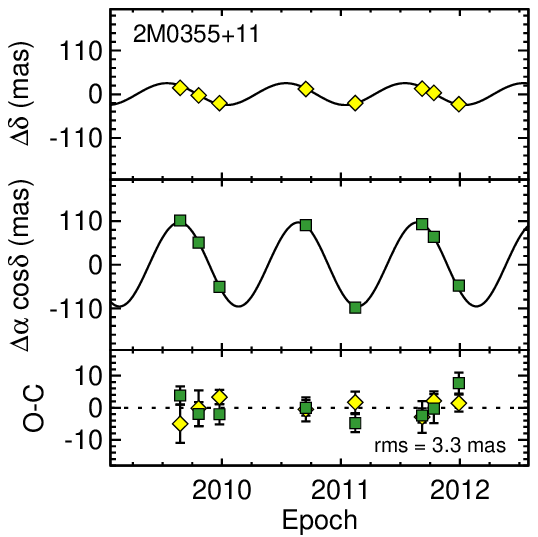}
    \includegraphics[width=2.0in,angle=0]{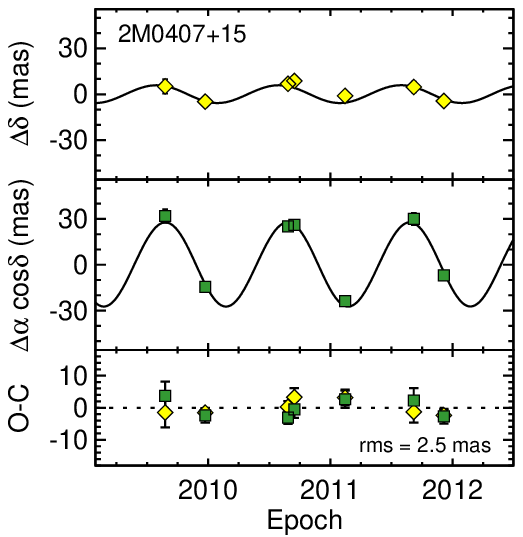}
    \includegraphics[width=2.0in,angle=0]{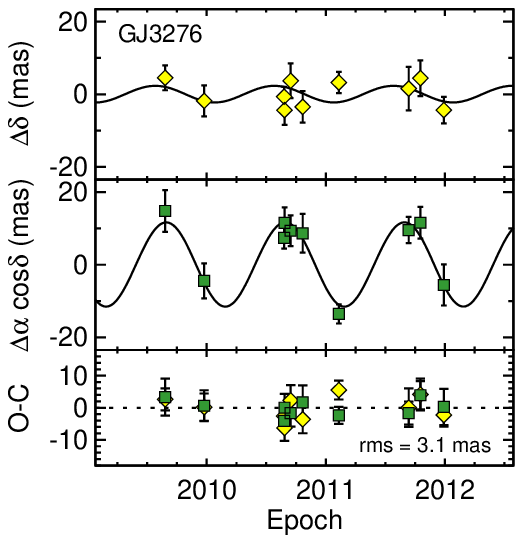}
    \includegraphics[width=2.0in,angle=0]{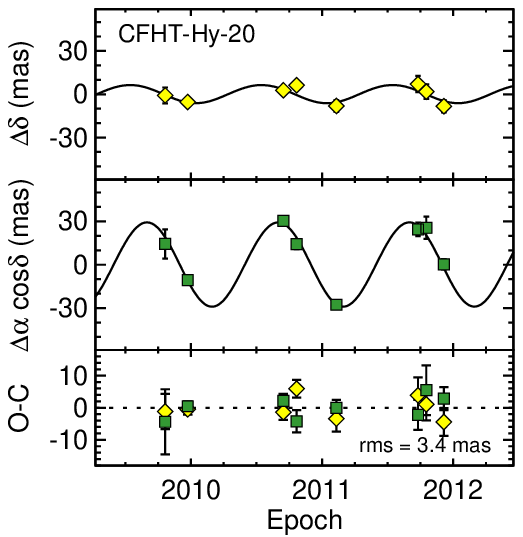}
  } 
  \centerline{
    \includegraphics[width=2.0in,angle=0]{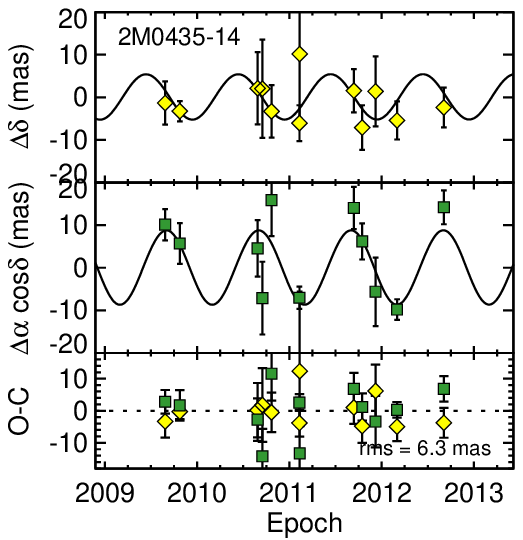}
    \includegraphics[width=2.0in,angle=0]{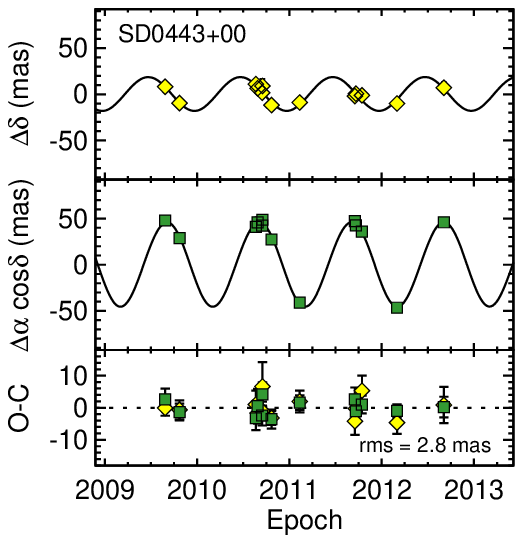}
    \includegraphics[width=2.0in,angle=0]{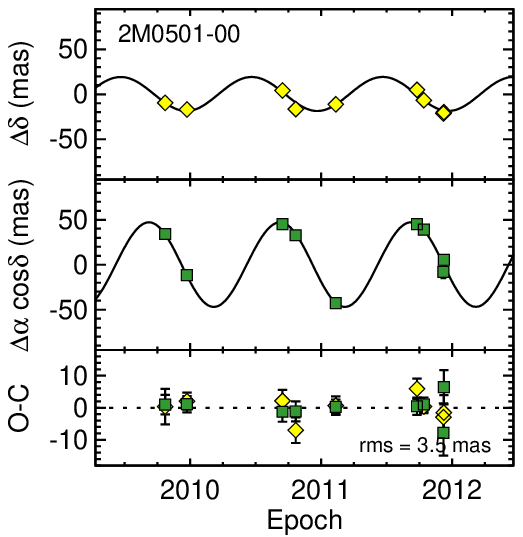}
    \includegraphics[width=2.0in,angle=0]{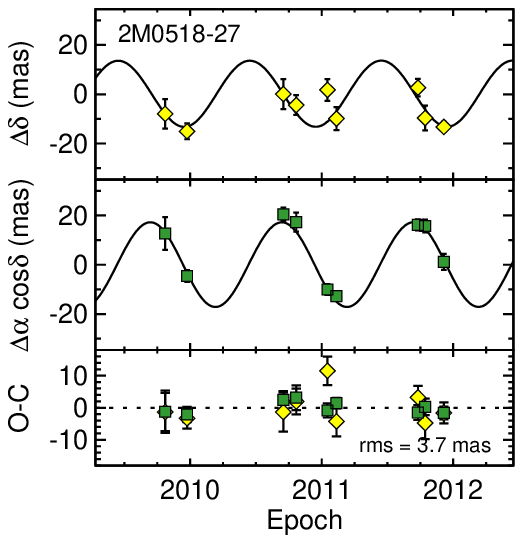}
  }
  \centerline{
    \includegraphics[width=2.0in,angle=0]{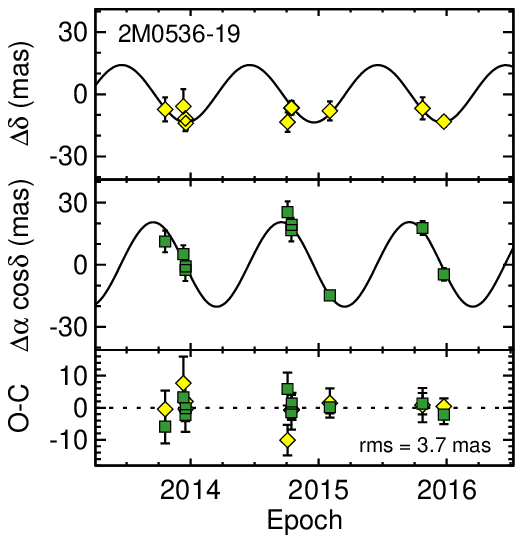}
    \includegraphics[width=2.0in,angle=0]{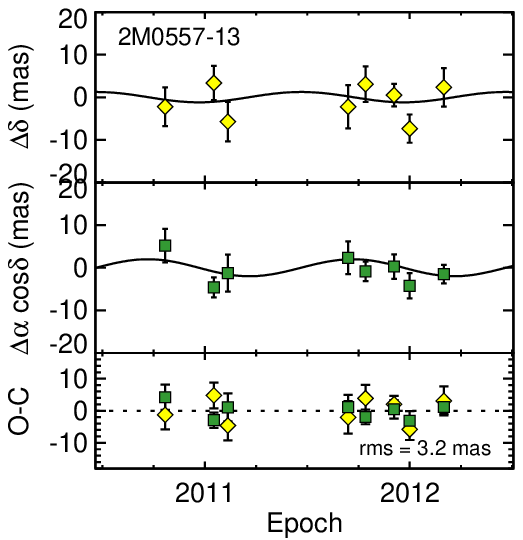}
    \includegraphics[width=2.0in,angle=0]{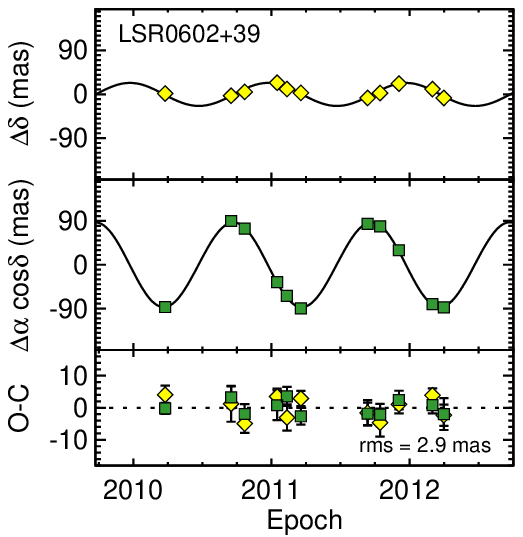}
    \includegraphics[width=2.0in,angle=0]{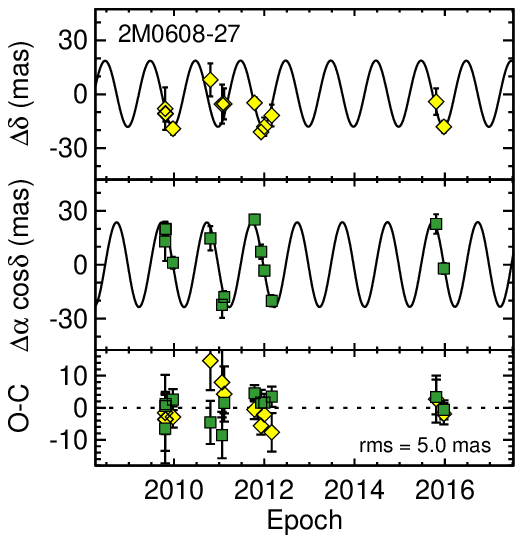}
  } 
  \caption{\normalsize (Continued)} 
\end{figure}

\setcounter{figure}{0}   
\begin{figure}
  \centerline{
    \includegraphics[width=2.0in,angle=0]{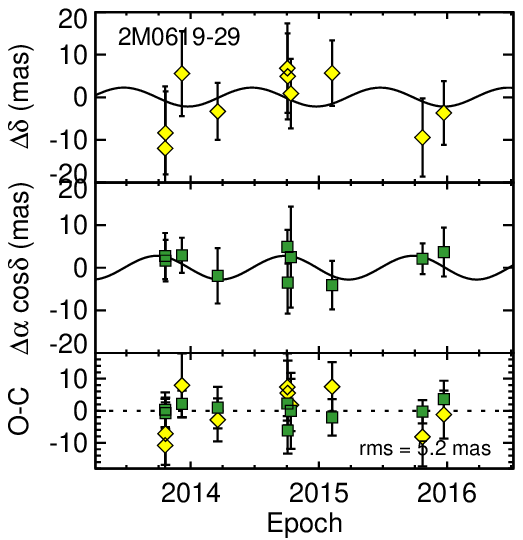}
    \includegraphics[width=2.0in,angle=0]{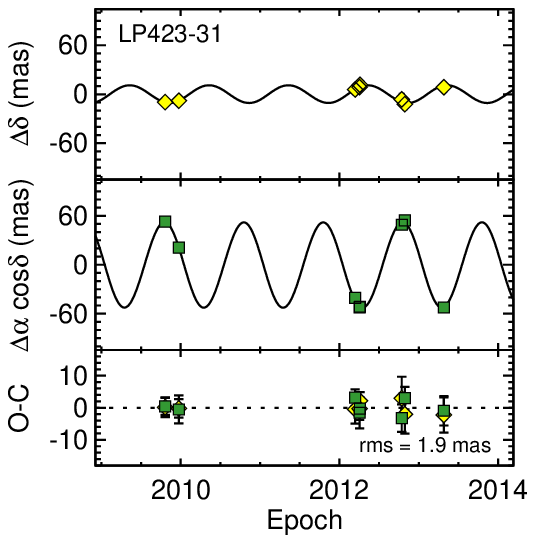}
    \includegraphics[width=2.0in,angle=0]{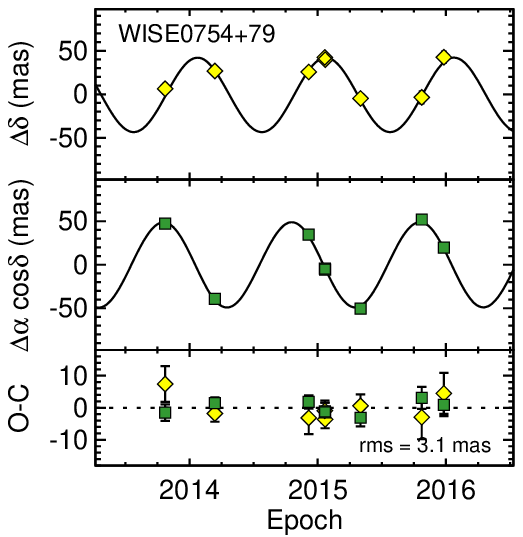}
    \includegraphics[width=2.0in,angle=0]{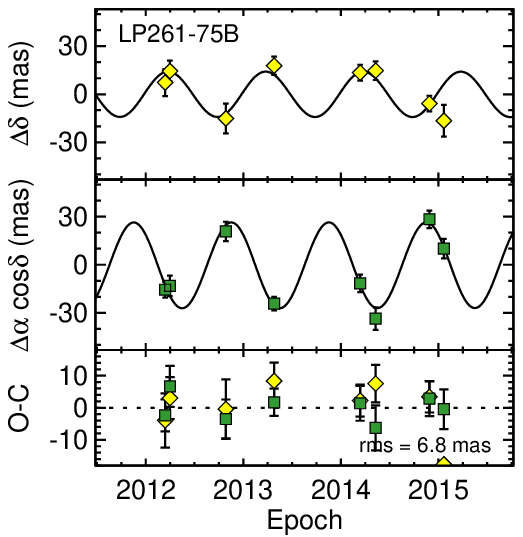}
  } 
  \centerline{
    \includegraphics[width=2.0in,angle=0]{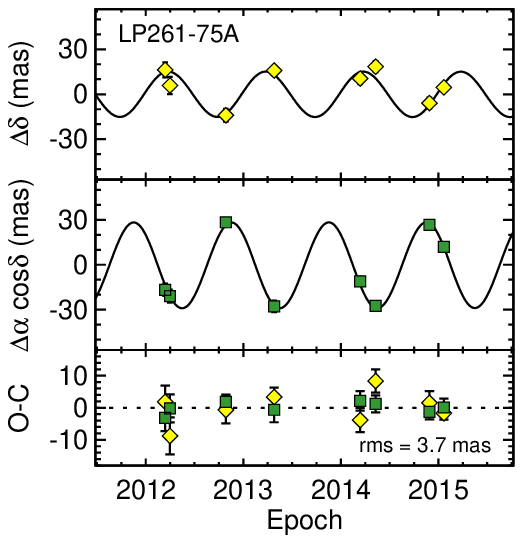}
    \includegraphics[width=2.0in,angle=0]{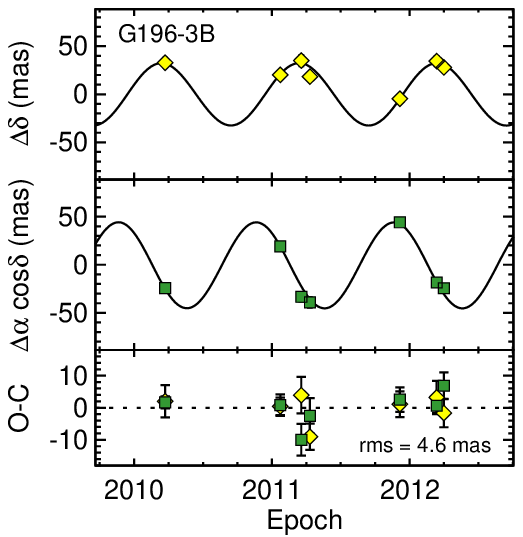}
    \includegraphics[width=2.0in,angle=0]{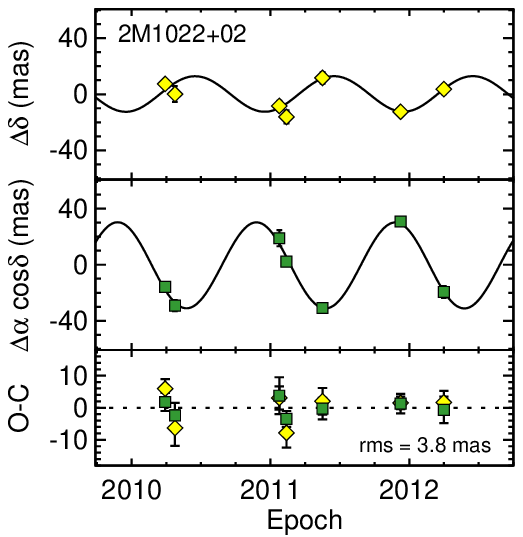}
    \includegraphics[width=2.0in,angle=0]{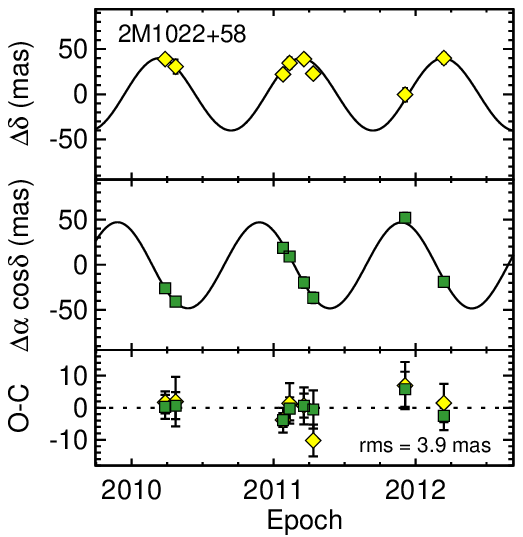}
  }
  \centerline{
    \includegraphics[width=2.0in,angle=0]{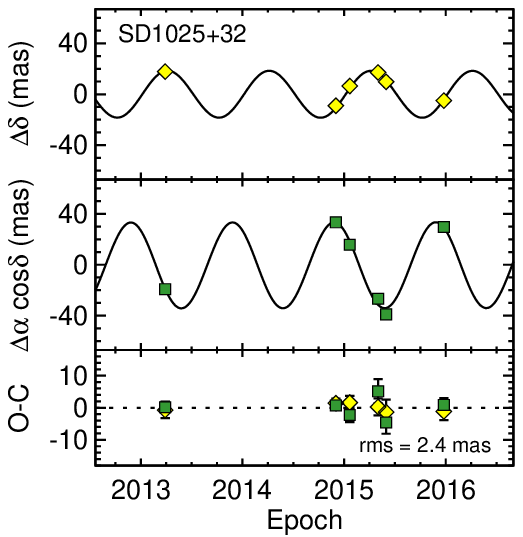}
    \includegraphics[width=2.0in,angle=0]{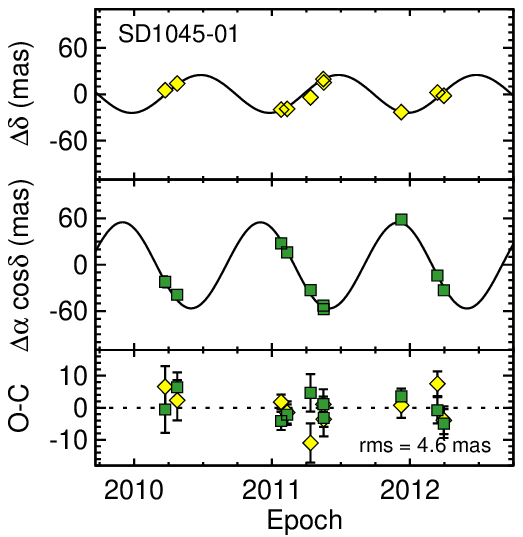}
    \includegraphics[width=2.0in,angle=0]{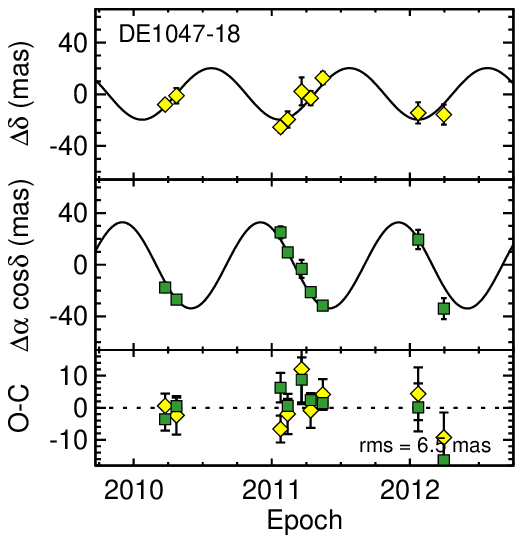}
    \includegraphics[width=2.0in,angle=0]{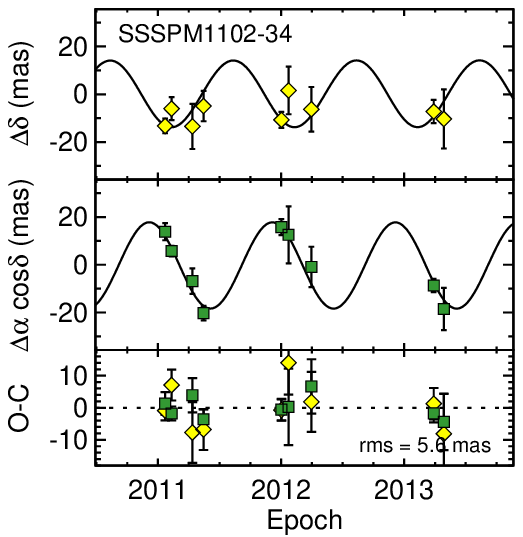}
  } 
  \caption{\normalsize (Continued)} 
\end{figure}

\setcounter{figure}{0}   
\begin{figure}
  \centerline{
    \includegraphics[width=2.0in,angle=0]{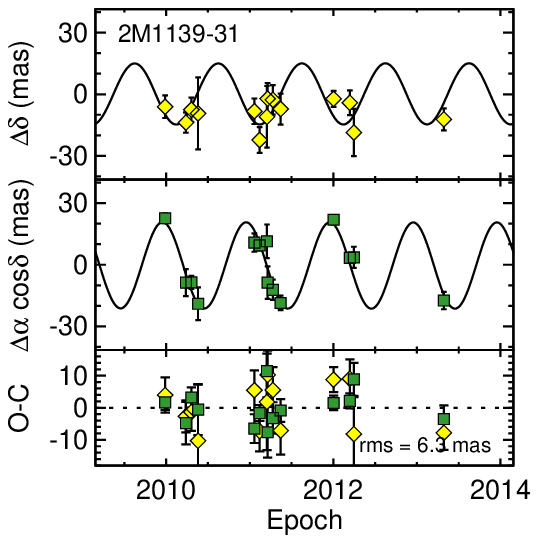}
    \includegraphics[width=2.0in,angle=0]{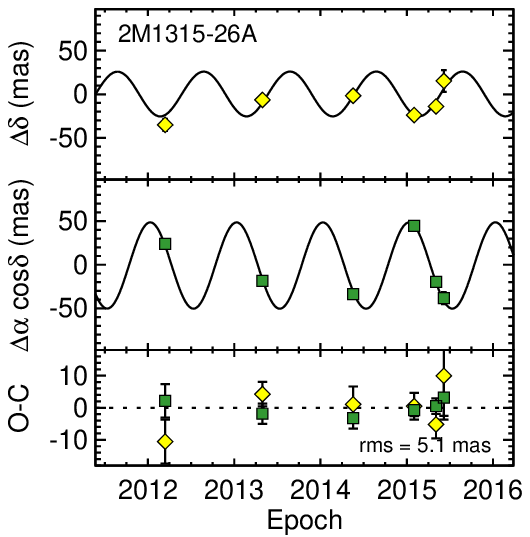}
    \includegraphics[width=2.0in,angle=0]{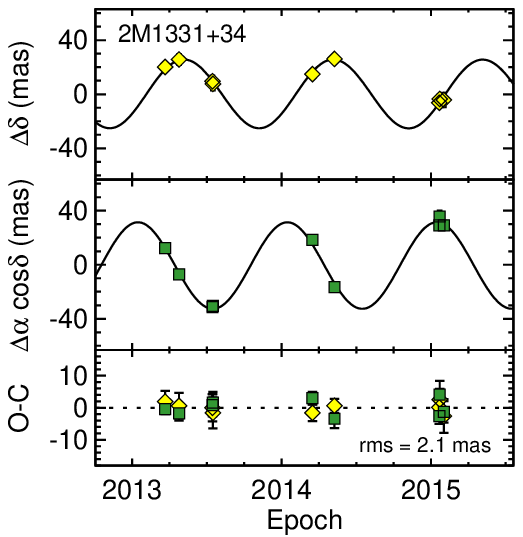}
    \includegraphics[width=2.0in,angle=0]{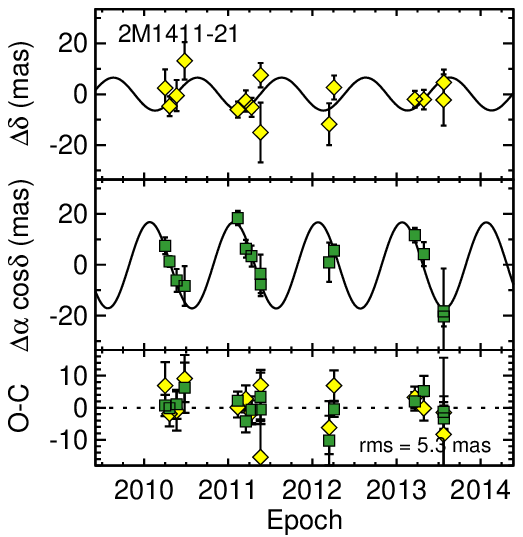}
  } 
  \centerline{
    \includegraphics[width=2.0in,angle=0]{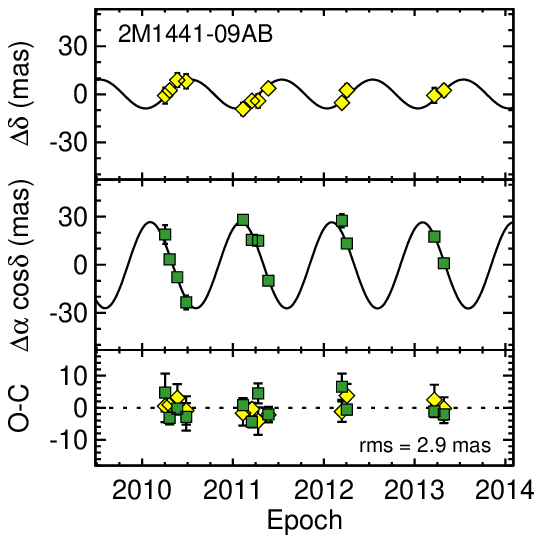}
    \includegraphics[width=2.0in,angle=0]{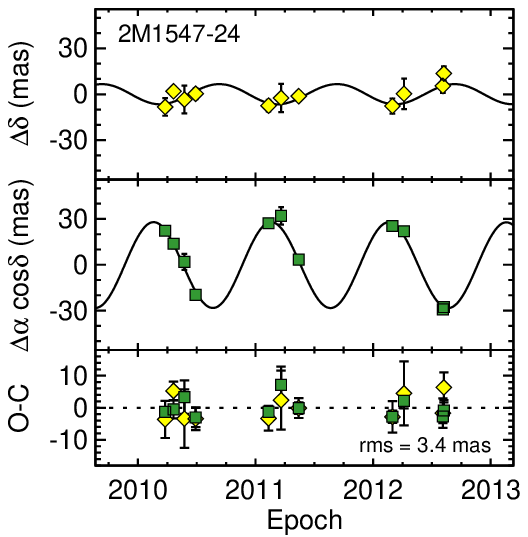}
    \includegraphics[width=2.0in,angle=0]{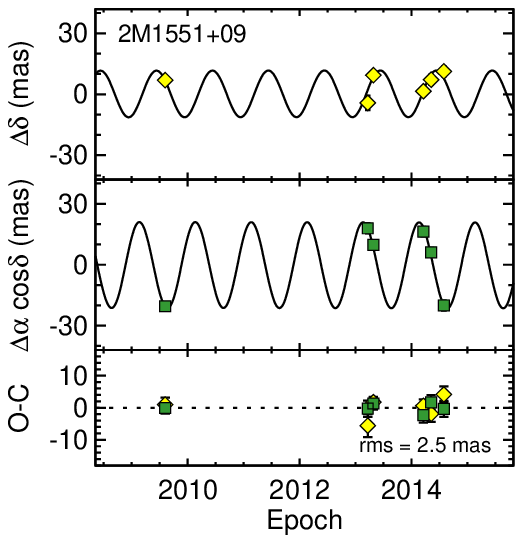}
    \includegraphics[width=2.0in,angle=0]{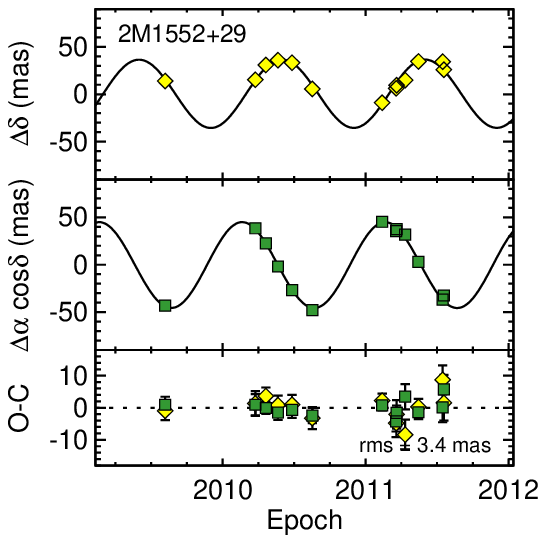}
  }
  \centerline{
    \includegraphics[width=2.0in,angle=0]{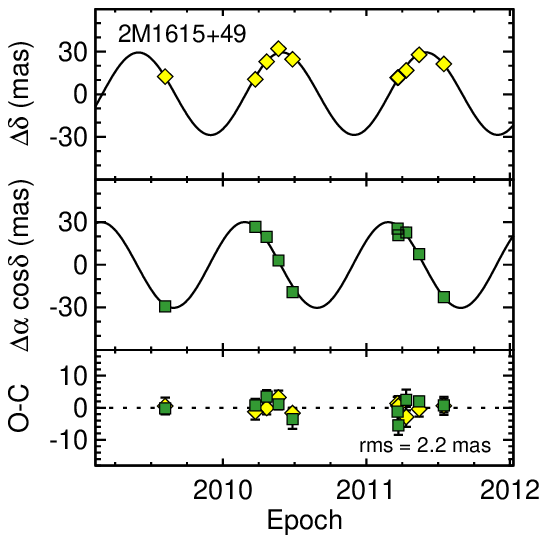}
    \includegraphics[width=2.0in,angle=0]{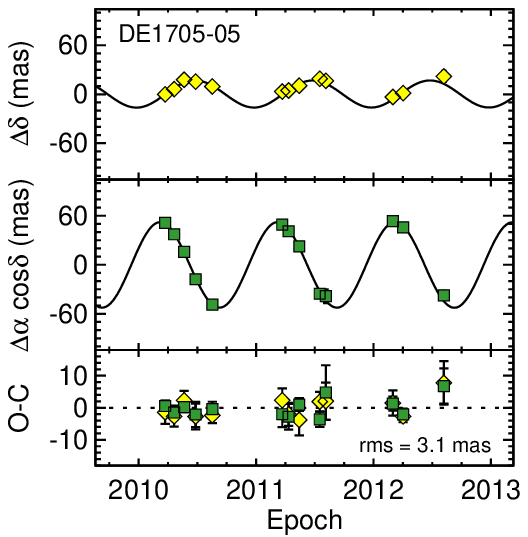}
    \includegraphics[width=2.0in,angle=0]{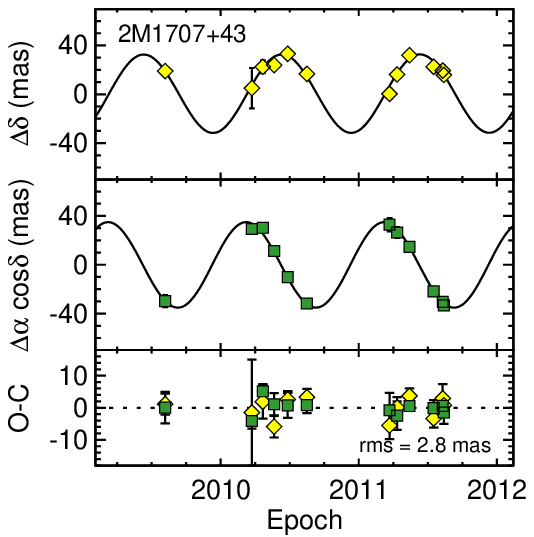}
    \includegraphics[width=2.0in,angle=0]{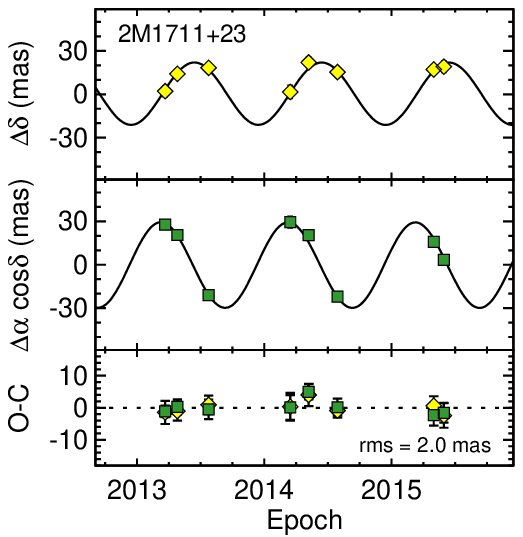}
  } 
  \caption{\normalsize (Continued) } 
\end{figure}

\setcounter{figure}{0}   
\begin{figure}
  \centerline{
    \includegraphics[width=2.0in,angle=0]{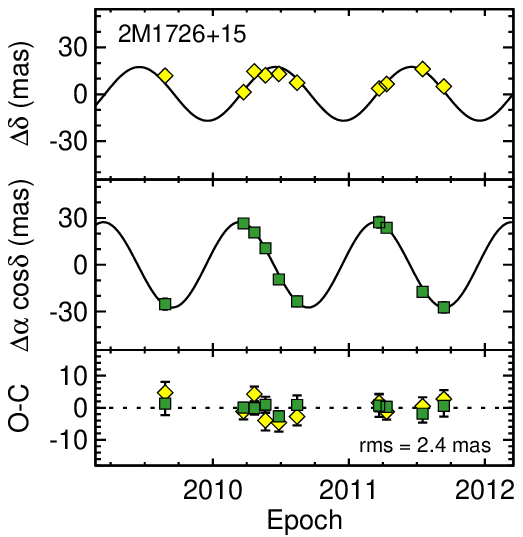}
    \includegraphics[width=2.0in,angle=0]{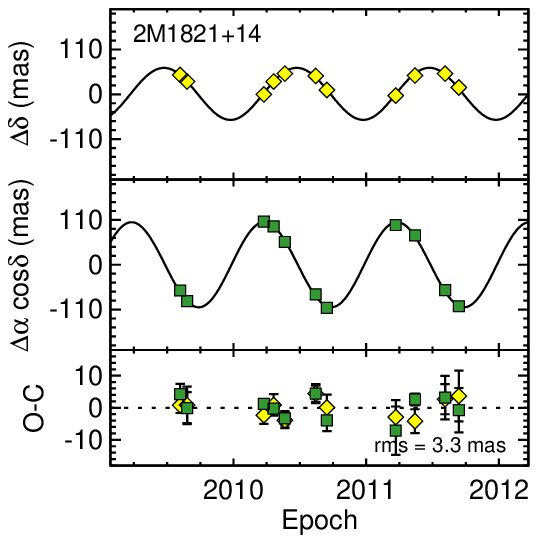}
    \includegraphics[width=2.0in,angle=0]{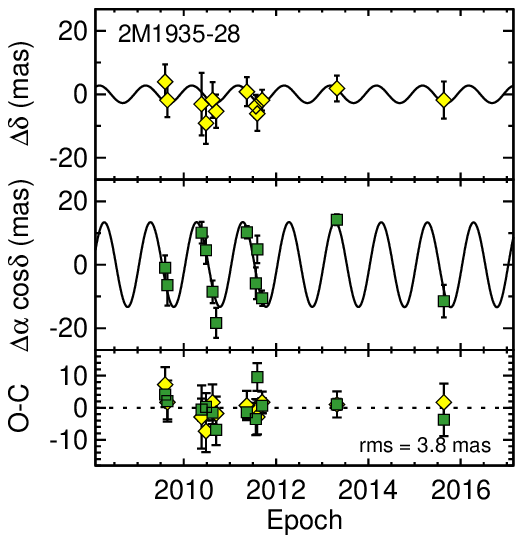}
    \includegraphics[width=2.0in,angle=0]{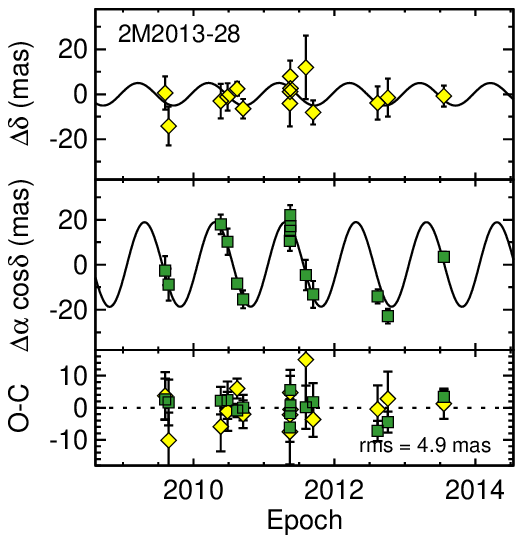}
  } 
  \centerline{
    \includegraphics[width=2.0in,angle=0]{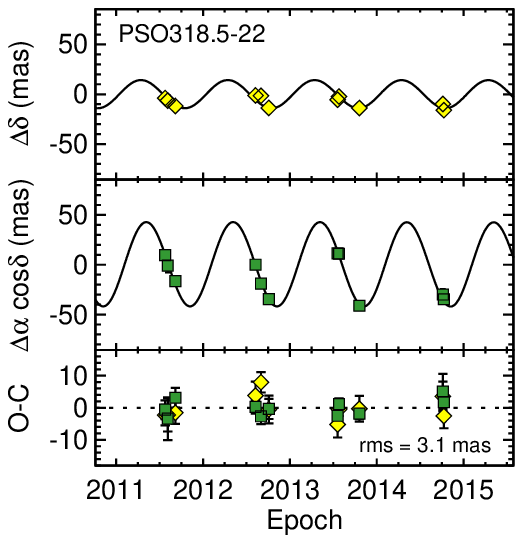}
    \includegraphics[width=2.0in,angle=0]{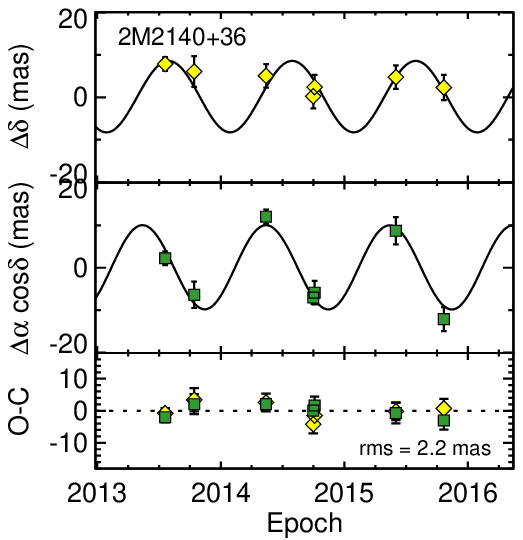}
    \includegraphics[width=2.0in,angle=0]{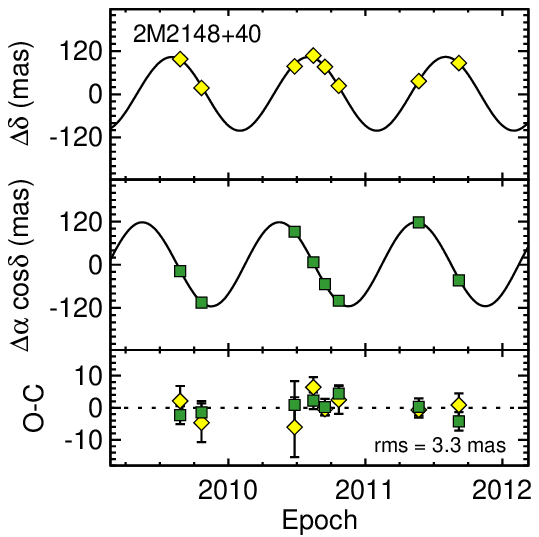}
    \includegraphics[width=2.0in,angle=0]{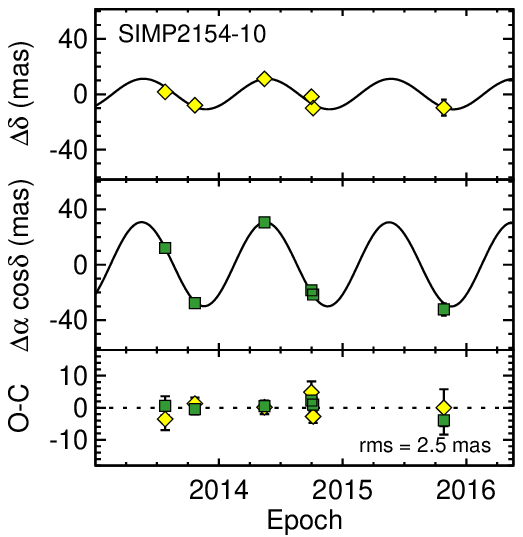}
  }
  \centerline{
    \includegraphics[width=2.0in,angle=0]{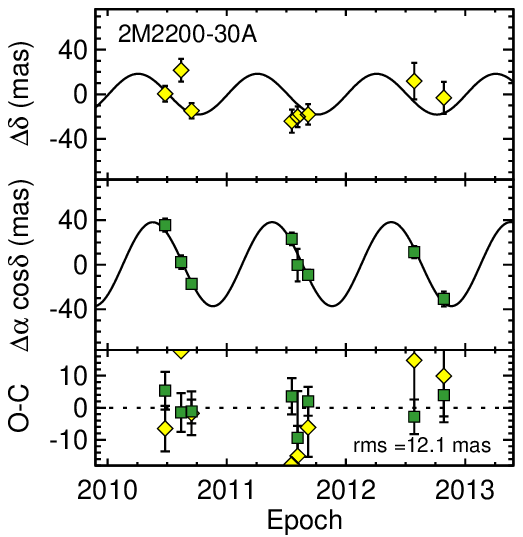}
    \includegraphics[width=2.0in,angle=0]{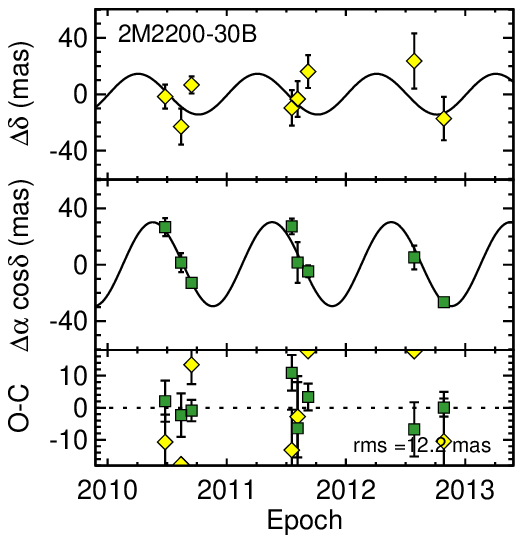}
    \includegraphics[width=2.0in,angle=0]{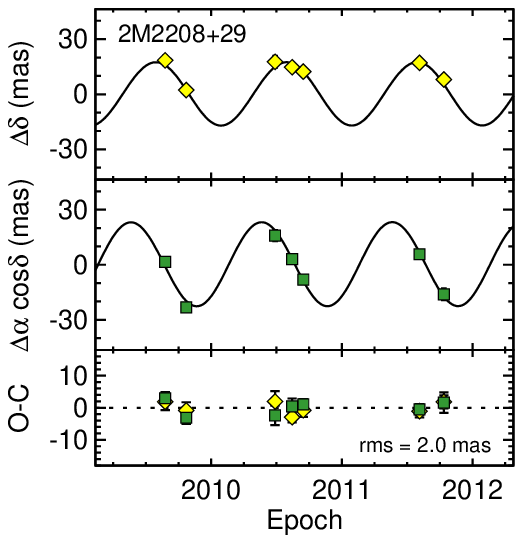}
    \includegraphics[width=2.0in,angle=0]{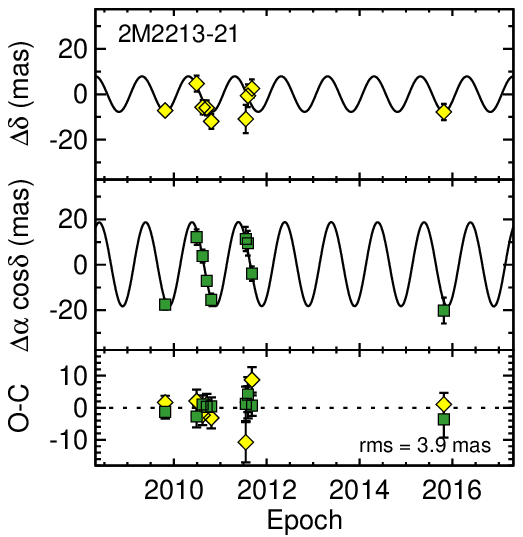}
  } 
  \caption{\normalsize (Continued)} 
\end{figure}

\setcounter{figure}{0}   
\begin{figure}
  \centerline{
    \includegraphics[width=2.0in,angle=0]{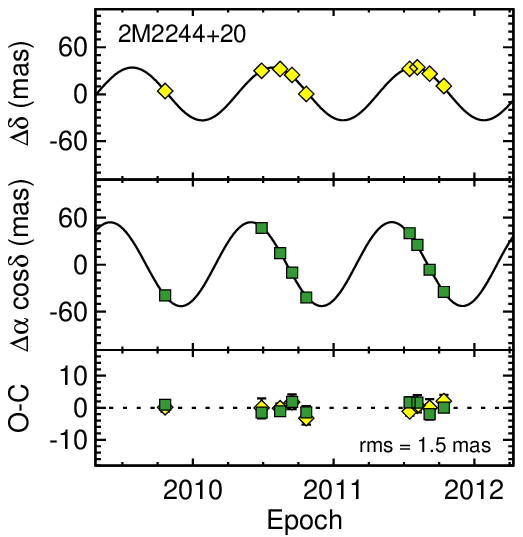}
    \includegraphics[width=2.0in,angle=0]{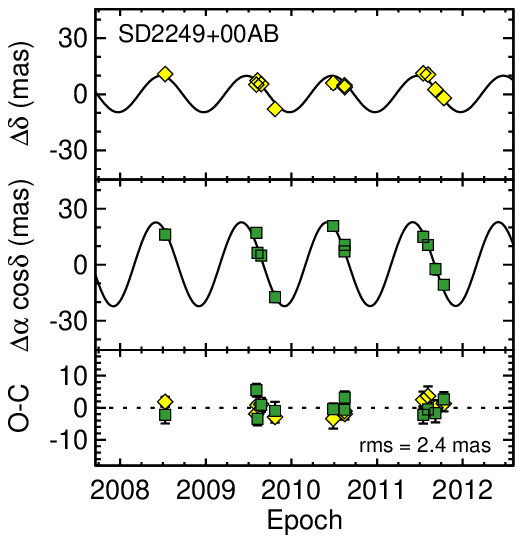}
    \includegraphics[width=2.0in,angle=0]{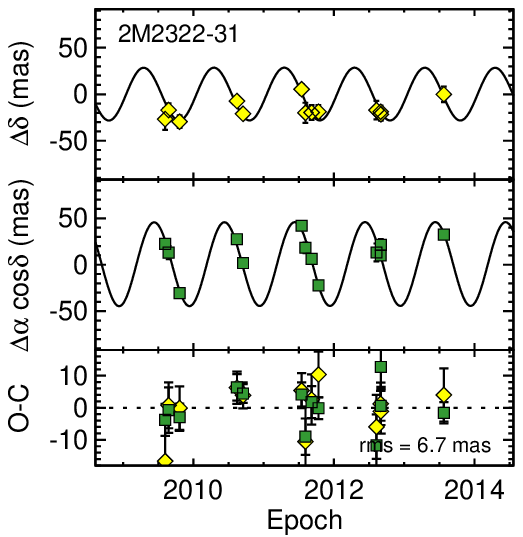}
    \includegraphics[width=2.0in,angle=0]{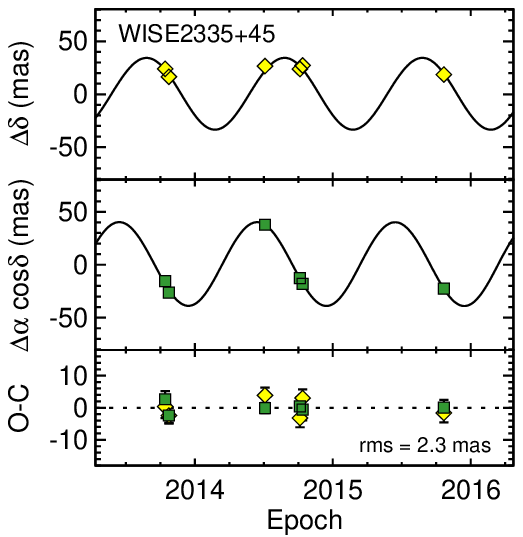}
  } 
  \centerline{
    \includegraphics[width=2.0in,angle=0]{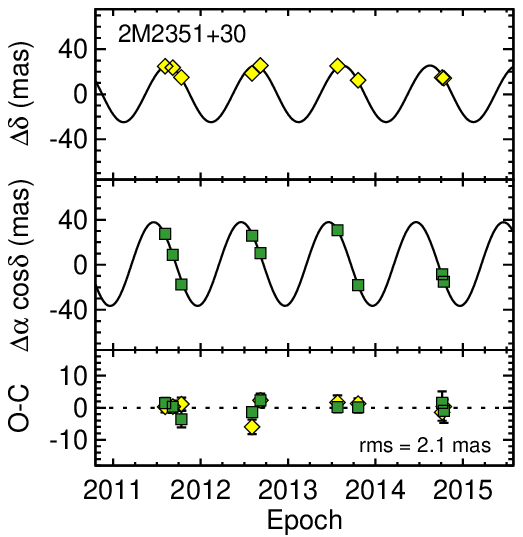}
    \includegraphics[width=2.0in,angle=0]{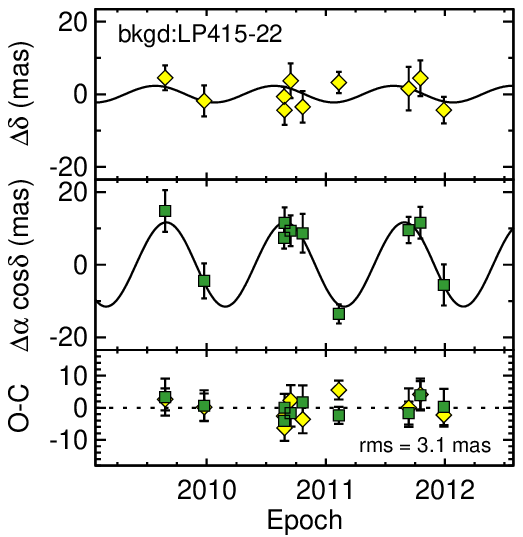}
    \includegraphics[width=2.0in,angle=0]{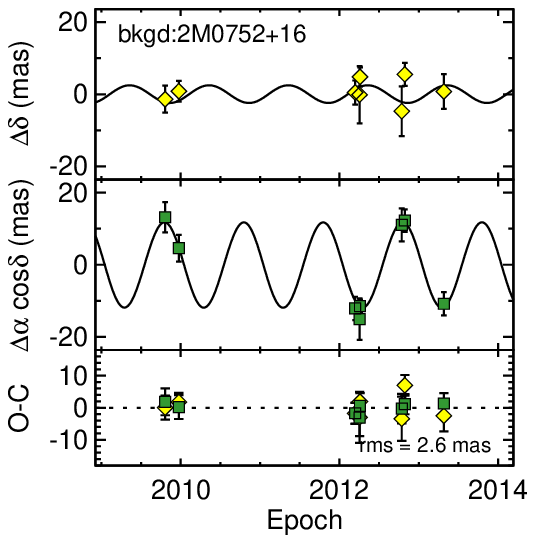}
  }
  \caption{\normalsize (Continued)} 
\end{figure}
\end{landscape}

\begin{landscape}
\begin{figure}

  \centerline{
    \includegraphics[width=8.5in,angle=0]{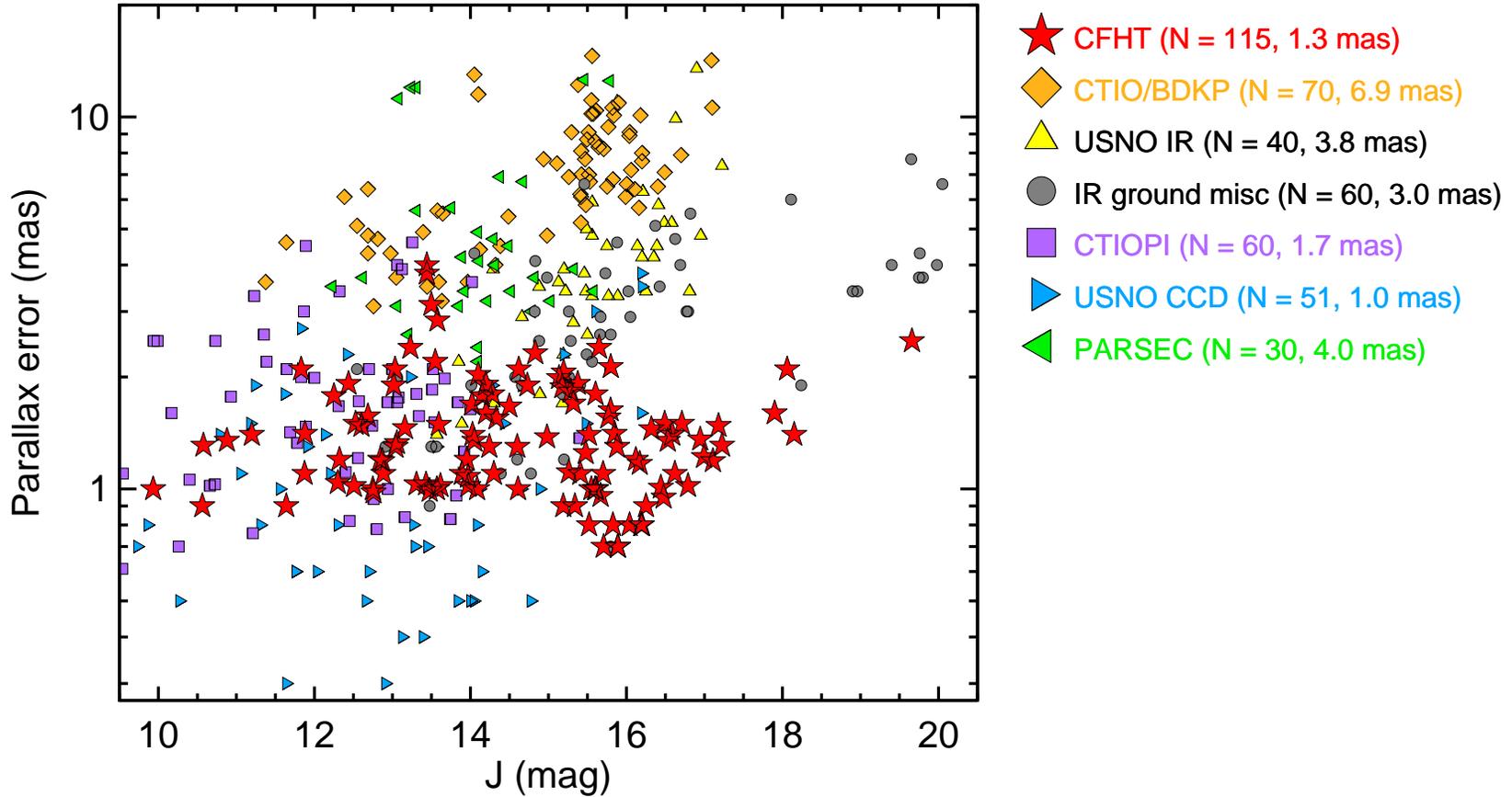}
  }
  \caption{\normalsize Our CFHT/WIRCam parallax errors plotted as a
    function of $J$-band magnitude alongside the results of other
    infrared parallax programs and larger optical programs that target
    ultracool dwarfs.  Plotted CFHT results include the parallaxes
    published here as well as our previous work
    \citep{2012ApJS..201...19D, 2015ApJ...805...56D}, in total
    115~parallax measurements with a median precision of 1.3\,mas.
    Other plotted parallaxes are from the Brown Dwarf Kinematics Project
    at CTIO \citep{2011AJ....141...71F, 2012ApJ...752...56F,
      2013AJ....145....2F}; the USNO IR program
    \citep{2004AJ....127.2948V, 2008ApJ...672.1159B}; the optical CTIO
    Parallax Investigation program \citep{2005AJ....130..337C,
      2006AJ....132.1234C, 2006AJ....132.2360H, 2007ApJ...669L..45G,
      2014AJ....147...85R, 2014AJ....147...94D, 2014AJ....148...91L};
    the optical USNO CCD program \citep{1992AJ....103..638M,
      2002AJ....124.1170D, 2008ApJ...686..548D, 2003AJ....125..354R,
      2015ApJ...799..203G}; and the optical PARSEC program
    \citep{2011AJ....141...54A, 2013AJ....146..161M}.  Other smaller
    individual samples of ground-based infrared parallaxes are plotted
    as a group, including results from NTT/SOFI
    \citep{2003AJ....126..975T}, Calar Alto 3.5-m/Omega-2000
    \citep{2009A&A...493L..27S, 2013A&A...560A..52M,
      2014A&A...568A...6Z}, UKIRT/WFCAM \citep{2013MNRAS.433.2054S}, and
    Magellan/FourStar \citep{2014ApJ...796...39T}.  Note that the
    extremely high-precision VLT optical parallaxes from
    \citet{2014arXiv1403.1275S}, with $\sigma_{\pi} = 0.06$--0.14\,mas
    at $J=11.1$--12.7\,mag, lie outside the plotted
    area.  \label{fig:j-eplx}}

\end{figure}
\end{landscape}

\begin{figure}
  \includegraphics[width=6in]{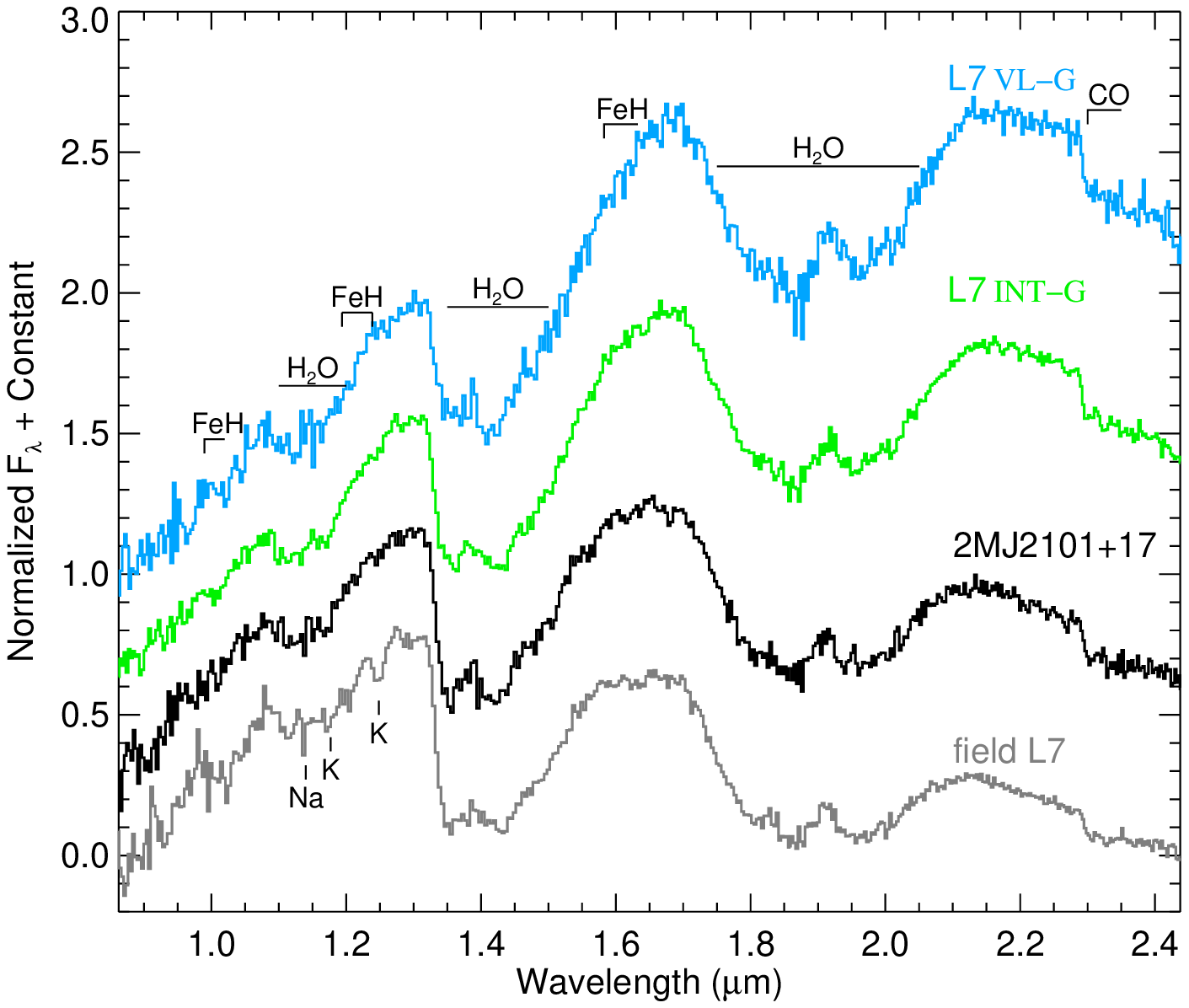}
  \caption{\normalsize Comparison of the spectrum of 2MASS~J2101+1756 to
    L7 dwarfs with a range of gravity classifications from
    \citet{2013ApJ...772...79A} and listed in the caption of
    Figure~\ref{fig:L7comp}. We assign a spectral type of L7 and do not
    find any strong indication of youth. \label{fig:2m2101}}
\end{figure}

\begin{figure}

  \centerline{
    \includegraphics[height=4.0in,angle=0]{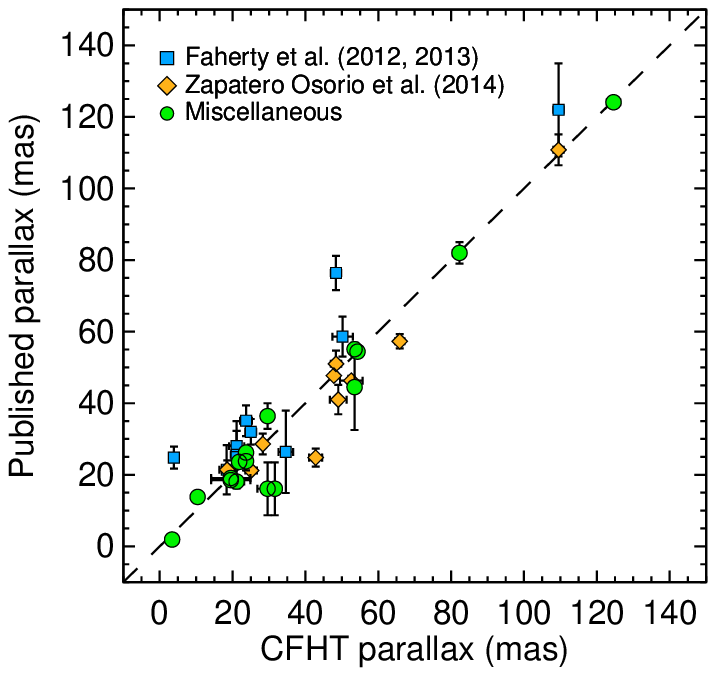} 
    \hskip 0.1in
    \includegraphics[height=3.8in,angle=0]{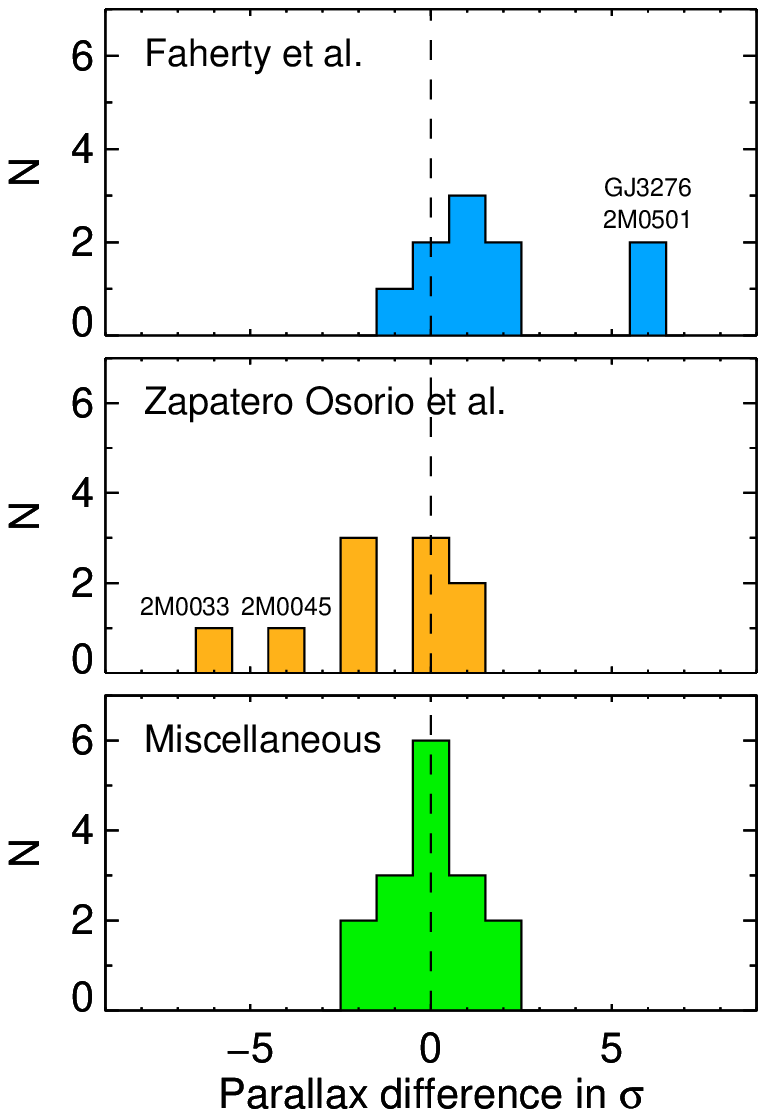} 
  }

  \caption{\normalsize \emph{Left:} Comparison of our CFHT parallaxes
    with other results in the literature. The published parallaxes are
    primarily from \citet{2012ApJ...752...56F} and
    \citet{2014A&A...568A...6Z}, with some measurements coming from
    other miscellaneous sources \citep{2002AJ....124.1170D,
      2004AJ....127.2948V, 2006AJ....132.1234C, 2008A&A...489..825T,
      2009AJ....137..402G, 2011AJ....141...54A, 2012ApJ...758...56S,
      2013ApJ...762..118W, 2014A&A...563A.121D, 2015ApJ...799..203G}.
    Data points without visible error bars have uncertainties smaller
    than the plotting symbols.  Most of our parallaxes are consistent at
    1$\sigma$ or less with previous results, and 90\% are consistent at
    2.5$\sigma$. However, there are a number of more extreme
    $\approx$4--6$\sigma$ outliers.  \emph{Right:} The histograms show
    differences in parallaxes computed as
    $(\pi_{\rm other}-\pi_{\rm CFHT})$.  The largest outliers come from
    the samples of \cite{2012ApJ...752...56F} and
    \citet{2014A&A...568A...6Z}. The histograms suggest that parallaxes
    from \citet{2012ApJ...752...56F} are systematically larger than our
    CFHT values and parallaxes from \citet{2014A&A...568A...6Z} are
    systematically lower, while no systematic offset is apparent for
    other literature results. \label{fig:compare}}
\end{figure}

\begin{figure}

  \centerline{
    \includegraphics[height=4.0in,angle=0]{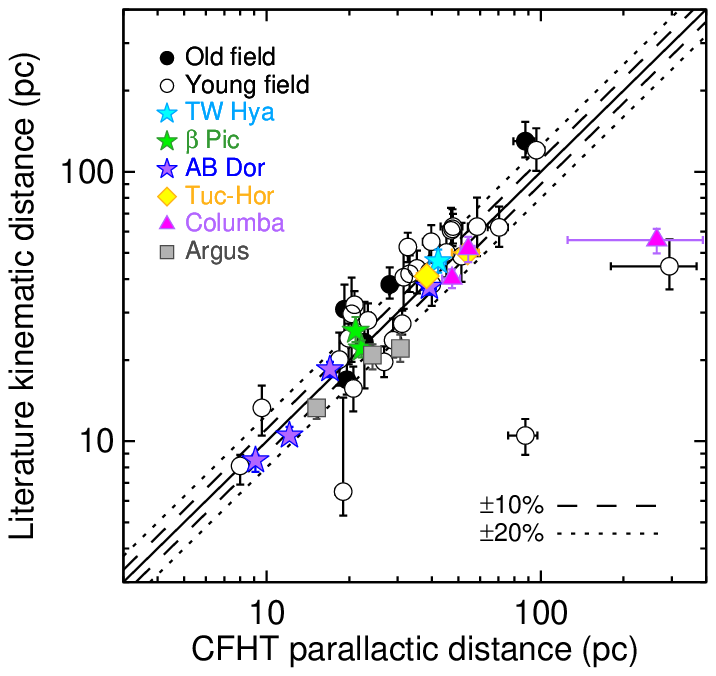} 
    \hskip 0.1in
    \includegraphics[height=3.8in,angle=0]{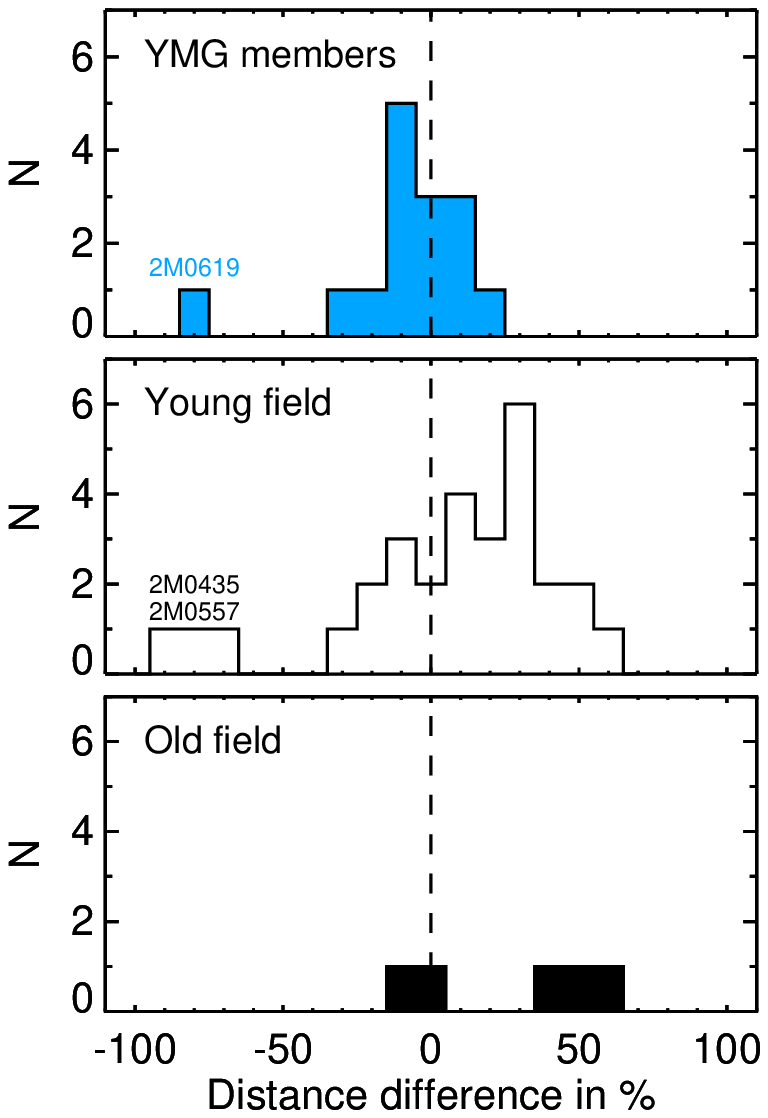}
  }

  \caption{\normalsize Comparison of our CFHT parallactic distances to
    kinematic distance estimates from BANYAN~II
    \citep{2014ApJ...783..121G} and \citet{2005ApJ...634.1385M}. Most
    objects have distances that agree to within 20\%, which is
    dominated by the uncertainty in the kinematic distances since the
    parallax errors are typically $<$6\%. Although the most visible
    outliers are those with CFHT parallaxes placing them at much
    larger distances, overall the kinematic distances for the ``young
    field'' population show a slight preference for being
    systematically 11\% larger than our parallax distances. Data
    points without obvious error bars have uncertainties smaller than
    the plotting symbol.  The histograms in the righthand panels show
    percentages computed as $(d_{\rm kin}-d_{\rm CFHT})/d_{\rm
      CFHT}$. \label{fig:dkin}}

\end{figure}


\begin{landscape}
\begin{figure}
  \includegraphics[width=2.4in]{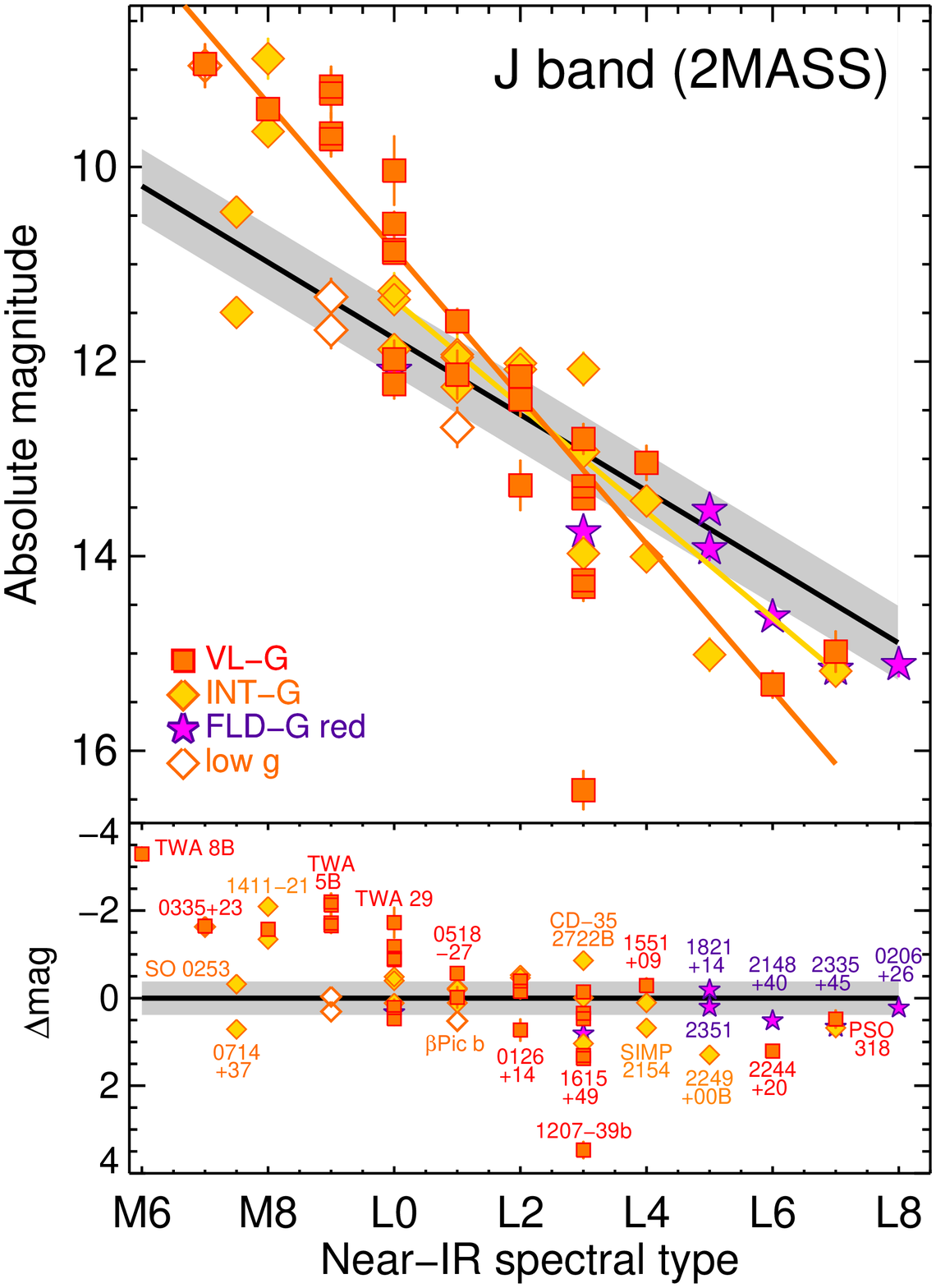}
  \hskip 0.5in
  \includegraphics[width=2.4in]{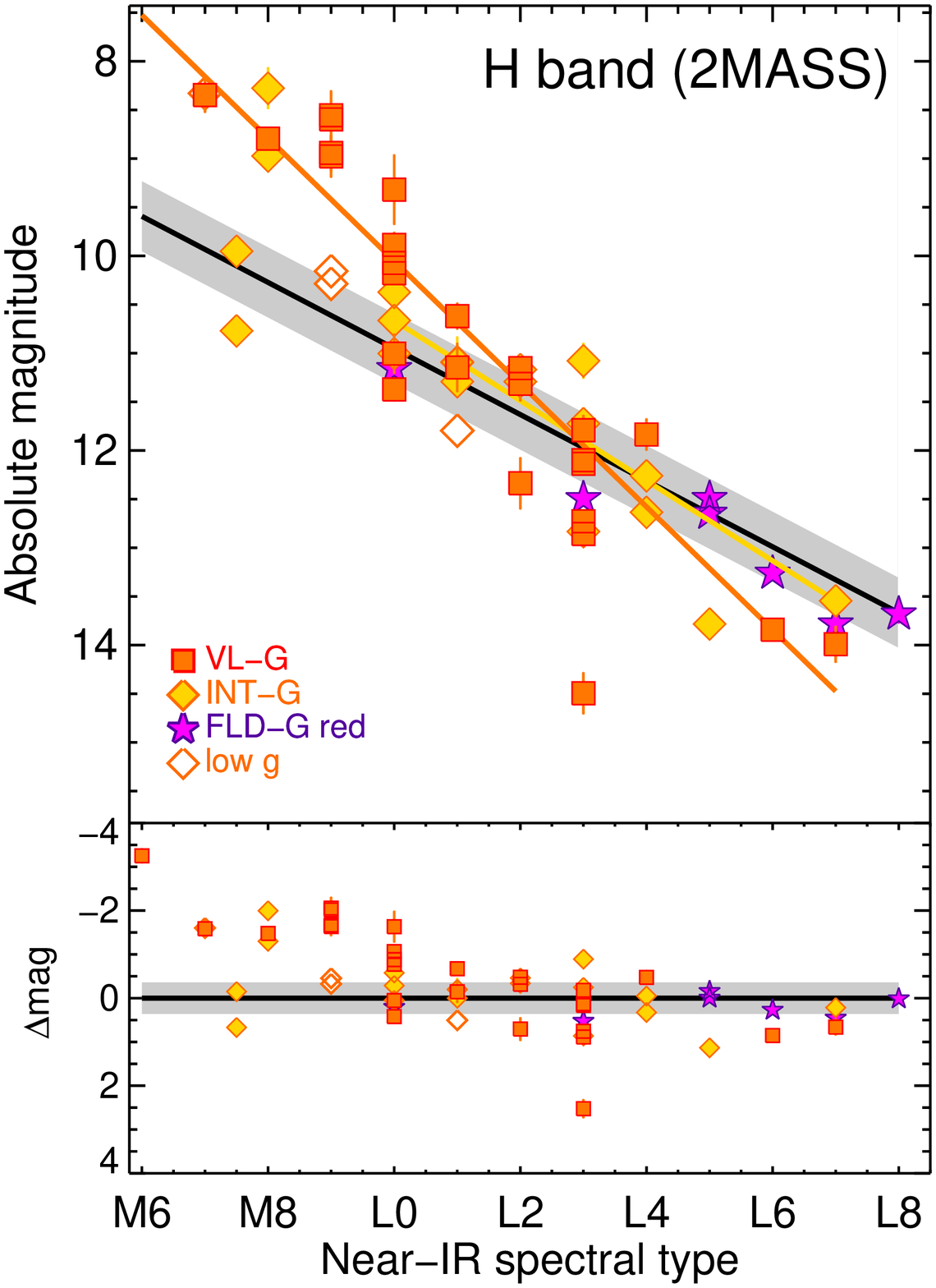}
  \hskip 0.5in
  \includegraphics[width=2.4in]{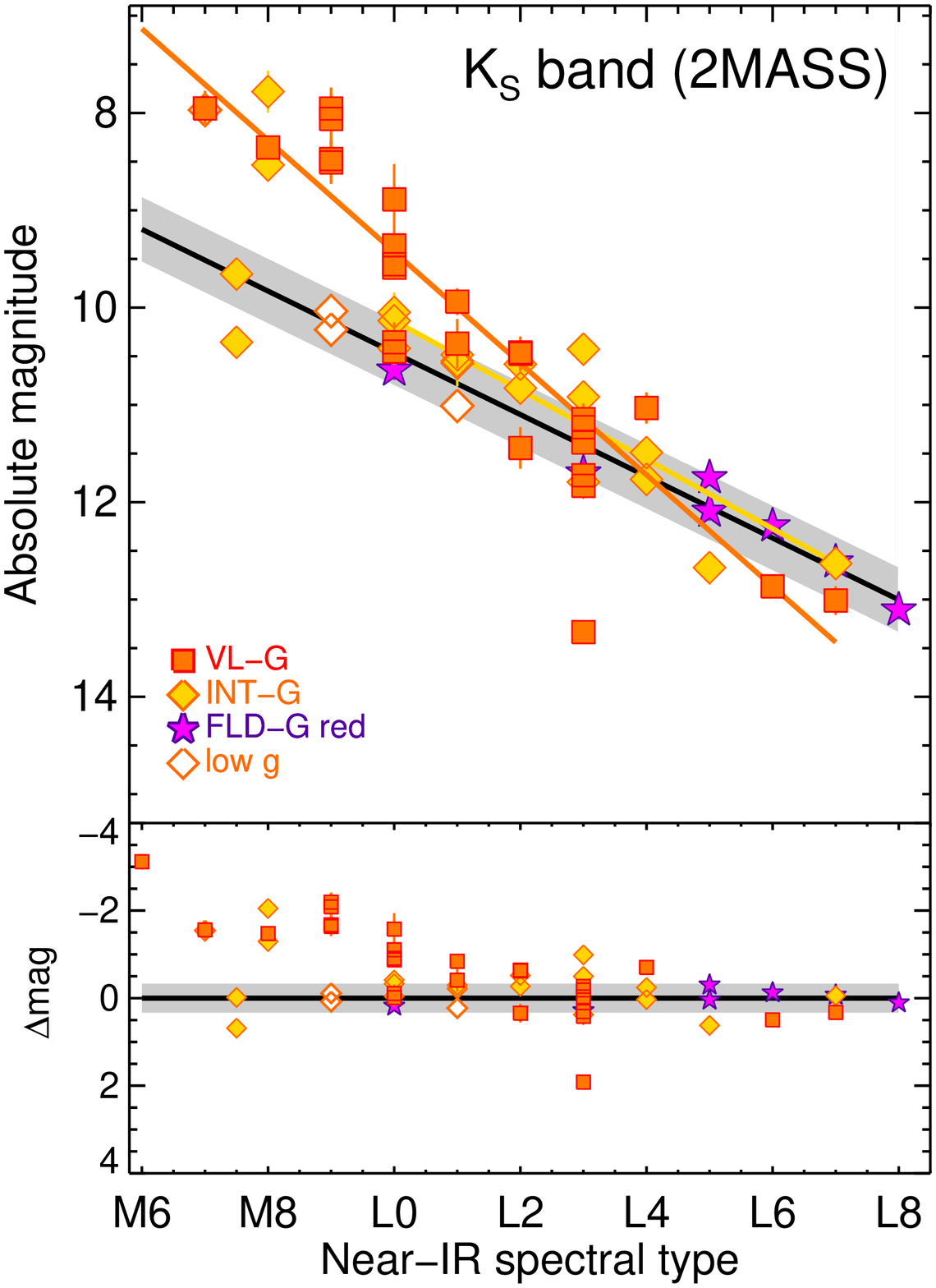}
  \vskip 6ex
  \caption{\normalsize 2MASS $JH\Ks$ absolute magnitudes for all known
    field late-M and L dwarfs with parallaxes, near-IR spectral types,
    and spectroscopic evidence of youth. (About 80\% of the objects
    are from this paper's CFHT parallaxes.) 
    Spectral types and gravity classifications are mostly based on the
    \citet{2013ApJ...772...79A} system (see
    Table~\ref{table:sample}). Objects labeled as ``low~g''
    (HR~6037~Ba and~Bb [both M9] and $\beta$~Pic~b [L1]) appear young
    spectroscopically, but their spectra have insufficient wavelength
    coverage for a formal gravity classification. Objects labeled as
    ``\fldg~red'' have field gravities but unusually red colors
    highlighted in the literature. %
    The errors on the absolute magnitudes are typically smaller than
    the plotting symbol. The thick black line shows our fit for field
    (old) ultracool dwarfs, and the grey swath shows the 1$\sigma$
    scatter about that fit. The colored lines show the linear fits for
    the \vlg\ and \intg\ samples (Table~\ref{table:coeff}); the 
    anomalously faint L3 object 2MASS~J1207$-$3932b has been excluded
    from the line fits. %
    Four low-gravity objects in our sample are not plotted here and
    not used for the line fits: three have low-S/N parallaxes due to
    their large distances ($S/N<4$, $d\gtrsim250$~pc; GJ~3276,
    2MASS~J0557$-$1359, and 2MASSI~J0619$-$2903), and the fourth
    object (2MASSI~J0435$-$1414, M7~\vlg) appears to have a
    circumstellar disk and its high luminosity places it off the plots
    (see Appendix). 
    The lower panels show difference with respect to the field sequence.
    Note that the extent of the $y$~axis is the same for all our
    plots of absolute magnitude versus spectral type 
    (Figures~\ref{fig:absmag-2mass}, \ref{fig:absmag-mko},
    \ref{fig:absmag-wise}, \ref{fig:absmag-optical}, and
    \ref{fig:absmag-misc}).
    \label{fig:absmag-2mass}}
\end{figure}
\end{landscape}

\begin{figure}
  \hbox{
    \hskip 0.2in
    \includegraphics[width=2.7in]{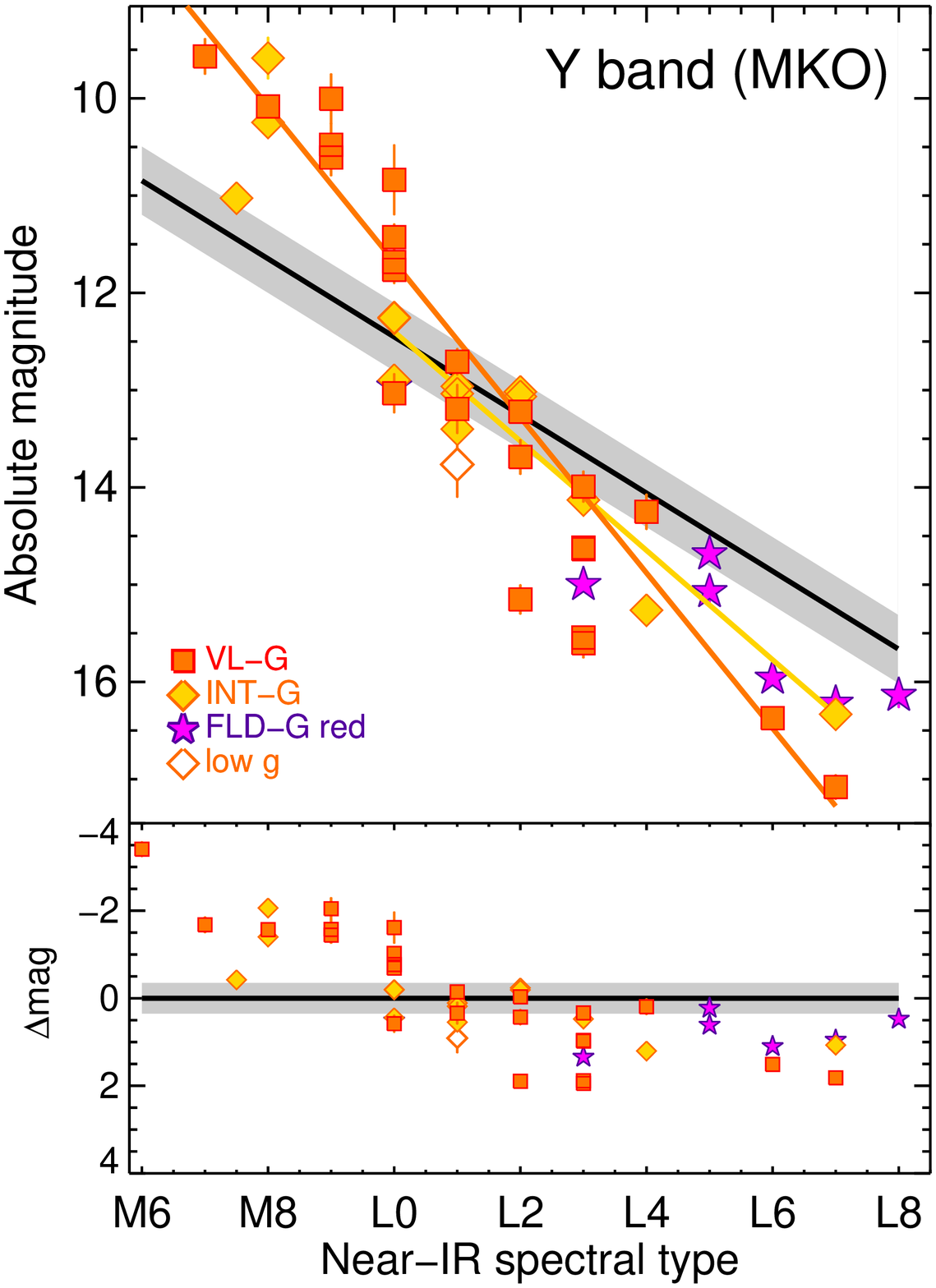}
    \hskip 0.5in
    \includegraphics[width=2.7in]{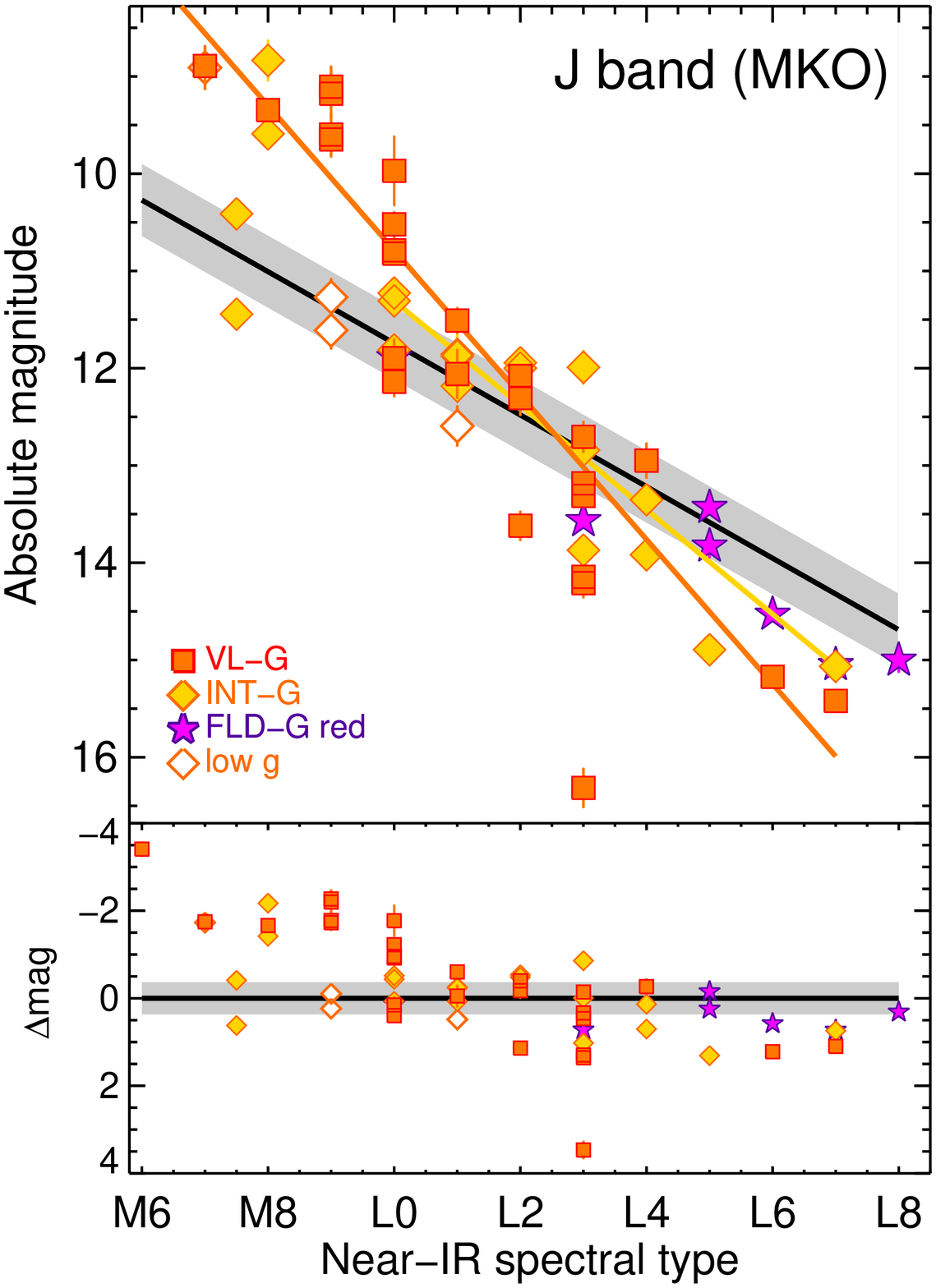}
  }
  \vskip 0.6in
  \hbox{
    \hskip 0.2in
    \includegraphics[width=2.7in]{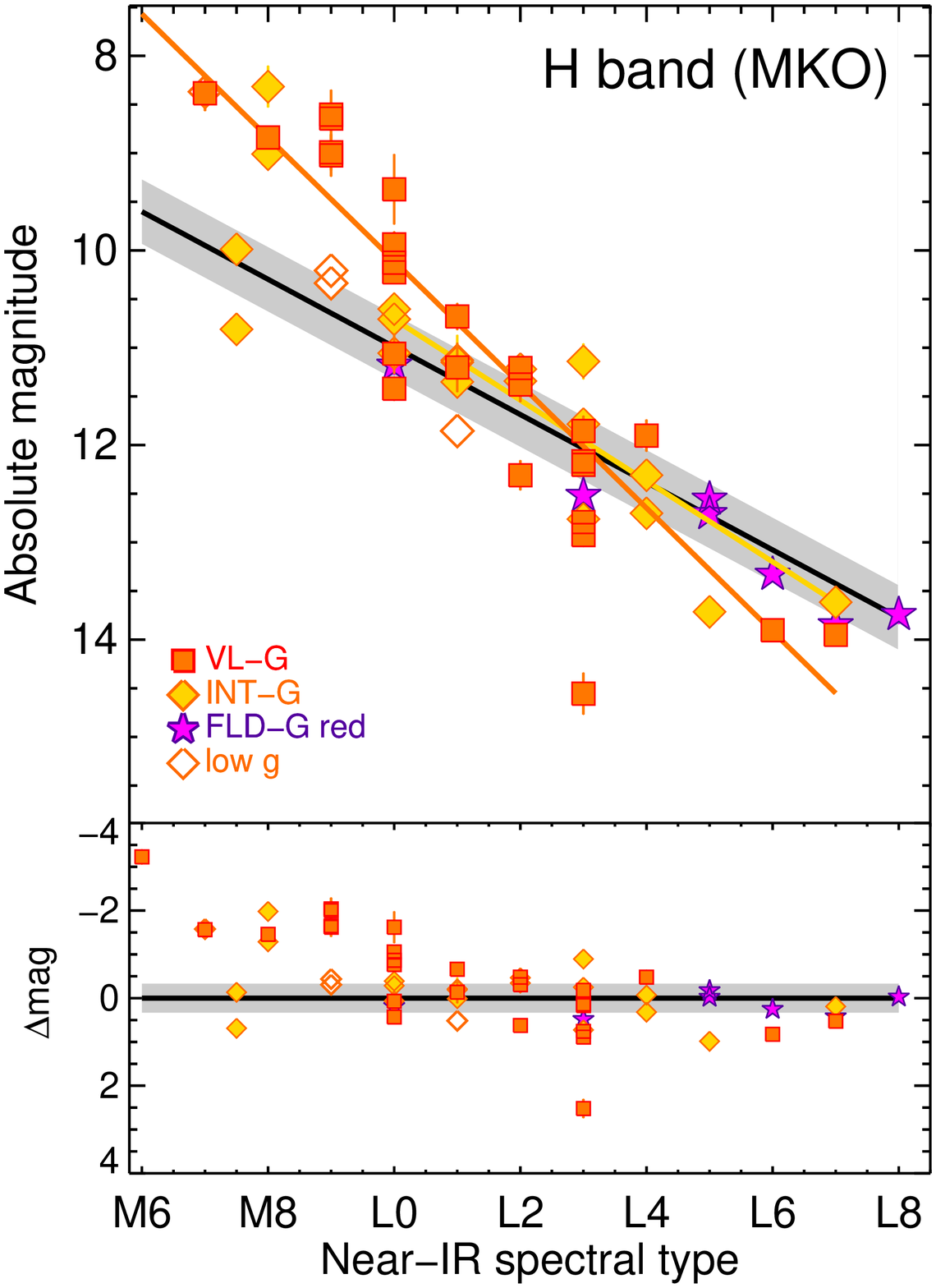}
    \hskip 0.5in
    \includegraphics[width=2.7in]{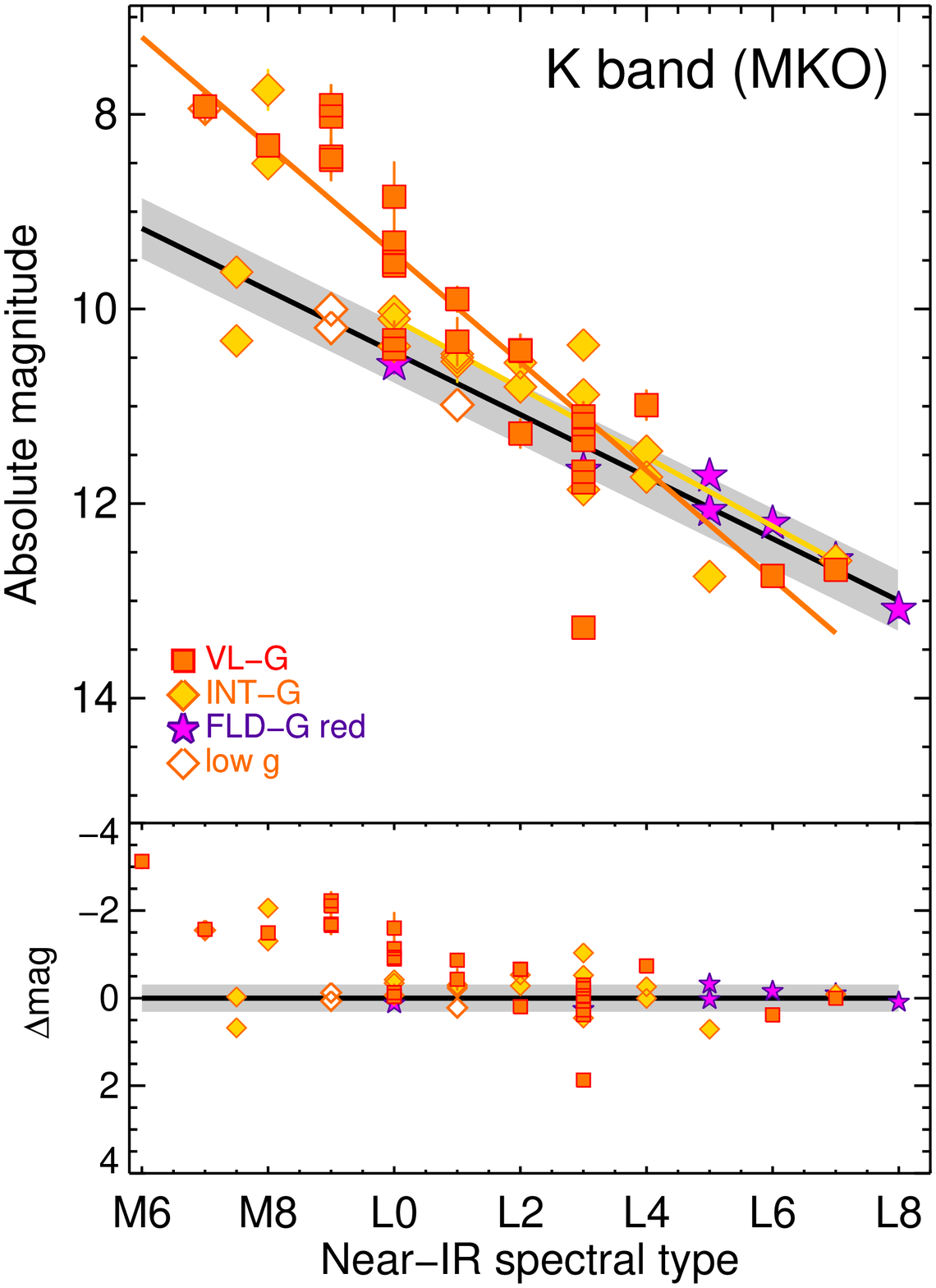}
  }
  \vskip 6ex
  \caption{\normalsize Absolute magnitudes in the MKO $YJHK$ filters as a
    function of near-IR spectral type for our young sample. See
    Figure~\ref{fig:absmag-2mass} caption for further
    details.\label{fig:absmag-mko}}
\end{figure}

\begin{figure}
  \begin{center}
    \hbox{
      \includegraphics[width=3in]{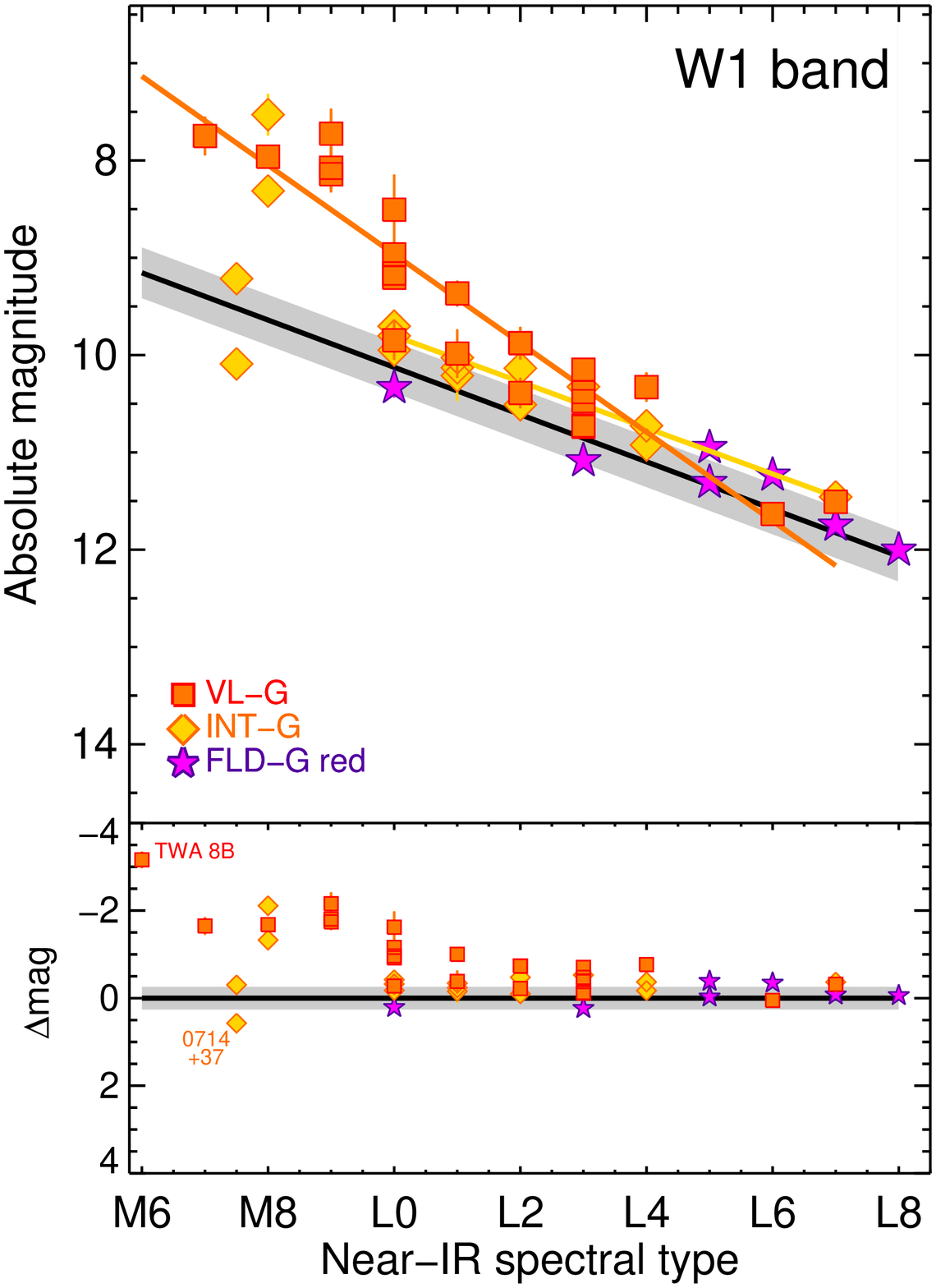}
      \hskip 0.5in
      \includegraphics[width=3in]{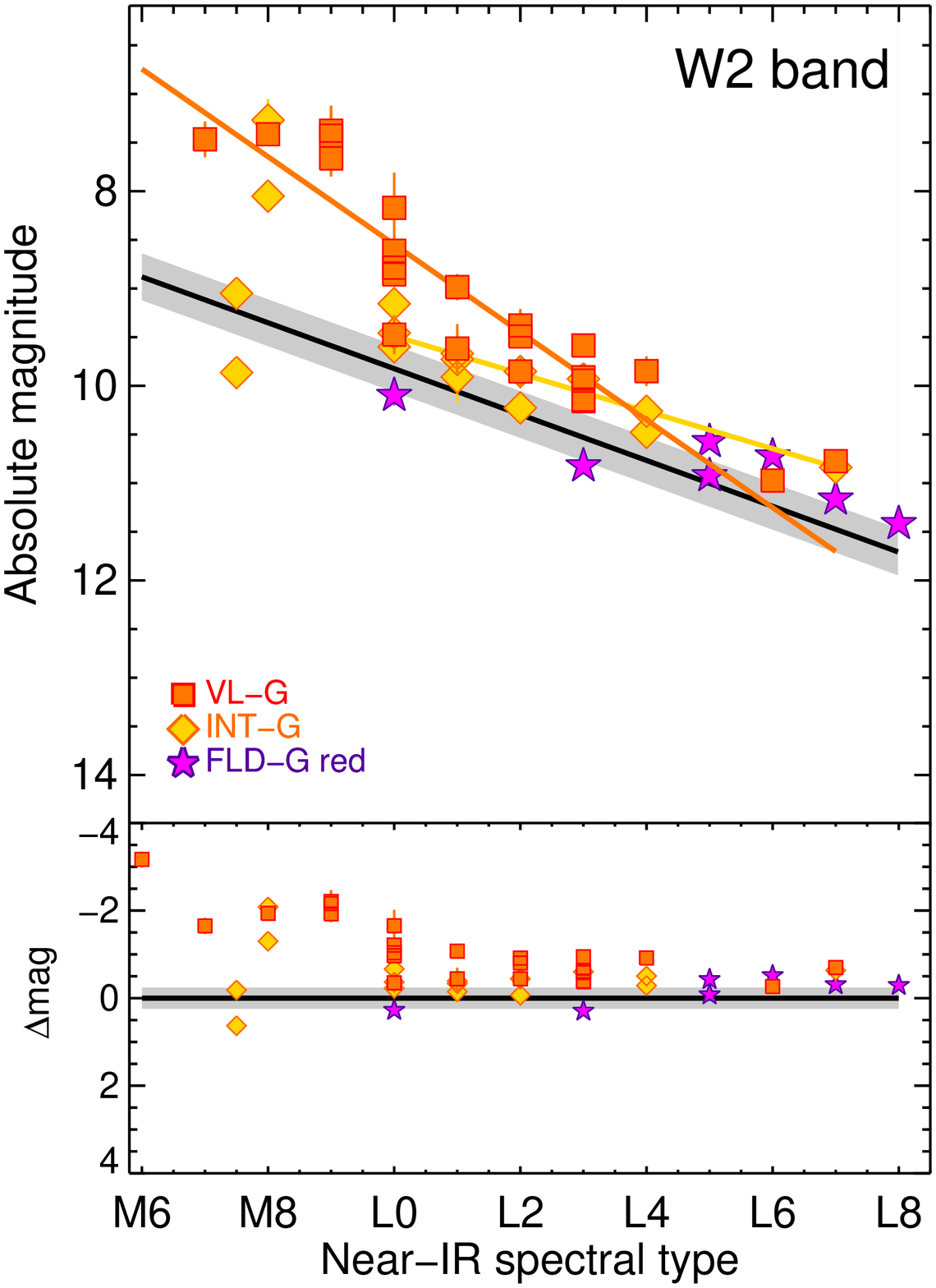}
    }
  \end{center}
  \vskip 6ex
  \caption{\normalsize Absolute magnitudes in the \WISE\ $W1$ and $W2$
    filters as a function of spectral type for our young sample. See
    Figure~\ref{fig:absmag-2mass} caption for further
    details. \label{fig:absmag-wise}}
\end{figure}

\begin{landscape}
\begin{figure}
  \includegraphics[width=2.4in]{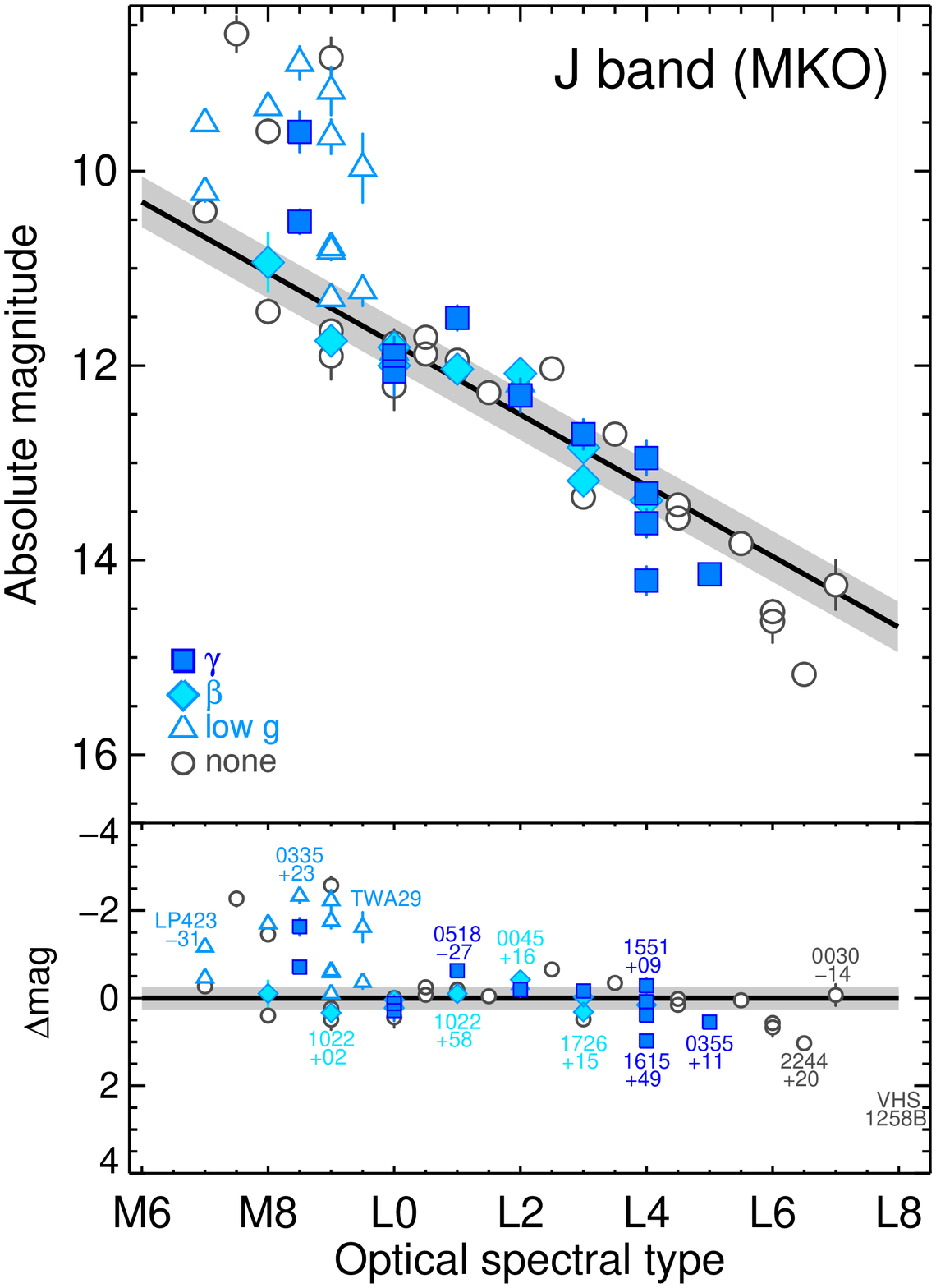}
  \hskip 0.5in
  \includegraphics[width=2.4in]{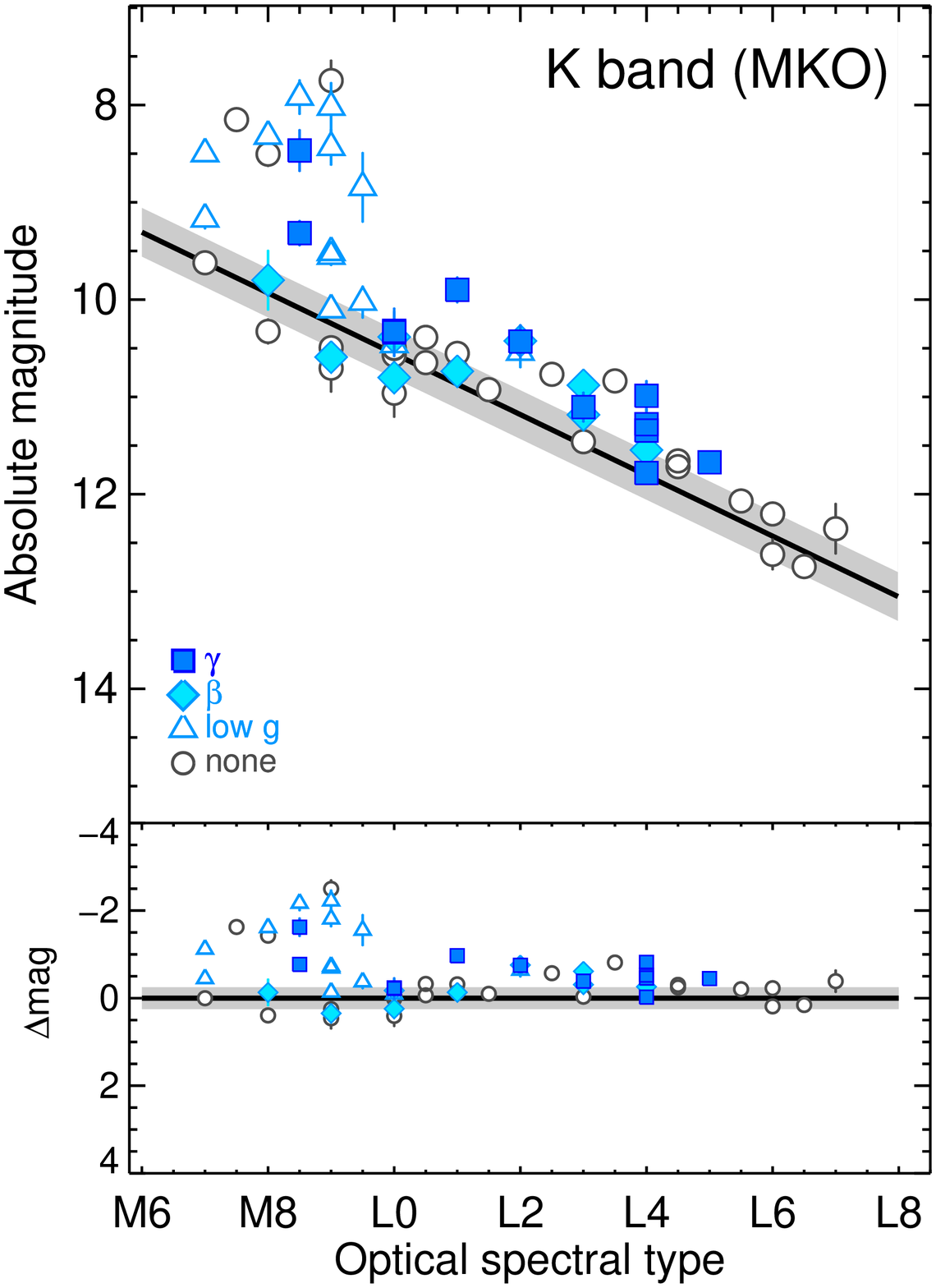}
  \hskip 0.5in
  \includegraphics[width=2.4in]{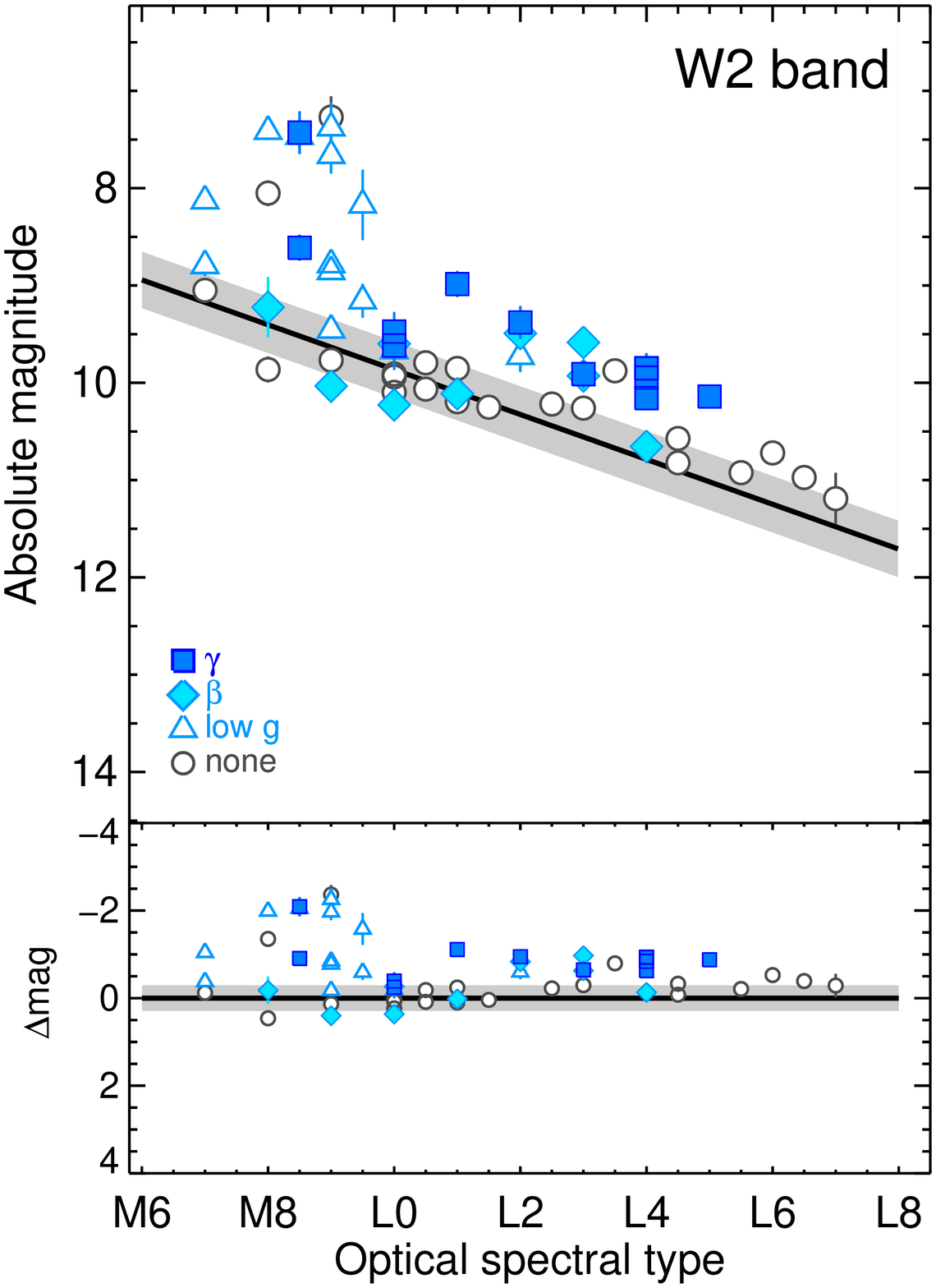}
  \vskip 6ex
  \caption{\normalsize Absolute magnitudes in the $J_{MKO}$, $K_{MKO}$,
    and $W2$ bandpasses, as a function of optical spectral type and
    gravity classifications, with the latter following the system of \citet{2009AJ....137.3345C}.
    Objects labeled as ``low~g'' are reported in the literature as being
    low gravity but do not have a formal gravity classification.  Objects
    labeled as ``none'' have optical spectral types but no gravity
    classification.  See Figure~\ref{fig:absmag-2mass} caption for further
    details. \label{fig:absmag-optical}}
\end{figure}
\end{landscape}

\begin{landscape}
\begin{figure}
  \includegraphics[width=2.4in]{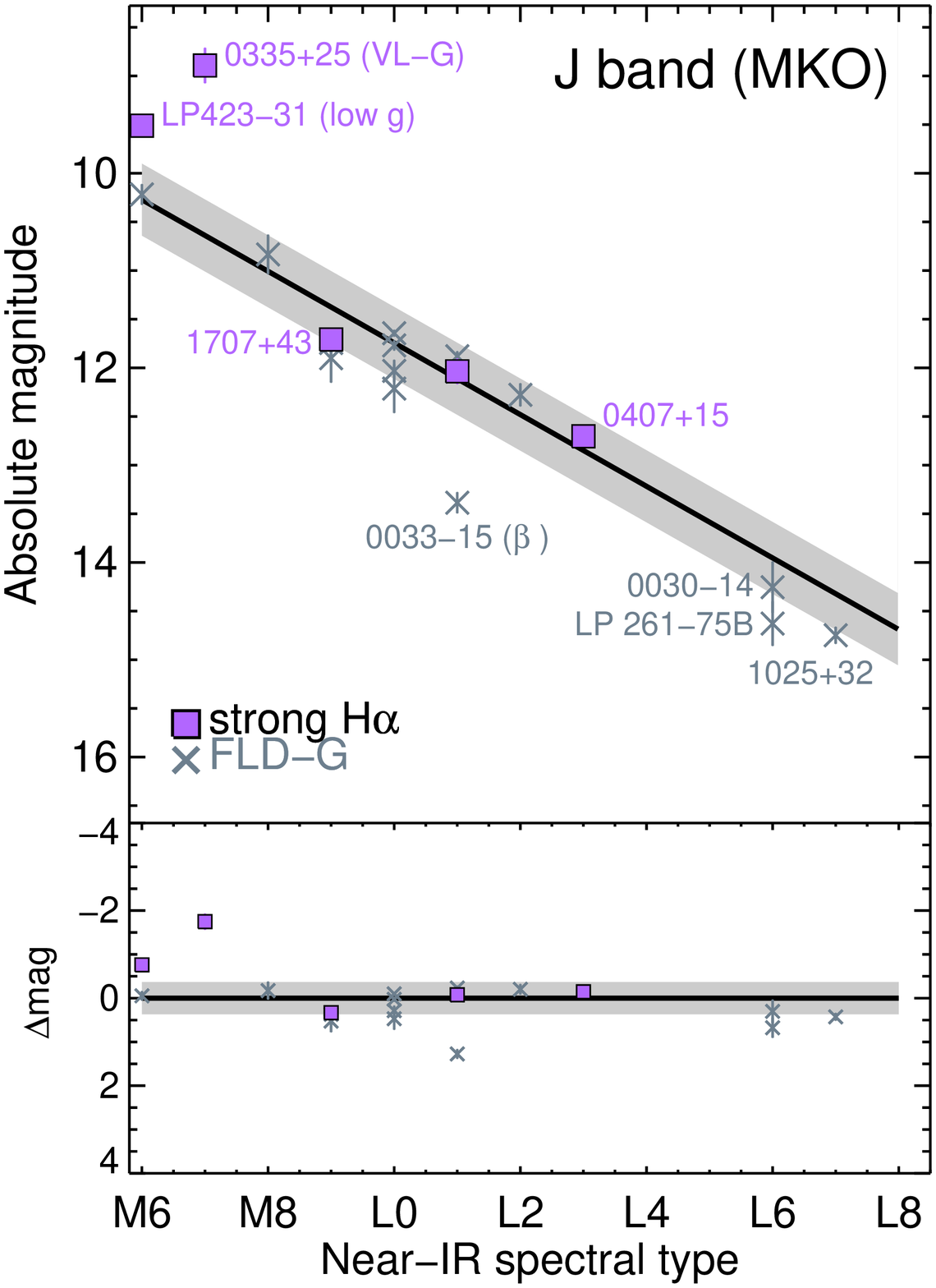}
  \hskip 0.5in
  \includegraphics[width=2.4in]{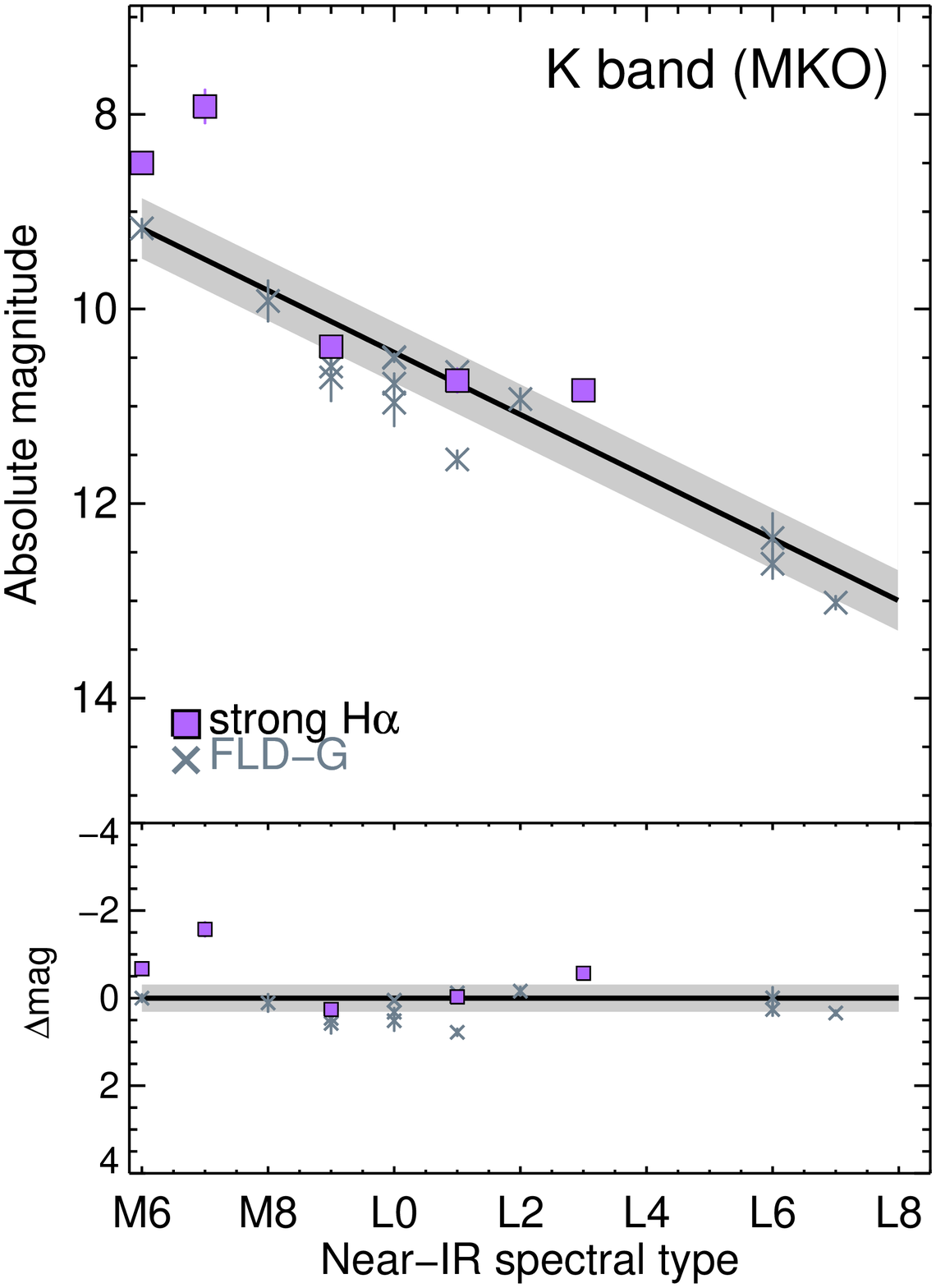}
  \hskip 0.5in
  \includegraphics[width=2.4in]{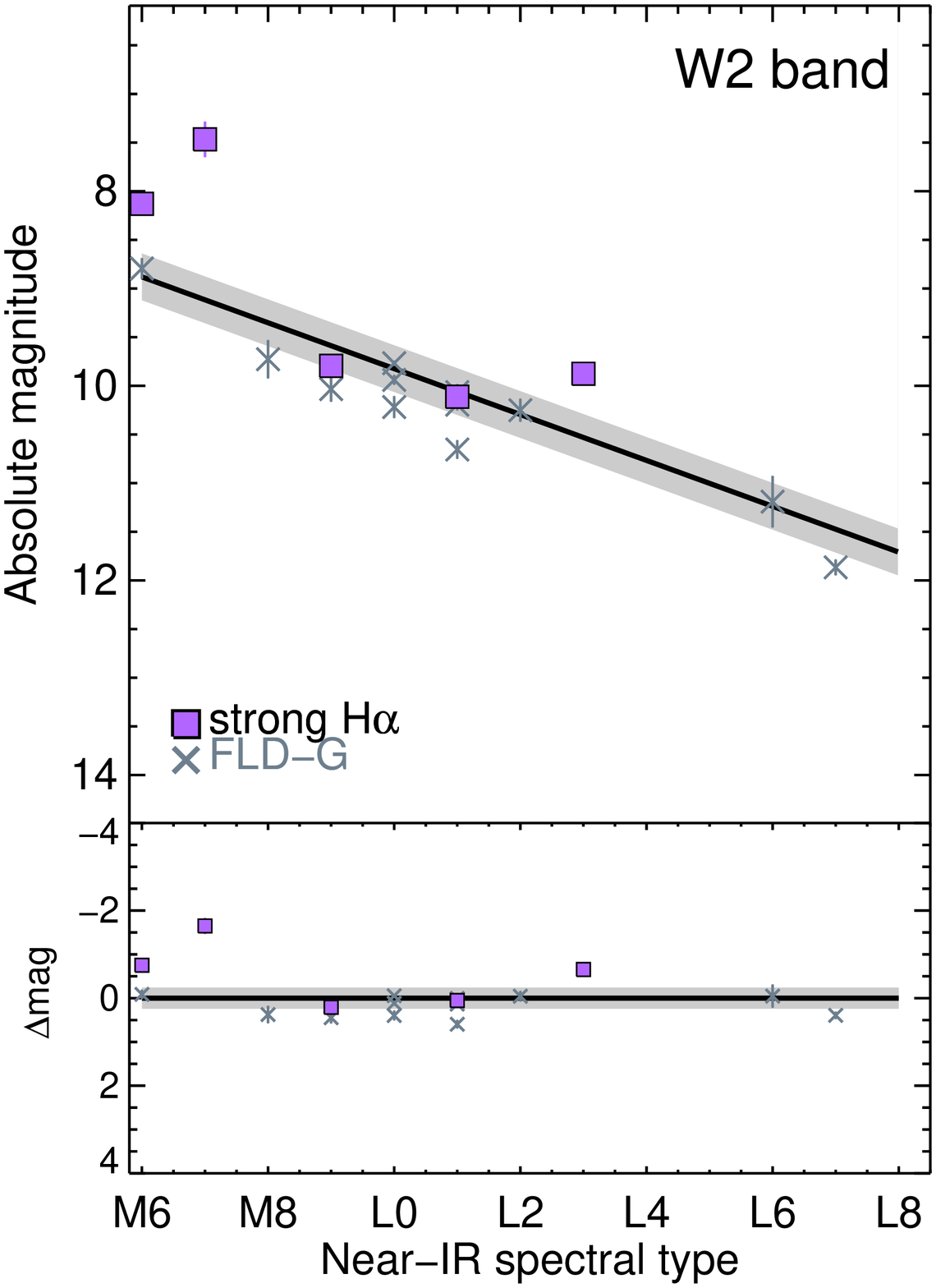}
  \vskip 6ex
  \caption{\normalsize Absolute magnitudes in the $J_{MKO}$, $K_{MKO}$,
    and $W2$ bandpasses for our CFHT targets with \fldg\ gravity
    classifications or with strong H$\alpha$ emission. Note that
    2MASS~J0033$-$1521 is classified as L1~\fldg\ in the near-IR but
    L4$\beta$ in the optical. See Figure~\ref{fig:absmag-2mass} caption
    for further details. (The H$\alpha$ emitting L~dwarf
    2MASS~J1315$-$2649 is not shown here as it does not have component
    spectral types on the \citealp{2013ApJ...772...79A}
    system.) \label{fig:absmag-misc}}
\end{figure}
\end{landscape}

\begin{figure}
  \hbox{
    \includegraphics[width=3.3in]{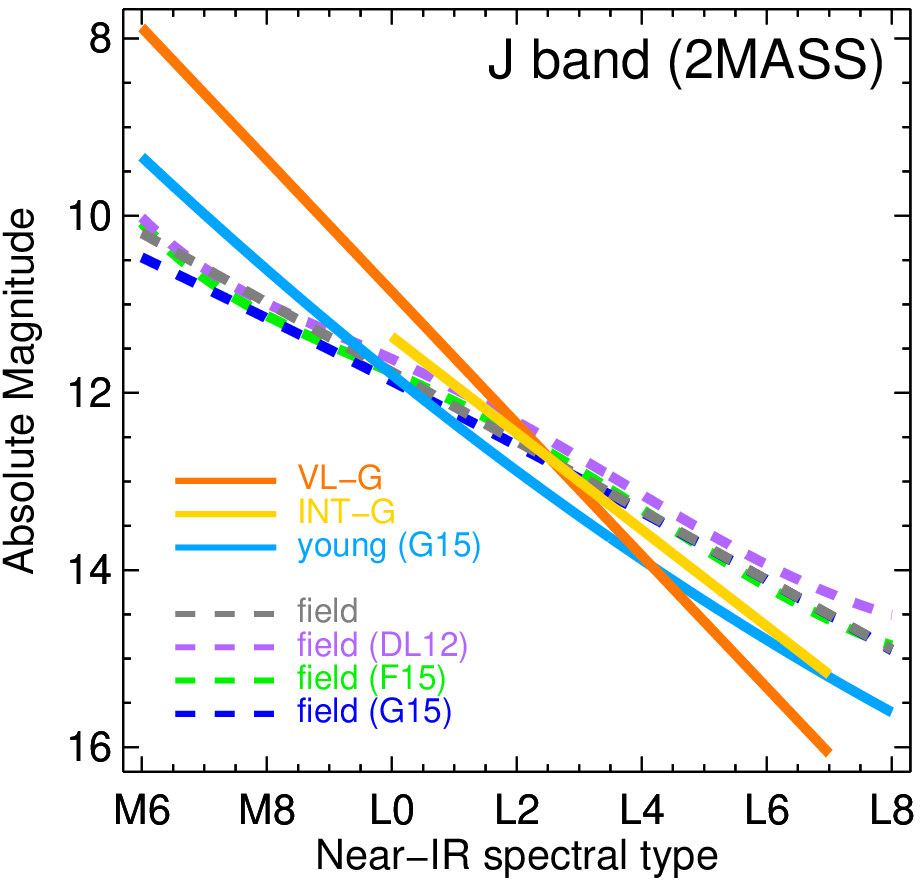}
    \includegraphics[width=3.3in]{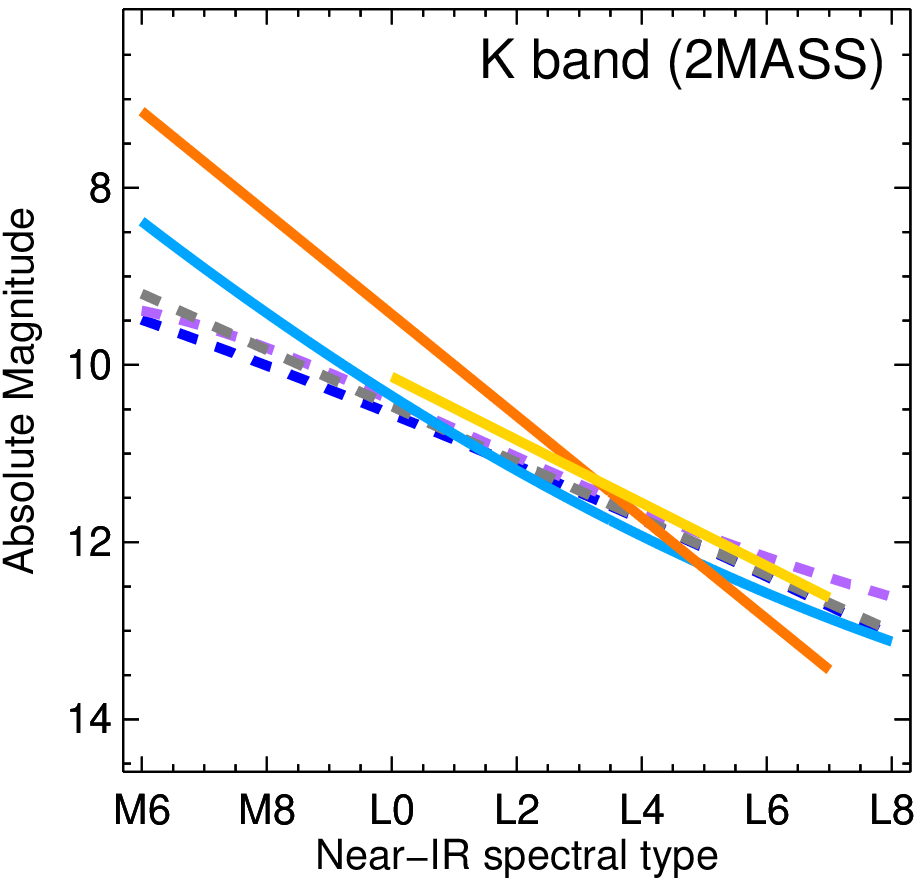}
  }
  \vskip -0.2in
  \hbox{
    \includegraphics[width=3.3in]{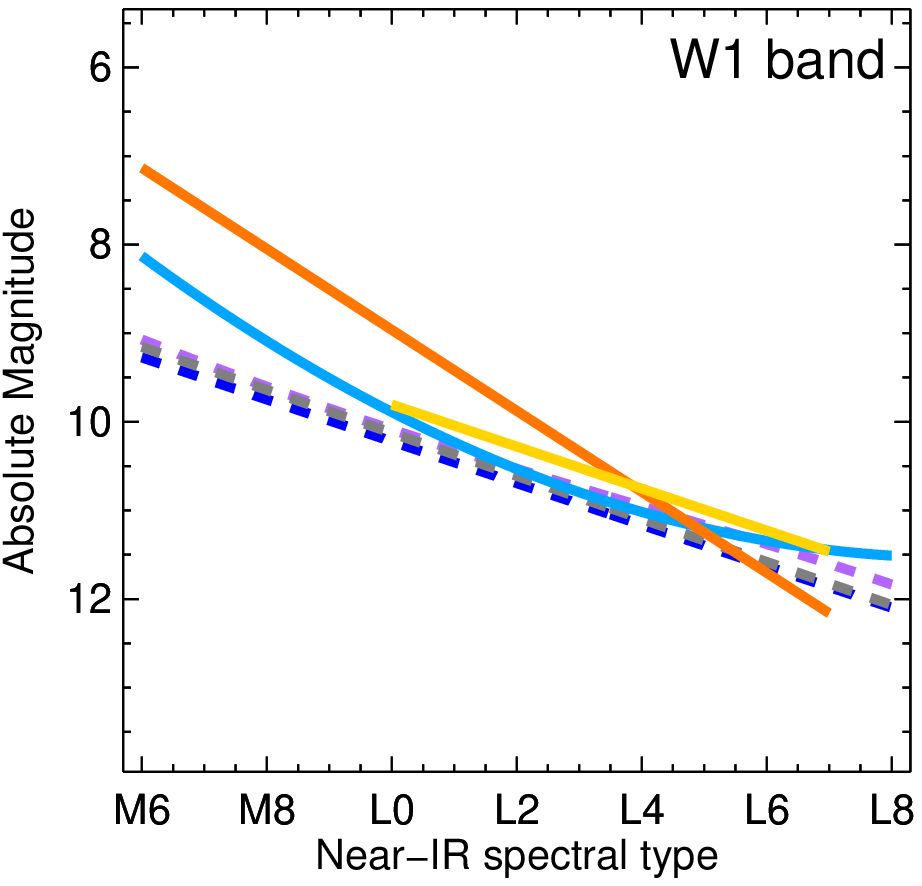}
    \includegraphics[width=3.3in]{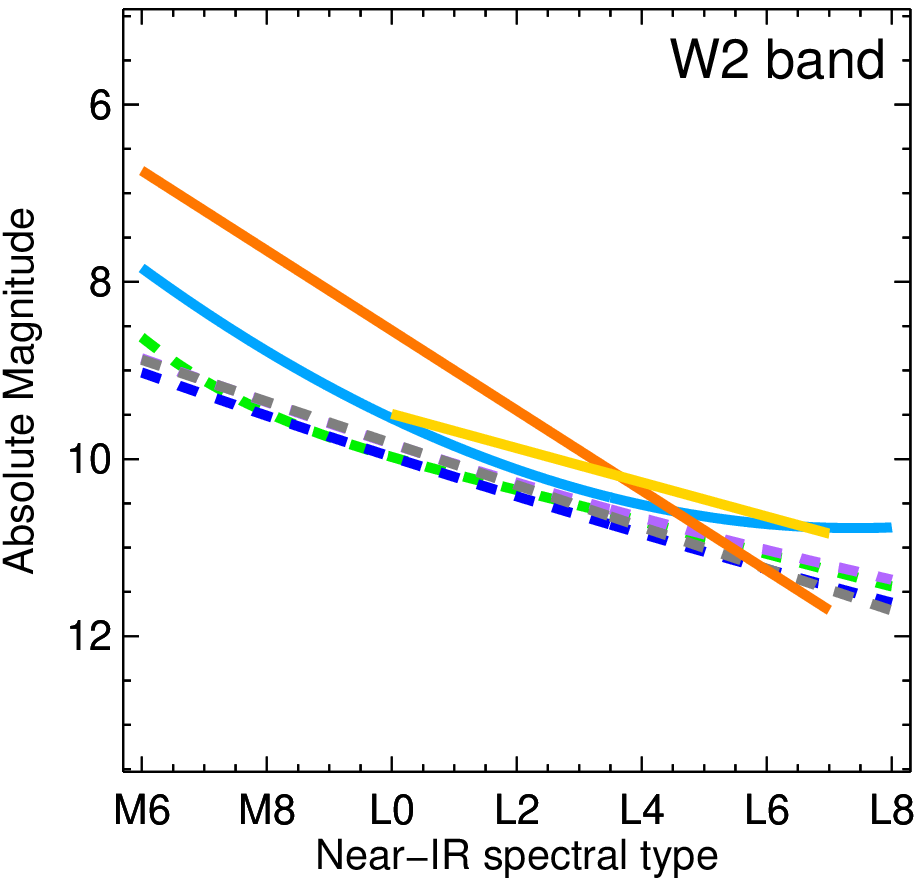}
  }
  \caption{\normalsize Comparison of polynomial fits for absolute
    magnitude as a function of near-IR spectral type.  
    Fits for low-gravity objects are plotted as solid lines, with the
    \vlg\ (solid orange) and \intg\ (solid yellow) relations from this
    work and the young relation from \citet{2015ApJS..219...33G} also
    plotted (solid medium blue).
    Fits for field objects are plotted as dashed lines, showing the
    ones for late-M and L dwarfs from this work (dashed grey) and
    \citet[][dashed dark blue]{2015ApJS..219...33G}, and the ones for
    all ultracool dwarfs from \citet[][dashed
    purple]{2012ApJS..201...19D} and \citet[][dashed
    green]{2015ApJ...810..158F}.
    The field dwarf relations are all in good agreement, while the
    low-gravity relations differ.\label{fig:absmag-compare}}
\end{figure}

\begin{figure}

  \centerline{
    \includegraphics[width=3.3in,angle=0]{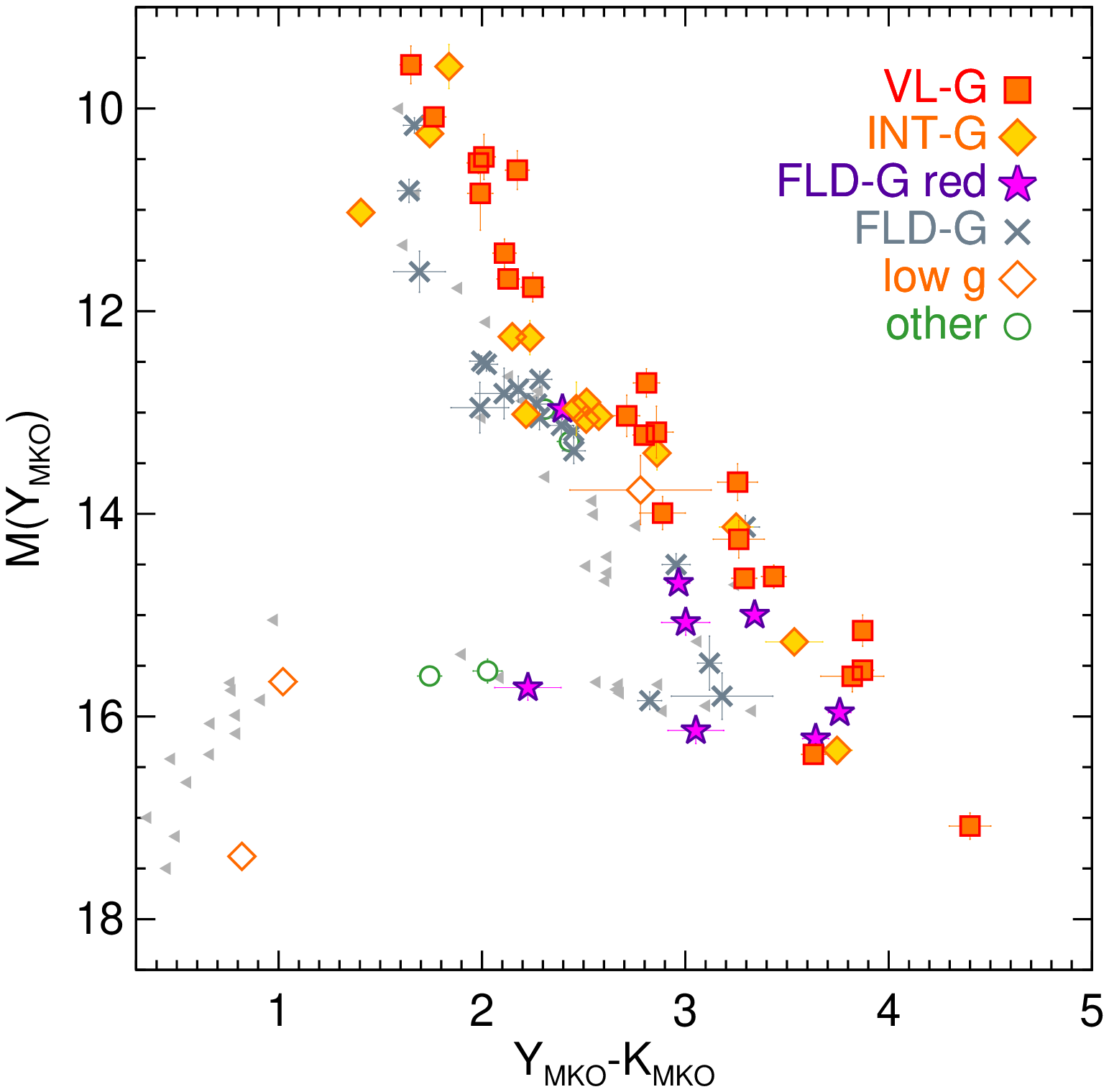}
    \includegraphics[width=3.3in,angle=0]{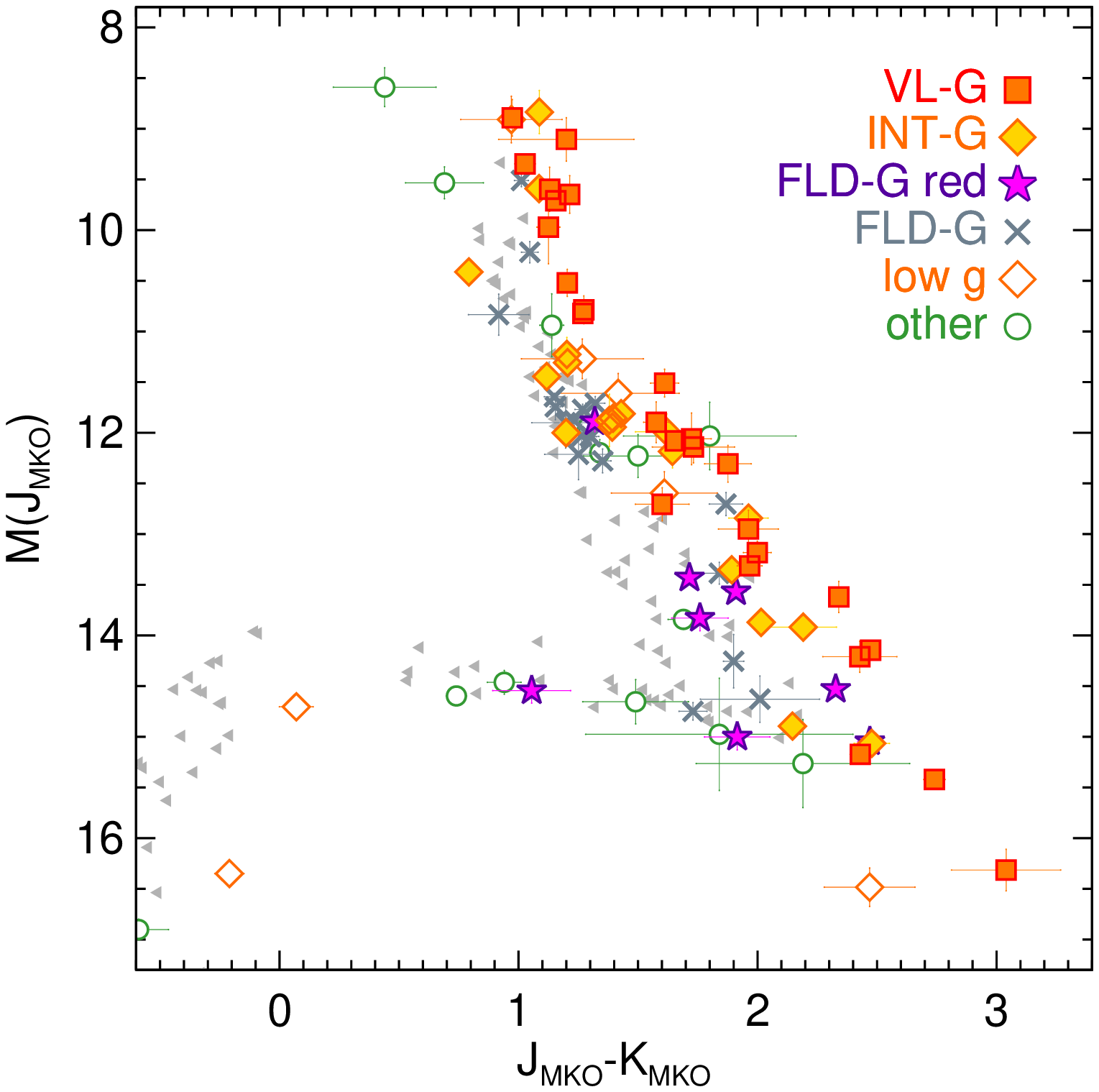}
  }
  \centerline{
    \includegraphics[width=3.3in,angle=0]{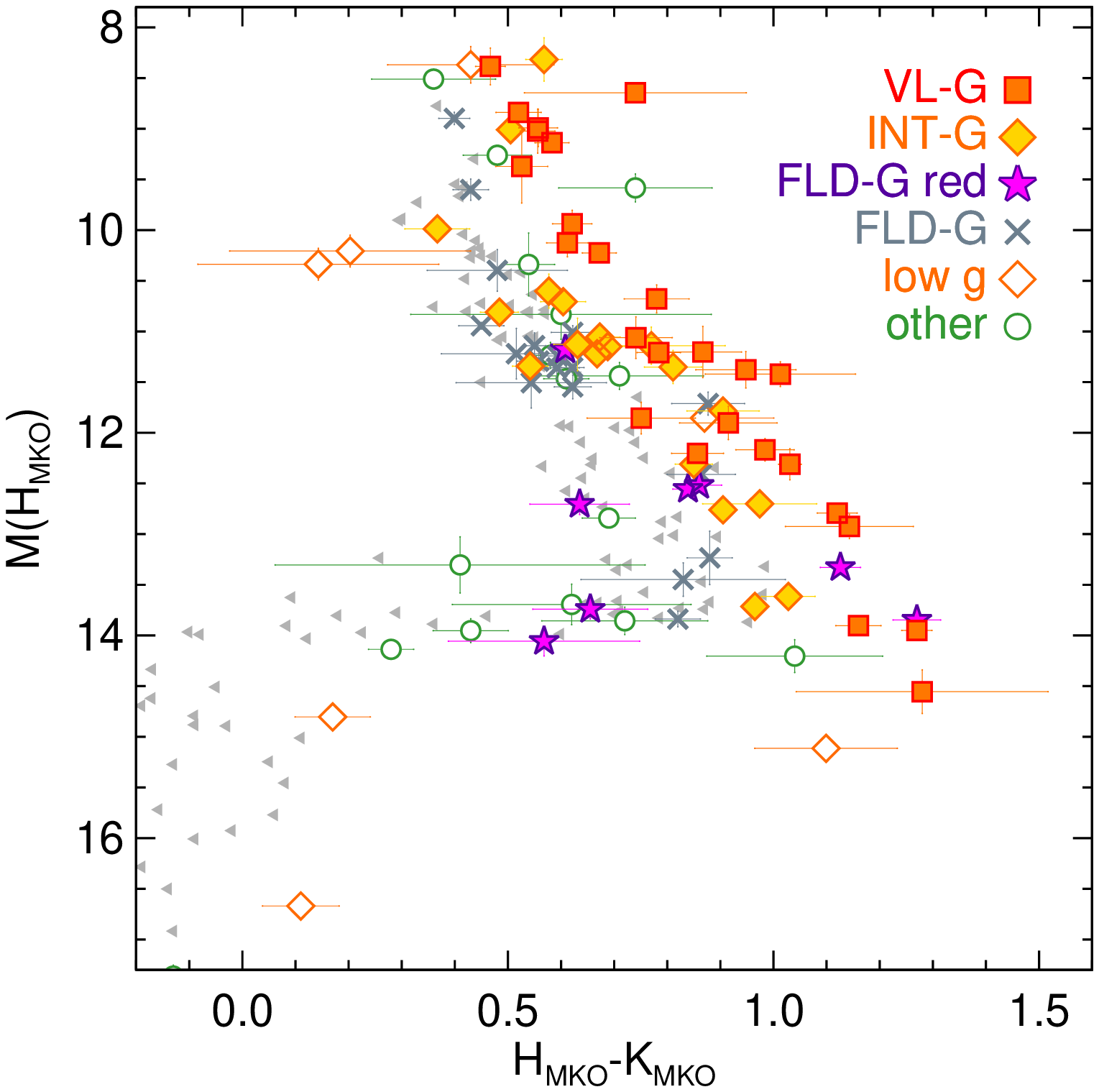}
    \includegraphics[width=3.3in,angle=0]{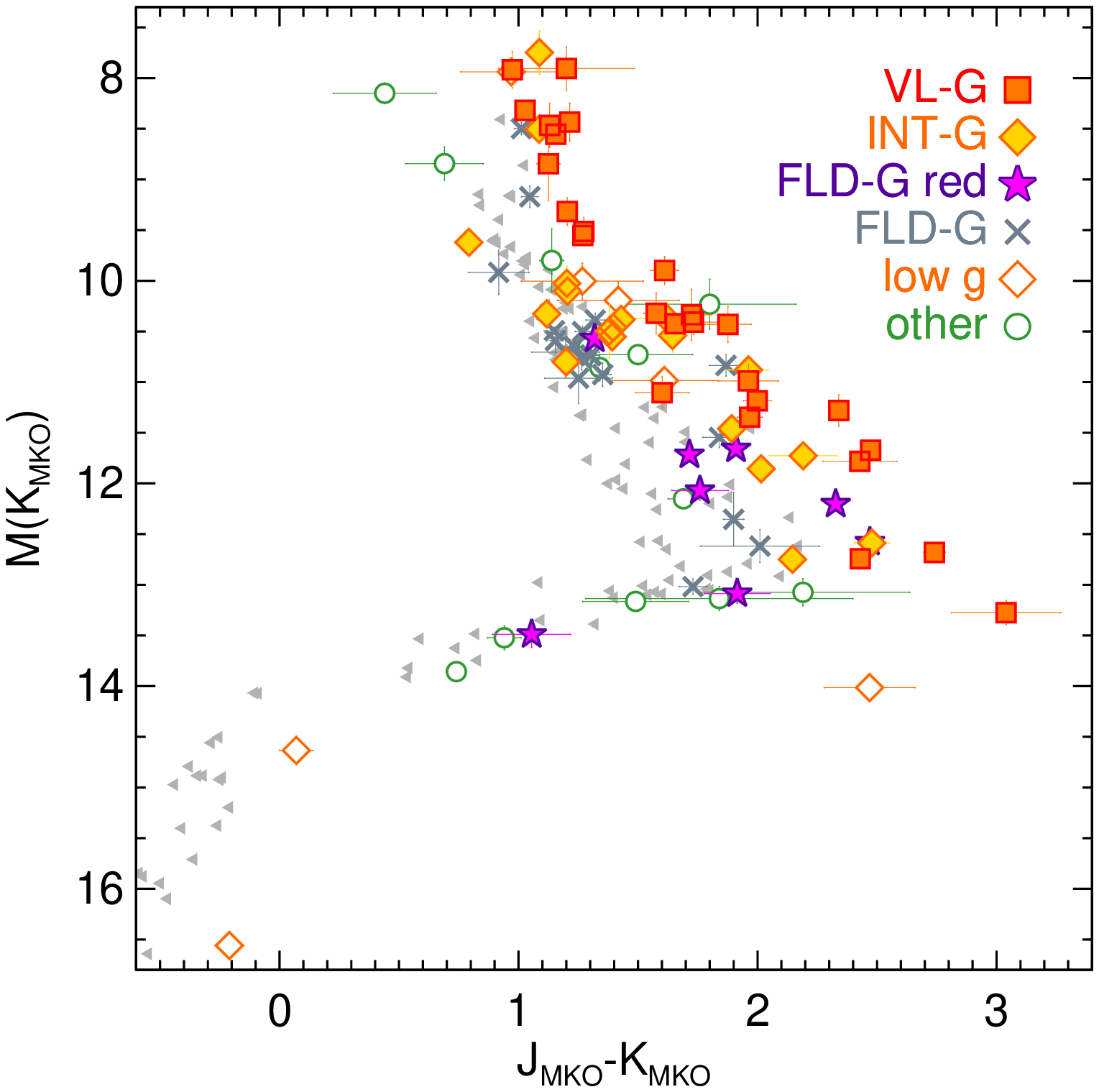}
  }
  \caption{\small Color--magnitude diagrams showing our full sample of
    potentially young ultracool ($\ge$M6) dwarfs with parallaxes from
    CFHT or the literature. All CMD plots span 9.5\,mag in absolute
    magnitude.  %
    Objects are plotted according to their \citet{2013ApJ...772...79A}
    gravity classification (\vlg, \intg\ and \fldg) when
    available. Objects having low-gravity IR spectra but without a
    formal classification are labeled ``low g.''  
    Field-gravity objects noted as being unusually red in the near-IR
    are labeled ``\fldg\ red.'' Open circles represent objects without
    any spectral gravity information (``other''); these are mostly
    companions and objects outside the M6--L7 spectral type range
    covered by the \citet{2013ApJ...772...79A} gravity system.  
    For comparison, normal field objects (as described in
      Section~\ref{sec:absmags}) are shown as small gray triangles
    (from \citealp{2012ApJS..201...19D}, updated at
    \hbox{\protect\url{http://www.as.utexas.edu/~tdupuy/plx}}).
    We only plot (old) field objects with apparent magnitude errors
    $<$0.10\,mag in the two bands used to compute the color 
    and absolute magnitude errors $<$0.10\,mag, and for clarity we do not plot error bars.
    For all the CMD plots, we exclude the young objects with ${\rm S/N} < 5$ parallaxes
    (GJ~3276, 2MASS~J0557$-$1359, and 2MASS~J0619$-$2903; all likely to
    be very young, as discussed in the Appendix) and also 
    LP~261-75A (M4.5), given its earlier spectral type.
    Two very bright objects are off all the CMD plots
    (2MASS~J0435$-$1414 [M7~\vlg] and TWA~8B [M6~\vlg]), and likewise the companion
    Gl~504b is too faint to appear.
    \label{fig:cmd1}}
\end{figure}

\addtocounter{figure}{-1}   
\begin{figure}

  \centerline{
    \includegraphics[width=3.3in,angle=0]{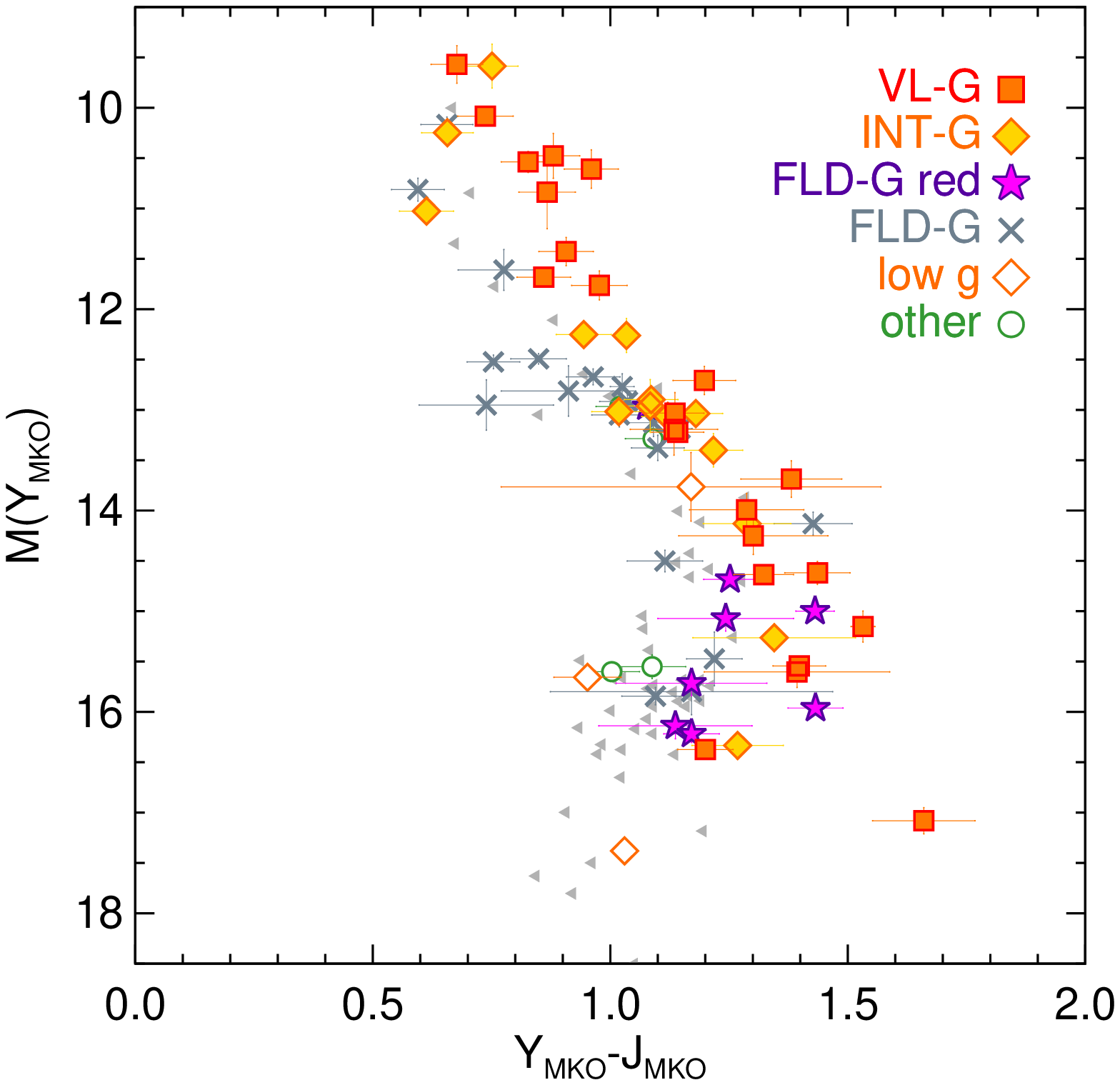}
    \includegraphics[width=3.3in,angle=0]{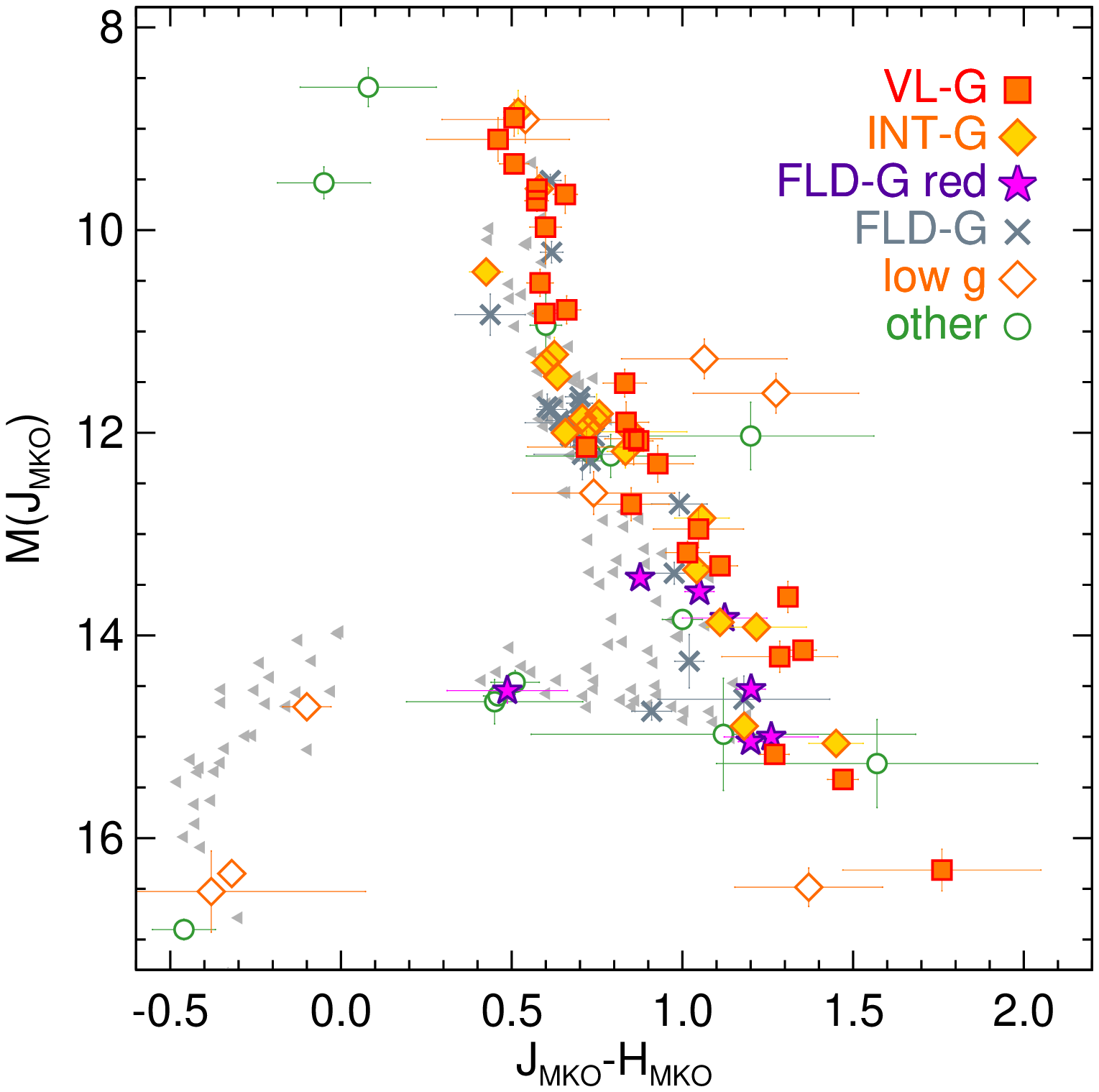}
  }
  \centerline{
    \includegraphics[width=3.3in,angle=0]{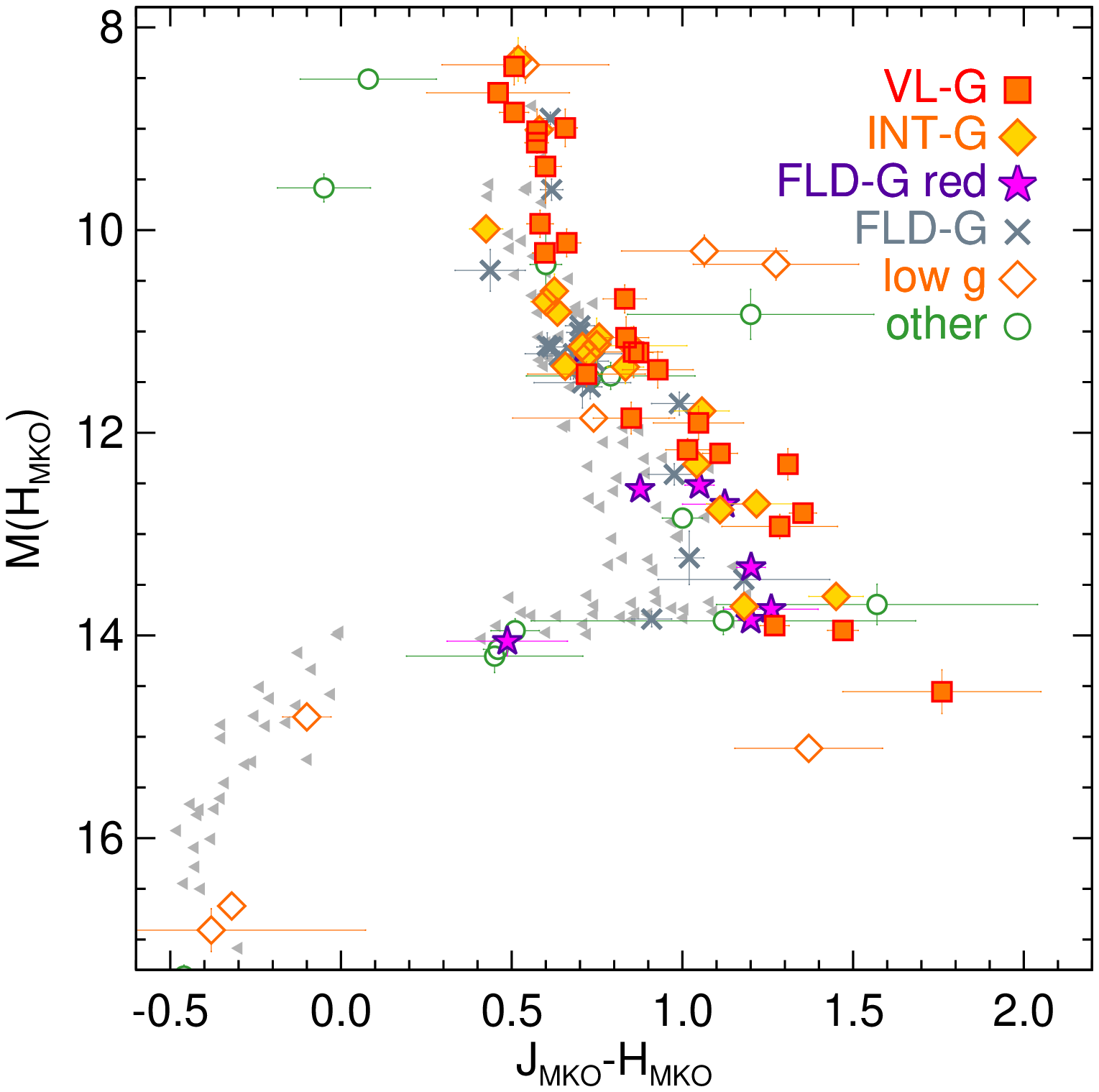}
    \includegraphics[width=3.3in,angle=0]{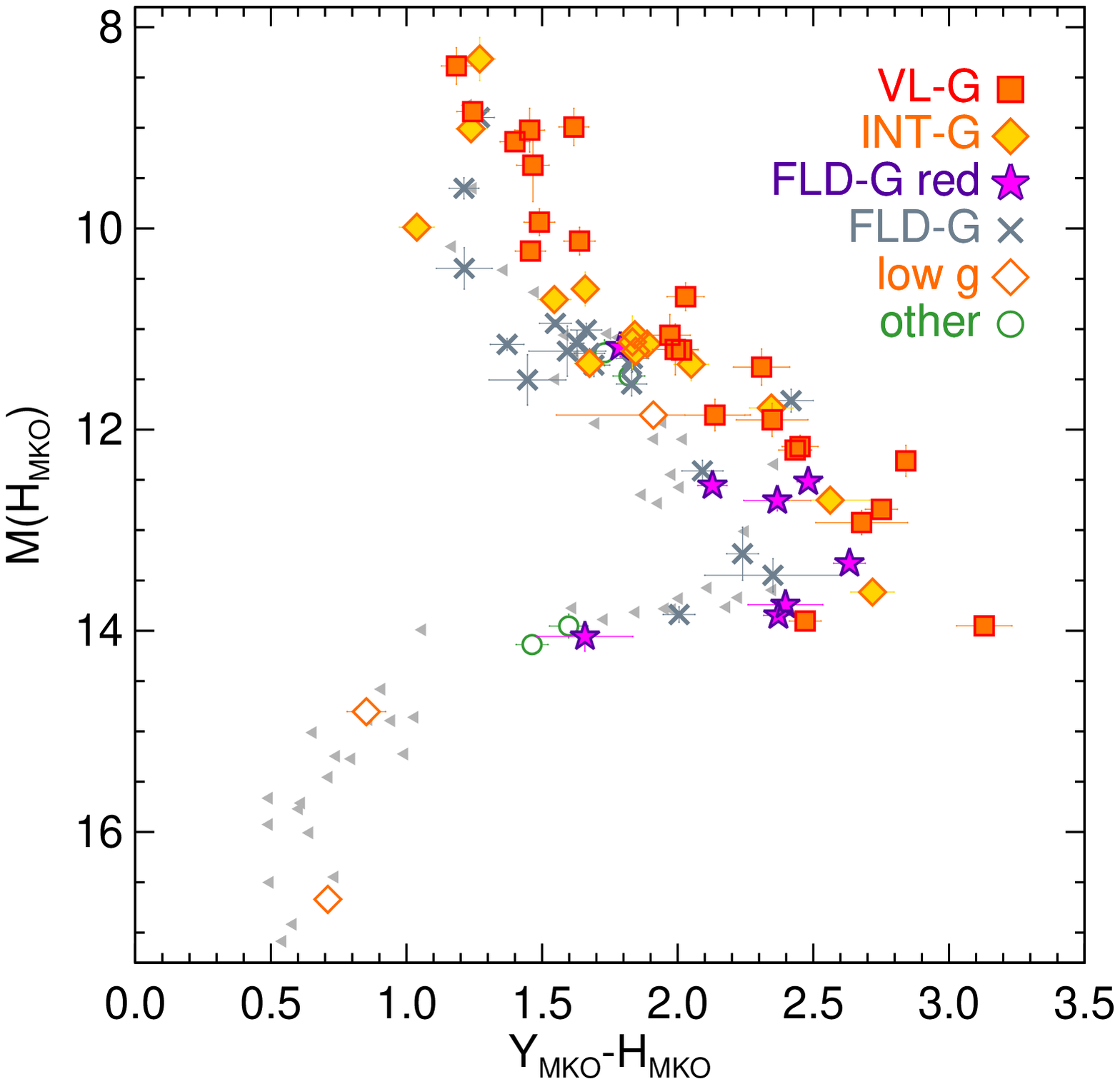}
  }
  \caption{\normalsize (continued) The faintest, bluest objects on the
    CMD plots involving $J-H$ are Ross~458C and 51~Eri~b (the one with
    the larger color uncertainties).} 
\end{figure}

\clearpage

\begin{figure}

  \centerline{
    \includegraphics[width=3.3in,angle=0]{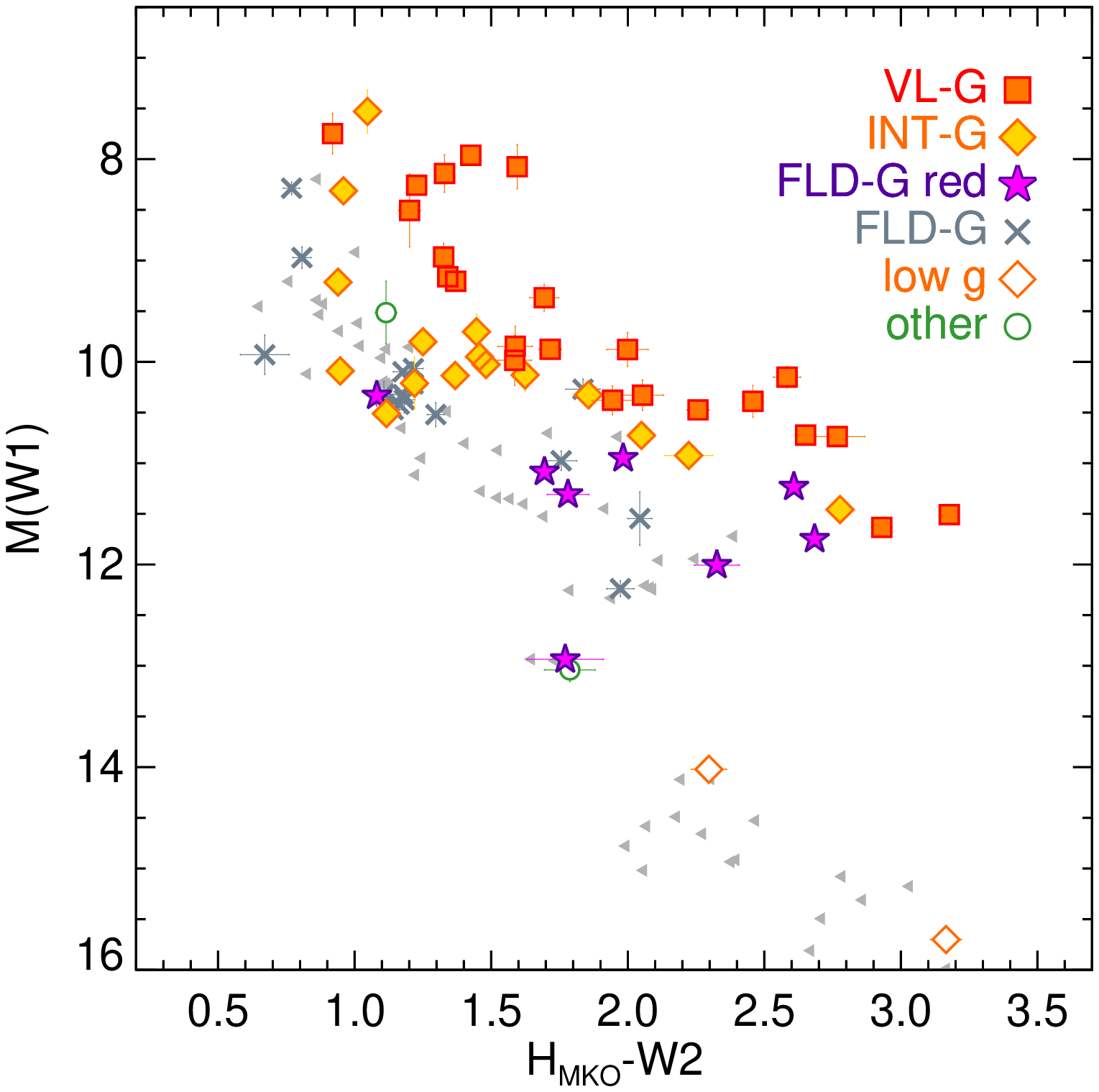}
    \includegraphics[width=3.3in,angle=0]{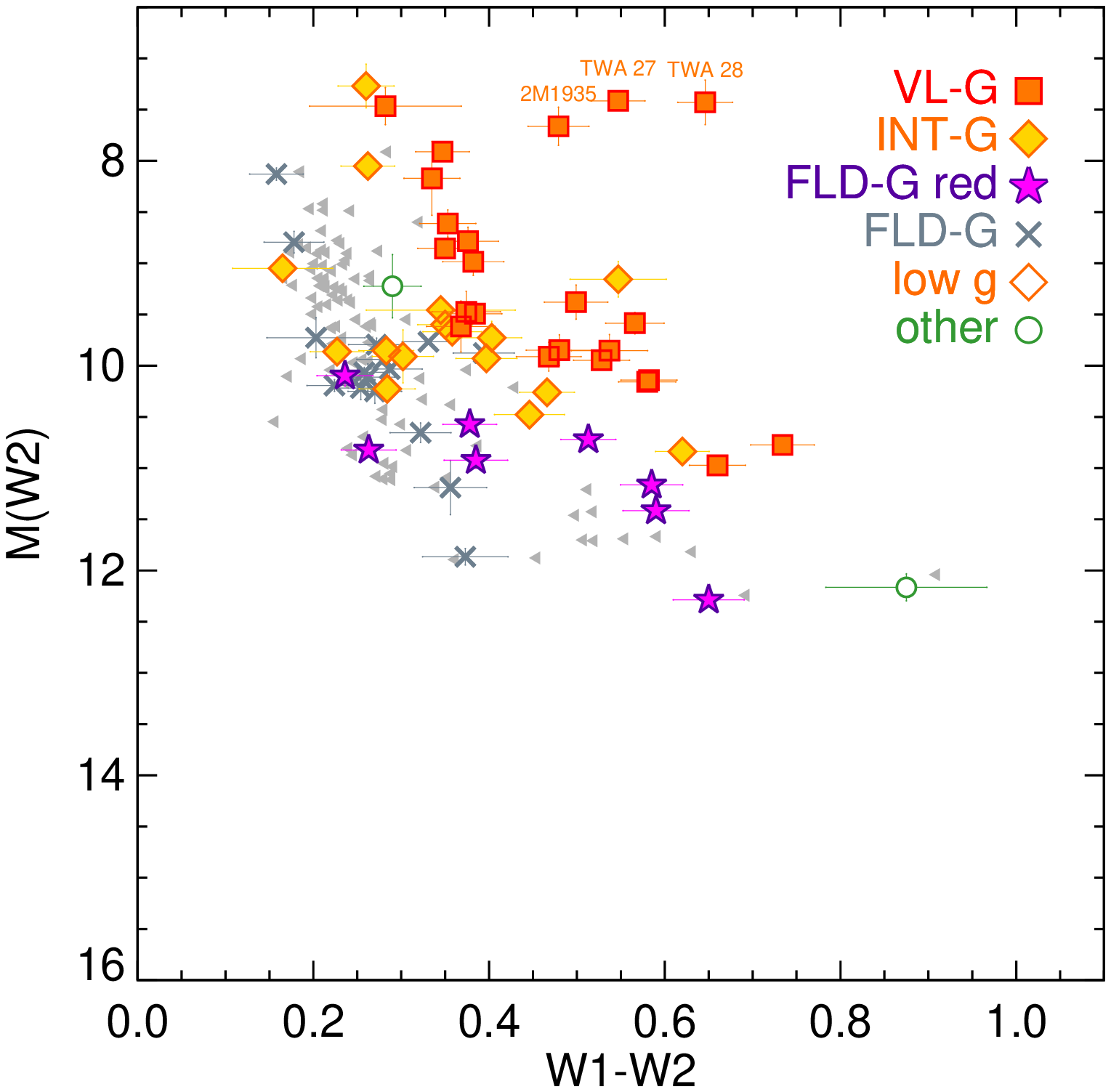}
  }
  \caption{\normalsize Same as Figure~\ref{fig:cmd1} but for \WISE\
    magnitudes. In order to more clearly show the gravity sequence, we
    restricted the $W1-W2$ color range, which causes the T dwarfs
    SDSS~J1110+0116 and Ross~458C be off the right plot. Likewise, the
    right plot labels the three young objects with evidence for a mid-IR
    excess (TWA~27 [2MASS~J1207$-$3932], TWA~28 [SSSPM~J1102$-$3431],
    2MASS~J1935$-$2846), likely due to circumstellar material (see
    Appendix).  \label{fig:cmd2}}

\end{figure}

\begin{landscape}
\begin{figure}

  \centerline{
    \includegraphics[width=7.5in,angle=0]{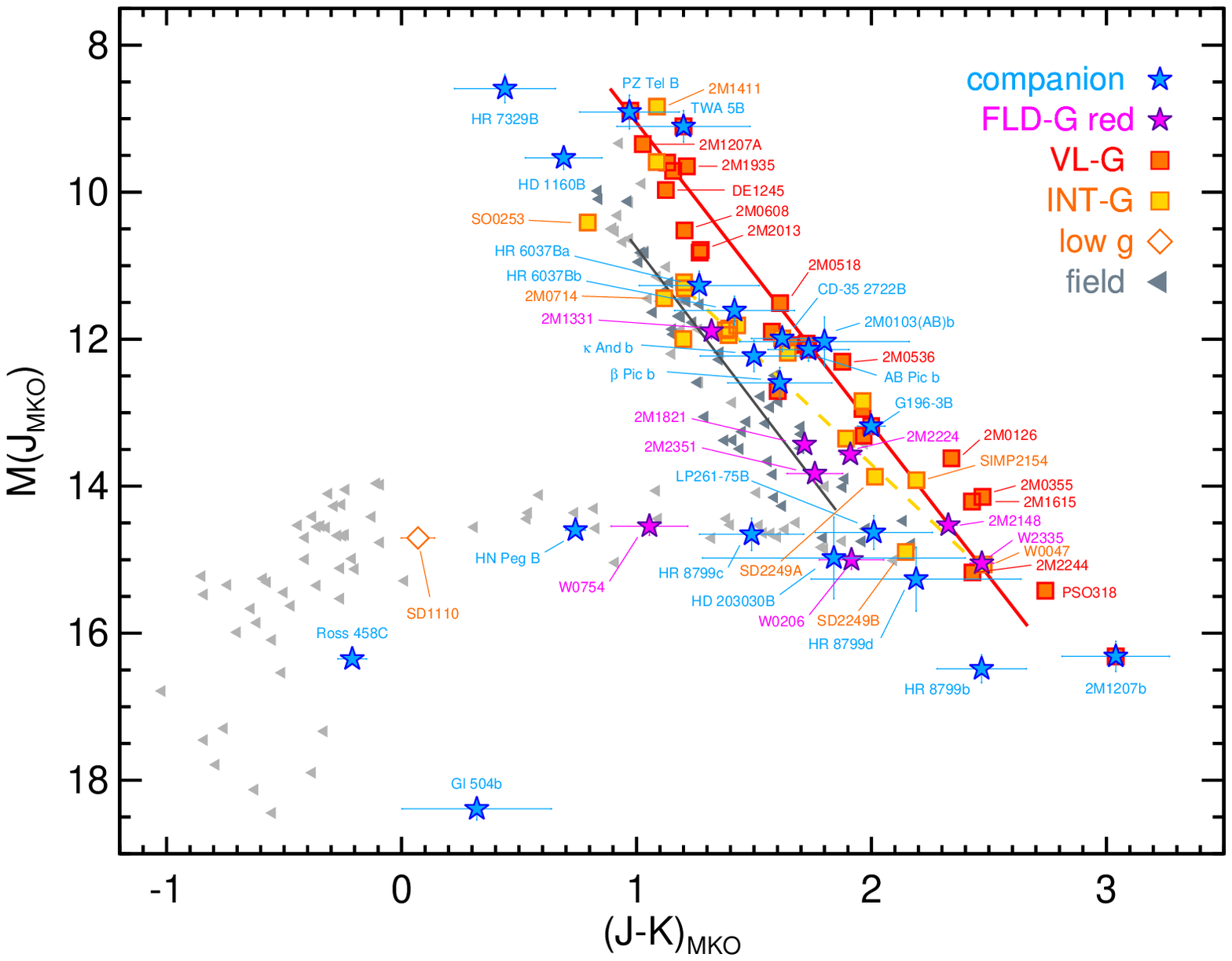}
  }
  \caption{\normalsize Color--magnitude diagram showing young ultracool
    companions with parallaxes in comparison to our low-gravity parallax
    sample (\textsc{vl-g} and \textsc{int-g}).  Normal field
      objects (gray triangles) are the same as in
      Figure~\ref{fig:cmd1}.  (The young companions GSC 08047$-$00232~B
    (M9.5$\pm$1; \citealp{2003A&A...404..157C, 2004A&A...420..647N,
      2005A&A...430.1027C}),
    1RXS~J235133.3+312720 (L0$^{+2}_{-1}$; \citealp{2012ApJ...753..142B}),
    2MASS J01225093$-$2439505~B (L3.7$\pm$1.0;
    \citealp{2013ApJ...774...55B, 2015ApJ...805L..10H}),
    and GU~Psc~b (T3.5$\pm$1; \citealp{2014ApJ...787....5N}) are not
    plotted, as they do not have parallactic
    distances.) \label{fig:cmd-comp}}
\end{figure}
\end{landscape}

\begin{figure}

  \centerline{
    \includegraphics[width=3.3in,angle=0]{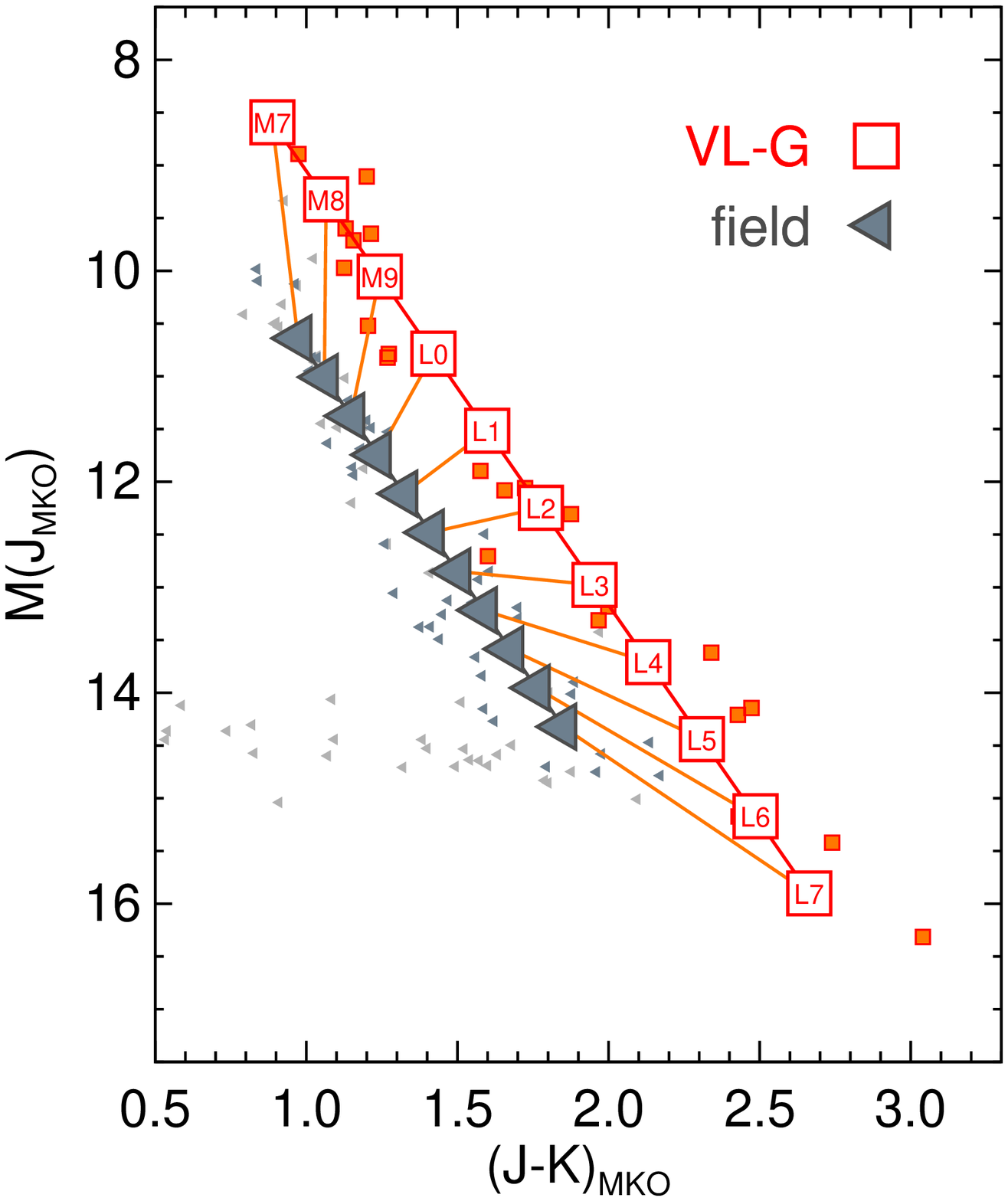}
    \includegraphics[width=3.3in,angle=0]{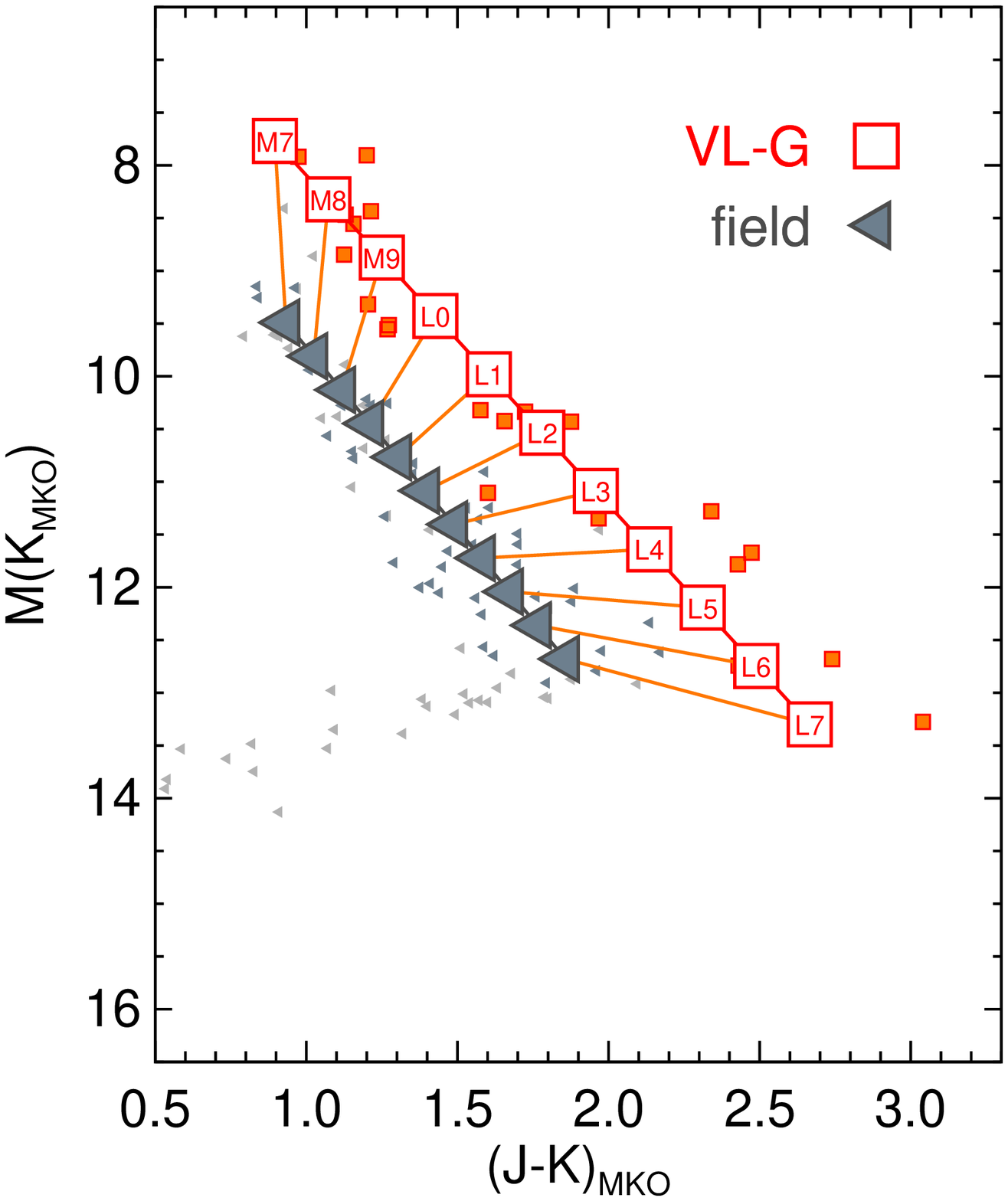}
  }
  \caption{\normalsize Color--magnitude diagram showing our
    lowest-gravity parallax sample (\textsc{vl-g}; small orange squares)
    alongside normal field objects (small gray triangles, same as in
    Figure~\ref{fig:cmd1}). For each population we fitted a line to
    absolute magnitude as a function of color for objects with spectral
    types M7--L7. Combining these fits with the linear spectral type to
    absolute magnitude relations from Table~\ref{table:coeff} allows us
    to plot the mean location of each spectral type for each population
    on the color--magnitude diagram.  Mean colors and magnitudes for
    low-gravity objects are marked by large, labeled open squares with
    orange lines connecting them to the mean field population location
    of the same spectral type. On average, the low-gravity sequence is
    brighter in both $J$ and $K$ at roughly the same $J-K$ color as the
    field for early spectral types. Then, going to later spectral types,
    the low-gravity sequence gradually becomes redder in $J-K$ as it
    becomes much fainter in $J$ and somewhat fainter in $K$.  (Only
    field objects with near-IR spectral types were used in that fit,
    indicated by darker gray triangles.) \label{fig:cmd-vlgseq}}

\end{figure}

\begin{figure}

  \centerline{
    \includegraphics[width=3.3in,angle=0]{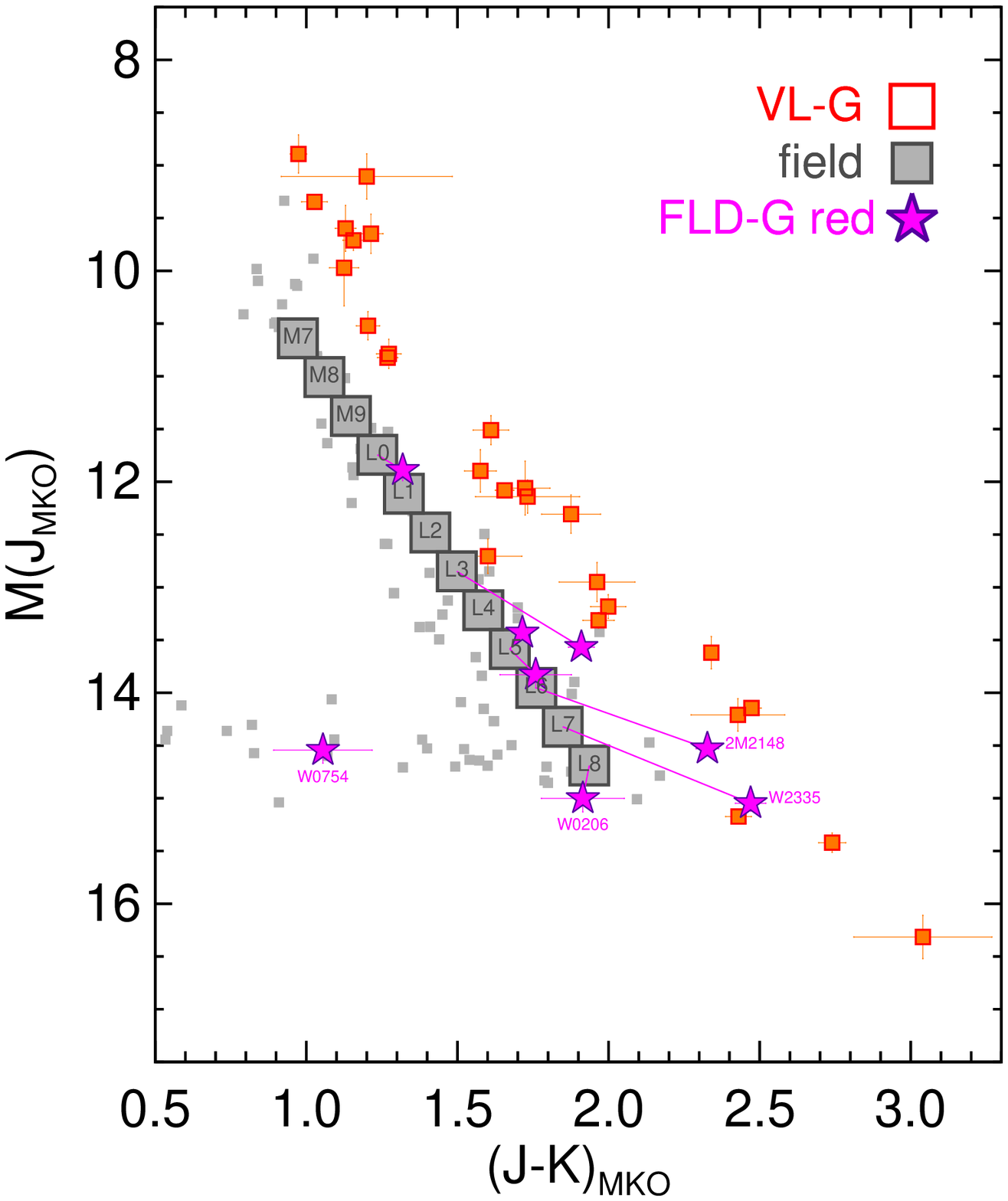}
    \includegraphics[width=3.3in,angle=0]{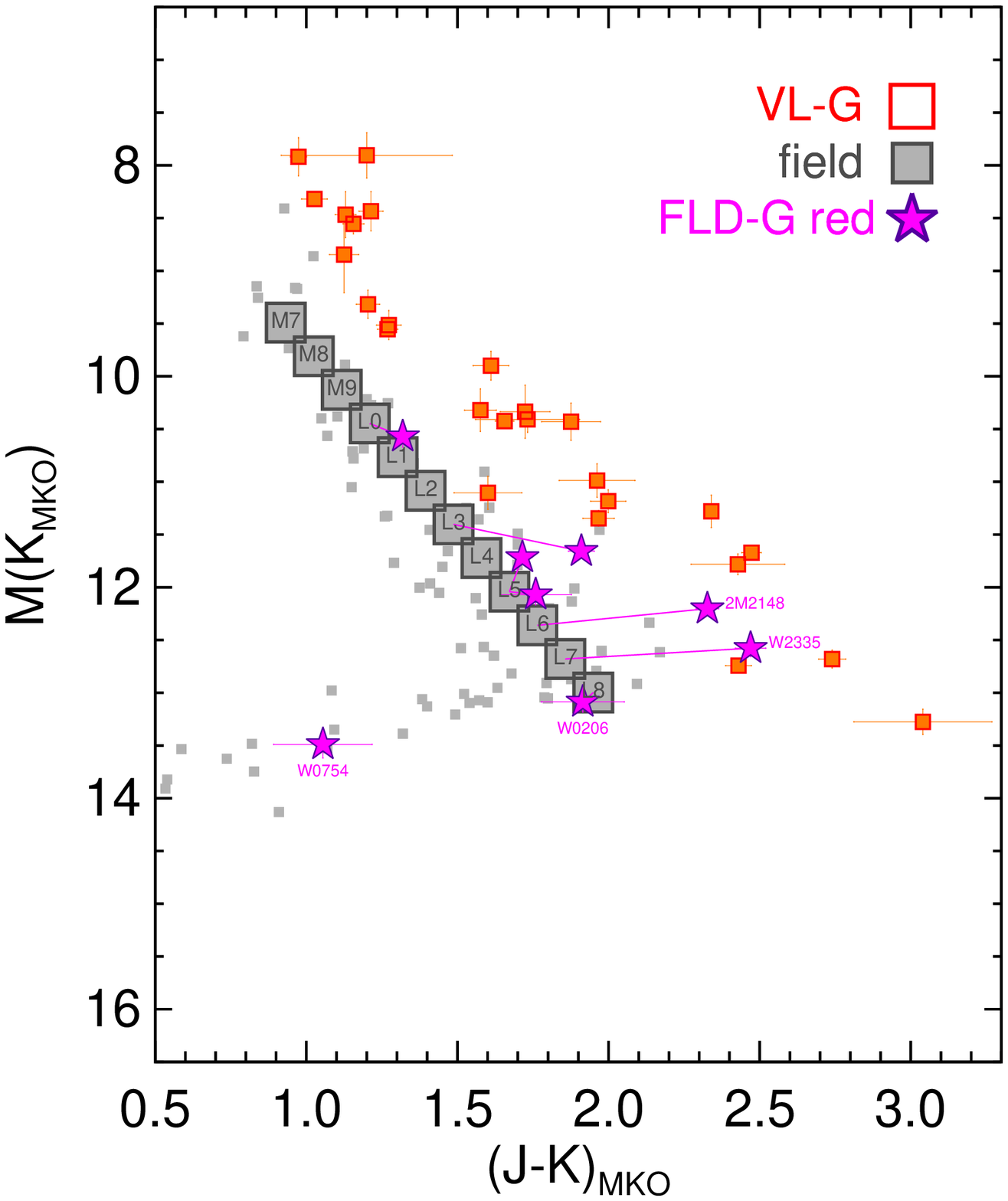}
  }
  \caption{\normalsize Color--magnitude diagram showing the CMD
      location of our unusually red field-gravity objects compared to
      our lowest-gravity parallax sample (\vlg; small orange squares)
      and normal field objects (small gray triangles, same as in
      Figure~\ref{fig:cmd1}).  Similar to Figure~\ref{fig:cmd-vlgseq},
      the mean field sequence as a function of near-IR spectral type is
      shown with labeled grey squares.  For each red object, a line
      connects its CMD position with the corresponding field value for
      the same near-IR spectral type. \label{fig:cmd-redseq}}
\end{figure}

\begin{figure}

  \centerline{
    \includegraphics[width=6.5in,angle=0]{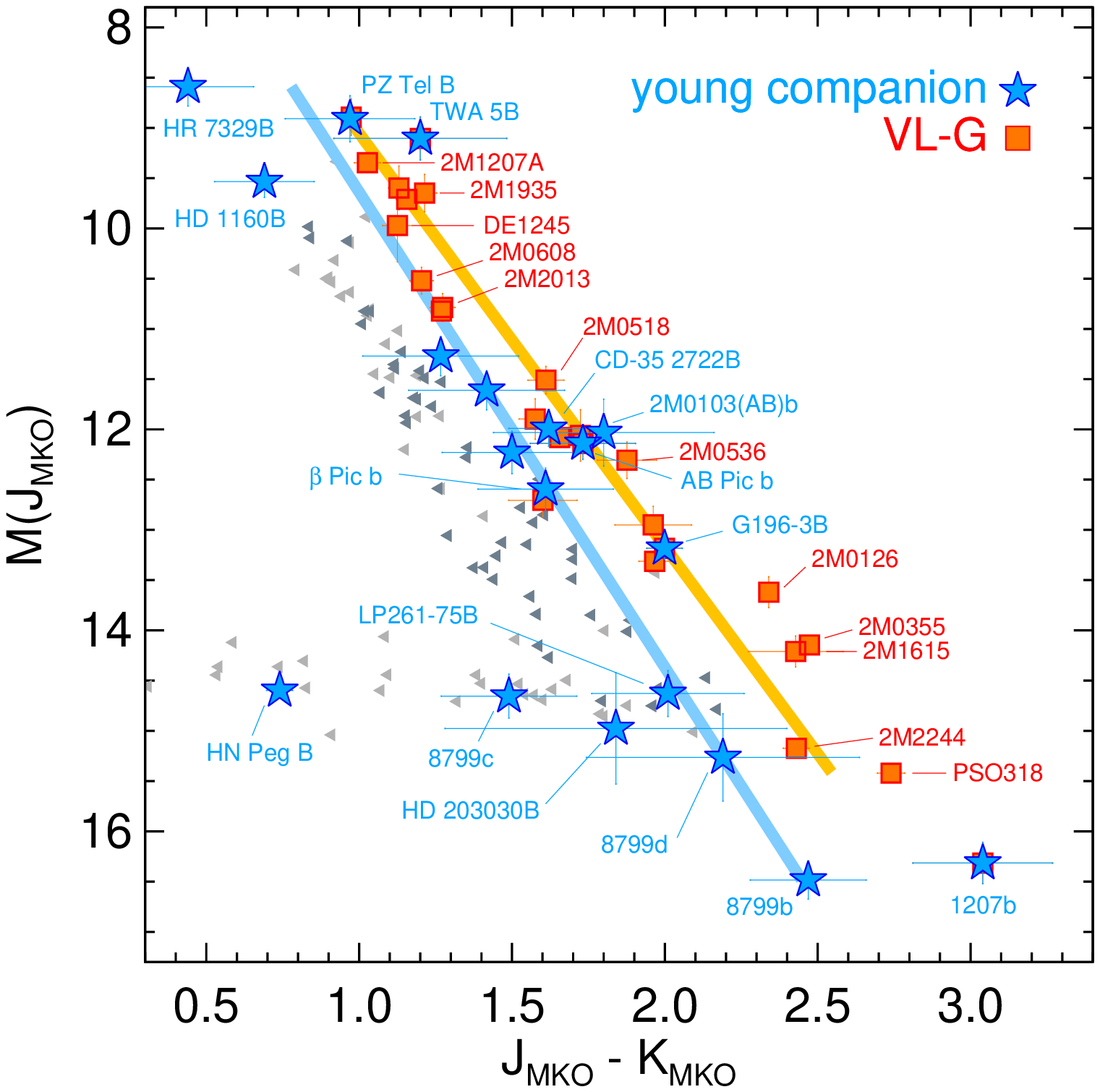}
  }
  \vskip -2ex
  \caption{\normalsize Color--magnitude diagram highlighting the late-M
    to late-L \vlg\ sequence compared to young ultracool companions with
    parallaxes, using the same data as in Figure~\ref{fig:cmd-comp}.
    The solid lines show robust linear fits through the two datasets.
    (Given its extreme nature, 2MASS~J1207$-$3932b is excluded from the
    fitting, as is the T2.5 companion HN~Peg~B given its spectral type
    is later than the \vlg\ sample here.)  Note that three companions
    have \vlg\ classifications so they are plotted as a blue star
    overlaid on a red square: TWA~5B, AB~Pic~b, and
    G~196-3B. \label{fig:cmd-comp-vs-VLG}}
\end{figure}

\begin{figure}

  \centerline{
    \hskip -0.2in
    \includegraphics[height=3.3in,angle=0]{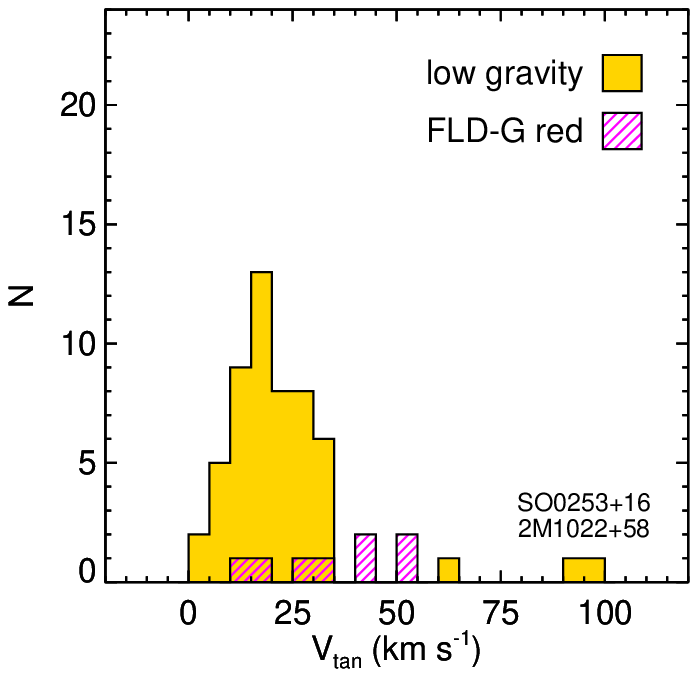}
    \hskip -0.2in
    \includegraphics[height=3.3in,angle=0]{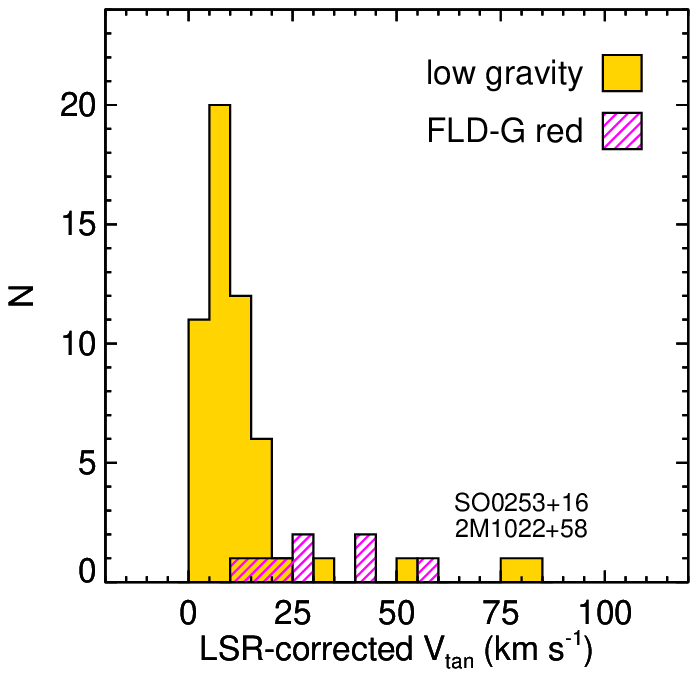}
  }

  \caption{\normalsize Tangential velocities are shown for all objects
    with some spectral signatures of low gravity (orange) and for
    unusually red field objects (hatched magenta).  The two plots
      show the observed tangential velocity ({\em left}) and the
      LSR-corrected tangential velocity ({\em right}).
    Two low-gravity objects 2MASS~J1022+5825 (optical L1$\beta$, near-IR
    L1~\fldg) and SO~J0253+1652 (near-IR M7.5~\intg; a.k.a. Teegarden's
    Star) have observed tangential velocities of $\approx$100\,\kms\ in
    contrast to almost all other low-gravity objects, which have
    $\Vtan\lesssim35$\,kms. This suggests either truly young objects
    acquire large space motions soon after birth or cases of old objects
    that somehow display spectral signatures of youth. The red \fldg\
    objects with $\Vtan>35$\,\kms\ are either at the earliest spectral
    types (2MASS~J1331+3407) or latest (WISE~J0206+2640,
    WISE~J0754+7909).  \label{fig:vtan}}

\end{figure}

\begin{landscape}
\begin{figure}

  \centerline{
    \includegraphics[width=7.5in,angle=0]{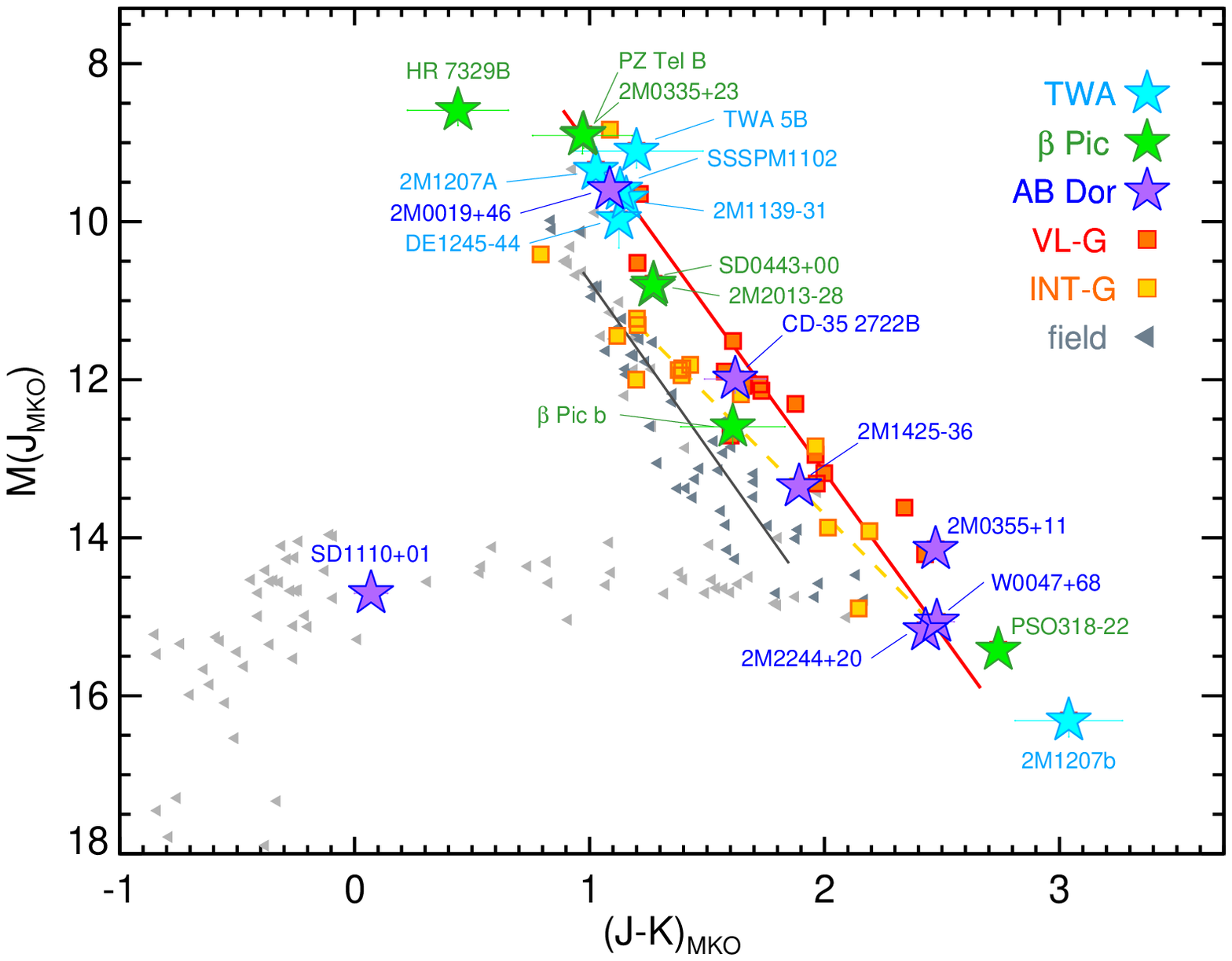}
  }
  \caption{\normalsize Color--magnitude diagram showing members of the
    three young moving groups with the most ultracool dwarfs
    identified to date: TWA ($\approx$10\,Myr), $\beta$~Pic
    ($\approx$25\,Myr), and AB~Dor ($\approx$150\,Myr). Our
    low-gravity parallax sample (\textsc{vl-g} and \textsc{int-g}) and
    the field population are shown for comparison, along with their
    linear fits (Section~\ref{sec:cmd}).  The darker triangles show
    the subset of field objects used for the new M7--L7 linear fit
    (Section~\ref{sec:absmags}).  \label{fig:cmd-ymg}}

\end{figure}
\begin{figure}

  \centerline{
    \includegraphics[width=7.5in,angle=0]{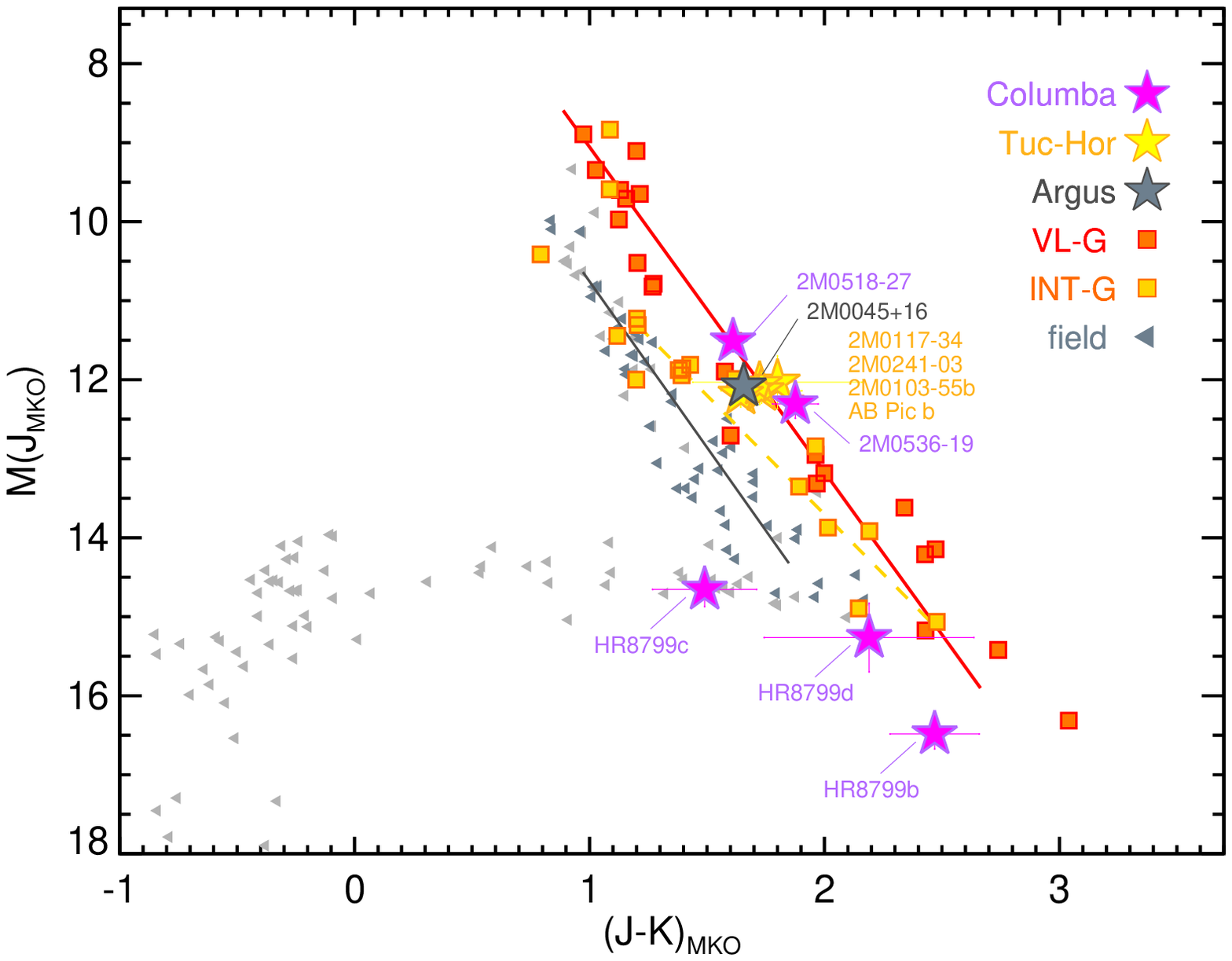}
  }
  \caption{\normalsize Same as Figure~\ref{fig:cmd-ymg}, for the more
    sparsely populated Columba ($\approx$40~Myr), Tuc-Hor
    ($\approx$50~Myr), and Argus (uncertain age) groups, with ages from
    \citet{2015MNRAS.454..593B}.  \label{fig:cmd-ymg2}}

\end{figure}
\end{landscape}

\begin{figure}
  \includegraphics[width=5in,angle=90]{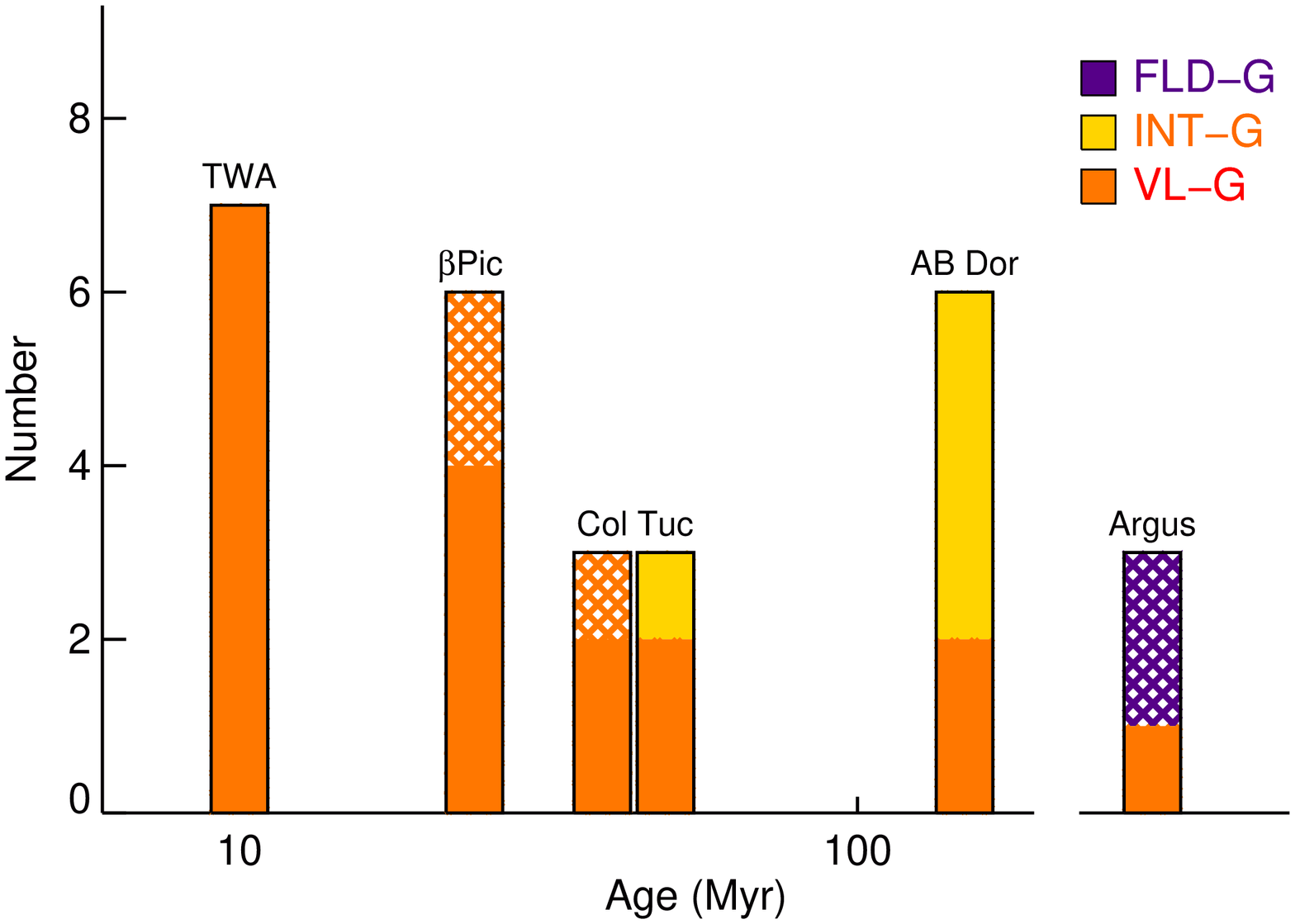}
  \caption{\normalsize Distribution of near-IR gravity classifications
    for different moving groups. The x-axis is logarithmically spaced,
    with assumed ages of 10, 24, 42, 45, and 149~Myr for the TWA, \bPic,
    Columba, Tuc-Hor, and AB~Dor groups \citep{2015arXiv150800898B} and
    with an indeterminate age for Argus. (The ages of Columba and
    Tuc-Hor have been nudged by 8\% to avoid overlapping bars.) Objects
    with uncertain group membership are represented by hatched regions.
    (We have excluded 2MASS~J0058$-$0651 [L1~\intg] from this plot,
    given its possible membership in both the \bPic\ and AB~Dor groups.)
    The \vlg\ objects dominate the younger groups but are a minority by
    the age of AB~Dor. The groups with intermediate ages may have an
    intermediate fraction of \vlg\ objects but larger samples of group
    members are needed for an assessment. \label{fig:ymg-gravities}}
\end{figure}

\begin{figure}

  \centerline{
    \includegraphics[width=6.5in,angle=0]{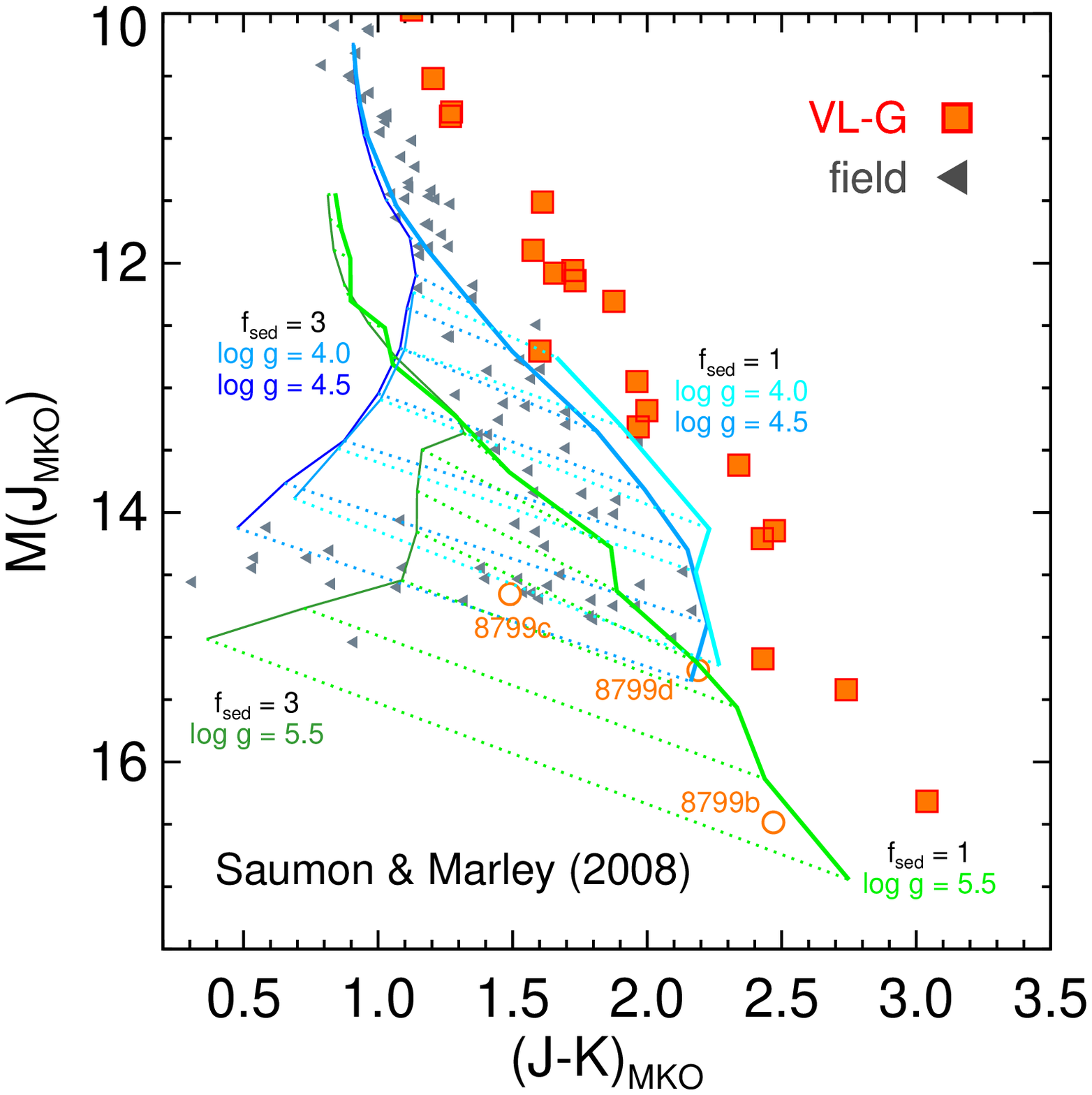}
  }
  \caption{\normalsize Color--magnitude diagram showing model-atmosphere
    grids from \citet{2008ApJ...689.1327S} alongside our low-gravity
    parallax sample (\textsc{vl-g}) and the field population. We plot
    model sequences having thick clouds and redder near-IR colors ($f_{\rm sed}
    = 1$) as well as thin clouds and bluer near-IR colors ($f_{\rm sed} = 3$).
    Dotted lines connect models of the same \Teff\
    and \logg\ but different $f_{\rm sed}$, showing the
    effect of cloud thickness. 
    We plot the highest gravity models ($\logg = 5.5$\,dex in green) for
    comparison to lower gravities ($\logg = 4.0$, 4.5\,dex in blue),
    where the lowest gravities are only complete for
    $\Teff = 1100$--1500\,K. Although these models successfully
    reproduce planetary-mass objects such as HR~8799b, they do not
    accurately follow the low-gravity sequence defined by our
    sample. \label{fig:cmd-sm08}}

\end{figure}

\begin{figure}

  \centerline{
    \includegraphics[width=6.5in,angle=0]{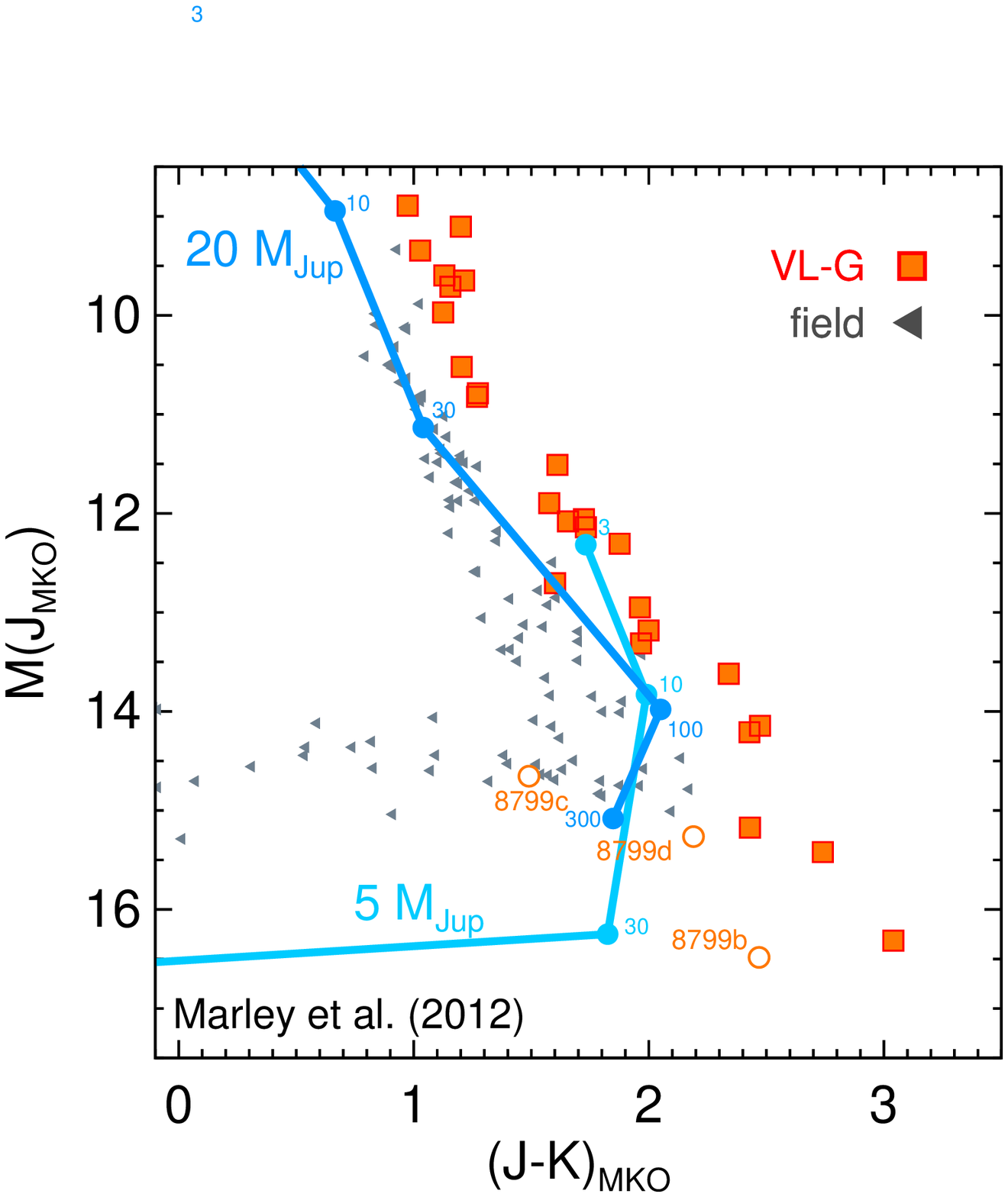}
  }
  \caption{\normalsize Color--magnitude diagram showing
    \citet{2012ApJ...754..135M} evolutionary models for 5\,\Mjup\ and
    20\,\Mjup\ objects.  The models use a simple prescription for the
    gravity-dependence of the L/T transition.  Beads plotted along each
    track demarcate ages in Myr, as young as 3~Myr.  Even these models
    intended for comparison to young planetary-mass objects such as
    HR~8799b do not accurately follow the low-gravity (\vlg) sequence
    defined by our sample.  (Note that our \vlg\ sample is expected to
    include objects above 20~\Mjup.  Plotting of these iso-mass model
    tracks is meant to illustrate the current theoretical locus relative
    to the data, not to assign actual masses to the \vlg\ sample using
    the models.)  \label{fig:cmd-m12}}

\end{figure}

\begin{landscape}
\begin{figure}

  \centerline{
    \includegraphics[width=2.7in,angle=0]{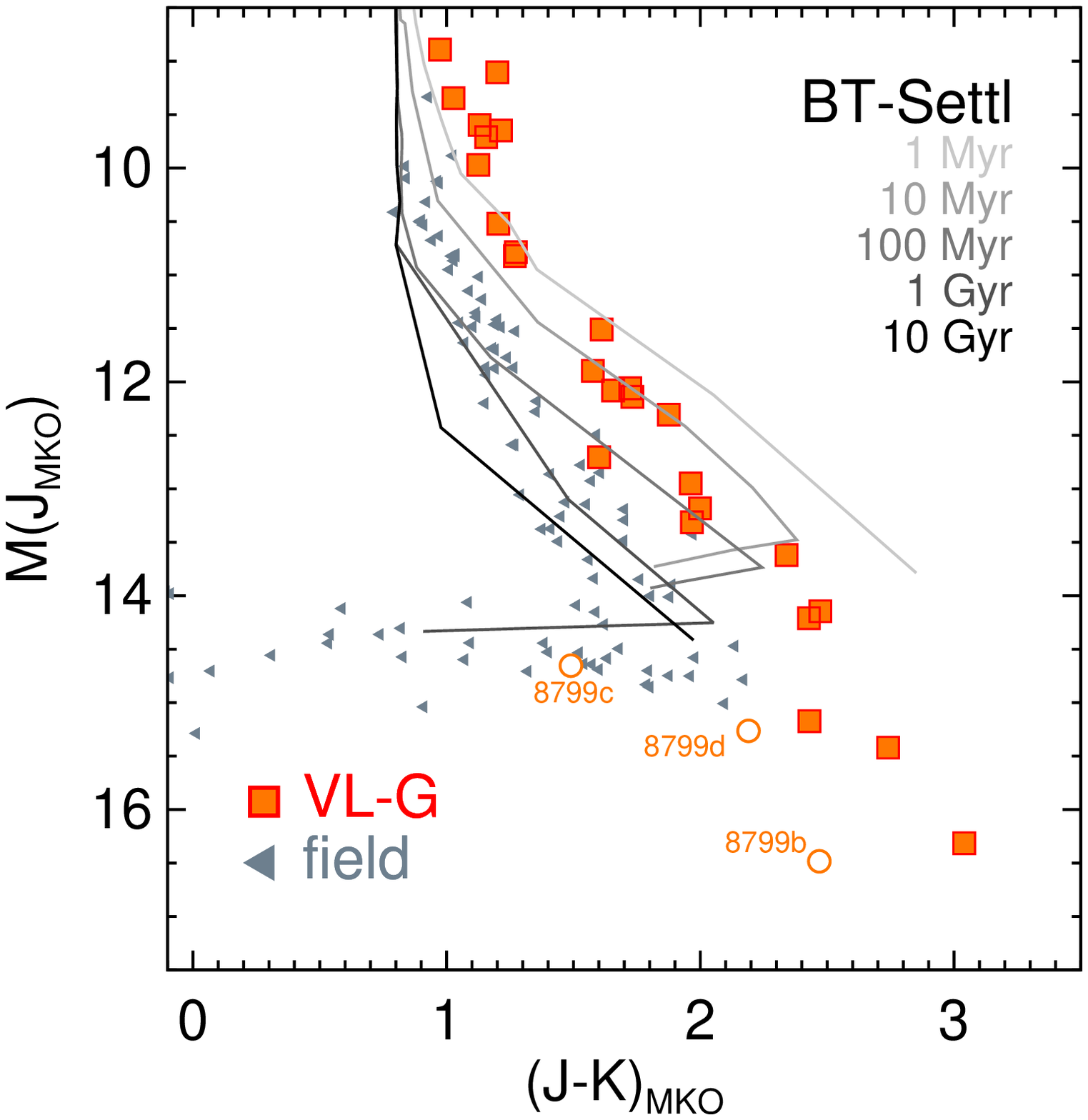}
    \includegraphics[width=2.7in,angle=0]{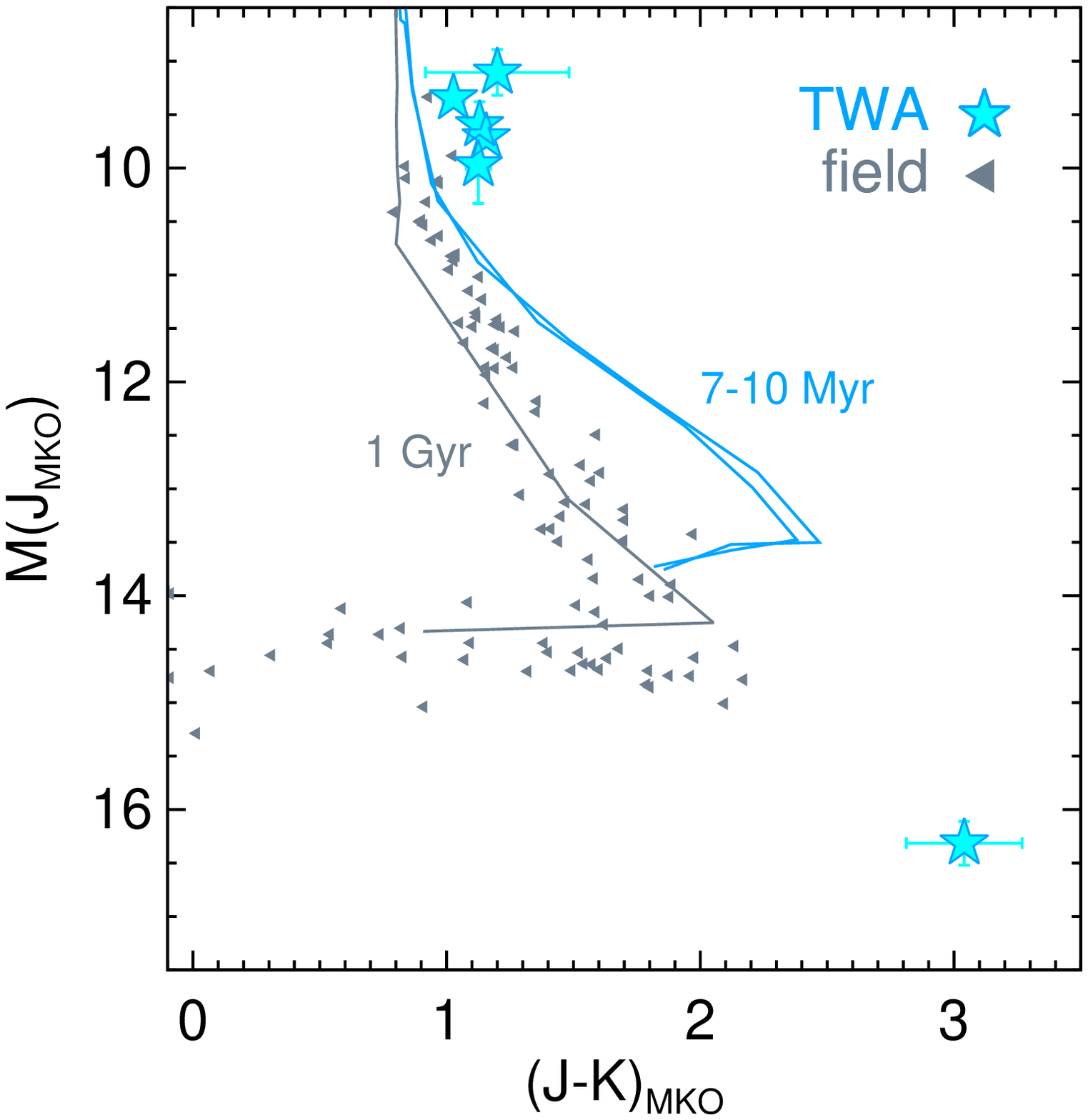}
    \includegraphics[width=2.7in,angle=0]{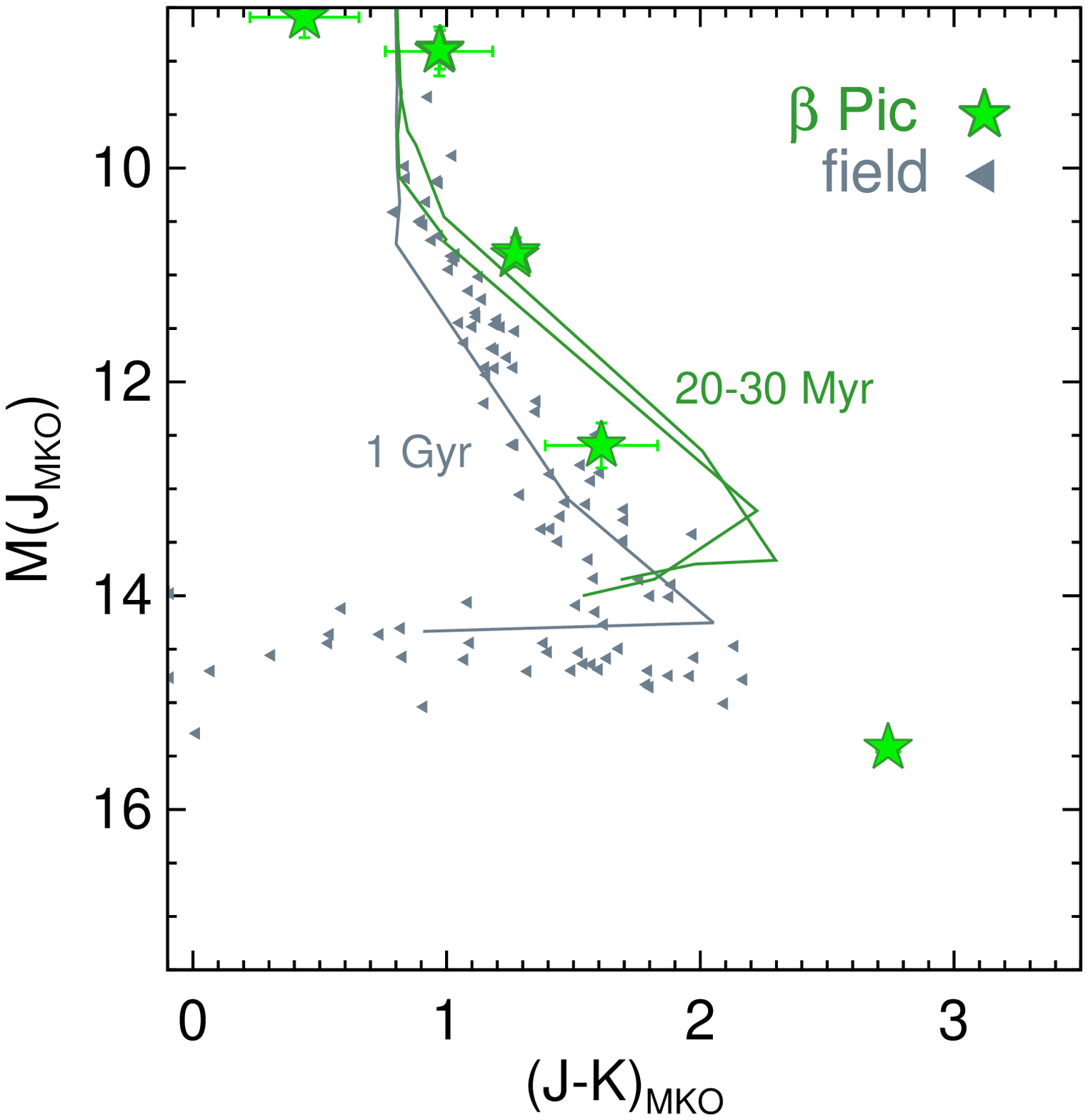}
  }
  \centerline{
    \includegraphics[width=2.7in,angle=0]{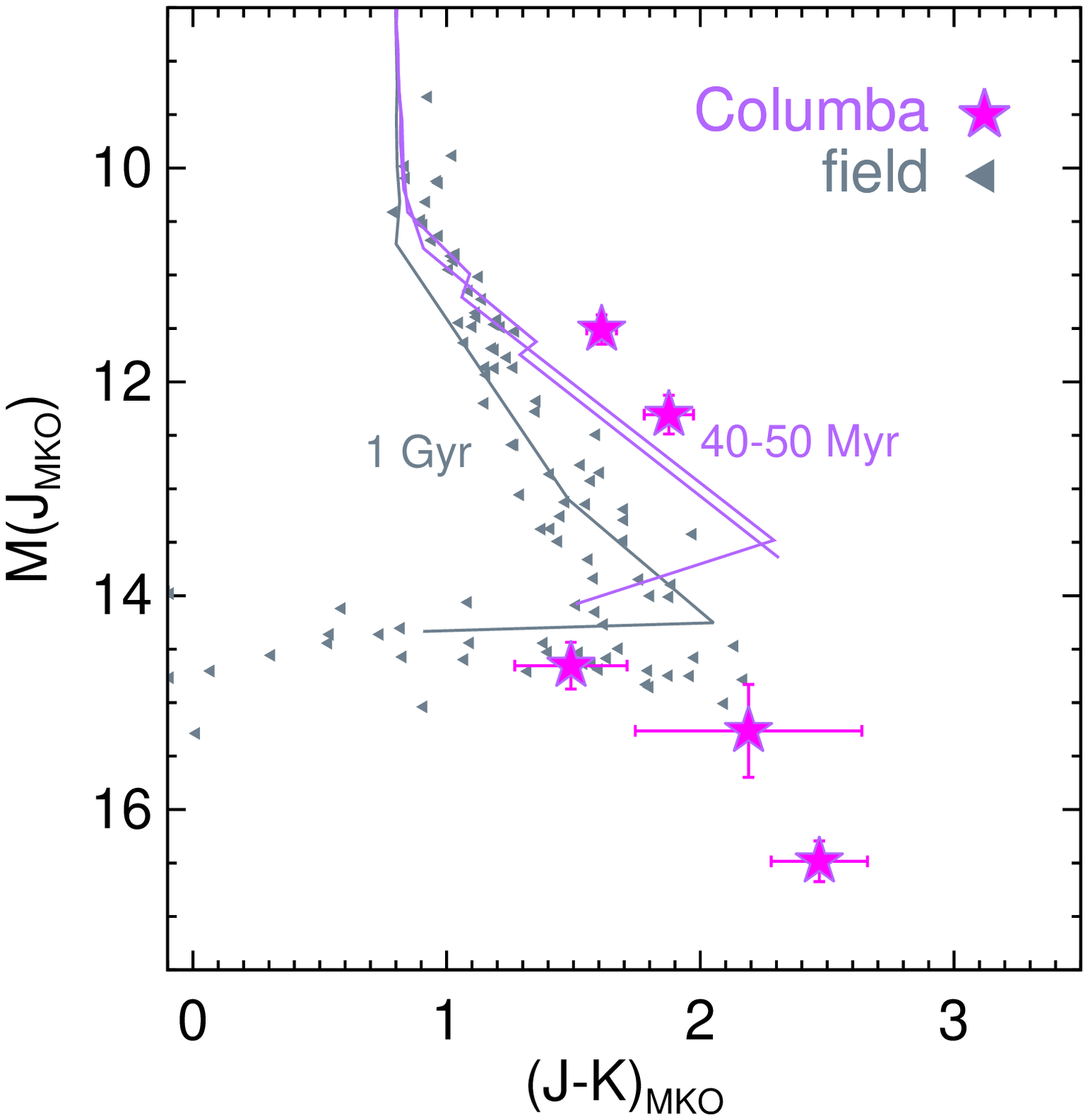}
    \includegraphics[width=2.7in,angle=0]{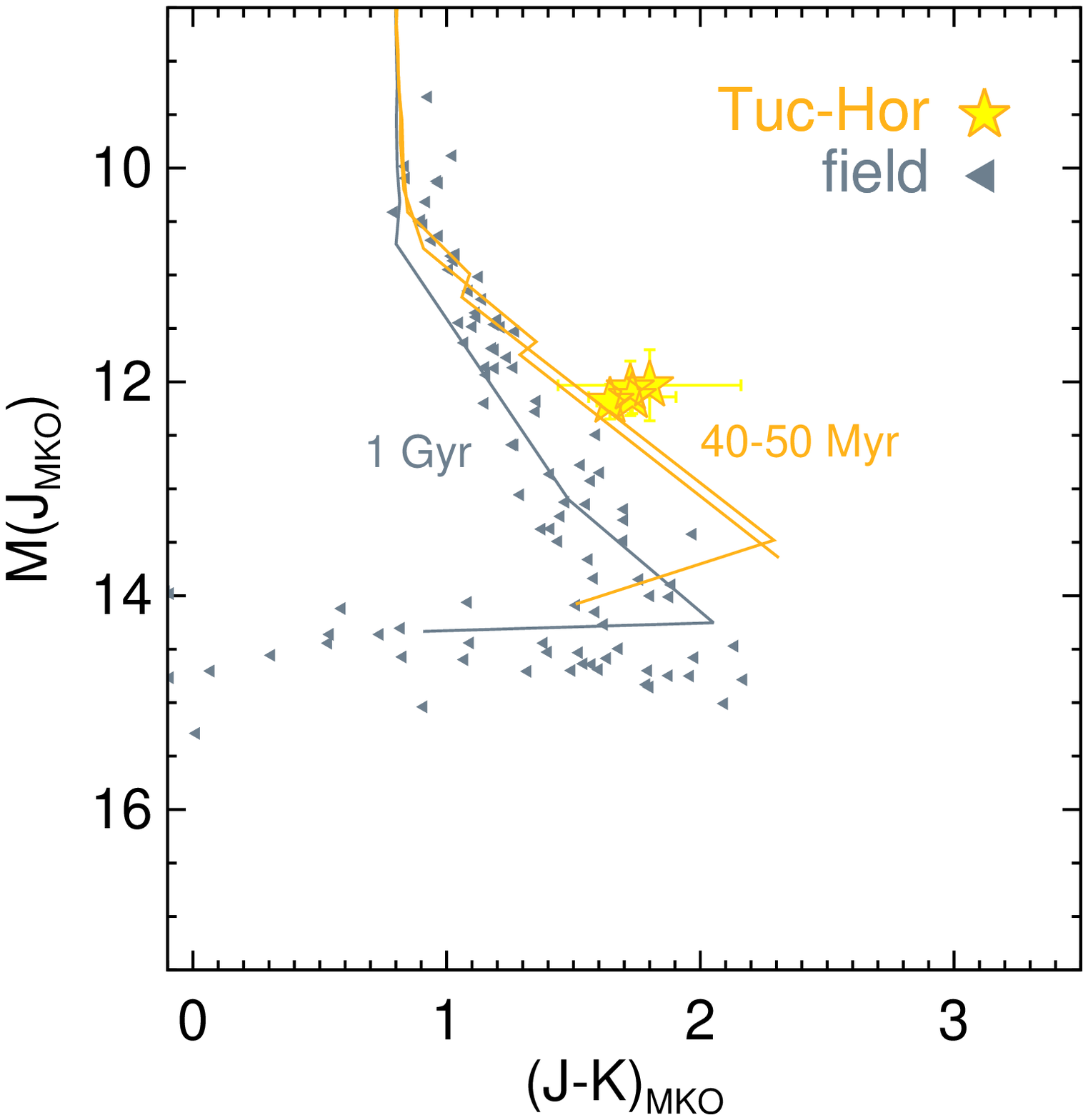}
    \includegraphics[width=2.7in,angle=0]{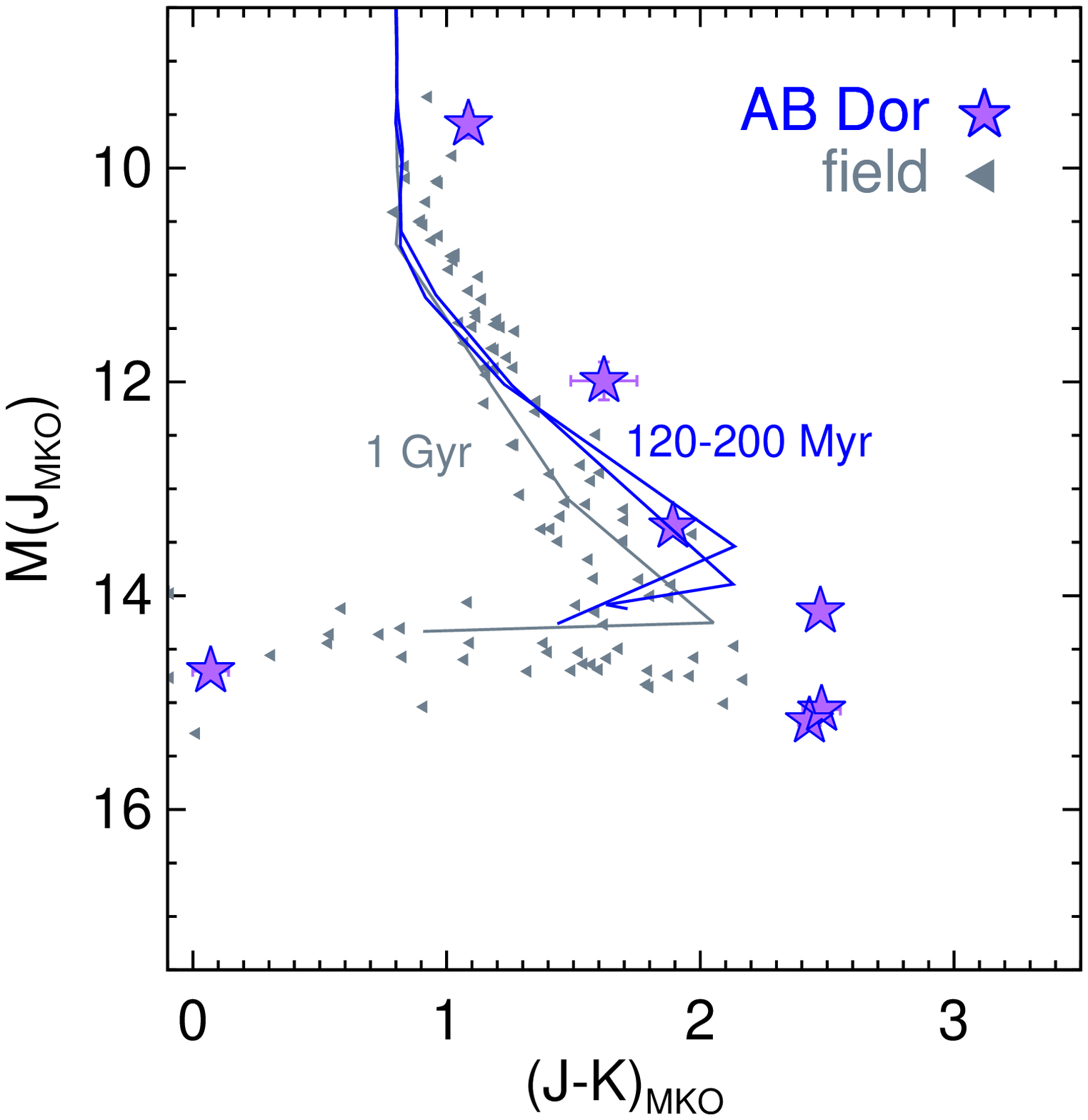}
  }
  \caption{\normalsize BT-Settl evolutionary model isochrones
    (\citealp{2015A&A...577A..42B}; Allard et al.~2015, in preparation)
    shown alongside our sample. {\em Top left:} Our empirical \vlg\
    sequence, the field sequence from \citet{2012ApJS..201...19D}, and
    BT-Settl isochrones ranging from 1\,Myr to 10\,Gyr. Although older
    isochrones provide a reasonable match to the field sequence, our
    \vlg\ sequence tends to be redder than the models. No model
    isochrones match the faint ($M_J\lesssim14$\,mag), red portion of
    the sequence. {\em Top middle to bottom right:} Subsets of our
    sample that belong to young moving groups (10--200\,Myr) shown
    alongside isochrones encompassing each group's age
    \citep{2015MNRAS.454..593B}. Overall, model isochrones do not tend
    to match the locus of the observations at any age. \label{fig:cmd-btsettl}}

\end{figure}
\end{landscape}


\begin{landscape}
\begin{figure}
  \includegraphics[width=2.4in]{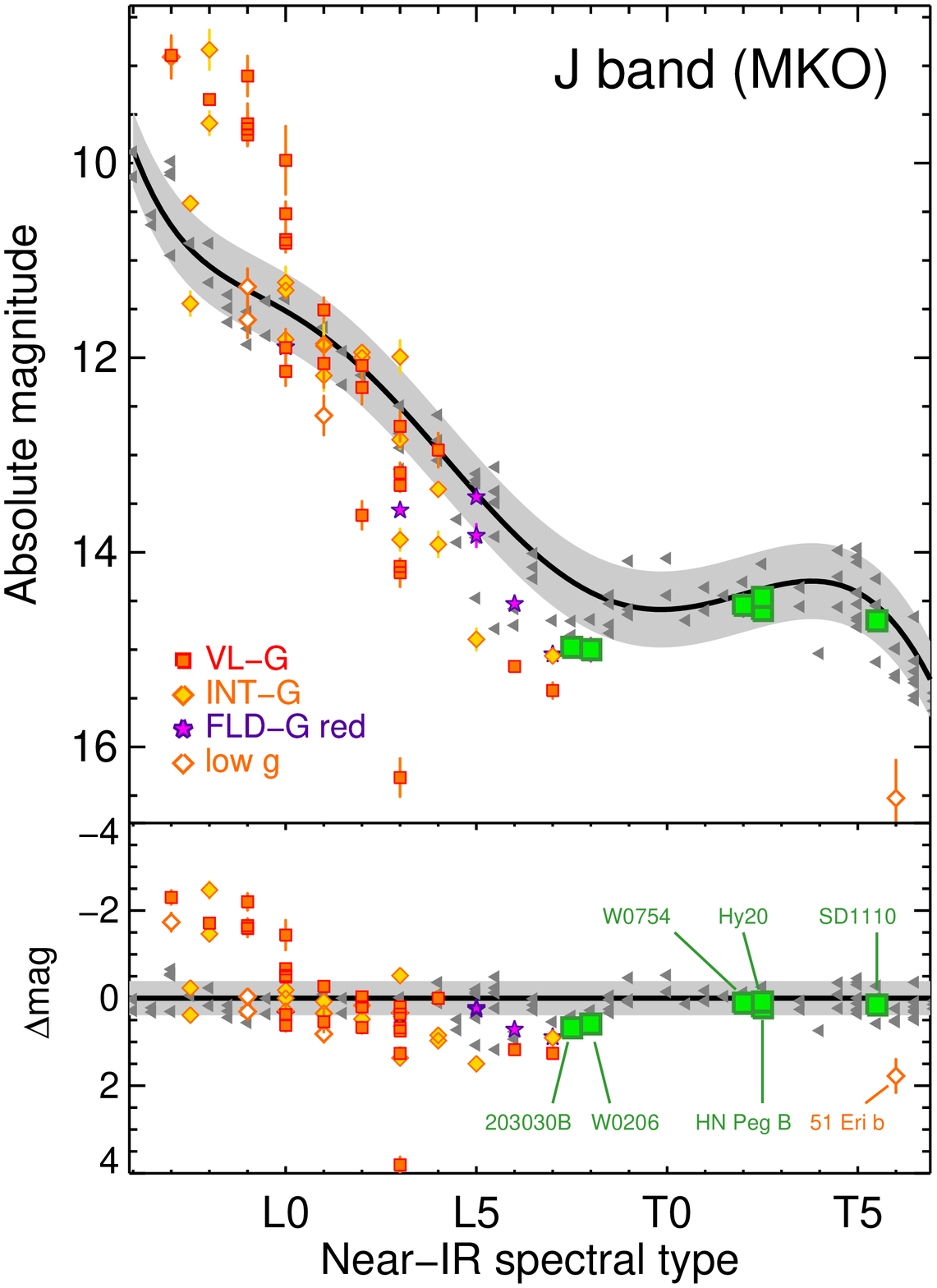}
  \hskip 0.5in
  \includegraphics[width=2.4in]{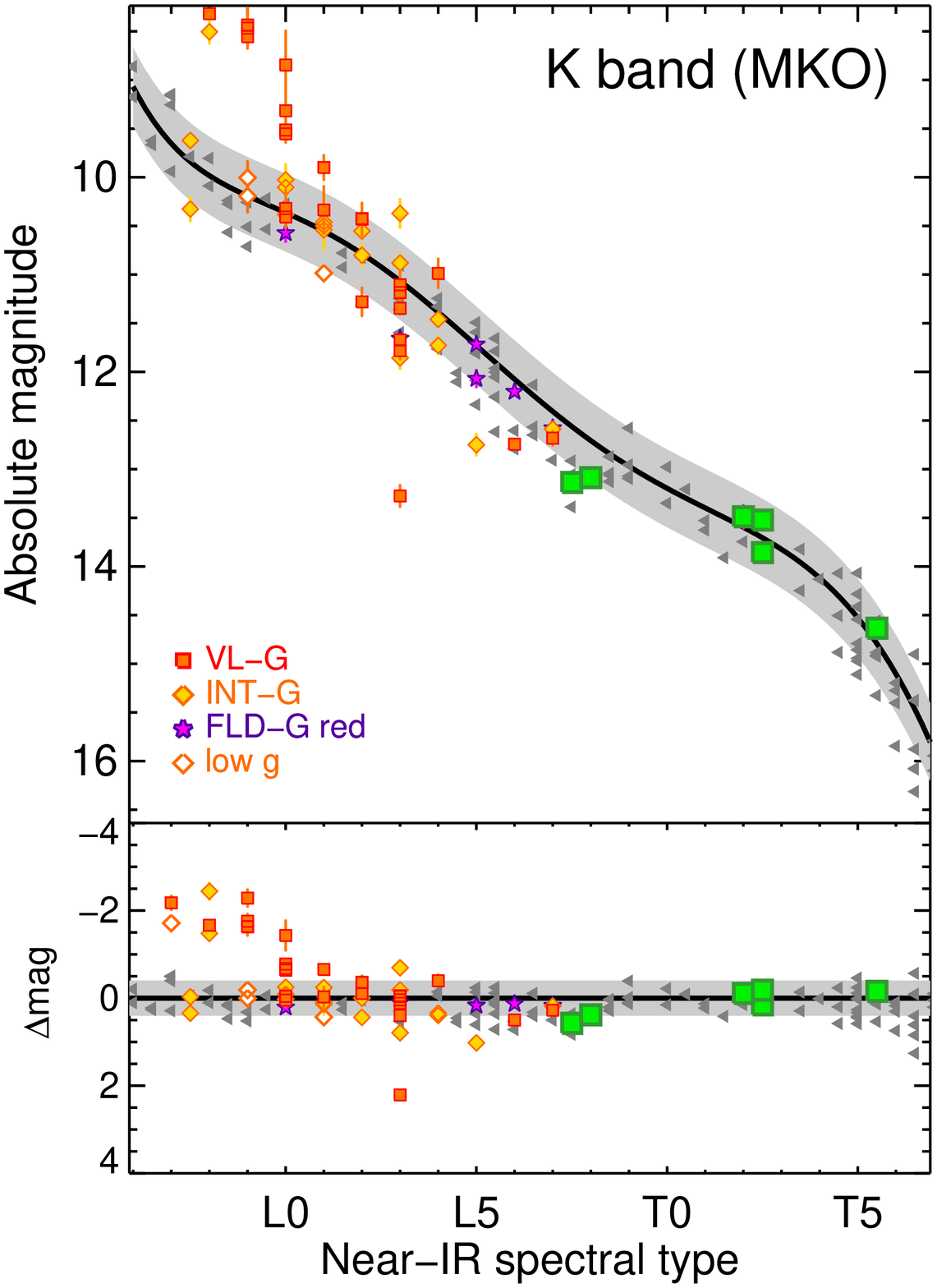}
  \hskip 0.5in
  \includegraphics[width=2.4in]{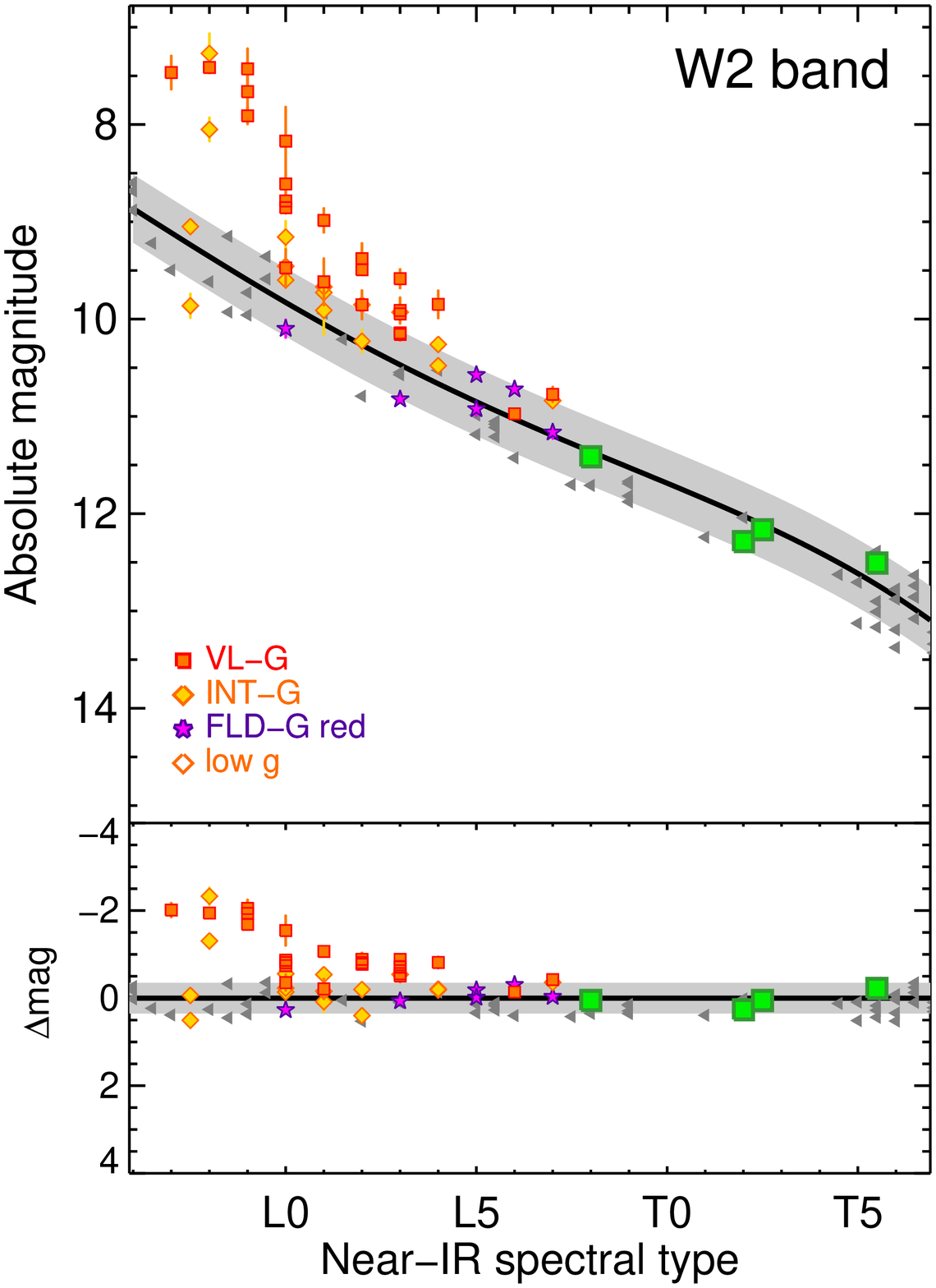}
  \vskip 6ex
  \caption{\normalsize Absolute magnitudes versus spectral type
      for our sample with an expanded range of spectral types, in order
      to highlight the latest-type objects.  The L/T objects are shown
      as green squares, with the measurement errors comparable to the
      symbol size.  Normal field objects (small grey triangles) are the
      same as in Figure~\ref{fig:cmd1}. The heavy black lines are the
      polynomial fits from \citet{2012ApJS..201...19D}, and the grey
      region shows the RMS about the fits. See
      Figure~\ref{fig:absmag-2mass} caption for further
      details.} \label{fig:absmag-LT}
\end{figure}
\end{landscape}


\begin{figure}
  \includegraphics[width=6in]{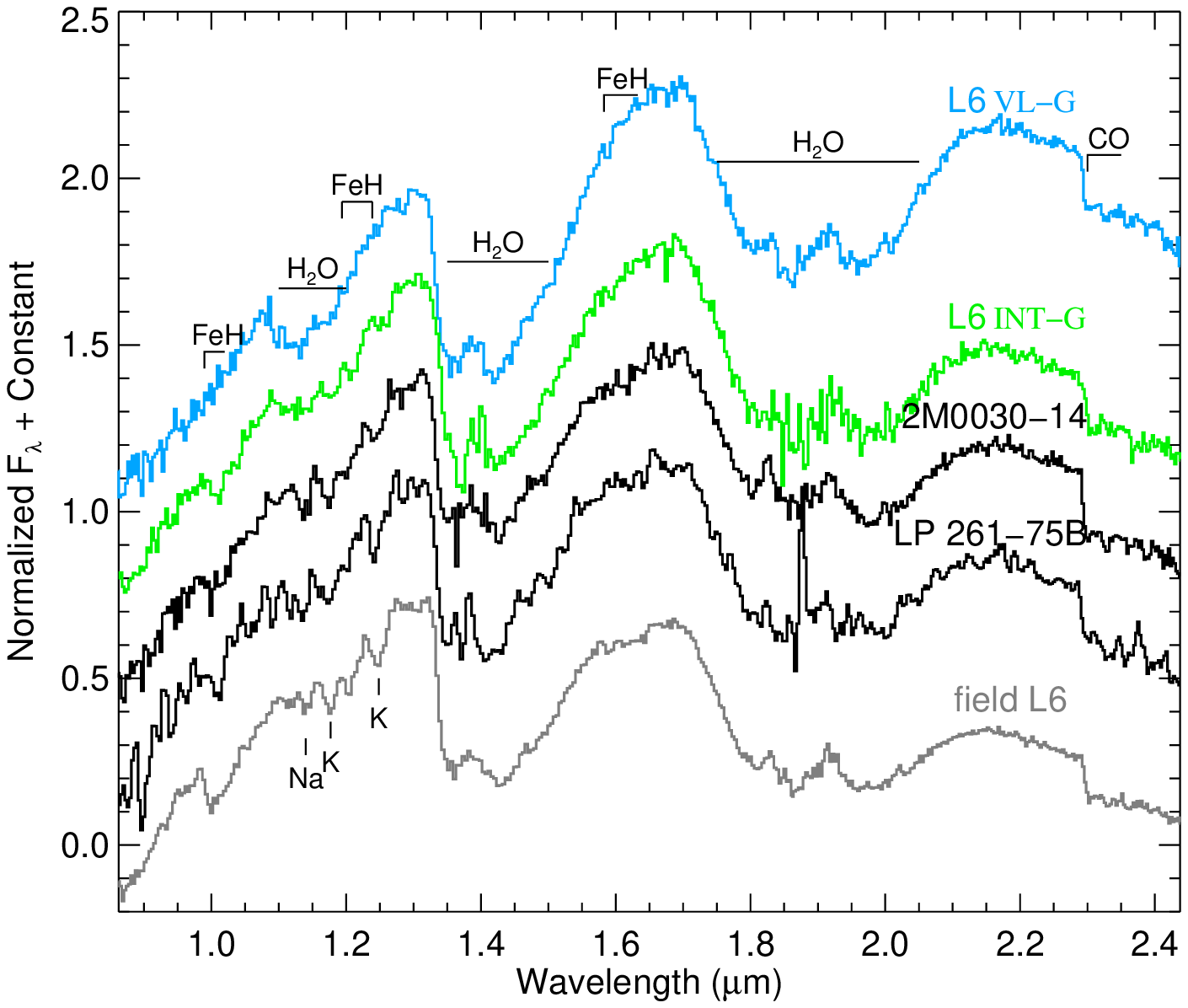}
  \caption{\normalsize Comparison of the spectra of 2MASS~J0030$-$1450
    \citep{2010ApJ...710.1142B} and LP~261-75B
    \citep{2013ApJ...774...55B} to L6 dwarfs with a range of gravity
    classifications from \citet{2013ApJ...772...79A}. The L6 \vlg,
    \intg, and \fldg\ spectra are 2MASS~J2244+2043
    \citep{2008ApJ...686..528L}, 2MASS~J0103+1935
    \citep{2013ApJ...772...79A}, and 2MASS~J1010$-$0406
    \citep{2010ApJS..190..100K}, respectively.
    \citet{2015ApJS..219...33G} type 2MASS~0030$-$1450 as L4--L6
    $\beta$, while we assign a spectral type of L6~\fldg.
    \citet{2013ApJ...774...55B} classify LP~261-75B as L4.5~\fldg, while
    we assign L6~\fldg\ (see Appendix). \label{fig:L6comp}}
\end{figure}

\begin{figure}
  \includegraphics[width=5in,angle=90]{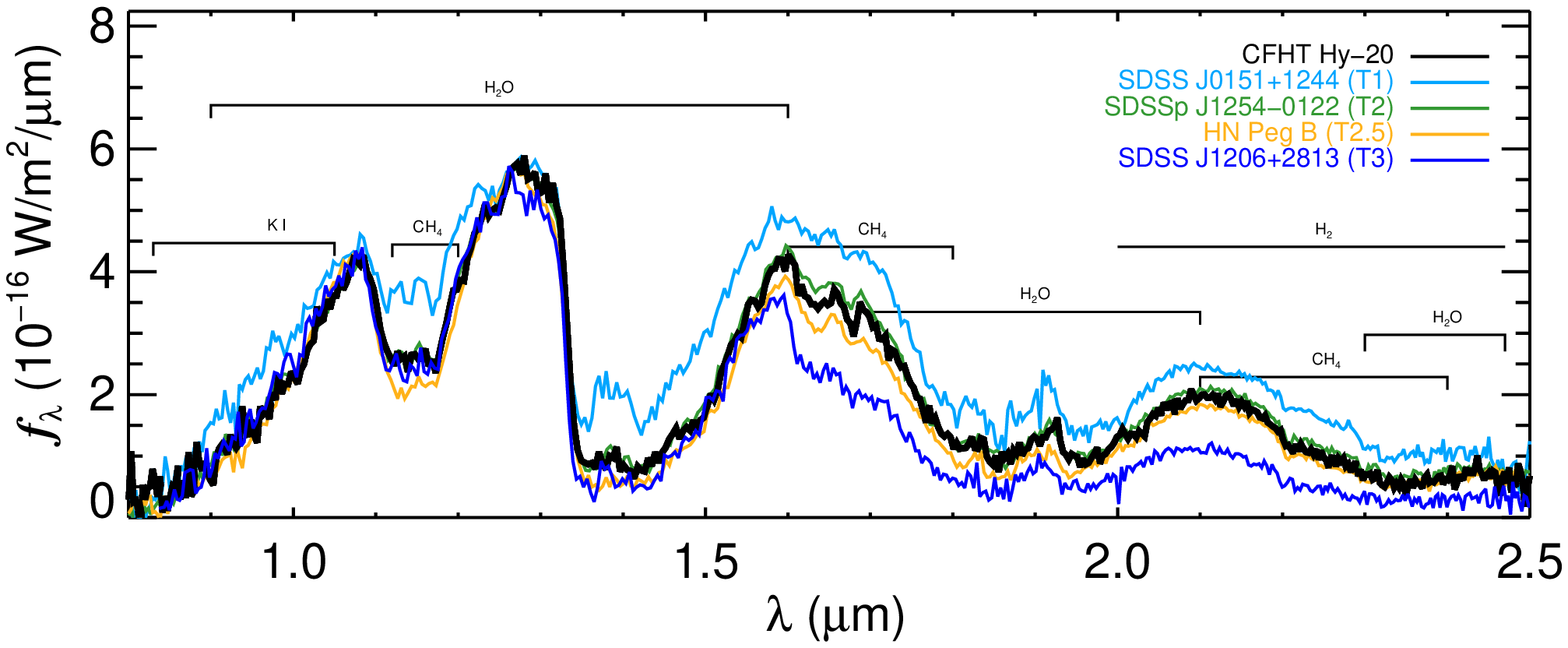}
  \vskip -1in
  \caption{\normalsize Near-IR spectrum of the Hyades member CFHT-Hy-20
    compared to the T~dwarf spectral standards
    SDSS~J015141.69+124429.6 (T1), SDSSpJ125453.90$-$012247.4 (T2), and
    SDSS~J120602.51+281328.7 (T3), along with the T2.5 companion HN~Peg~B.
    \label{fig:Hy20}}
\end{figure}

\begin{figure}
  \vbox{
    \centerline{\includegraphics[width=4in]{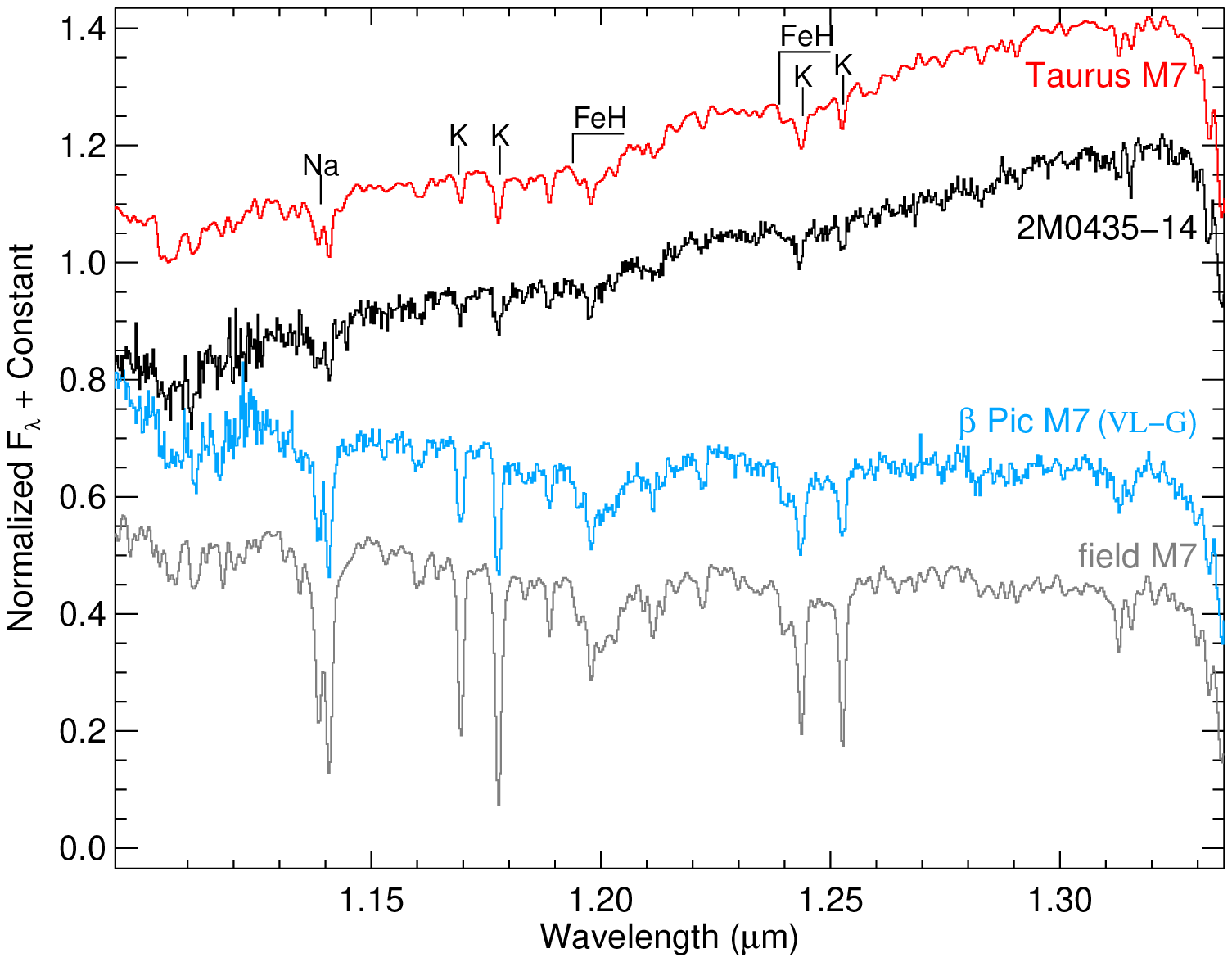}}
    \vskip 2ex
    \centerline{\includegraphics[width=4in]{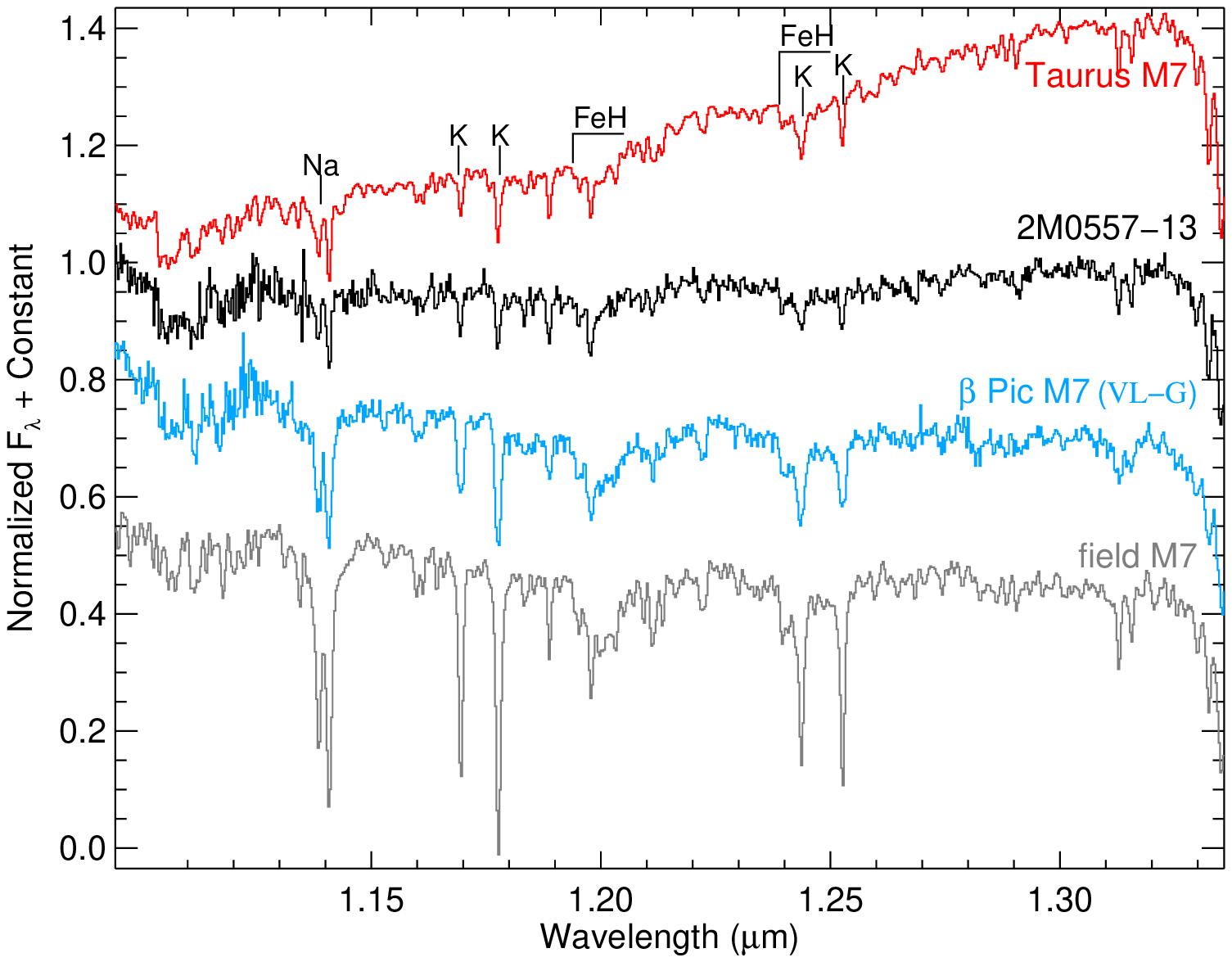}}
  }
  \caption{\normalsize Gravity-sensitive features in the
      \citet{2013ApJ...772...79A} spectra of 2MASS~J0435$-$1414 (top;
      $R=750$) and 2MASS~J0557$-$1359 (bottom; $R=1200$) compared to a
      Taurus M7 dwarf \citep[CFHT-BD-Tau~4;][]{2009ApJ...697..824A}, a
      $\beta$~Pic moving group M7 dwarf
      \citep[2MASS~J0335+2342;][]{2013ApJ...772...79A}, and a field M7
      dwarf \citep[GJ~644C;][]{2005ApJ...623.1115C}.  %
      For each panel, all the plotted spectra have been smoothed to a
      common spectral
      resolution.  
      The Taurus M7 spectrum shows slightly deeper \ion{K}{1},
      \ion{Na}{1}, and FeH absorption, indicating that
      2MASS~J0435$-$1414 and 2MASS~J0557$-$1359 are likely very young
      ($\lesssim$3~Myr old).  (The difference in the overall continuum
      slopes of 2MASS~J0435$-$14 and CFHT-BD-Tau~4 compared to the older
      objects can be minimize for assumed extinctions of $A_{V} = 4.5$
      and 3.0, respectively, though we do not attempt to robustly
      determine the extinction here.)
      \label{fig:2m0435+2m0557}}
\end{figure}

\begin{figure}
  \includegraphics[width=6in]{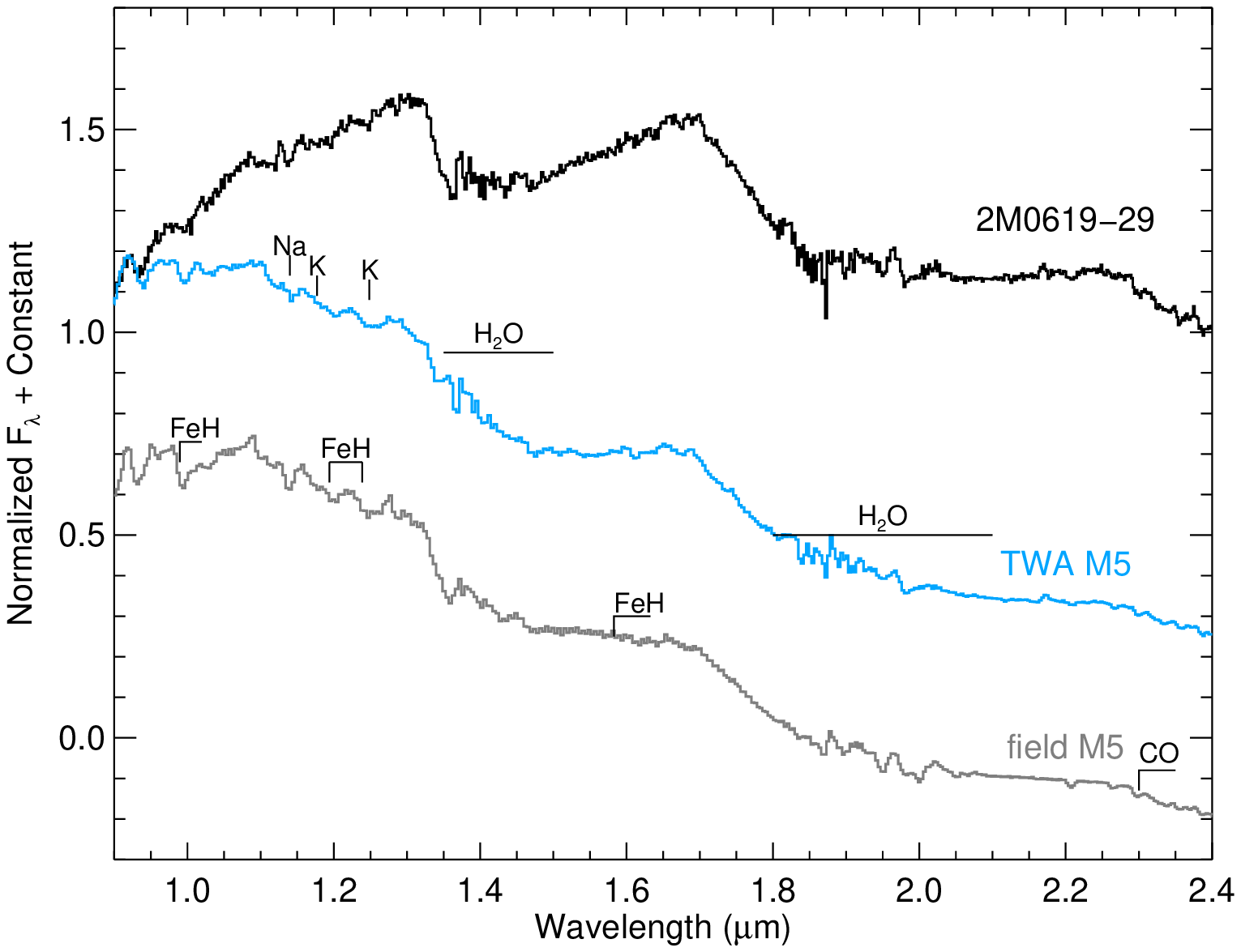}
  \caption{\normalsize The spectrum of 2MASS~J0619$-$2903
      \citep{2013ApJ...772...79A} compared to the TWA member,
      2MASS~J1235$-$3950 \citep[M5~\vlg;][]{2013ApJ...772...79A}, and a
      field M5 \citep[Gl~51;][]{2010ApJS..190..100K}.  The spectrum of
      2MASS~J0619$-$2903 is significantly reddened and has the
      triangular $H$-band shape and weak FeH absorption features typcial
      of low gravity objects. \label{fig:2m0619}}
\end{figure}

\begin{figure}
  \includegraphics[width=6in]{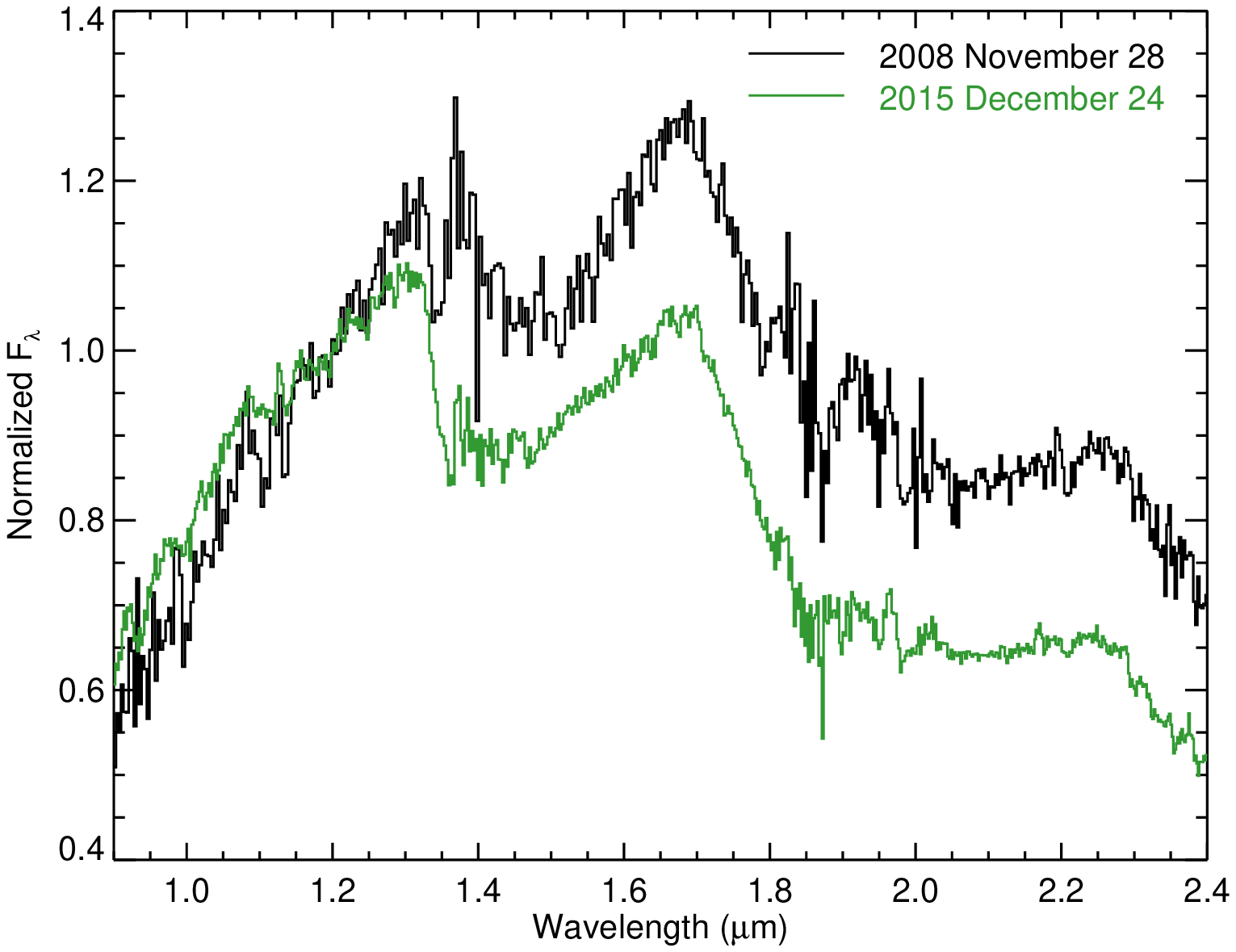}
  \caption{\normalsize Comparison of our two epochs of near-IR
      prism spectra of 2MASS~J0619$-$2903. The green-colored spectra was
      obtained by us in 2015, while the black spectra was obtained in
      2008 \citep{2013ApJ...772...79A}. Spectra are normalized to the
      median flux from 1.1--1.3~\micron.  The two epochs show
      significant differences in overall spectral shape, suggesting that
      2MASS J0619-2903 may be variable. \label{fig:2m0619-2epochs}}
\end{figure}

\begin{figure}
  \includegraphics[width=6in]{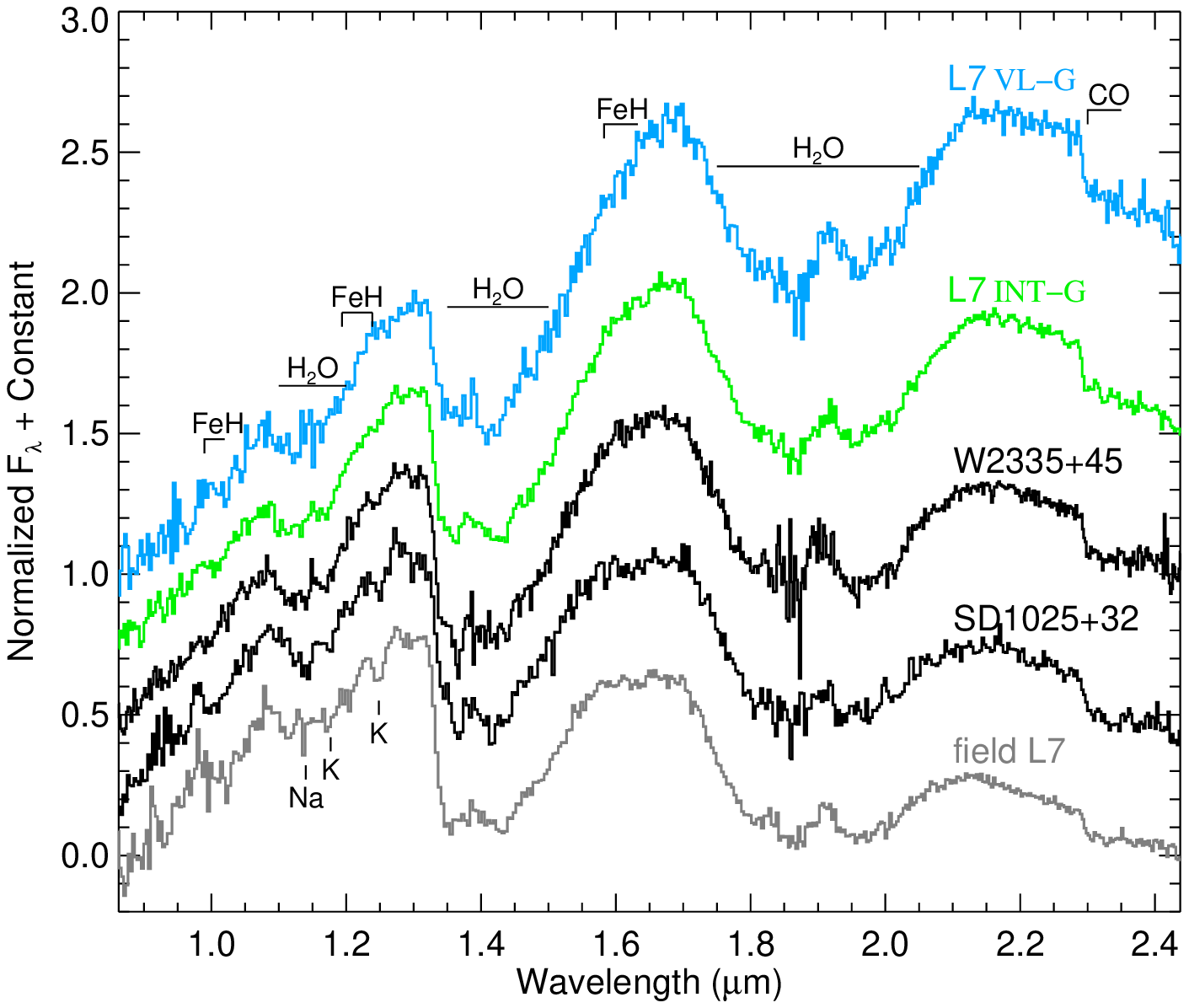}
  \caption{\normalsize Near-IR spectra of SDSS~J1025+3212
    \citet{2006AJ....131.2722C} and WISE~J2335+4511 (this work) compared
    to L7 dwarfs with a range of gravity classications
    \citep{2013ApJ...772...79A}. The L7 \vlg, \intg, and \fldg\ spectra
    are PSO~J318.5-22 \citep{2013ApJ...777L..20L}, WISE~J0047+6803
    \citep{2012AJ....144...94G}, and DENIS~J0205$-$1159
    \citep{2010ApJ...710.1142B}, respectively. The gravity
    classifications of PSO~J318.5-22 and WISE~0047+6803
    are consistent with their membership in the $\beta$~Pictoris
    ($\approx$25~Myr; \citealp{2015MNRAS.454..593B}) and AB~Doradus
    ($\approx$150~Myr; \citealp{2015MNRAS.454..593B}) moving groups,
    respectively. Based on visual comparison, we assign gravity
    designations of \fldg\ to both SDSS~J1025+3212 and WISE~2335+4511,
    but note that higher-resolution spectroscopy would allow for a more
    thorough analysis. \label{fig:L7comp}}
\end{figure}

\begin{figure}
  \includegraphics[width=6in]{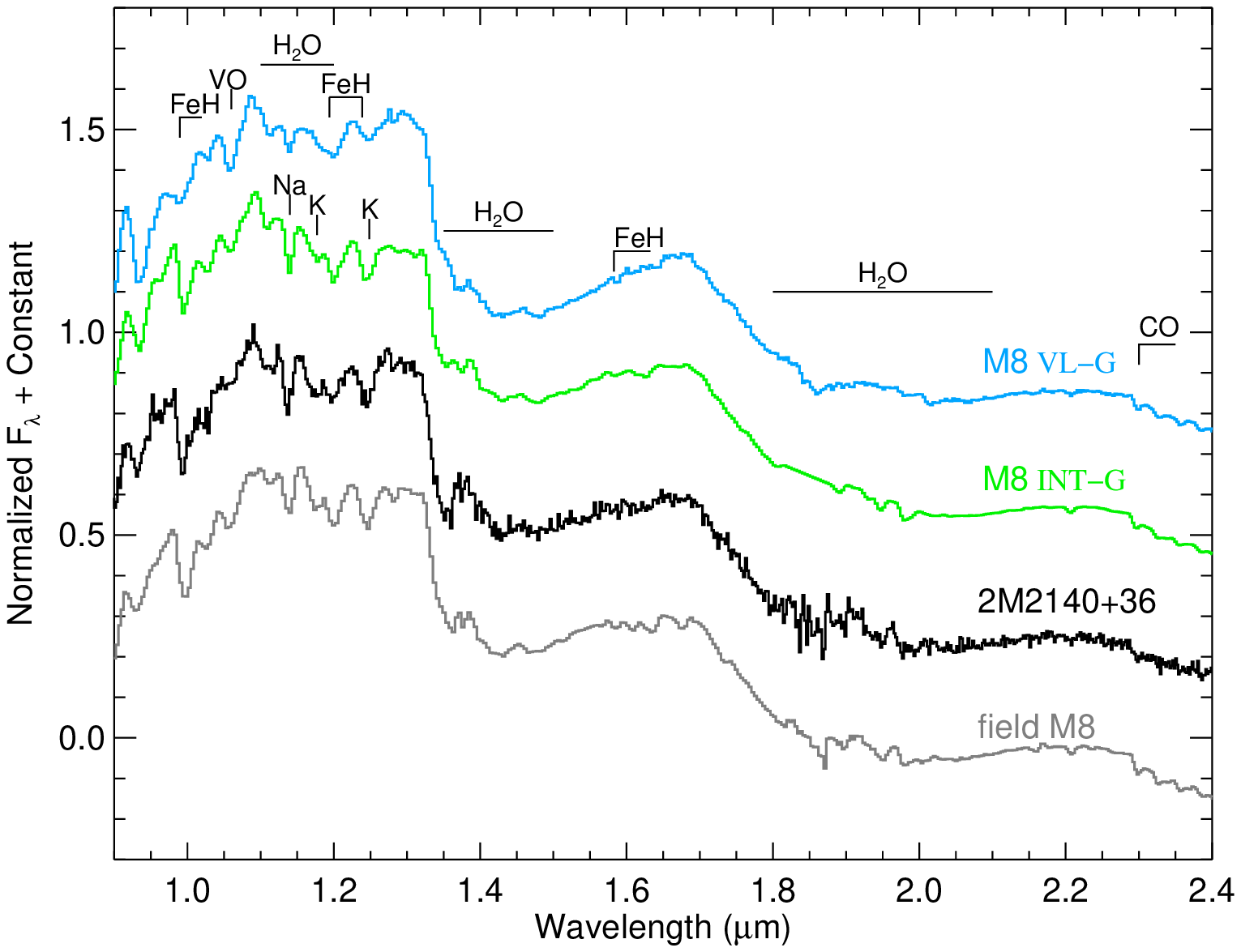}
  \caption{\normalsize Our IRTF/SpeX near-IR spectrum of
    2MASS~J2140+3655 (July 2015 epoch) compared to objects spanning a
    range of gravity classifications from
    \citet{2013ApJ...772...79A}. The M8 \vlg, \intg, and \fldg\ spectra
    are 2MASS~J1207$-$3932 \citep{2007ApJ...669L..97L}, 2MASS~J0019+4614
    \citep{2013ApJ...772...79A}, and vB~10 \citep{2004AJ....127.2856B},
    respectively. From this comparison, we type 2MASS~J2140+3655 as
    M8p~\fldg. Its triangular $H$-band continuum is peculiar given the
    lack of other low-gravity indicators in its
    spectrum. \label{fig:2mass2140}}
\end{figure}

\begin{figure}
  \includegraphics[width=6in]{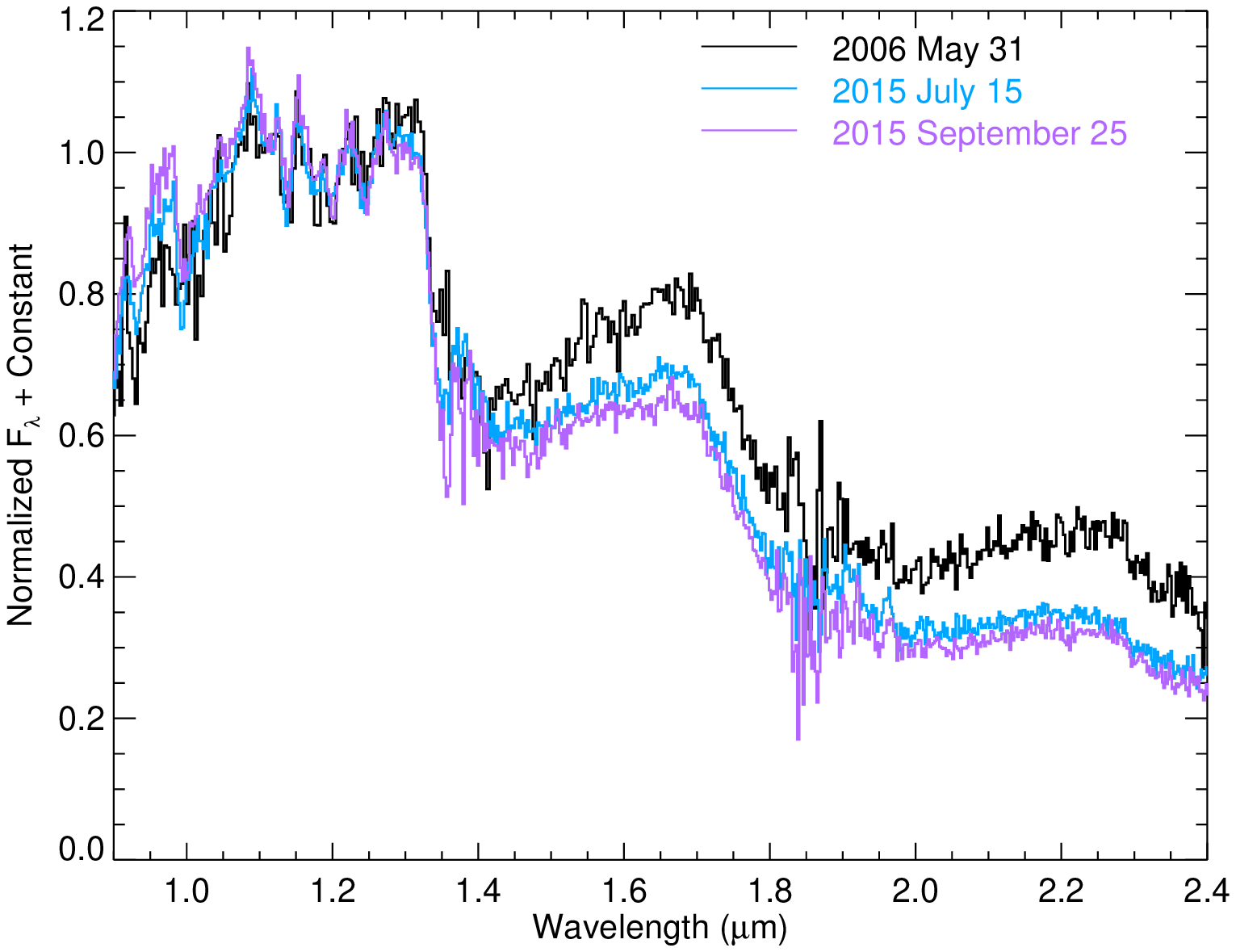}
  \caption{\normalsize Three epochs of near-IR prism spectra for
    2MASS~J2140+3655. The two colored spectra were obtained by us in
    2015 (Table~\ref{table:spex}), while the black one was obtained in
    2006 by \citet{2010ApJS..190..100K}. Spectra are normalized to the
    median flux from 1.1--1.3~\micron. \label{fig:2mass2140-compare}}
\end{figure}

\clearpage
\begin{deluxetable}{lccccccccccl}
\setlength{\tabcolsep}{0.035in}
\tabletypesize{\tiny}
\tablecaption{Summary of CFHT/WIRCam Observed Sample \label{table:sample}}
\tablewidth{0pt}
\tablehead{
  \colhead{Target} &
  \colhead{Spec.\ Type} &
  \colhead{Filter} &
  \colhead{FWHM} &
  \colhead{$\Delta$AM$_{\rm max}$} &
  \colhead{$N_{\rm fr}$} &
  \colhead{$N_{\rm ep}$} &
  \colhead{$\Delta{t}$} &
  \colhead{$N_{\rm ref}$} &
  \colhead{$N_{\rm cal}$} &
  \colhead{$\pi_{\rm abs}-\pi_{\rm rel}$} &
  \colhead{Ref.\ for attributes} \\
  \colhead{} &
  \colhead{Optical/NIR} &
  \colhead{} &
  \colhead{(\arcsec)} &
  \colhead{} &
  \colhead{} &
  \colhead{} &
  \colhead{(yr)} &
  \colhead{} &
  \colhead{} &
  \colhead{(mas)} &
  \colhead{(youth, etc.)} 
}

\startdata

\multicolumn{12}{c}{\bf Objects with spectroscopic signatures of youth (low gravity, lithium)} \\
\cline{1-12}
2MASSI~J0019262+461407      & M8/M8~\textsc{int-g}              & \Kn & $0.53\pm0.09$ & 0.040 &  242 &   12 & 2.95 &   63 &   61 & $1.17\pm0.11$ & 3, 41                                  \\  
PC~0025+0447                & M9.5$\ddagger$/L0~\textsc{int-g}  & $J$ & $0.63\pm0.10$ & 0.043 &  179 &    9 & 2.24 &   65 &   61 & $0.23\pm0.02$ & 3, 29                                  \\  
2MASSW~J0033239$-$152131    & L4$\beta$/L1~\textsc{fld-g}       & $J$ & $0.58\pm0.15$ & 0.046 &  186 &   12 & 2.18 &   29 &   14 & $ 1.6\pm0.2 $ & 3, 8, 40                               \\  
2MASSW~J0045214+163445      & L2$\beta$/L2~\textsc{vl-g}        & \Kn & $0.71\pm0.12$ & 0.053 &  130 &   12 & 2.19 &   32 &   29 & $1.56\pm0.20$ & 3, 8, 40                               \\  
WISEP~J004701.06+680352.1   & \nodata/L7~\textsc{int-g}         & $J$ & $0.61\pm0.09$ & 0.046 &   96 &   11 & 3.18 & 1192 &  248 & $0.99\pm0.03$ & 16, 24                                 \\  
2MASSI~J0117474$-$340325    & L2?$\ddagger$/L1~\textsc{int-g}   & $J$ & $0.69\pm0.08$ & 0.015 &  106 &    7 & 2.32 &   48 &   16 & $1.42\pm0.17$ & 3, 9                                   \\  
2MASS~J01262109+1428057     & L4$\gamma$/L2~\textsc{vl-g}       & $J$ & $0.64\pm0.22$ & 0.062 &  133 &   12 & 2.41 &   66 &   63 & $0.43\pm0.05$ & 3, 11, 34                              \\
\enddata
\end{deluxetable}

\begin{landscape}
\begin{deluxetable}{lcccclllclllcc}
\setlength{\tabcolsep}{0.050in}
\tabletypesize{\tiny}
\tablecaption{CFHT/WIRCam Parallax and Proper Motion Results \label{table:plx}}
\tablewidth{0pt}
\tablehead{                            
\colhead{} &
\colhead{} &
\colhead{} &
\colhead{} &
\colhead{} &
\multicolumn{3}{c}{Relative} & 
\colhead{} &
\multicolumn{3}{c}{Absolute} & 
\colhead{} &
\colhead{} \\
\cline{6-8}
\cline{10-12}
\colhead{Target} &
\colhead{$\alpha_{J2000}$} &
\colhead{$\delta_{J2000}$} &
\colhead{Epoch} &
\colhead{} &
\colhead{$\pi_{\rm rel}$} &
\colhead{$\mu_{\alpha, {\rm rel}}\cos{\delta}$} &
\colhead{$\mu_{\delta, {\rm rel}}$} &
\colhead{} &
\colhead{$\pi_{\rm abs}$} &
\colhead{$\mu_{\alpha, {\rm abs}}\cos{\delta}$} &
\colhead{$\mu_{\delta, {\rm abs}}$} &
\colhead{} &
\colhead{$\chi^2$/dof} \\
\colhead{} &
\colhead{(deg)} &
\colhead{(deg)} &
\colhead{(MJD)} &
\colhead{} &
\colhead{(\arcsec)} &
\colhead{(\arcsec\ yr$^{-1}$)} &
\colhead{(\arcsec\ yr$^{-1}$)} &
\colhead{} &
\colhead{(\arcsec)} &
\colhead{(\arcsec\ yr$^{-1}$)} &
\colhead{(\arcsec\ yr$^{-1}$)} &
\colhead{} &
\colhead{}}
\startdata

2MASSI~J0019262+461407         & 004.8602716 & $+$46.2353705 & 55423.55 & & 0.0243(15)    &  \phs0.1194(9)\phn &   $-$0.0754(9)\phn & & 0.0255(15)    &  \phs0.1193(13)    &   $-$0.0777(11)    & &     24.8/19     \\
PC~0025+0447                   & 006.9251141 & $+$05.0617997 & 56497.62 & & 0.0102(8)\phn &  \phs0.0166(11)    &   $-$0.0014(11)    & & 0.0104(8)\phn &  \phs0.0161(11)    &   $-$0.0033(12)    & &     12.6/13     \\
2MASSW~J0033239$-$152131       & 008.3501667 & $-$15.3585656 & 55050.54 & & 0.0412(18)    &  \phs0.3025(10)    &  \phs0.0360(15)    & & 0.0428(18)    &  \phs0.3061(24)    &  \phs0.0324(26)    & &     22.1/19     \\
2MASSW~J0045214+163445         & 011.3387503 & $+$16.5794601 & 55050.55 & & 0.0644(13)    &  \phs0.3513(11)    &   $-$0.0475(12)    & & 0.0659(13)    &  \phs0.3544(22)    &   $-$0.0511(20)    & &     16.9/19     \\
WISEP~J004701.06+680352.1      & 011.7555865 & $+$68.0645252 & 55780.58 & & 0.0816(19)    &  \phs0.3846(11)    &   $-$0.2005(14)    & & 0.0823(18)    &  \phs0.3807(11)    &   $-$0.2042(14)    & &     22.4/17     \\
2MASSI~J0117474$-$340325       & 019.4479413 & $-$34.0572078 & 55050.59 & & 0.0247(19)    &  \phs0.1060(14)    &   $-$0.0498(35)    & & 0.0261(19)    &  \phs0.1115(21)    &   $-$0.0524(38)    & &     11.0/9 \phn \\
2MASS~J01262109+1428057        & 021.5883123 & $+$14.4682530 & 56500.62 & & 0.0167(12)    &  \phs0.0760(11)    &   $-$0.0362(10)    & & 0.0171(12)    &  \phs0.0749(12)    &   $-$0.0394(12)    & &     17.1/19     \\
WISEP~J020625.26+264023.6      & 031.6058917 & $+$26.6733413 & 56585.42 & & 0.0514(14)    &  \phs0.4439(19)    &   $-$0.0377(23)    & & 0.0521(14)    &  \phs0.4427(21)    &   $-$0.0419(23)    & & \phn 9.7/11     \\
\enddata

\tablecomments{This table gives all the astrometric parameters derived
  from our Markov Chain Monte Carlo (MCMC) analysis for each
  target. For parameters in units of arcseconds, errors are given in
  parentheses in units of $10^{-4}$~arcsec. ($\alpha$, $\delta$,
  MJD):~Coordinates that correspond to the epoch listed, which is the
  first epoch of our observations for that target.  ($\pi$,
  $\mu_{\alpha}\cos{\delta}$, $\mu_{\delta}$):~Parallax and proper
  motion parameters are listed as both as relative and absolute
  values.  Relative values are our directly fitted results.  Absolute
  values include a correction for the mean parallax and proper motion
  of our reference stars, which includes an additional uncertainty
  added in quadrature to our direct fitting results (see
  Section~\ref{sec:parallaxes}).  $\chi^2$/dof:~The lowest $\chi^2$ in each set
  of MCMC chains along with the degrees of freedom (dof).}

\end{deluxetable}
\end{landscape}

\begin{deluxetable}{llccccl}
\tablecaption{IRTF/SpeX Prism Near-IR Spectroscopy \label{table:spex}}
\tabletypesize{\footnotesize}
\tablewidth{0pt}
\tablehead{
 \colhead{Object} &
 \colhead{UT Date} &
 \colhead{Slit (\arcsec)} &
 \colhead{$\sec z$} &
 \colhead{$N_{exp} \times t$ (s)} &
 \colhead{$T_{int}$ (s)} &
 \colhead{$\langle S/N \rangle\ (YJHK)$} 
}

\startdata
CFHT-Hy-20                 & 2009-11-03  &   0.5 $\times$ 15 &   1.21 &   26 $\times$ 120   &   3120.0 &   20,  54,  34,  20 \\
2MASS J0619$-$2903         & 2015-12-24  &   0.5 $\times$ 15 &   1.52 &   12 $\times$ 119.5 &   1434.5 &   70,  98,  98, 102 \\
DENIS-P J1441$-$0945AB     & 2015-01-28  &   0.5 $\times$ 15 &   1.19 &    9 $\times$ 9.7   &     87.6 &   40,  65,  56,  46 \\
2MASS J2140+3655           & 2015-07-15  &   0.5 $\times$ 15 &   1.06 &    9 $\times$ 59.8  &    537.9 &   59,  67,  49,  36 \\ 
                           & 2015-09-25  &   0.5 $\times$ 15 &   1.05 &    9 $\times$ 119.5 &   1075.9 &   58,  65,  45,  35 \\ 
WISE J2335+4511            & 2015-07-23  &   0.5 $\times$ 15 &   1.21 &    9 $\times$ 179.8 &   1618.0 &   20,  35,  38,  50 \\
\enddata 

\tablecomments{The last column gives the median S/N of the spectrum
  within the standard near-IR bandpasses.}

\end{deluxetable}

\begin{landscape}
\begin{deluxetable}{llclcllllcl}
\setlength{\tabcolsep}{0.050in}
\tabletypesize{\scriptsize}
\tablecaption{Near-IR Spectral Classification of Late-M and L~Dwarfs \label{table:spectra}}
\tablewidth{0pt}
\tablehead{                            
  \colhead{Object} &
  \multicolumn{2}{c}{Published SpT} &
  \multicolumn{2}{c}{SpT (Visual)} &
  \multicolumn{4}{c}{SpT (Index)} &
  \colhead{Gravity} &
  \colhead{SpT} \\
  \colhead{} &
  \colhead{Opt./NIR} &
  \colhead{Refs} &
  \colhead{$J$-band\tablenotemark{a}} &
  \colhead{$K$-band\tablenotemark{a}} &
  \colhead{H$_2$O} &
  \colhead{H$_2$OD} &
  \colhead{H$_2$O-1} &
  \colhead{H$_2$O-2} &
  \colhead{Scores\tablenotemark{a}} &
  \colhead{Adopted} \\
}
\startdata

2MASSI~J0030300$-$145033         & L7/L4-L6$\beta$  &3, 4, 5& L6$\pm$1   & L7$\pm$1   & \nodata      & L4.6$\pm$0.9 & \nodata      & \nodata      & nnn1 & L6~\textsc{fld-g}\tablenotemark{b}\\
2MASSI~J0619526$-$290359         & M6/M5            & 14, 1 & M6$\pm$1   & M6 $\pm$1  & M7.0$\pm$0.5 & \nodata      & M3.8$\pm$1.1 & M5.9$\pm$0.5 & 2n22 & M6~\textsc{vl-g}\tablenotemark{b}\\
LP~261-75B                       & L6/L4.5          & 1, 5  & L7$\pm$1   & L6$\pm$1   & \nodata      & L4.5$\pm$0.9 & \nodata      & \nodata      & nnn1 & L6~\textsc{fld-g}\tablenotemark{b}\\
2MASS~J13153094$-$2649513AB      & L5.5/L5          & 6, 13 & L7$\pm$1   & L5$\pm$1   & \nodata      & L5.9$\pm$0.8 & \nodata      & \nodata      & nnn0 & L6~\textsc{fld-g} \\
2MASS~J13313310+3407583          & L0/L1p           &   7   & M9$\pm$1   & L0$\pm$1   & L0.1$\pm$0.5 & \nodata      & M7.0$\pm$1.1 & M9.7$\pm$0.6 & 0000 & L0~\textsc{fld-g} \\
DENIS~J144137.2$-$094558AB       & L0.5/\nodata     &   6   & L1$\pm$1   & L1$\pm$1   & L1.3$\pm$0.5 & L1.1$\pm$0.9 & L0.0$\pm$1.1 & L1.0$\pm$0.6 & 00?0 & L1~\textsc{fld-g} \\
2MASS~J18212815+1414010          & L4.5/L5p         &  7, 8 & L4$\pm$1   & L5$\pm$1   & \nodata      & L5.3$\pm$0.9 & L4.6$\pm$1.1 & \nodata      & 0n0? & L5~\textsc{fld-g} \\
2MASS~J21403907+3655563          & \nodata /M8p     &   7   & M7.5$\pm$1 & M8.5$\pm$1 & M7.2$\pm$0.6 & \nodata      & M9.7$\pm$1.1 & M7.8$\pm$0.8 & ?n01 & M8p~\textsc{fld-g} \\
\enddata

\tablenotetext{a}{Gravity scores are listed in the following order: FeH,
  VO, alkali lines, and $H$-band continuum shape. See
  \citet{2013ApJ...772...79A} for details.}

\tablenotetext{b}{Gravity classification based on visual inspection.}

\tablerefs{
(1)~\citet{2013ApJ...774...55B};
(2)~\citet{2006AJ....131.1007B};
(3)~\citet{2010ApJ...710.1142B};
(4)~\citet{2015ApJ...798...73G};
(5)~\citet{2000AJ....120..447K};
(6)~\citet{2008ApJ...689.1295K};
(7)~\citet{2010ApJS..190..100K};
(8)~\citet{2008ApJ...686..528L};
(9)~\citet{2013PASP..125..809T};
(10)~\citet{2004AJ....127.3553K};
(11)~\citet{2003AJ....126.2421C};
(12)~\citet{2014AJ....147...34S};
(13)~\citet{2011ApJ...739...49B};
(14)~\citet{2007AJ....133..439C}
}
\end{deluxetable}
\end{landscape}

\begin{deluxetable}{lllllllllllllllllccc}
\tablecaption{Near-IR Spectral Classification of Late-L and T~Dwarfs\label{table:spectra-lateL+T}}
\tabletypesize{\footnotesize}
\rotate
\tablewidth{0pt}
\tablehead{
 \colhead{Object} &
 \colhead{\htwoo-$J$} &
 \colhead{\meth-$J$ } &
 \colhead{\htwoo-$H$} &
 \colhead{\meth-$H$ } &
 \colhead{\meth-$K$ } &
 \colhead{\htwoo-$K$} &
 \colhead{SpT: Index}  &
 \colhead{SpT: Visual} &
 \colhead{SpT: Final}  
}

\startdata
 WISE~J0206+2640  &  0.675 (L8.0) & 0.806 ($<$T0) & 0.695 (L6.7) & 1.075 ($<$T1) & 0.922 (L6.9) & 0.726 ( -- ) &  L7.2$\pm$0.7  &  L8 (red)  &  L8 (red)   \\
 CFHT-Hy-20       &  0.461 (T2.9) & 0.670 ($<$T0) & 0.459 (T3.3) & 0.896  (T1.6) & 0.544 (T2.3) & 0.539 ( -- ) &  T2.5$\pm$0.7  &  T2        &  T2.5       \\
 WISE~J0754+7909  &  0.471 (T2.7) & 0.652 (T0.1)  & 0.465 (T3.2) & 1.072 ($<$T1) & 0.715 (T0.7) & 0.478 ( -- ) &  T1.7$\pm$1.5  &  T2p (red) &  T2p (red)  \\

\enddata

\end{deluxetable}

\begin{landscape}
\begin{deluxetable}{lcccccccccccccccl}
\setlength{\tabcolsep}{0.020in}
\tabletypesize{\tiny}
\tablecaption{Photometry for Our Sample and Other Potentially Young Objects \label{table:photom}}
\tablewidth{0pt}
\tablehead{                            
\colhead{} &
\colhead{} &
\colhead{} &
\colhead{} &
\multicolumn{4}{c}{MKO} & 
\colhead{} &
\multicolumn{3}{c}{2MASS} & 
\colhead{} &
\multicolumn{2}{c}{AllWISE} & 
\colhead{} &
\colhead{} \\
\cline{5-8}
\cline{10-12}
\cline{14-15}
\colhead{Object} &
\colhead{Parallax} &
\colhead{Ref.} &
\colhead{} &
\colhead{$Y$} &
\colhead{$J$} &
\colhead{$H$} &
\colhead{$K$} &
\colhead{} &
\colhead{$J$} &
\colhead{$H$} &
\colhead{\Ks} &
\colhead{} &
\colhead{W1} &
\colhead{W2} &
\colhead{} &
\colhead{Refs.} \\
\colhead{} &
\colhead{(mas)} &
\colhead{} &
\colhead{} &
\colhead{(mag)} &
\colhead{(mag)} &
\colhead{(mag)} &
\colhead{(mag)} &
\colhead{} &
\colhead{(mag)} &
\colhead{(mag)} &
\colhead{(mag)} &
\colhead{} &
\colhead{(mag)} &
\colhead{(mag)} &
\colhead{} &
\colhead{}}
\startdata

HD~984B                                     &     $ 21.2\pm0.6  $     & 52 &&       \nodata          &       \nodata          &     \{$12.63\pm0.05$\} &     \{$12.15\pm0.04$\} &&       \nodata          &       $12.58\pm0.05$   &       $12.19\pm0.04$   &&    \nodata         &    \nodata         && 1, 44             \\  
HD~1160B                                    & \phn$  9.7\pm0.5  $     & 52 &&       \nodata          &       $14.60\pm0.11$   &       $14.65\pm0.08$   &     \{$13.91\pm0.12$\} &&     \{$14.66\pm0.11$\} &     \{$14.60\pm0.08$\} &       $13.95\pm0.12$   &&    \nodata         &    \nodata         && 1, 38, 47         \\  
2MASSIJ0019262+461407                       &     $ 25.5\pm1.5  $     & 1  &&      [$13.22\pm0.05$]  &      [$12.56\pm0.02$]  &      [$11.98\pm0.02$]  &      [$11.47\pm0.01$]  &&       $12.60\pm0.02$   &       $11.94\pm0.02$   &       $11.50\pm0.01$   &&     $11.28\pm0.02$ &     $11.02\pm0.02$ && 1, 2              \\  
PC~0025+0447                                &     $ 10.4\pm0.8  $     & 1  &&       $17.18\pm0.02$   &       $16.14\pm0.02$   &       $15.52\pm0.02$   &       $14.94\pm0.02$   &&       $16.19\pm0.09$   &       $15.29\pm0.10$   &       $14.96\pm0.12$   &&     $14.62\pm0.03$ &     $14.07\pm0.05$ && 1, 2, 3           \\  
2MASSWJ0030300$-$145033                     &     $   37\pm5    $     & 53 &&      [$17.61\pm0.05$]  &       $16.39\pm0.03$   &       $15.37\pm0.03$   &       $14.49\pm0.03$   &&       $16.28\pm0.11$   &       $15.27\pm0.10$   &       $14.48\pm0.10$   &&     $13.68\pm0.03$ &     $13.33\pm0.03$ && 1, 2, 29          \\  
2MASSWJ0033239$-$152131                     &     $ 42.8\pm1.8  $     & 1  &&      [$16.34\pm0.06$]  &      [$15.23\pm0.06$]  &      [$14.25\pm0.05$]  &      [$13.39\pm0.04$]  &&       $15.29\pm0.06$   &       $14.21\pm0.05$   &       $13.41\pm0.04$   &&     $12.82\pm0.02$ &     $12.50\pm0.03$ && 1, 2              \\
\enddata
\end{deluxetable}
\end{landscape}

\begin{landscape}
\begin{deluxetable}{lcccccccccccccccc}
\setlength{\tabcolsep}{0.020in}
\tabletypesize{\tiny}
\tablecaption{Absolute Magnitudes for Our Sample and Other Potentially Young Objects \label{table:absmag}}
\tablewidth{0pt}
\tablehead{                            
\colhead{} &
\colhead{} &
\colhead{} &
\colhead{} &
\multicolumn{4}{c}{MKO} & 
\colhead{} &
\multicolumn{3}{c}{2MASS} & 
\colhead{} &
\multicolumn{2}{c}{AllWISE} & 
\colhead{} &
\colhead{} \\
\cline{5-8}
\cline{10-12}
\cline{14-15}
\colhead{Object} &
\colhead{Spec.\ Type} &
\colhead{Ref.} &
\colhead{} &
\colhead{$M_Y$} &
\colhead{$M_J$} &
\colhead{$M_H$} &
\colhead{$M_K$} &
\colhead{} &
\colhead{$M_J$} &
\colhead{$M_H$} &
\colhead{$M_{\Ks}$} &
\colhead{} &
\colhead{$M_{\rm W1}$} &
\colhead{$M_{\rm W2}$} &
\colhead{} &
\colhead{Notes} \\
\colhead{} &
\colhead{Optical/NIR} &
\colhead{} &
\colhead{} &
\colhead{(mag)} &
\colhead{(mag)} &
\colhead{(mag)} &
\colhead{(mag)} &
\colhead{} &
\colhead{(mag)} &
\colhead{(mag)} &
\colhead{(mag)} &
\colhead{} &
\colhead{(mag)} &
\colhead{(mag)} &
\colhead{} &
\colhead{}}
\startdata

\multicolumn{17}{c}{\bf Very low gravity objects (\vlg)} \\
\cline{1-17}
2MASSJ0045+1634                       & L2$\beta$/L2~\textsc{vl-g}      & 5, 22, 73     &&      [$13.22\pm0.07$]  &      [$12.08\pm0.05$]  &      [$11.21\pm0.06$]  &      [$10.43\pm0.05$]  &&       $12.15\pm0.05$   &       $11.15\pm0.06$   &       $10.46\pm0.05$   && \phn$ 9.88\pm0.05$ & \phn$ 9.49\pm0.05$ && Argus                         \\  
2MASSJ0126+1428                       & L4$\gamma$/L2~\textsc{vl-g}     & 5, 30         &&       $15.15\pm0.17$   &       $13.62\pm0.16$   &       $12.31\pm0.15$   &       $11.27\pm0.15$   &&       $13.27\pm0.26$   &       $12.33\pm0.27$   &       $11.44\pm0.21$   &&     $10.38\pm0.15$ & \phn$ 9.85\pm0.16$ && \nodata                       \\  
2MASSJ0241$-$0326                     & L0$\gamma$/L1~\textsc{vl-g}     & 5, 22         &&      [$13.19\pm0.26$]  &      [$12.06\pm0.26$]  &      [$11.20\pm0.26$]  &      [$10.33\pm0.26$]  &&       $12.13\pm0.26$   &       $11.15\pm0.26$   &       $10.37\pm0.26$   && \phn$ 9.98\pm0.25$ & \phn$ 9.62\pm0.25$ && (Tuc-Hor)                     \\  
2MASSJ0335+2342                       & M8.5/M7~\textsc{vl-g}           & 5, 33, 71, 79 && \phn [$ 9.57\pm0.18$]  & \phn [$ 8.89\pm0.18$]  & \phn [$ 8.38\pm0.18$]  & \phn [$ 7.92\pm0.18$]  && \phn  $ 8.94\pm0.18$   & \phn  $ 8.35\pm0.18$   & \phn  $ 7.95\pm0.18$   && \phn$ 7.75\pm0.20$ & \phn$ 7.47\pm0.18$ && $\beta$~Pic,~strong~H$\alpha$ \\  
2MASSJ0355+1133                       & L5$\gamma$/L3~\textsc{vl-g}     & 5, 22, 73     &&      [$15.54\pm0.06$]  &      [$14.15\pm0.04$]  &      [$12.79\pm0.04$]  &      [$11.67\pm0.03$]  &&       $14.25\pm0.04$   &       $12.73\pm0.04$   &       $11.72\pm0.03$   &&     $10.72\pm0.04$ &     $10.14\pm0.03$ && AB~Dor                        \\
\enddata
\end{deluxetable}
\end{landscape}

\begin{deluxetable}{lccrrccccc}
\setlength{\tabcolsep}{0.050in}
\tabletypesize{\scriptsize}
\tablecaption{Comparison of CFHT Parallaxes to Literature Values \label{table:compare}}
\tablewidth{0pt}
\tablehead{
\colhead{Object} &
\multicolumn{4}{c}{Absolute Parallax (mas)} & 
\colhead{} &
\multicolumn{2}{c}{$\Delta$Proper Motion (mas/yr)\tablenotemark{a}} & 
\colhead{} &
\colhead{Ref.} \\
\cline{2-5}
\cline{7-8} 
\colhead{} &
\colhead{CFHT} &
\colhead{Lit.} &
\multicolumn{2}{c}{Difference} & 
\colhead{} &
\colhead{RA} &
\colhead{Dec.} &
\colhead{} &
\colhead{}}
\startdata

\multicolumn{10}{c}{\citet{2012ApJ...752...56F,2013AJ....145....2F}} \\
\cline{1-10}
2MASS J03552337+1133437         &     $109.5\pm 1.4$\phn &     $122.0\pm13.0$     & $ 1.0\sigma$\phs & $  11\%$\phs & &     \phs$  5.6\pm 5.2$     &     \phs$  0.4\pm 5.2$     & & 1   \\
GJ~3276                         & \phn$  3.9\pm 1.0$     &     $ 24.8\pm 3.1$     & $ 6.4\sigma$\phs & $ 536\%$\phs & &     \phs$ 20.8\pm 3.1$\phn &         $ -8.8\pm 3.1$     & & 2   \\
2MASS J05012406$-$0010452       &     $ 48.4\pm 1.4$     &     $ 76.4\pm 4.8$     & $ 5.6\sigma$\phs & $  58\%$\phs & &     \phs$  7.3\pm 4.5$     &         $-11.2\pm 4.5$\phn & & 2   \\
2MASSI J0518461$-$275645        &     $ 18.4\pm 1.1$     &     $ 21.4\pm 6.9$     & $ 0.4\sigma$\phs & $  16\%$\phs & &     \phs$  6.3\pm 4.5$     &     \phs$ 11.1\pm 4.4$\phn & & 2   \\
2MASSI J0536199$-$192039        &     $ 21.1\pm 1.6$     &     $ 25.6\pm 9.4$     & $ 0.5\sigma$\phs & $  21\%$\phs & &     \phs$  0.9\pm 5.5$     &     \phs$  4.2\pm 5.3$     & & 2   \\
2MASSI J0608528$-$275358        &     $ 25.0\pm 1.5$     &     $ 32.0\pm 3.6$     & $ 1.8\sigma$\phs & $  28\%$\phs & &     \phs$  3.5\pm 3.6$     &         $ -3.4\pm 3.6$     & & 2   \\
2MASS J10220489+0200477         &     $ 34.6\pm 2.0$     & \phn$ 26.4\pm11.5$     & $-0.7\sigma$\phs & $ -24\%$\phs & &         $-22.1\pm 7.1$\phn &     \phs$ 10.5\pm 7.1$\phn & & 2   \\
\enddata

\tablerefs{(1)~\citet{2013AJ....145....2F},
  (2)~\citet{2012ApJ...752...56F}, (3)~\citet{2014A&A...568A...6Z},
  (4)~\citet{2002AJ....124.1170D}, (5)~\citet{2015ApJ...799..203G},
  (6)~\citet{2012ApJ...758...56S}, (7)~\citet{2009AJ....137..402G},
  (8)~\citet{2004AJ....127.2948V}, (9)~\citet{2008A&A...489..825T},
  (10)~\citet{2014A&A...563A.121D}, (11)~\citet{2013ApJ...762..118W},
  (12)~\citet{2008A&A...477L...1D}, (13)~\citet{2007ApJ...669L..45G},
  (14)~\citet{2006AJ....132.1234C}, (15)~\citet{2011AJ....141...54A}, 
  (16)~\citet{2014AJ....147...94D}.}

\tablenotetext{a}{The difference between our absolute proper motions
  and literature values that are typically reported as relative.}

\end{deluxetable}

\begin{deluxetable}{lccrrc}
\setlength{\tabcolsep}{0.050in}
\tabletypesize{\small}
\tablecaption{Comparison of CFHT Parallax Distances to Literature Kinematic Distances \label{table:dkin}}
\tablewidth{0pt}
\tablehead{
\colhead{Object} &
\colhead{CFHT Distance (pc)} &
\colhead{Kinematic Distance (pc)} &
\multicolumn{2}{c}{Difference} &
\colhead{Ref.}}
\startdata

\multicolumn{6}{c}{AB Dor} \\
\cline{1-6}
2MASSI J0019262+461407          & $ 39.2^{+ 2.2}_{- 2.4}$ & $ 37.4^{+ 2.8}_{- 2.4}$ & $ -0.5\sigma$ & $  -5$\% & 1 \\ 
WISEP J004701.06+680352.1       & $ 12.1\pm 0.3       $ & $ 10.5^{+ 0.8}_{- 0.4}$ & $ -2.0\sigma$ & $ -14$\% & 1 \\ 
2MASS J03552337+1133437         & $  9.1\pm 0.1       $ & $  8.5^{+ 0.4}_{- 0.8}$ & $ -1.5\sigma$ & $  -7$\% & 1 \\ 
2MASSW J2244316+204343          & $ 17.0^{+ 0.3}_{- 0.3}$ & $ 18.5\pm 1.2       $ & $  1.2\sigma$ & $   9$\% & 1 \\ 

\cline{1-6}
\multicolumn{6}{c}{} \\                                                                                                                               
\multicolumn{6}{c}{Argus} \\
\cline{1-6}
2MASSW~J0045214+163445          & $ 15.2\pm 0.3       $ & $ 13.3^{+ 0.8}_{- 1.2}$ & $ -2.2\sigma$ & $ -12$\% & 1 \\
\enddata

\tablerefs{(1)~\citet{2014ApJ...783..121G}, (2)~\citet{2014ApJ...792L..17G}, (3)~\citet{2005ApJ...634.1385M}.}

\tablecomments{The group memberships listed here are those proposed in
  the literature. Objects with $\dag$ have changes in their memberships,
  as summarized in Table~\ref{table:membership}.}

\end{deluxetable}

\begin{landscape}
\begin{deluxetable}{llccccccccccccccccc}
\tablecaption{Coefficients of Linear Relations in the Paper \label{table:coeff}}
\tablewidth{0pt}
\tablehead{                            
\colhead{} &
\colhead{} &
\colhead{} &
\multicolumn{3}{c}{Field (IR types)} & 
\colhead{} &
\multicolumn{3}{c}{Field (Opt.\ types)} & 
\colhead{} &
\multicolumn{3}{c}{\vlg} & 
\colhead{} &
\multicolumn{3}{c}{\intg\tablenotemark{a}} \\
\cline{4-6}
\cline{8-10}
\cline{12-14}
\cline{16-18}
\colhead{$x$} &
\colhead{$y$} &
\colhead{} &
\colhead{$c_0$} &
\colhead{$c_1$} &
\colhead{rms} &
\colhead{} &
\colhead{$c_0$} &
\colhead{$c_1$} &
\colhead{rms} &
\colhead{} &
\colhead{$c_0$} &
\colhead{$c_1$} &
\colhead{rms} &
\colhead{} &
\colhead{$c_0$} &
\colhead{$c_1$} &
\colhead{rms} }
\startdata

SpT          & $Y_{\rm{MKO}}$      &&  8.437 &  0.402 & 0.35 &&  8.677 &  0.404 & 0.29     &&  3.689  &  0.799 & 0.77 &&  6.778 &  0.562 & 0.36 \\
SpT          & $J_{\rm{MKO}}$      &&  8.062 &  0.368 & 0.37 &&  8.131 &  0.364 & 0.26     &&  3.475  &  0.731 & 0.72 &&  5.941 &  0.537 & 0.50 \\
SpT          & $H_{\rm{MKO}}$      &&  7.516 &  0.348 & 0.33 &&  7.820 &  0.331 & 0.24     &&  3.778  &  0.633 & 0.64 &&  6.557 &  0.415 & 0.43 \\
SpT          & $K_{\rm{MKO}}$      &&  7.258 &  0.319 & 0.31 &&  7.433 &  0.312 & 0.25     &&  3.898  &  0.553 & 0.56 &&  6.555 &  0.355 & 0.39 \\
SpT          & $J_{\rm{2MASS}}$    &&  7.848 &  0.391 & 0.38 &&  8.021 &  0.380 & 0.30     &&  3.406  &  0.745 & 0.72 &&  5.914 &  0.545 & 0.50 \\
SpT          & $H_{\rm{2MASS}}$    &&  7.555 &  0.340 & 0.36 &&  7.723 &  0.332 & 0.30     &&  3.734  &  0.632 & 0.64 &&  6.552 &  0.411 & 0.47 \\
SpT          & $K_{S,{\rm{2MASS}}}$ &&  7.292 &  0.317 & 0.33 &&  7.491 &  0.308 & 0.29    &&  3.699   &  0.573 & 0.55 &&  6.573 &  0.356 & 0.36 \\
SpT          & W1                &&  7.698 &  0.243 & 0.26 &&  7.610 &  0.255  & 0.29    &&  4.392  &  0.457 & 0.46 &&  7.436 &  0.237 & 0.13 \\
\enddata

\tablecomments{These fits are defined as $y = c_0 + c_1x$, and the rms
  of the data about the fit is also given. Spectral type (SpT) is
  defined to be 10.0 for L0, 15.0 for L5, and so on. The field
  relations are valid for objects of spectral types M6 to L8. The
  \vlg\ relations are valid from M6 to L7. The \intg\ relations are
  valid from L0 to L7. For the \vlg\ sample, 2MASS~J1207$-$3932b was
  excluded from the fits. Its impact on the fitted coefficients is
  neglible, but its extreme outlying position would artificially
  inflate the resulting RMS values if it were included.
  Note that for some spectral type ranges for some filters, the linear
  fit is not a particularly good representation (\eg, \WISE\ photometry
  for late-L dwarfs) and more accurate photometric distances may be
  obtained by averaging over the specific spectral subclasses of interest.}

\tablenotetext{a}{For the \intg\ sample, these line fits for the late-M
  dwarfs are not very good, given the small number of objects and large
  dispersion (\eg, see Figure~\ref{fig:absmag-2mass}).}

\end{deluxetable}
\end{landscape}

\begin{deluxetable}{lrrrr}
  \setlength{\tabcolsep}{0.050in}
\tabletypesize{\footnotesize}
\tablecaption{Tangential Velocities \label{table:vtan}}
\tablewidth{0pt}
\tablehead{                            
  \colhead{} &
  \colhead{} &
  \multicolumn{3}{c}{LSR Corrected} \\
  \cline{3-5}
  \colhead{Object} &
  \colhead{\Vtan\ (\kms)} &
  \colhead{$\Delta\mu_{\alpha}\cos{\delta}$ (\masyr)} &
  \colhead{$\Delta\mu_{\delta}$ (\masyr)} &
  \colhead{$V_{\rm tan}^{\prime}$ (\kms)}
}
\startdata

\multicolumn{5}{c}{Low-gravity objects} \\
\cline{1-5}
2MASSI~J0019262+461407         & $ 26.5\pm  1.6$ & $  77.9\pm  4.6$ & $ -36.3\pm  2.1$ & $ 10.9\pm  1.6$ \\
PC~0025+0447                   & $  7.5\pm  0.8$ & $  31.6\pm  2.4$ & $ -17.5\pm  1.3$ & $  9.6\pm  0.7$ \\
2MASSW~J0030300$-$145033       & $ 31.3\pm  4.0$ & $ 113.4\pm 13.6$ & $ -54.3\pm  6.5$ & $ 17.0\pm  3.9$ \\
2MASSW~J0033239$-$152131       & $ 34.1\pm  1.5$ & $ 129.5\pm  5.4$ & $ -61.3\pm  2.6$ & $ 22.2\pm  1.4$ \\
2MASSW~J0045214+163445         & $ 25.8\pm  0.5$ & $ 197.1\pm  3.9$ & $-117.8\pm  2.3$ & $ 12.3\pm  0.5$ \\
WISEP~J004701.06+680352.1      & $ 24.9\pm  0.5$ & $ 245.7\pm  5.4$ & $-108.5\pm  2.4$ & $  9.5\pm  0.5$ \\
2MASSW~J0058425$-$065123       & $ 25.8\pm  3.2$ & $  99.4\pm 11.8$ & $ -51.6\pm  6.1$ & $ 11.3\pm  3.1$ \\
\enddata

\end{deluxetable}

\clearpage
\thispagestyle{empty}
\begin{deluxetable}{lcccccc}
\tablecaption{$UVW$ Determinations for Objects with Published Radial Velocities\label{table:uvw}}
\tabletypesize{\footnotesize}
\tablewidth{0pt}
\tablehead{
  \colhead{Object}  &
  \multicolumn{2}{c}{RV}  &
  \colhead{$U$} &
  \colhead{$V$} &
  \colhead{$W$} &
  \colhead{$v_{\rm{tot}}$} \\
\cline{2-3}
  \colhead{}  &
  \colhead{$v$ (\kms)} &
  \colhead{Ref}  &
  \colhead{(\kms)} &
  \colhead{(\kms)} &
  \colhead{(\kms)} &
  \colhead{(\kms)} 
}

\startdata
2MASSI~J0019262+461407      & $ -19.5\pm  2.0$                  & 8       & $  -7.2\pm  1.3$ & $ -30.1\pm  1.9$ & $ -11.0\pm  1.2$ & $  32.9\pm  1.7$\\ 
2MASSW~J0045214+163445      & $   3.3\pm  0.2$                  & 1       & $ -21.3\pm  0.4$ & $ -13.9\pm  0.4$ & $  -5.5\pm  0.2$ & $  26.0\pm  0.5$\\ 
WISEP~J004701.06+680352.1   & $ -20.0\pm  1.4$                  & 5       & $  -8.2\pm  0.9$ & $ -27.6\pm  1.2$ & $ -13.9\pm  0.3$ & $  31.9\pm  1.0$\\ 
SO~J025300.5+165258         & $  68.3\pm  0.1$\tablenotemark{a} & 16      & $ -69.5\pm  0.5$ & $ -71.9\pm  0.6$ & $ -58.8\pm  0.4$ & $ 116.0\pm  0.5$\\ 
2MASSI~J0253597+320637      & $ -36.3\pm  0.8$                  & 13      & $  17.8\pm  0.8$ & $ -43.2\pm  1.4$ & $   7.3\pm  0.5$ & $  47.3\pm  1.1$\\ 
2MASS~J03140344+1603056     & $  -8.0\pm  1.1$                  & 12      & $  15.9\pm  0.9$ & $   6.6\pm  0.3$ & $  -5.8\pm  0.7$ & $  18.2\pm  0.6$\\ 
2MASSI~J0335020+234235      & $  15.5\pm  1.7$                  & 15      & $ -16.6\pm  1.5$ & $ -12.6\pm  1.5$ & $  -9.5\pm  0.9$ & $  22.9\pm  1.6$\\ 
LP~944-20                   & $   7.6\pm  2.0$                  & 8       & $ -13.5\pm  0.6$ & $  -5.3\pm  1.0$ & $  -0.7\pm  1.6$ & $  14.6\pm  1.0$\\
\enddata

\tablenotetext{a}{Adopted radial velocity is the weighted mean of published values.}

\tablenotetext{b}{Radial velocity measurements are for the comoving
 M~dwarf companions to these objects \citep{1998Sci...282.1309R,
   2010ApJ...715..561A}.}

\tablerefs{
(1)~\citealp{2010ApJ...723..684B};   
(2)~\citealp{2012AJ....144...99D};   
(3)~\citealp{2013AJ....146..156D};   
(4)~\citealp{2015ApJ...808L..20G};   
(5)~\citealp{2015ApJ...799..203G};   
(6)~\citealp{2003ApJ...593L.109M};   
(7)~\citealp{2003MNRAS.342..837R};   
(8)~\citealp{2009ApJ...705.1416R};   
(9)~\citealp{2010ApJS..186...63R};   
(10)~\citealp{2010ApJ...715L.165R};  
(11)~\citealp{2010AJ....139.1808S};  
(12)~\citealp{2010A&A...512A..37S};  
(13)~E. Shkolnik, priv. comm.;
(14)~\citealp{2011ApJ...727....6S};  
(15)~\citealp{2012ApJ...758...56S};  
(16)~\citealp{2012ApJS..203...10T}; 
(17)~\citealp{2008AJ....135..785W};  
}

%
%

\end{deluxetable}

\clearpage
\thispagestyle{empty}
\begin{deluxetable}{lccccccccc}
\tablecaption{Membership Summary for Young Field Ultracool Dwarfs\label{table:membership}}
\tabletypesize{\tiny}
\tablecolumns{9}
\rotate
\tablewidth{0pt}
\setlength{\tabcolsep}{0.03in} 
\tablehead{
  \colhead{Object}  &
  \colhead{Spec.\ Type} &
  \multicolumn{2}{c}{BANYAN II: Kinematics + SEDs}  &
  \colhead{} &
  \multicolumn{2}{c}{BANYAN II: Kinematics only}  &
  \colhead{} &
  \colhead{\rchisq($UVW$, $XYZ$)} &
  \colhead{Membership} \\
  \colhead{} &
  \colhead{Optical/NIR} &
  \multicolumn{2}{c}{\citep{2014ApJ...783..121G, 2015ApJS..219...33G}} &
  \colhead{} &
  \multicolumn{2}{c}{(This work)}  &
  \colhead{} &
  \colhead{} &
  \colhead{} \\
  \cline{3-4}
  \cline{6-7}
  \colhead{}  &
  \colhead{}  &
  \colhead{RV $\pi$ yng} &
  \colhead{YMG}  &
  \colhead{} &
  \colhead{RV $\pi$ yng} &
  \colhead{YMG}   &
  \colhead{} &
  \colhead{} &
  \colhead{} 
}

\startdata

\cutinhead{\bf \em AB Dor}
2MASSI J0019262+461407                       & M8/M8~\textsc{int-g}             & y \enskip - \enskip y & ABDMG(92.1\%)                            && y \enskip y \enskip y & ABDMG(96\%)                       && ABDMG[2.6/0.6]                               & ABDMG \\
WISEP J004701.06+680352.1                    &     .../L7~\textsc{int-g}        & - \enskip - \enskip y & ABDMG(98.2\%)                            && y \enskip y \enskip y & ABDMG(100\%)                      && ABDMG[0.4/0.5]                               & ABDMG \\
2MASSW J0058425$-$065123                     & L0/L1~\textsc{int-g}             & - \enskip y \enskip y & ABDMG(63.7\%)/BPMG(31.8\%)               && - \enskip y \enskip y & ABDMG(75\%)/BPMG(22\%)            && ABDMG[$>$1.9/0.2], BPMG[$>$2.3/1.0]          & (ABDMG)/(BPMG) \\
2MASS J03552337+1133437                      & L5$\gamma$/L3~\textsc{vl-g}      & y \enskip y \enskip y & ABDMG(99.5\%)                            && y \enskip y \enskip y & ABDMG(99.7\%)                     && ABDMG[1.4/0.3]                               & ABDMG \\
SDSS J111010.01+011613.1                     &     .../T5.5                     & y \enskip y \enskip y & ABDMG(97\%)\tablenotemark{1}             && y \enskip y \enskip y & ABDMG(97\%)                       && ABDMG(1.2/1.2)                               & ABDMG \\
DENIS J142527.9$-$365023                     & L3:/L4~\textsc{int-g}            & y \enskip y \enskip y & ABDMG(99.9\%)                            && y \enskip y \enskip y & ABDMG(99.9\%)                     && ABDMG[1.9/0.6]                               & ABDMG \\
2MASSW J2244316+204343                       & L6.5/L6~\textsc{vl-g}            & - \enskip - \enskip y & ABDMG(99.1\%)                            && - \enskip y \enskip y & ABDMG(99.9\%)                     && ABDMG[$>$1.0/0.3]                            & (ABDMG) \\
\enddata
\end{deluxetable}

\clearpage
\begin{deluxetable}{lccccccc}
\tablecaption{Stellar Moving Groups \label{table:ymg}}
\tabletypesize{\footnotesize}
\tablewidth{0pt}
\tablehead{
 \colhead{Group} &
 \colhead{$U$} &
 \colhead{$V$} &
 \colhead{$W$} &
 \colhead{$X$} &
 \colhead{$Y$} &
 \colhead{$Z$} &
 \colhead{Ref} 
}

\startdata

\cutinhead{Young ($\lesssim$100~Myr) Groups}
AB Dor\tablenotemark{b}            & $-$7.6$\pm$1.1  & $-$27.3$\pm$1.5 & $-$14.9$\pm$1.0& $-$6.0$\pm$32.3 & $-$14.1$\pm$38.5  & $-$20.5$\pm$19.5 & 1, 2 \\
Argus\tablenotemark{a}             & $-$22.0$\pm$0.3 & $-$14.4$\pm$1.3 & $-$5.0$\pm$1.3 &   5.2$\pm$19.7  & $-$115.1$\pm$35.5 & $-$17.5$\pm$12.5 & 2 \\
$\beta$~Pic\tablenotemark{a}       & $-$10.9$\pm$1.5 & $-$16.0$\pm$1.4 & $-$9.2$\pm$1.8 &   8.4$\pm$31.9  &   -5.0$\pm$15.4   & $-$15.0$\pm$8.0  & 3 \\
Columba\tablenotemark{a}           & $-$13.2$\pm$1.3 & $-$21.8$\pm$0.8 & $-$5.9$\pm$1.2 & $-$42.6$\pm$25.7& $-$56.5$\pm$33.6  & $-$48.0$\pm$23.2 & 2 \\
Tuc-Hor\tablenotemark{b}           & $-$10.6$\pm$1.1 & $-$21.0$\pm$1.1 & $-$2.1$\pm$1.1 &   2.6$\pm$24.3  & $-$24.3$\pm$9.3   & $-$35.3$\pm$3.5  & 2, 4 \\
TWA\tablenotemark{b}               & $-$10.9$\pm$0.8 & $-$18.2$\pm$0.8 & $-$5.3$\pm$0.8 &  18.3$\pm$14.7  & $-$52.2$\pm$18.5  &  21.9$\pm$5.6    & 5, 6 \\

\cutinhead{Older ($\gtrsim$100~Myr) Groups}
Castor Stream\tablenotemark{a}     & $-$10.7$\pm$6.5 & $-$7.5$\pm$4.4  & $-$8.8$\pm$4.2 &     \nodata    &      \nodata     &      \nodata   & 7 \\
IC2391 Stream\tablenotemark{a}     & $-$23.9$\pm$3.8 & $-$18.9$\pm$4.9 & $-$8.0$\pm$5.6 &     \nodata    &      \nodata     &      \nodata   & 7 \\
\enddata 

\tablenotetext{a}{Uncertainties on $UVW$ are the observed dispersion in
  measured $UVW$ of group members.}

\tablenotetext{b}{Uncertainties on $UVW$ includes uncertainties on the
  $UVW$ of the group as well as the groups' internal dispersion.}

\tablerefs{
(1)~\citet{2013ApJ...766....6B};  
(2)~\citet{2008hsf2.book..757T};  
(3)~\citet{2014MNRAS.445.2169M};  
(4)~\citet{2014AJ....147..146K};  
(5)~\citet{2013ApJ...762..118W};  
(6)~\citet{2014A&A...563A.121D};  
(7)~\citet{2014A&A...567A..52K};  
(8)~\citet{2003AJ....125.1980K}   
}

\end{deluxetable}

\clearpage
\thispagestyle{empty}
\begin{deluxetable}{lcc}
\tablecolumns{3}
\tablecaption{Summary of Membership Changes Relative to BANYAN~II Results\label{table:changes-membership}}
\tabletypesize{\small}
\tablewidth{0pt}
\tablehead{
  \colhead{Object}  &
  \colhead{Previous} &
  \colhead{This Work} 
}

\startdata

\cutinhead{Changed Memberships}
2MASSW J0030300$-$145033             &  ARG(26.5\%)                           &    ARG? \\    
2MASSW J0033239$-$152131             &  ARG(32\%)/Y.FLD(68\%)                 &    Y.FLD \\
2MASS J01262109+1428057              &  BPMG(3\%)/Y.FLD(97\%)                 &    Y.FLD \\
2MASSI J0253597+320637               &  BPMG(21\%)/Y.FLD(75\%)                &    Y.FLD \\
2MASSI J0435145$-$141446             &  Y.FLD(100\%)                          &    MBM20 \\
2MASS J05012406$-$0010452            &  COL(49\%)/CAR(17\%)                   &    Y.FLD \\
2MASS~J05575096$-$1359503            &  Y.FLD(99.7\%)                         &    n/a (too far) \\
2MASSI J0608528$-$275358             &  COL(3\%)/Y.FLD(96\%)                  &    COL? \\
\enddata

\tablecomments{See Section~\ref{sec:membership} and the notes of
  Table~\ref{table:membership} for an explanation of results and the
  adopted notation. Objects designated as "Improved Membership" are
  mostly previously identified candidates (based on analyses without
  parallaxes) whose membership is now corroborated by our new parallax
  measurements.  See Appendix for details of individual objects.}

\tablenotetext{1}{\citet{2015ApJ...798...73G}}   
\tablenotetext{2}{\citet{2014ApJ...792L..17G}}   

\end{deluxetable}

\end{document}